\documentclass[conference,compsoc]{IEEEtran}
\usepackage{graphicx}
\usepackage{subcaption}
\usepackage{tikz}
\usepackage{amsmath}
\usepackage{amsfonts}
\usepackage{colortbl}
\usepackage{threeparttable}
\usepackage{booktabs}
\usepackage{adjustbox}
\usepackage{multirow}
\newtheorem{definition}{Definition}
 
\ifCLASSOPTIONcompsoc
  \usepackage[nocompress]{cite}
\else
  \usepackage{cite}
\fi
\ifCLASSINFOpdf
\else
\fi
\IEEEoverridecommandlockouts
\begin{document}
\title{Data Lineage Inference: Uncovering Privacy Vulnerabilities of Dataset Pruning}

\author{\IEEEauthorblockN{Qi Li \\
National University of Singapore\\
liqi@u.nus.edu
}
\and
\IEEEauthorblockN{Cheng-Long Wang* \\
KAUST\\
chenglong.wang@kaust.edu.sa}
\and
\IEEEauthorblockN{Yinzhi Cao \\
Johns Hopkins University\\
yinzhi.cao@jhu.edu
}
\and
\IEEEauthorblockN{Di Wang\\
KAUST\\
di.wang@kaust.edu.sa}
}

\IEEEpubid{\makebox[\columnwidth]{*Corresponding author\hfill}%
\hspace{\columnsep}\makebox[\columnwidth]{}}


% compsoc conferences compared to traditional conferences), use this
%Telephone: (800) 555--1212, Fax: (888) 555--1212}


\maketitle

\begin{abstract}
Dataset Pruning, or Coreset Selection, is one of the core topics in data-centric AI. It removes redundant portions of a dataset to select a smaller, more efficient subset for downstream training, improving efficiency without significantly impacting model performance. While dataset pruning explicitly aligns with GDPR’s data minimization principle by limiting redundant data exposure in training, its data-side privacy risks remain underexplored. Existing privacy attacks and defenses in machine learning typically focus on training samples, overlooking and unable to address the privacy risks of data used prior to model training. Despite being excluded from training by an optimized algorithm, the redundant set plays a ‘dark side’ role in contributing implicitly to the machine learning system. They share similar privacy-sensitive information as the selected training samples, since both are often collected for the same purposes and undergo the same data preprocessing. Their exposure can also pose significant privacy risks.

In this work, we systematically explore the data privacy issues of dataset pruning in machine learning systems. Our findings reveal, for the first time, that even if data in the redundant set is solely used before model training, its pruning-phase membership status can still be detected through attacks. Since this is a fully upstream process before model training, traditional model output-based privacy inference methods are completely unsuitable. To address this, we introduce a new task called \textit{Data-Centric Membership Inference} and propose the first ever data-centric privacy inference paradigm named \textit{\underline{Da}ta} \textit{\underline{L}ineage} \textit{\underline{I}nference} \textit{(DaLI)}. Under this paradigm, four threshold-based attacks are proposed, named \textit{WhoDis}, \textit{CumDis}, \textit{ArraDis} and \textit{SpiDis}. We show that even without access to downstream models, adversaries can accurately identify the redundant set with only limited prior knowledge. Furthermore, we find that different pruning methods involve varying levels of privacy leakage, and even the same pruning method can present different privacy risks at different pruning fractions. We conducted an in-depth analysis of these phenomena and introduced a metric called the \textit{Brimming score} to offer guidance for selecting pruning methods with privacy protection in mind.
\end{abstract}

\IEEEpeerreviewmaketitle

\section{Introduction}
\begin{figure}
    \centering
    \vspace{-4mm}        \includegraphics[width=0.9\linewidth]{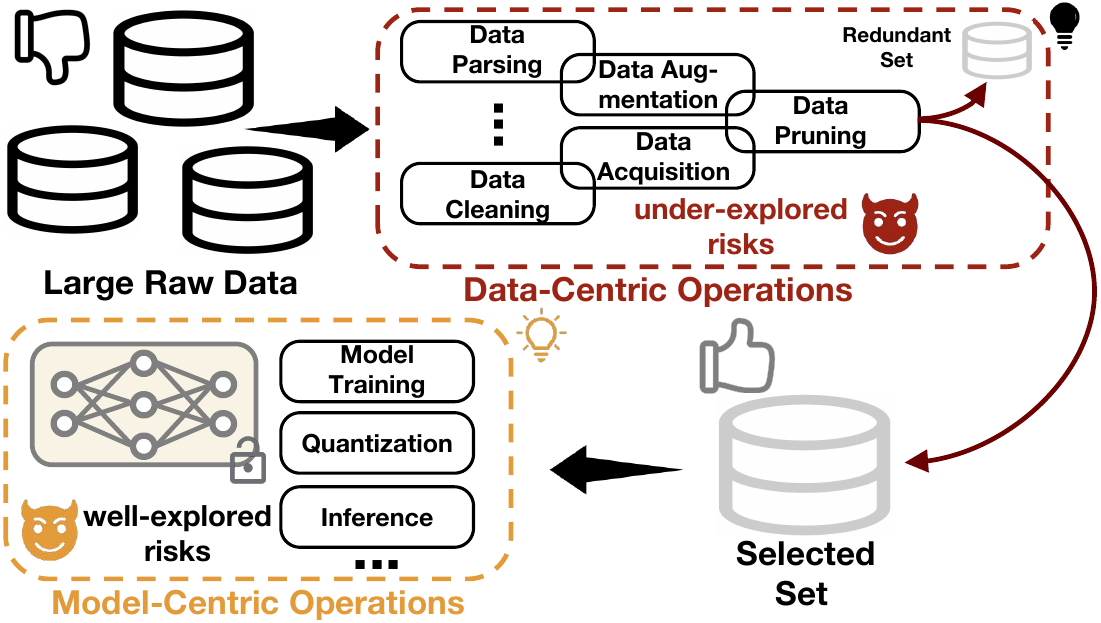}
    \caption{A typical machine learning system, among which the privacy risks of data-centric operations are overlooked.}
    \label{fig:datacentric}
    \vspace{-5mm}
\end{figure}

During the past decades, deep neural networks have achieved significant success across various research fields. As they evolve, the scale of the datasets has expanded considerably~\cite{hu2020randla,rong2020self,li2024encapsulating}, which presents challenges when training and deploying advanced neural networks on resource-constrained devices. One promising idea to alleviate the resource consumption caused by large datasets is Dataset Pruning~\cite{qin2023infobatch,yang2022dataset,tan2024data,maharana2023mathbb}, also known as Coreset Selection~\cite{guo2022deepcore,yang2024mind}. As shown in Figure \ref{fig:datacentric}, this technique identifies a smaller yet informative selected set from a larger dataset for training while leaving the redundant set aside, enabling models to achieve similar generalization performance as those trained on the full dataset~\cite{chen2010super,agarwal2020contextual,sener2017active}. Due to its exceptional scalability and versatility, it has been widely applied in scenarios requiring extreme efficiency\cite{radenovic2023filtering,abbas2023semdedup,gadre2024datacomp}. Numerous tech giants have also adopted it widely for scaling model training, including companies like Google and Meta~\cite{vo2024automatic,evans2024data,sorscher2022beyond}.

Dataset pruning prevents the exposure of redundant data in downstream model training, effectively aligning with GDPR's data minimization principle~\cite{politou2018backups, gdpr_data_minimization,wang2024has}.
For example, consider a scenario where some data from an HIV patient is initially collected for the purpose of training a medical model~\cite{shokri2015privacy, kaissis2020secure} but is subsequently excluded during the pruning phase. As a result, these data are not included in the actual model training. In such cases, even if an adversary (e.g., a third-party auditor or the data entity itself) attempts to infer information from the model, they would not detect the presence of this patient. 
Consequently, data service providers can seemingly bypass scrutiny and evade responsibility by concealing the fact that data in the redundant set was collected and processed while claiming compliance with data regulations.

However, if we delve deeper into the data processing stream of an ML system, as shown in Figure \ref{fig:datacentric}, we can discern a privacy `dark side' associated with the upstream pruning process. Although pruned data does not directly participate in the model training, it is still collected, stored, and processed for the same purpose, and it serves as an implicit contributor to model training by influencing the pruning process. The privacy risks associated with the pruned data’s pruning-phase membership status are effectively the same as those of the selected set used for training.
As we will demonstrate in this paper, an adversary can still accurately infer this pruning-phase membership even without any access to downstream models, 
thereby compromising the privacy of data in the pruned redundant set.

\begin{figure}[t]
    \centering
    \begin{minipage}[b]{\linewidth}
        \centering
        \begin{subfigure}[b]{0.63\linewidth}
            \centering
            \includegraphics[width=\linewidth]{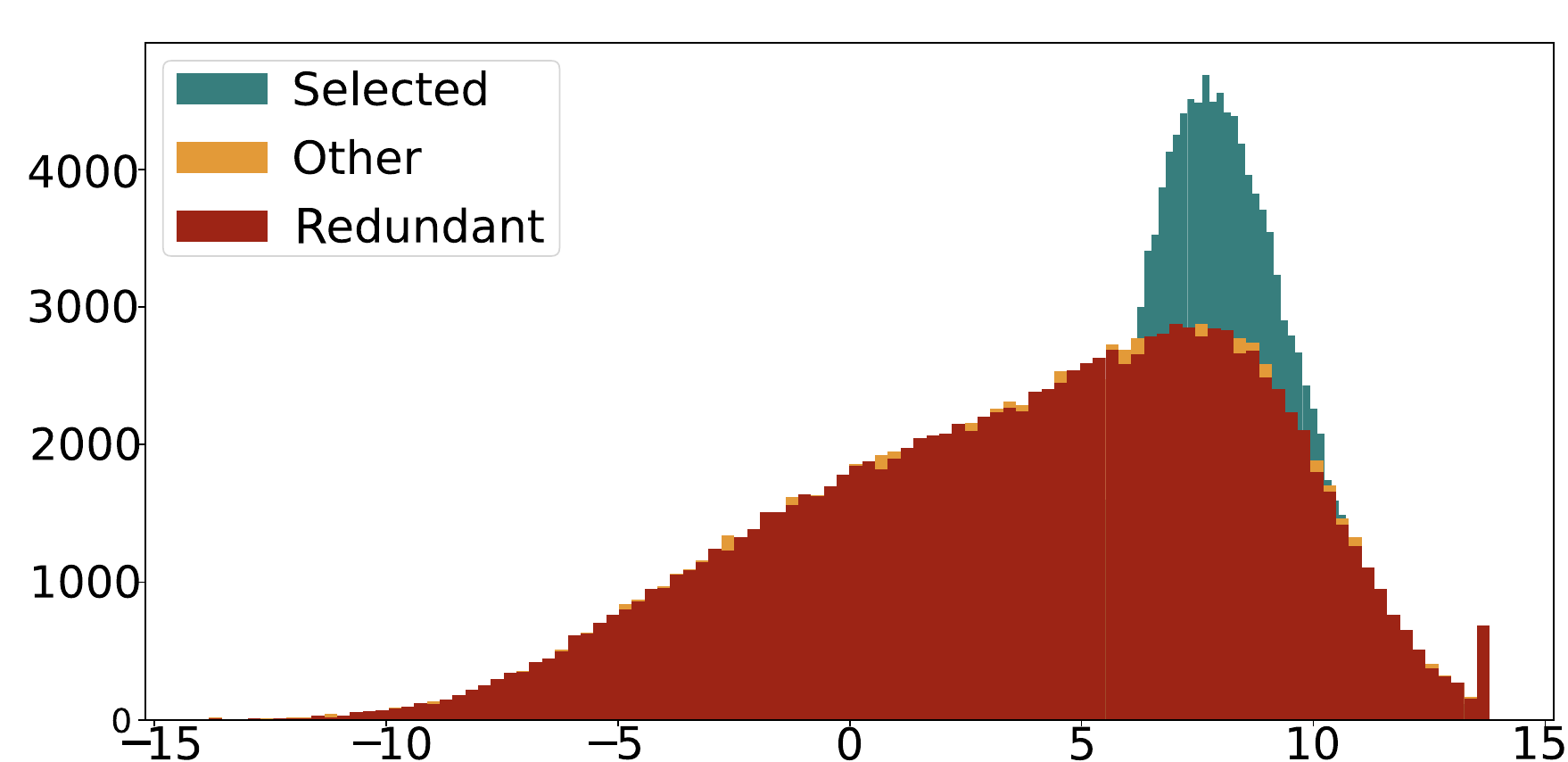}
            \label{fig:method1}
        \end{subfigure}
        \hfill 
        \begin{subfigure}[b]{0.35\linewidth}
            \centering
            \includegraphics[width=\linewidth]{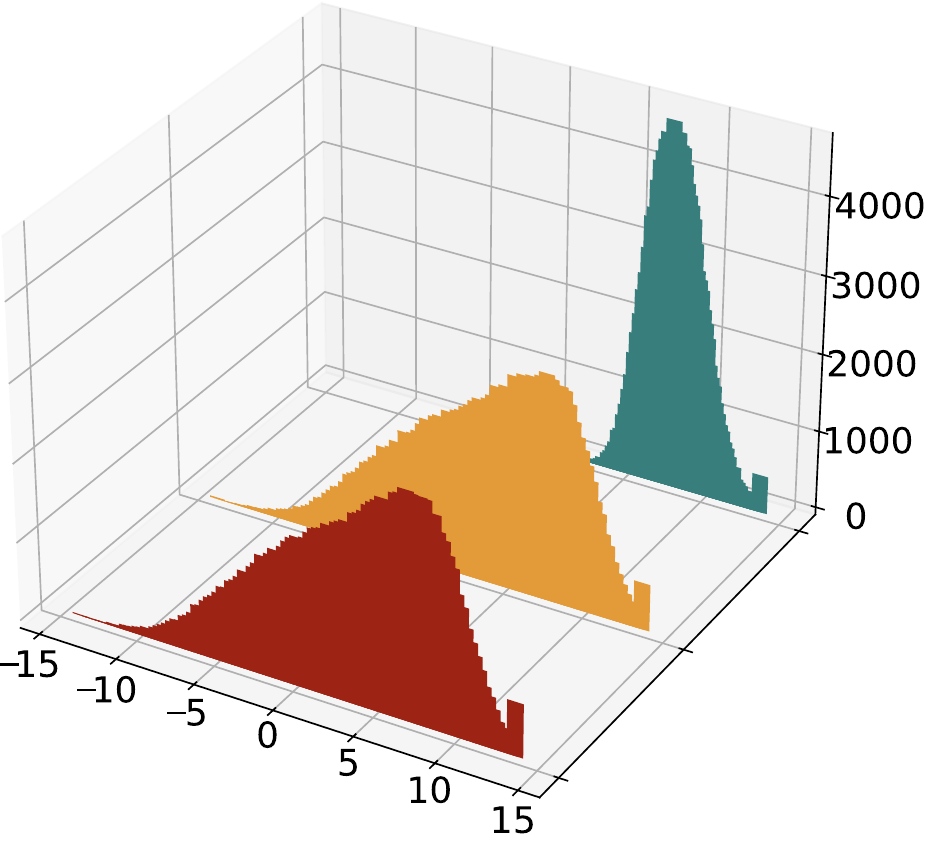}
            \label{fig:method2}
        \end{subfigure}
        \vspace{-7mm}
        \subcaption{Model Confidence under Random Selection.}
    \end{minipage}
    
    \begin{minipage}[b]{\linewidth}
        \centering
        \begin{subfigure}[b]{0.63\linewidth}
            \centering
            \includegraphics[width=\linewidth]{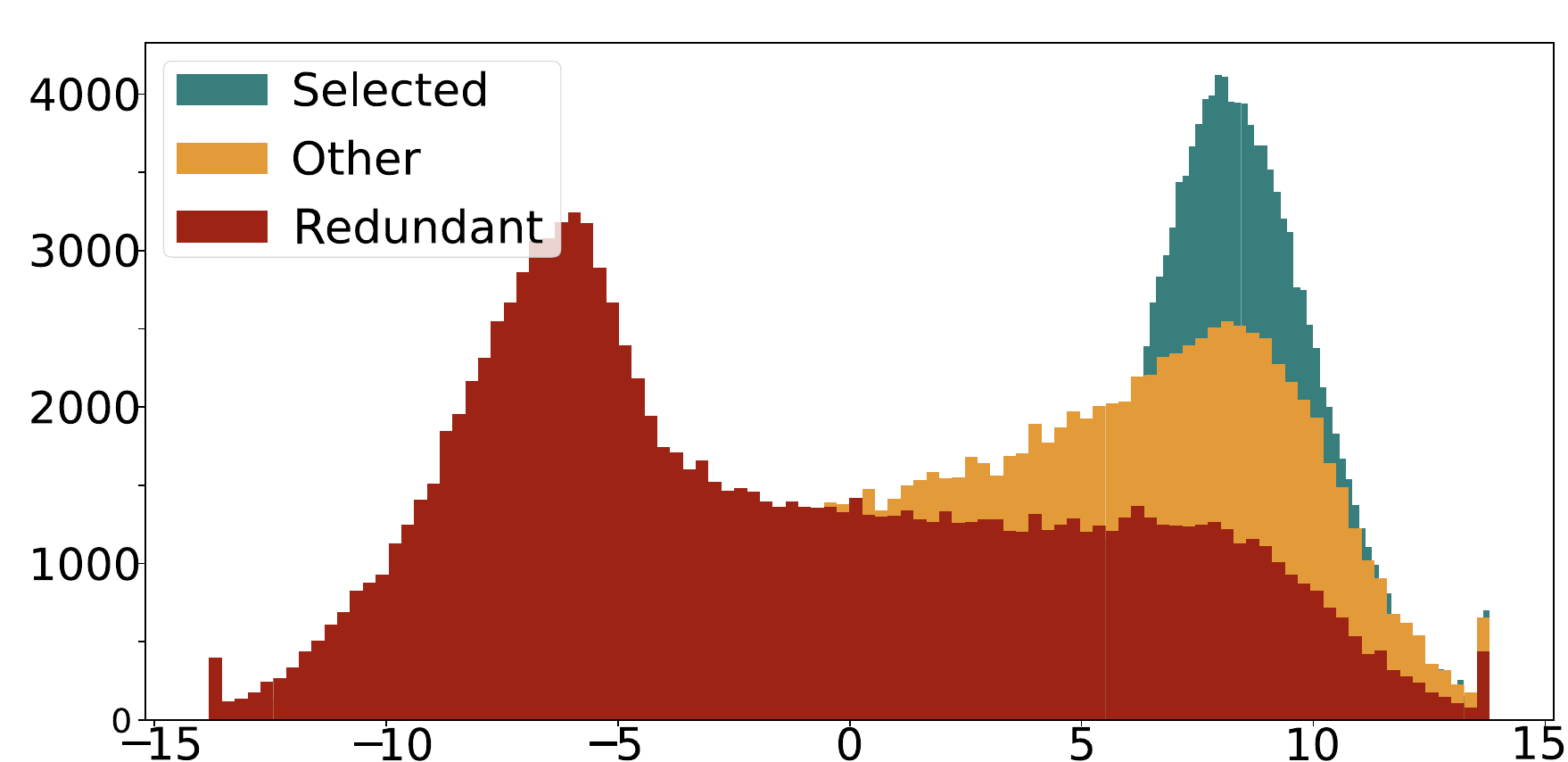}
        \end{subfigure}%
        \hfill 
        \begin{subfigure}[b]{0.35\linewidth}
            \centering
            \includegraphics[width=\linewidth]{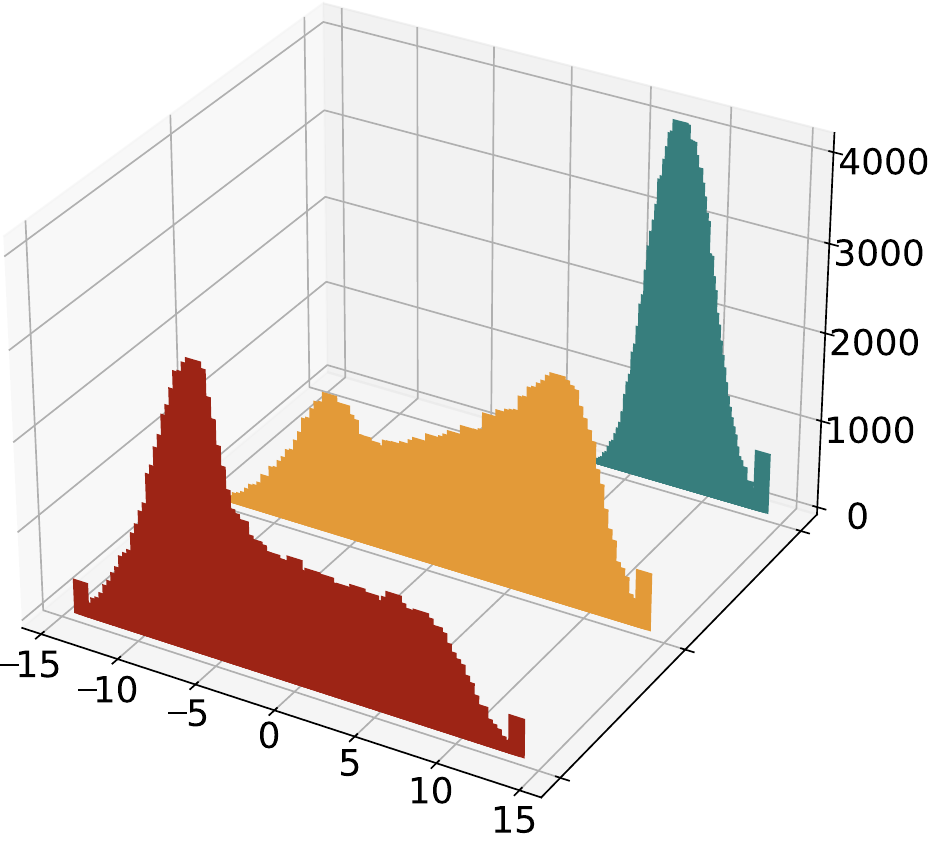}
            
        \end{subfigure}
        \subcaption{Model Confidence under Dataset Pruning.}
    \end{minipage}
    \vspace{-5mm}
    \caption{A toy example on model confidence after logit scaling, where x-axis is the logits after scaling, y-axis is the frequency. Random selection only causes a slight difference in model confidence between the redundant set and the other non-members. However, a carefully designed dataset pruning process causes significant differences.}
    \label{fig:initial}
    \vspace{-5mm}
\end{figure}

\noindent\textbf{Motivation. }In this work, we consider a scenario where a data service provider attempts to evade responsibility and risk associated with data collection and usage by concealing the redundant set after upstream dataset pruning. An adversary aims to infer the pruning-phase membership status of the redundant set. Specifically, the goal is to distinguish which of the data that were not used for model training (including both the redundant set and other non-members) belong to the redundant set that underwent pruning. This ensures that data within the redundant set, which was collected and used, receives the appropriate credit and holds the service provider accountable for its processing of those auditable samples. As an initial attempt, we conducted a downstream analysis to assess whether a carefully curated pruning process can implicitly pose the privacy risk of redundant samples. 

As shown in Figure \ref{fig:initial}, we performed a toy experiment to examine model confidence through logit scaling~\cite{carlini2022membership} after training on the selected set. Here, we use a benchmark dataset pruning method Glister~\cite{killamsetty2021glister} to compare with random selection. It can be observed that, with random selection from the entire training set, there is no noticeable difference in the model’s confidence distribution between the redundant set and other non-members. In contrast, the confidence distribution for a model trained on the selected set produced by Glister reveal significant differences between the redundant set and other non-members. This finding motivates us to trace and inference the data privacy back to the dataset pruning process.

\noindent\textbf{Contributions. }
Unlike traditional membership inference, which assesses the membership risk of data used in model training, we focus on an upstream objective: evaluating the pruning-phase membership of data collected but ultimately excluded by pruning algorithms. We name it as a \textit{Data-Centric Membership Inference (DCMI)} task. In this scenario, the adversary is limited to intercepting the selected set upstream without access to the downstream model, distinguishing it from previous model-centric attacks that infer membership by querying the target model.

To this end, we propose the first data-centric membership inference paradigm, named \textit{\underline{Da}ta} \textit{\underline{L}ineage} \textit{\underline{I}nference} \textit{(DaLI)}. This paradigm attempts to accurately identify which samples belong to the redundant set from a datapool containing both the redundant set and other non-members without access to the model trained on the selected set. 
To overcome the challenge, DaLI utilizes the dataset pruning process to reveal the differences between the carefully designed `occurrence distribution' (see Sections \ref{occ_dis_vic} and \ref{shadow_occ}) of redundant set and other non-members. 
Based on these differences, thresholds are constructed to infer membership within the datapool. Throughout this entire process, the target model is not directly involved. 
Our main contributions are summarized as follows:

\begin{itemize}
    \item For the first time,  we identify the privacy risks in the dataset pruning process and introduce the problem of Data-Centric Membership Inference (DCMI). We propose the first attack paradigm with broad attack capabilities in this new scenario named DaLI. Within the DaLI paradigm, we propose four different threshold-based attacks based on occurrence distribution differences observed from the datapool.
    \item We use DaLI on twelve pruning methods, three datasets, and four pruning fractions under different adversary knowledge levels. Results demonstrate DaLI's resilience under different settings. 
    \item Based on the experimental results, we designed a metric called the Brimming score to measure the privacy risk of different dataset pruning methods.
    Experimental results show that the Brimming score effectively provides a solid basis for choosing pruning methods with privacy protection in mind.
    \item We propose a defense strategy named ReDoMi, marking an initial attempt to defend against DCMI. It enhances the indistinguishability between redundant set and other non-members, making it more challenging for adversaries to conduct inference.
\end{itemize}

\section{Related Work}
\subsection{Data Side-Channels in ML Systems} 
\label{dp_mlsys}
In machine learning, privacy related research often treats models as isolated entities, ignoring data pre-processing and post-training processes. However, model training is typically part of a broader ML system with components like data filtering and post-training output/query filtering.~\cite{debenedetti2023privacy} experimentally demonstrated that system-level attacks can extract private information more effectively than attacks on standalone models.

Another line of research aims to explore the privacy issues in dataset condensation (DC) or dataset distillation (DD). 
\cite{dong2022privacy} established a theoretical connection between dataset condensation and differential privacy (DP), proving dataset condensation's superiority in privacy preservation over traditional methods. However,~\cite{carlini2022no} pointed out flaws in their experimental and theoretical evidence.~\cite{liu2023backdoor} successfully conduct backdoor attacks by embedding triggers during the distillation process and iteratively optimizing them to enhance attack performance. Current works all conduct attacks or develop defenses against attacks within the model-centric scenario. Additionally, the limitations of DC/DD are quite apparent. The condensation/distillation process itself places a high demand on computational resources~\cite{liu2023slimmable,sun2024diversity}, 
which contradicts its original goal of efficiency and greatly limits its applicability.
In contrast, dataset pruning is much more resource-friendly and has been widely applied in the ML community. Some pruning services in diverse fields are already  publicly available. For example, 
Data Brokers like Acxiom~\cite{acxiom} and Experian~\cite{experian} collect and curate data from a variety of sources and offer tailored datasets that meet specific user requirements, including de-duplication and filtering processes that enhance dataset quality for downstream tasks. Additionally, the `Global Database of Events, Language, and Tone (GDELT)'\cite{gdelt} autonomously gathers data from news and social media, organizing and categorizing them by topic, sentiment, and geographic region, providing pre-processed datasets for research and ML applications.

The data-centric trend in ML community has also caught the attention of security researchers. Some have started recognizing privacy risks beyond model training~\cite{ganesh2024data, debenedetti2024privacy,wen2024understanding}. Compared to these works, this paper is the first to explore the privacy risks associated with data collected but excluded from training at the upstream level.

% 下面这个subsection你也应该强调是model centric
\subsection{Model-Centric Membership Inference}
Previous mainstream Membership Inference Attacks (MIAs) can be categorized into two types: Binary Classifier-based and Metric-based\cite{hu2022membership}. The former involves training a binary classifier to differentiate between the behaviors of the target model's training members and non-members. 
% The challenge lies in training this binary classifier effectively.
\cite{shokri2017membership} proposed an influential technique known as shadow training, which remains widely used. The core idea is that an adversary can train multiple shadow models to mimic the target model's behavior, allowing them to construct attack datasets. These attack datasets are then used to train a binary classifier for the final inference.
In contrast, metric-based MIAs\cite{salem2018ml,song2021systematic,yeom2018privacy} are simpler and less computationally intensive, yet can still achieve comparable or even superior performance. Metric-based MIAs infer membership by calculating metrics on the target model’s prediction vectors and comparing these metrics against a predefined threshold to determine the membership status of data records. This type of method bypasses the need to train a separate binary classifier, relying instead on straightforward metric comparisons.

Regardless of the type of attack mentioned above, all require a target downstream model to provide information for the attack, making them closely tied to a specific model. In this paper, we introduce DaLI, the first paradigm for data-centric membership inference. It occurs during the dataset pruning process before model training and does not require the involvement of trained downstream models.

\section{Preliminaries}
\subsection{Data-Centric AI}
A significant driver of the remarkable advancements in Artificial Intelligence (AI) over the past decade has been the availability of abundant and high-quality data, which give rise to Data-Centric AI (DCAI)\cite{mazumder2024dataperf,zha2023data,jakubik2024data,data-centric-ai}. As Figure \ref{fig:datacentric} infers, in contrast to traditional Model-Centric AI (MCAI), DCAI places greater emphasis on enhancing the quality and quantity of data, with relatively fixed models\cite{zha2023data,jakubik2024data}. 
Below is a formal definition of MCAI and DCAI:
\begin{itemize}
    \item \textbf{Model-Centric AI:} MCAI is a paradigm that focuses on selecting the appropriate model type, hyperparameters, learning algorithms, etc., from a variety of options to build effective and efficient AI systems.
    \item \textbf{Data-Centric AI:} DCAI is a paradigm that emphasizes the importance of systematically designing and engineering data to build effective and efficient AI systems.
\end{itemize}

While the transformation from MCAI to DCAI is ongoing, several achievements already underscore its benefits.
For instance, the advancements in large-scale language models largely depend on the utilization of extensive high-quality datasets\cite{kenton2019bert,radford2018improving}. Compared to GPT-2\cite{radford2019language}, GPT-3\cite{brown2020language} made only minor modifications to the network architecture while investing considerable effort into gathering significantly larger and higher-quality datasets for training. 
\subsection{Dataset Pruning}
\noindent\textbf{Workflow.}
Consider a classification task with a training dataset containing \(N\) examples drawn i.i.d. from an underlying distribution \(\mathcal{P}\). We denote the training dataset as \(D_{\textrm{tra}} = \{(x_i, y_i)\}_{i=1}^{N}\), where \(x_i\) is the feature vector, and \(y_i\) is the ground-truth label. The goal of dataset pruning is to choose a selected set with a pruning fraction \(\alpha\) before model training that should have a minimal impact on the model, i.e., the test performances of the models learned on the training sets before and after pruning should be very close. After the pruning process, \(D_{\textrm{tra}}\) will be divided into two subsets: the selected set \(D_{\textrm{sel}}\) for further model training and the redundant set \(D_{\textrm{red}}\) which is excluded from model training. Specifically, for a pruning method \(\mathbb{M}(\cdot)\), the pruning process can be formulated as \( D_{\textrm{sel}}, D_{\textrm{red}} = \mathbb{M}(D_{\textrm{train}}, \alpha) \), with the following optimization objective:
\begin{equation}
\resizebox{\columnwidth}{!}{
$
\begin{aligned}
 \min_{D_{\textrm{sel}} \subseteq D_{\textrm{tra}}: \frac{|D_{\textrm{sel}}|}{|D_{\textrm{tra}}|} = \alpha} |\mathbb{E}_{z \sim \mathcal{P}} [l(z; \hat\theta_{D_{\textrm{tra}}})] - \mathbb{E}_{z \sim \mathcal{P}} [l(z; \hat\theta_{D_{\textrm{sel}}})]|, 
\end{aligned}
$
}
\end{equation}
where \(l\) is the loss function for model training, \(\hat\theta_{D_{\textrm{sel}}}\) represents the model parameters trained with \(D_{\textrm{sel}}\): \(\hat{\theta}_{\textrm{sel}} = \arg\min_{\theta \in \Theta} \frac{1}{|D_{\textrm{sel}}|} \sum_{z_i \in D_{\textrm{sel}}} l(z_i, \theta)\), while \(\hat\theta_{D_{\textrm{tra}}}\) represents the model parameters trained with \(D_{\textrm{tra}}\).  

Note that besides $D_{\textrm{tra}}$ we always have a non-member dataset \(D_{\textrm{non}}\) whose samples are also i.i.d. sampled from the same distribution, and both \(D_{\textrm{red}}\) and other non-members \(D_{\textrm{non}}\) are not used in model training. Thus, in the view of traditional model-centric MIAs, data in \(D_{\textrm{sel}}\) are viewed as members while those in \(D_{\textrm{red}}\) and \(D_{\textrm{non}}\) are all non-members. 

\noindent\textbf{Method Taxonomy.} Popular dataset pruning methods can be categorized into seven types\cite{guo2022deepcore}. To avoid redundancy, we will provide specific explanations for different types of methods in Appendix \ref{taxonomy}. All the seven types of methods are considered in our experiments.

\noindent\textbf{Remark. }There are two orthogonal fields related to dataset pruning: (1) Few-shot learning\cite{wang2020generalizing,snell2017prototypical} focuses on enhancing model performance with limited training data. In contrast, dataset pruning aims to minimize the amount of training data used for training without significantly compromising model performance. (2) Dataset condensation/distillation\cite{zhao2021dataset,zhao2023dataset} aims at creating a small set of synthetic data. A model trained on this synthetic dataset can attain comparable performance with a model trained on the full dataset. The synthetic data are newly generated, while dataset pruning methods do not generate but select real samples from $D_{\textrm{tra}}$.

\section{Threat Model}
\subsection{Attack Scenario}
We define our attack scenario as a game between an upstream data service provider (e.g., the data brokers mentioned in Section \ref{dp_mlsys}) and an adversary (e.g., a third-party auditor or data entity). Upon receiving a service request from the adversary, the service provider optimizes, filters, and processes data under their management according to specified requirements, then delivers the resulting selected set to the adversary. The adversary then uses this selected set to infer the pruning-phase membership status of a victim datapool (containing both redundant data that have undergone pruning and other non-members that have not), aiming to identify redundant data that was involved in pruning but ultimately excluded from model training. This scenario assumes no knowledge or access to downstream models, aligning with the upstream nature of dataset pruning in the ML system data flow~\cite{yang2022dataset,maharana2023mathbb}.

\subsection{Adversary’s Capability}
As described in the aforementioned attack scenario, we assume no knowledge of or access to downstream models trained on the selected set. 
Furthermore, under the strictest service agreements (representing the most challenging attack scenario), besides the inference subjects (i.e., the victim datapool, analogous to the target dataset containing both members and non-members in traditional MIAs), the adversary receives only the selected set delivered by the service provider to infer the membership of data in the victim datapool without any additional information such as the pruning method, pruning fraction, or data source/distribution that could aid in building local auxiliary data—although in practice, these details may sometimes be disclosed depending on the service agreement~\cite{acxiom,experian,gdelt}. We will explore all practical scenarios in our experiments and demonstrate that none of these various pieces of information or prior knowledge are essential.

\subsection{Adversary's Goal}
\begin{figure}[h]
    \centering
        \includegraphics[width=\linewidth]{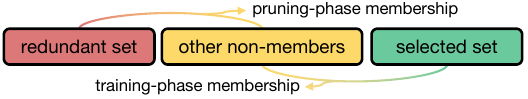}
    \caption{A comparison between traditional membership inference and our proposed data-centric membership inference: the former focuses on inferring the training-phase membership status of the selected set, while the latter targets the pruning-phase membership status of the redundant set.}
    \vspace{-4mm}
    \label{fig:mias}
\end{figure}

The adversary's goal is to distinguish the pruning-phase membership status of data in the victim datapool. As the term suggests, `membership' here is based on whether the data participated in the pruning process. Data involved in the pruning process (the known selected set and the redundant set to be inferred) are members, while all other data are regarded as non-members (we name them as other non-members).
Since the redundant set inherently contributes a positive impact on downstream model training, this distinction can help protect data entity rights and prevent service providers from evading responsibility. Figure \ref{fig:mias} presents a comparison between the adversary’s goal in our scenario and that of traditional model-centric membership inference. They target at different aspects of data flow within the ML system, yet both focus on the privacy issues of critical data that affect the model utility.

\subsection{Adversary's Challenge}
Since the adversaries are facing an upstream data privacy tracing issue, previous data privacy inference frameworks based on model outputs are inapplicable here. These frameworks essentially compress and store the knowledge inherent in the data to a model~\cite{hu2022membership,li2021membership,shokri2017membership}, allowing adversaries to extract this knowledge through model access. In contrast, we operate directly on the data itself, lacking a model-like knowledge-storing counterpart. Furthermore, the adversary has no control over the dataset pruning process and can only utilize the received selected set. Therefore, a paradigm shift in attack is necessary to tackle this new scenario.

\section{Attack Overview}
\label{at_ov}
Before diving into the detailed components of DaLI, this section provides an overview of the attack framework. We will first introduce the general idea behind DaLI and discuss the primary information source for conducting the attack: the occurrence distributions. 
\subsection{General Idea} 
\begin{figure}[h!]
    \centering
    \vspace{-4mm}
        \includegraphics[width=\linewidth]{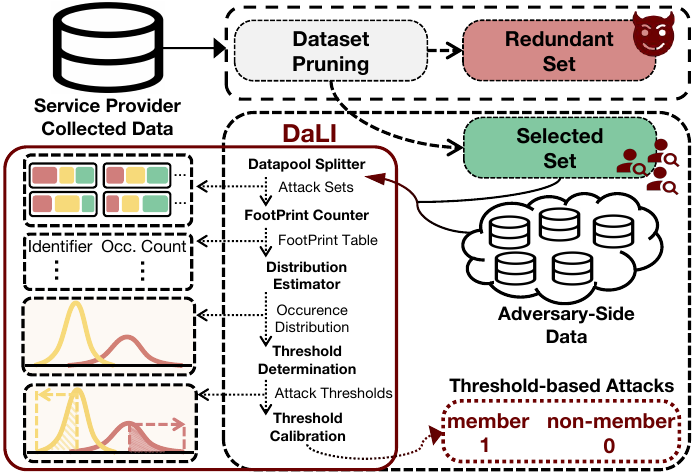}
    \caption{Overall pipeline of DaLI.}
    \label{fig:pipeline}
    \vspace{-2mm}
\end{figure}
Figure \ref{fig:pipeline} shows the general pipeline of DaLI. In simple terms, the intuition behind our inference attack is the observation that the occurrence distribution (the definition will be given later) of the redundant set and other non-members shows a significant difference. 
In DaLI, the adversary constructs several shadow datapools and gradually builds a shadow occurrence distribution pair for each (i.e., \(\mathbb{S}^{\textrm{shadow}}_{\textrm{red}}(\textrm{i})\) for the redundant set and \(\mathbb{S}^{\textrm{shadow}}_{\textrm{non}}(\textrm{i})\) for other non-members, where \(\textrm{i}\) denotes the index of a shadow datapool) through the Datapool Splitter, Footprint Counter, and Distribution Estimator. Attack thresholds are then constructed based on these occurrence distribution pairs.
Meanwhile, a mixed occurrence distribution is constructed for the victim datapool (i.e., \(\mathbb{S}^{\textrm{victim}}_{\textrm{red}\cup\textrm{non}}\)). Finally, a threshold calibration process is conducted to derive the final thresholds that can be applied to the victim datapool. Throughout the overall pipeline, constructing the occurrence distribution for the victim and shadow datapools is key to executing the attack. We now present the formal definition of occurrence distribution.

\subsection{Occurrence Distribution of Victim Datapool}
\label{occ_dis_vic}
For the construction of the victim occurrence distribution \(\mathbb{S}^{\textrm{victim}}_{\textrm{red}\cup\textrm{non}}\) of the victim datapool \(Q_{\textrm{v}}\) composed of the victim redundant set \(D^{\textrm{victim}}_{\textrm{red}}\) and other non-members \(D^{\textrm{victim}}_{\textrm{non}}\) (i.e., \(Q_{\textrm{v}} = D^{\textrm{victim}}_{\textrm{red}} \cup D^{\textrm{victim}}_{\textrm{non}}\)), we perform extensive sampling from \(Q_{\textrm{v}}\) to form a query set collection \(\{q^{\textrm{victim}}_\textrm{j}\}_{\textrm{j}\in \textrm{J}}\) via Datapool Splitter (see Section \ref{splitter}). Each of the query sets is processed along with the victim selected set \(D^{\textrm{victim}}_{\textrm{sel}}\) using the underlying dataset pruning method to obtain the corresponding dataset split. To avoid confusion, we denote the selected set and the redundant set in the dataset split as the picking set \(D^{\textrm{victim}}_{\textrm{pic}}(\textrm{j})\) and the culling set \(D^{\textrm{victim}}_{\textrm{cul}}(\textrm{j})\), respectively (i.e., \(D^{\textrm{victim}}_{\textrm{pic}}(\textrm{j}), D^{\textrm{victim}}_{\textrm{cul}}(\textrm{j}) = \mathbb{M}(D^{\textrm{victim}}_{\textrm{sel}} \cup q^{\textrm{victim}}_\textrm{j}, \alpha)\)). 

After constructing the dataset split for each query set, several \(D^{\textrm{victim}}_{\textrm{cul}}(\textrm{j})\) are produced. 
These sets are used to build the victim occurrence distribution \(\mathbb{S}^{\textrm{victim}}_{\textrm{red}\cup\textrm{non}}\). For each sample in the victim datapool, if it appears in any \(D^{\textrm{victim}}_{\textrm{cul}}(\textrm{j})\), it is counted as an \textit{occurrence}. The occurrence count for each data is then determined, ranging from 0 to the total number of culling sets. Finally, the occurrence distribution for the victim datapool is constructed, which can be formally defined as follows. 

\begin{definition}[Occurrence Distribution of Victim Datapool]\label{def:1}
    Given the occurrence count $t_s$ of each sample $x_s$ in the victim datapool, the occurrence distribution of the victim datapool can be represent as a function $\mathbb{S}^{\textrm{victim}}_{\textrm{red}\cup\textrm{non}}(\cdot)$:  
\begin{equation}
\mathbb{S}^{\textrm{victim}}_{\textrm{red}\cup\textrm{non}}(t) = \sum_{s=1}^{|Q_{\textrm{v}}|} \mathbb{I}(t_s = t), 
\label{eq2}
\end{equation}
where $\mathbb{S}^{\textrm{victim}}_{\textrm{red}\cup\textrm{non}}(t)$ denotes the number of samples with an occurrence count of $t$, $\mathbb{I}$ is the indicator function.\footnote{\label{fn1}
In practice, there are instances where no samples appear for \(t\) above a
certain high value. Consequently, the occurrence distribution may contain a
segment of all-zero values. In the following text, we assume that all-zero
segments are truncated, meaning the actual range of \(t\) in the occurrence
distribution spans from 0 to the highest non-zero value.} 
\end{definition}

\subsection{Shadow Occurrence Distribution} 
\label{shadow_occ}
For the shadow occurrence distribution construction, the adversary first constructs several shadow sets \(D_{\textrm{tra}}^{\textrm{shadow}}(\textrm{i})\) via the auxiliary data \footnote{In Section \ref{no_aux}, we'll demonstrate that the attack can still succeed without auxiliary data.}
and runs a dataset pruning process on each of them, resulting in each shadow set's own shadow selected set \(D_{\textrm{sel}}^{\textrm{shadow}}(\textrm{i})\) and shadow redundant set \(D_{\textrm{red}}^{\textrm{shadow}}(\textrm{i})\). The adversary then builds several shadow datapools using the redundant set and other non-members, i.e., \(Q^{\textrm{shadow}}_{\textrm{i}} = D_{\textrm{red}}^{\textrm{shadow}}(\textrm{i}) \cup D_{\textrm{non}}^{\textrm{shadow}}(\textrm{i})\).  Each shadow datapool \(Q^{\textrm{shadow}}_{\textrm{i}}\) shares a same scale \(|Q_{\textrm{s}}|\).

Similar to the victim datapool, for each \(Q^{\textrm{shadow}}_{\textrm{i}}\), the adversary samples multiple shadow query sets to form a shadow query set collection \(\{q^{\textrm{shadow}}_\textrm{ij}\}\), where \(\textrm{i}\) denotes the index of the shadow datapool and \(\textrm{j}\) denotes the index of the query set of the \(\textrm{i}\)-th shadow datapool. Each shadow query set \(q^{\textrm{shadow}}_\textrm{ij}\) is processed along with the corresponding shadow selected set \(D_{\textrm{sel}}^{\textrm{shadow}}(\textrm{i})\) using the dataset pruning method, resulting in multiple shadow splits. Similarly, we refer to the shadow selected set and the shadow redundant set in a shadow split as the shadow picking set \(D^{\textrm{shadow}}_{\textrm{pic}}(\textrm{i,j})\) and the shadow culling set \(D^{\textrm{shadow}}_{\textrm{cul}}(\textrm{i,j})\), respectively, i.e., \(D^{\textrm{shadow}}_{\textrm{pic}}(\textrm{i,j}), D^{\textrm{shadow}}_{\textrm{cul}}(\textrm{i,j}) = \mathbb{M}(D_{\textrm{sel}}^{\textrm{shadow}}(\textrm{i}) \cup q^{\textrm{shadow}}_\textrm{ij}, \alpha)\). 

After constructing the shadow split for each shadow query set \(q^{\textrm{shadow}}_\textrm{ij}\), several \(D^{\textrm{shadow}}_{\textrm{cul}}(\textrm{i,j})\) are produced. 
These sets are used to build the shadow occurrence distribution pair \( (\mathbb{S}^{\textrm{shadow}}_{\textrm{red}}(\textrm{i}), \mathbb{S}^{\textrm{shadow}}_{\textrm{non}}(\textrm{i})) \). Similarly, for each sample in the current shadow datapool \(Q^{\textrm{shadow}}_{\textrm{i}}\), if it appears in any \(D^{\textrm{shadow}}_{\textrm{cul}}(\textrm{i,j})\), it is counted as an \textit{occurrence}. The occurrence count for each sample in \(Q^{\textrm{shadow}}_{\textrm{i}}\) is then determined, ranging from 0 to the total number of shadow culling sets of current shadow datapool. Finally, the occurrence distributions \(\mathbb{S}^{\textrm{shadow}}_{\textrm{red}}(\textrm{i})\) and \(\mathbb{S}^{\textrm{shadow}}_{\textrm{non}}(\textrm{i})\) are constructed for \(D_{\textrm{red}}^{\textrm{shadow}}(\textrm{i})\) and \(D_{\textrm{non}}^{\textrm{shadow}}(\textrm{i})\), respectively.
We give the formal definition of \(\mathbb{S}^{\textrm{shadow}}_{\textrm{red}}(\textrm{i})\) and \(\mathbb{S}^{\textrm{shadow}}_{\textrm{non}}(\textrm{i})\) as follows:

\begin{definition}[Occurrence Distribution of Shadow Datapool]\label{def:2}
    Given the occurrence $t_s$ of each sample $x_s$ in a shadow datapool, the occurrence distributions of the shadow redundant set $\mathbb{S}^{\textrm{shadow}}_{\textrm{red}}(\textrm{i})(\cdot)$ and other non-members   $\mathbb{S}^{\textrm{shadow}}_{\textrm{non}}(\textrm{i})(\cdot)$ are defined as  
    \begin{align}
&\mathbb{S}^{\textrm{shadow}}_{\textrm{red}}(\textrm{i})(t) = \sum_{s=1}^{|Q_{\textrm{s}}|} \mathbb{I}(x_s \in D^{\textrm{shadow}}_{\textrm{red}}(\textrm{i}))\mathbb{I}(t_s = t),
\label{eq3} \\
&\mathbb{S}^{\textrm{shadow}}_{\textrm{non}}(\textrm{i})(t) = \sum_{s=1}^{|Q_{\textrm{s}}|} \mathbb{I}(x_s \in D^{\textrm{shadow}}_{\textrm{non}}(\textrm{i}))\mathbb{I}(t_s = t). 
\label{eq4}
\end{align}
\end{definition}

The difference between the victim and shadow occurrence distributions lies in their roles: the former is the adversary's inference target—a mixed distribution—while the latter are already separated. This aligns with the concept in traditional membership inference, where the victim set is a mixture containing both training and test sets, serving as the inference target, and the shadow set is auxiliary data with known divisions to aid in the inference process~\cite{shokri2017membership}. In the next section, we will provide a detailed explanation of each component in DaLI.

\section{Data Lineage Inference Components}
\label{components}
\subsection{Datapool Splitter} 
\label{splitter}

\begin{figure}[h!]
    \centering
    \vspace{-4mm}
        \includegraphics[width=\linewidth]{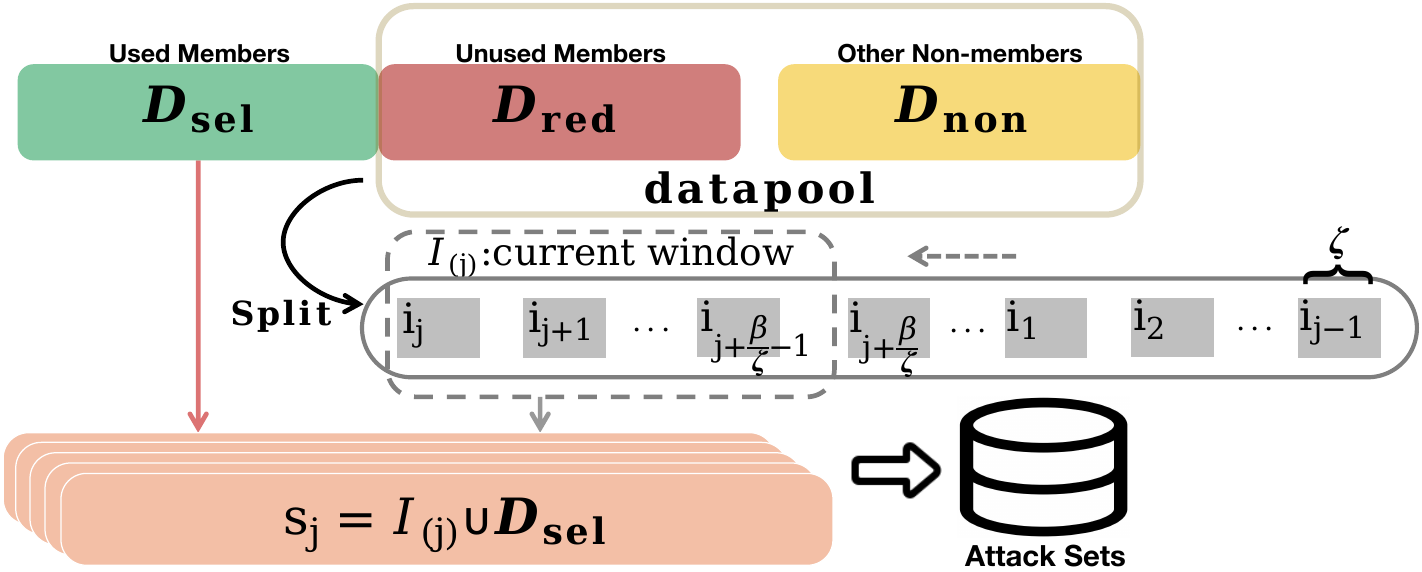}
    \caption{Overall architecture of datapool splitter.}
    \label{fig:splitter}
\end{figure}
The datapool splitter is designed for the adversary to conduct a sliding window sampling. After each sampling process, a query set sharing the same scale with \(D_{\textrm{red}}\) is produced, which is then combined with \(D_{\textrm{sel}}\) for further use\footnote{Since the datapool splitter here applies to both the victim datapool and the shadow datapool, we will use \(D_{\textrm{red}}\) to represent \(D^{\textrm{victim}}_{\textrm{red}}\) and all \(D^{\textrm{shadow}}_{\textrm{red}}(\textrm{i})\) without additional superscripts. Similarly, we have $Q, D_{\textrm{sel}}, \zeta, \textrm{q}_\textrm{j}, \textrm{s}_\textrm{j}$. }. To do so, we divide the datapool \(Q\) into small batches. According to the size of the redundant set, a batch size \(\zeta\) is pre-defined, which should be chosen such that the total size of several batches equals the size of the redundant set\footnote{In real scenarios, the size of the datapool \(|Q|\) may not always be divisible by the batch size \(\zeta\). In such cases, we simply leave the last batch incomplete.}.

After dividing \(Q\) to $\frac{|Q|}{\zeta}$ batches of data,
for the construction of the query sets, the adversary essentially conducts a sliding window sampling. We use \(\beta\) to represent the scale of the redundant set (i.e., \(\beta=|D_{\textrm{red}}|\)). Each time there are $\frac{\beta}{\zeta}$ number of batches covered in the sliding window. For example, as shown in Figure \ref{fig:splitter}, suppose currently the first batch in the sliding window is the $\textrm{j}^{\textrm{th}}$ batch (i.e., \(\textrm{i}_{\textrm{j}}\)), then the current state of the sliding window is: $\mathcal{I}_{(\textrm{j})} = \{\textrm{i}_{\textrm{j}}, \textrm{i}_{\textrm{j+1}},..., \textrm{i}_{\textrm{j}+\frac{\beta}{\zeta}-1}\}$, which together form a query set (i.e., \(q_\textrm{j}=\mathcal{I}_{(\textrm{j})}\)) and is further combined with $D_{\textrm{sel}}$. We name the combination of a query set \(q_\textrm{j}\) and $D_{\textrm{sel}}$ as an attack set \(\textrm{s}_\textrm{j}\) (i.e., \(\textrm{s}_\textrm{j} = q_\textrm{j} \cup D_{\textrm{sel}}\)). Every time a query set is formed, the first batch in the sliding window is moved to the end, allowing a new batch of data to enter the end of the sliding window to construct a new query set. This process will not end until the last batch is in the $1^{\textrm{st}}$ position of the window. Finally, $\frac{|Q|}{\zeta}$ number of attack sets are constructed: 
\begin{equation}
A(Q, D_{\textrm{sel}}, \zeta, \alpha) = \{\textrm{s}_1, \textrm{s}_2, ..., \textrm{s}_{\frac{|Q|}{\zeta}}\}.  
\end{equation}
To simplify, we will use \(A\) to denote the attack sets (i.e., \(A^{\textrm{v}}\) for victim datapool and \(A^{\textrm{s}}_{\textrm{i}}\) for each shadow datapool). Note that the sliding window sampling process ensures that all samples in the datapool are used \(\frac{\beta}{\zeta}\) times, thereby ensuring that they share the same occurrence count range.

After the construction of attack sets, we can now use them to reproduce the dataset pruning process. Specifically, given the attack sets, we apply the pruning algorithm to each of them using the pruning fraction \(\alpha\), obtaining the picking sets and culling sets as defined in Section \ref{occ_dis_vic}, which are subsequently used for footprint construction. In the next section, we will provide the details of the Footprint Counter for the shadow datapools and the victim datapool.

\noindent {\bf Unknown Pruning Fraction \(\alpha\).}
For adversaries who are informed the pruning fraction, inferring the exact size of the redundant set is straightforward. However, there are cases in which adversaries do not know the pruning fraction. In such cases, they can employ a mark-recapture method~\cite{southwood2009ecological,chao2001applications} to estimate the fraction. We found that that this approach can help the adversary identify a fraction that is nearly identical to the ground-truth one (see Section \ref{partial_know}).

\noindent {\bf Influence of Batch Size \(\zeta\).}
A simplistic strategy is to set \(\zeta\) to 1. While this strategy provides the adversary with the most fine-grained information, it also makes the inference inefficient. We find that setting \(\zeta\) to a relatively large value can also lead to a successful inference (see Appendix \ref{batch_size}). 

\subsection{FootPrint Counter}
The footprint counter leverages the culling sets to gather data-level statistics and create a footprint table. This table contains the index and the occurrence count of each data, supporting the construction of occurrence distributions.
\begin{figure}[ht!]
\vspace{-1mm}
    \centering
        \includegraphics[width=\linewidth]{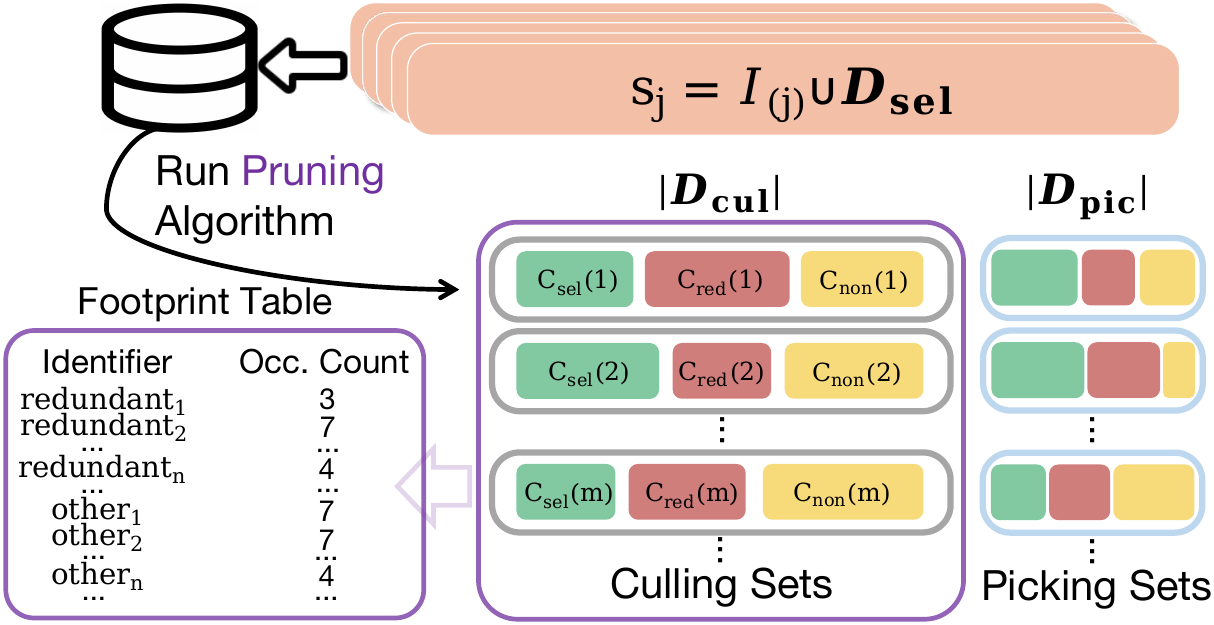}
    \caption{Overall architecture of Footprint Counter.}
    \vspace{-5mm}
    \label{fig:footprint}
\end{figure}
\subsubsection{Shadow Datapool}
\label{footprint_shadow}
For each shadow datapool \(Q^{\textrm{shadow}}_{\textrm{i}}\), we have obtained $\frac{|Q_{\textrm{s}}|}{\zeta_{\textrm{s}}}$ number of culling sets \(D^{\textrm{shadow}}_{\textrm{cul}}(\textrm{i,j})\), where \(\zeta_{\textrm{s}}\) is the batch size of the shadow datapool. For samples in a culling set that belong to \(D_{\textrm{sel}}^{\textrm{shadow}}(\textrm{i})\), $D_{\textrm{red}}^{\textrm{shadow}}(\textrm{i})$ and $D_{\textrm{non}}^{\textrm{shadow}}(\textrm{i})$, we termed them as $\textrm{C}^{\textrm{shadow}}_{\textrm{sel}}(\textrm{i,j})$, $\textrm{C}^{\textrm{shadow}}_{\textrm{red}}(\textrm{i,j})$ and $\textrm{C}^{\textrm{shadow}}_{\textrm{non}}(\textrm{i,j})$, respectively.
Finally, we have a footprint for the current shadow datapool: 
\begin{equation} 
\resizebox{0.85\columnwidth}{!}{
$
G^{\textrm{shadow}}_{\textrm{i}}(A_{\textrm{i}}^{\textrm{s}}, \mathbb{M}(\cdot)) = \{\textrm{C}^{\textrm{shadow}}_{\textrm{red}}(\textrm{i,j}), \textrm{C}^{\textrm{shadow}}_{\textrm{non}}(\textrm{i,j})\}_{\textrm{j}\in[1,
\frac{|Q_{\textrm{s}}|}{\zeta_{\textrm{s}}}
]}.
$
}
\end{equation}
As shown in Figure \ref{fig:footprint} (for simplicity, we omit the shadow datapool index \(\textrm{i}\)), from the footprint of the current shadow datapool, we can finally form its FootPrint Table, where each data record has its own entry. The entry of a specific data record includes its identifier and occurrence count.

\subsubsection{Victim Datapool}
\label{footprint_victim}
For the victim datapool \(Q_{\textrm{v}}\), we have $\frac{|Q_{\textrm{v}}|}{\zeta_{\textrm{v}}}$ number of culling sets \(D^{\textrm{victim}}_{\textrm{cul}}(\textrm{j})\), where \(\zeta_{\textrm{v}}\) is the batch size of the victim datapool. Similarly, we firstly term samples in the victim culling set \(D^{\textrm{victim}}_{\textrm{cul}}(\textrm{j})\) belong to \(D_{\textrm{sel}}^{\textrm{victim}}\), $D_{\textrm{red}}^{\textrm{victim}}$ and $D_{\textrm{non}}^{\textrm{victim}}$ as $\textrm{C}^{\textrm{victim}}_{\textrm{sel}}(\textrm{j})$, $\textrm{C}^{\textrm{victim}}_{\textrm{red}}(\textrm{j})$ and $\textrm{C}^{\textrm{victim}}_{\textrm{non}}(\textrm{j})$, respectively. Note that \(\textrm{C}^{\textrm{victim}}_{\textrm{red}}(\textrm{j}) \) and \(\textrm{C}^{\textrm{victim}}_{\textrm{non}}(\textrm{j})\) are mixed and the adversary can only obtain the intersection of the two, i.e., \(\textrm{C}^{\textrm{victim}}_{\textrm{red}\cup\textrm{non}}(\textrm{j}) = \textrm{C}^{\textrm{victim}}_{\textrm{red}}(\textrm{j})\cup\textrm{C}^{\textrm{victim}}_{\textrm{non}}(\textrm{j})\). Thus, the footprint of the victim datapool can be written as:
\begin{equation} 
\resizebox{0.63\columnwidth}{!}{
$
G^{\textrm{victim}}(A^{\textrm{v}}, \mathbb{M}(\cdot)) = \{\textrm{C}^{\textrm{victim}}_{\textrm{red}\cup\textrm{non}}(\textrm{j})\}_{\textrm{j}\in[1,
\frac{|Q_{\textrm{v}}|}{\zeta_{\textrm{v}}}
]}.
$
}
\end{equation}
Finally, the FootPrint Table of the victim datapool is formed, which differs slightly from that of the shadow datapool. While each entry still includes the identifier and occurrence count, the identifier only specifies the sample without indicating its type.

\subsection{Distribution Estimator}
With the footprint tables providing sample-level information, the occurrence distribution and its derivatives can now be formed. Referring to Definition \ref{def:2} and Section \ref{footprint_shadow}, for each shadow datapool, we count the \textit{occurrence} of each data. This results in two occurrence distributions $\mathbb{S}^{\textrm{shadow}}_{\textrm{red}}(\textrm{i})(\cdot)$ and $\mathbb{S}^{\textrm{shadow}}_{\textrm{non}}(\textrm{i})(\cdot)$. For the victim datapool, referring back to Definition \ref{def:1} and discussions in Section \ref{footprint_victim}, a mixed occurrence distribution \(\mathbb{S}^{\textrm{victim}}_{\textrm{red}\cup\textrm{non}}(\cdot)\) is formed.

Once we have the occurrence distributions of a shadow datapool, their cumulative distribution function (CDF)~\cite{wiki:cdf} \(F^{\textrm{cdf}}_{\textrm{red}}(\textrm{i})(t)\), \(F^{\textrm{cdf}}_{\textrm{non}}(\textrm{i})(t)\) and complementary cumulative distribution function (CCDF)~\cite{wiki:cdf} \(F^{\textrm{ccdf}}_{\textrm{red}}(\textrm{i})(t)\), \(F^{\textrm{ccdf}}_{\textrm{non}}(\textrm{i})(t)\) can be easily derived. Formal definitions can be found in Appendix \ref{CDF_def}. In addition, we define a flag function for \(\delta \in \{\textrm{cdf}, \textrm{ccdf}\}\) as follows:
\begin{equation} 
U_{\delta}(t) = \mathbb{I} \{F^{\delta}_{\textrm{red}}(\textrm{i})(t) > F^{\delta}_{\textrm{non}}(\textrm{i})(t)\}. 
\label{flag}
\end{equation}
Based on the CDFs, we introduce the \textit{cumulative disparity rates} (CDRs) \(R_{\textrm{cdr}}(\textrm{i})(t)\) of each shadow datapool. The CDRs quantify the relative difference between the CDFs of the redundant set and other non-members at different occurrence count \(t\). Accordingly, the \textit{complementary cumulative disparity rates} (CCDRs) \(R_{\textrm{ccdr}}(\textrm{i})(t)\) can also be defined. They can be formed as:
\begin{equation} 
R(\textrm{i})(t) = \frac{\textrm{max}\{F^{\delta}_{\textrm{red}}(\textrm{i})(t), F^{\delta}_{\textrm{non}}(\textrm{i})(t)\}}{F^{\delta}_{\textrm{red}}(\textrm{i})(t)+F^{\delta}_{\textrm{non}}(\textrm{i})(t)},
\label{cdr}
\end{equation}
where \(R(\textrm{i})(t)\) can be \(R_{\textrm{cdr}}(\textrm{i})(t)\) when \(\delta=\textrm{cdf}\) or \(R_{\textrm{ccdr}}(\textrm{i})(t)\) when \(\delta=\textrm{ccdf}\). 

Moreover, for CDFs, we define a function \(\Delta(p,q)\)  to quantify the absolute difference between different occurrence counts \(p < q\) of the same CDF 
(e.g., for \(F^{\textrm{cdf}}_{\textrm{red}}(\textrm{i})(t)\), \(\Delta_{\textrm{i}}^{\textrm{red}}(p,q)=F^{\textrm{cdf}}_{\textrm{red}}(\textrm{i})(q)-F^{\textrm{cdf}}_{\textrm{red}}(\textrm{i})(p)\)). The concept of both the flag function and CDR of CDFs can then be extended to a bivariate version:
\begin{align}
& U_{\textrm{cdf}}(p,q) = \mathbb{I} \{\Delta_{\textrm{i}}^{\textrm{red}}(p,q) > \Delta_{\textrm{i}}^{\textrm{non}}(p,q)\},
\label{bflag}\\
&R_{\textrm{cdr}}(\textrm{i})(p,q) = \frac{\textrm{max}\{\Delta_{\textrm{i}}^{\textrm{red}}(p,q), \Delta_{\textrm{i}}^{\textrm{non}}(p,q)\}}{\Delta_{\textrm{i}}^{\textrm{red}}(p,q)+\Delta_{\textrm{i}}^{\textrm{non}}(p,q)}. 
\label{bcdr}
\end{align}
Now we introduce the proposed attacks based on the previously determined concepts.

\subsection{Threshold Determination}

\subsubsection{Whole Distribution Difference Ratio (WhoDis)} 
In WhoDis, we are based on the following observation: \textit{The occurrence distributions of most dataset pruning methods share a phenomenon where the occurrence counts of other non-members are densely clustered around smaller values, while those of the redundant set are more dispersed.} A detailed illustration of this can be found in the experiments (see Section \ref{evidence}). In WhoDis, we aim at inferring all data in the victim datapool. To do so, for each shadow datapool, we first identify the occurrence count that maximizes the difference between the two CDFs (for the redundant set and other non-members). This specific occurrence count is defined as the W ho Di s threshold $\tau_{\textrm{w}}(\cdot)$ of each shadow datapool. Formally, $\tau_{\textrm{w}}(\textrm{i})$ of the \(\textrm{i}\)-th shadow datapool is: 
\begin{equation}
    \tau_{\textrm{w}}(\textrm{i}) = \arg\max_{t} \left| F^{\textrm{cdf}}_{\textrm{red}}(\textrm{i})(t) - F^{\textrm{cdf}}_{\text{non}}(\textrm{i})(t) \right|.
\end{equation}
Subsequently, we classify all samples with occurrence counts lower than $\tau_{\textrm{w}}(\cdot)$ as belonging to the other non-members, while those with occurrence counts higher than the threshold are considered to belong to the redundant set.

We then perform a majority vote on the \(\tau_{\textrm{w}}(\textrm{i})\) values from all shadow datapools. The most frequently occurring \(\tau_{\textrm{w}}(\textrm{i})\) is selected as the final \(\tau_{\textrm{w}}\) value. This can be expressed as: \( \tau_{\textrm{w}} = \text{mode}(\{\tau_{\textrm{w}}(\textrm{i})\}) \), where
\(\text{mode}(\cdot)\) denotes the most frequently occurring value in a set.

\subsubsection{Cumulative Difference Ratio (CumDis)} 
Here, two thresholds are determined to infer part of the datapool. Firstly, we find the occurrence count corresponding to the maximum CDR and set it as the lower threshold \(\tau_{\textrm{cL}}(\textrm{i})\): 
\begin{equation}
    \tau_{\textrm{cL}} (\textrm{i}) = \mathop{\arg\max}\limits_{t} R_{\textrm{cdr}}(\textrm{i})(t).
\end{equation}
Accordingly, a flag \(\kappa_{\textrm{cL}}(\textrm{i})\) is designed to determine the higher value among the two CDFs used in \(R_{\textrm{cdr}}(\textrm{i})(t)\): \(\kappa_{\textrm{cL}}(\textrm{i}) = U_{\textrm{cdf}}(\tau_{\textrm{cL}}(\textrm{i}))\). If \(\kappa_{\textrm{cL}}(\textrm{i})\) equals 1, redundant data constitutes the majority below the threshold \(\tau_{\textrm{cL}}(\textrm{i})\). Therefore, we infer the data below \(\tau_{\textrm{cL}}(\textrm{i})\) as redundant data. For the two CCDFs, we can determine the upper threshold \(\tau_{\textrm{cU}}(\textrm{i})\) and the flag \(\kappa_{\textrm{cU}}(\textrm{i})\) in the same manner. Finally, a majority vote is conducted, similar to the process in WhoDis, to get the final thresholds \( \tau_{\textrm{cL}}\), \( \tau_{\textrm{cU}}\) and flags \(\kappa_{\textrm{cL}}\), \(\kappa_{\textrm{cU}}\). 

\subsubsection{Arbitrary Range Difference Ratio (ArraDis)} Here, we aim at identifying the interval with the most significant difference between \(F^{\textrm{cdf}}_{\textrm{red}}(\textrm{i})(\cdot)\) and \(F^{\textrm{cdf}}_{\textrm{non}}(\textrm{i})(\cdot)\). We define two thresholds to represent the beginning and the end of the interval, respectively. These thresholds are determined as:
\begin{equation}
\tau_{\textrm{aL}}(\textrm{i}), \tau_{\textrm{aU}}(\textrm{i}) = \mathop{\arg\max}\limits_{p,q} R_{\textrm{cdr}}(\textrm{i})(p,q).
\end{equation}
Recall Equation \eqref{bflag}, the flag \(\kappa_{\textrm{a}}(\textrm{i})\) can thus be defined as: \(\kappa_{\textrm{a}}(\textrm{i}) = U_{\textrm{cdf}}(\tau_{\textrm{aL}}(\textrm{i}),\tau_{\textrm{aU}}(\textrm{i}))\).
Finally, we conduct a majority vote for the threshold pair and the flag, resulting in the final thresholds \(\tau_{\textrm{aL}}\) and \(\tau_{\textrm{aU}}\) and the final flag \(\kappa_{\textrm{a}}\). 
 
\subsubsection{Spike Difference Ratio (SpiDis)} In SpiDis, we aim at identifying the most venerable data records directly via the occurrence distributions, i.e., for each shadow datapool, we locate a specific occurrence count value \(t\), where the two types of data have the most significant occurrence distribution difference. We set this specific value \(t\) as the threshold \(\tau_{\textrm{sp}}(\textrm{i})\): 
\begin{equation}
    \tau_{\textrm{sp}}(\textrm{i}) = \arg\max_{t}\frac{\textrm{max}\{\mathbb{S}^{\textrm{shadow}}_{\textrm{red}}(\textrm{i})(t), \mathbb{S}^{\textrm{shadow}}_{\textrm{non}}(\textrm{i})(t)\}}{\mathbb{S}^{\textrm{shadow}}_{\textrm{red}}(\textrm{i})(t)+\mathbb{S}^{\textrm{shadow}}_{\textrm{non}}(\textrm{i})(t)}.
\end{equation}
The flag \(\kappa_{\textrm{sp}}(\textrm{i})=\mathbb{I} \{\mathbb{S}^{\textrm{shadow}}_{\textrm{red}}(\textrm{i})(\tau_{\textrm{sp}}(\textrm{i}))>\mathbb{S}^{\textrm{shadow}}_{\textrm{non}}(\textrm{i})(\tau_{\textrm{sp}}(\textrm{i}))\}\) is then derived. Finally, a majority vote is done on all the shadow datapools to get the final threshold \(\tau_{\textrm{sp}}\) and flag \(\kappa_{\textrm{sp}}\). 

\subsection{Threshold Calibration}
According to previous discussions, 
the occurrence count range and the batch size of the victim datapool and the shadow datapool set by the adversary may be different. Thus, to align the shadow datapool with the victim datapool, a threshold calibration step should be conducted. Specifically, the calibration is applied on all of the designed thresholds \(\tau\) in the previous section. We utilize the batch numbers of the two datapools, as they contain information related to both the datapool scale and the batch size. The formal definition is as follows:
\begin{equation}
\tau' = \left\lfloor \tau \frac{|Q_{\textrm{v}}|}{|Q_{\textrm{s}}|} \frac{\zeta_{\textrm{s}}}{\zeta_{\textrm{v}}} \right\rfloor.
\end{equation}

We can now use the derived thresholds and flags to perform threshold-based attacks on the victim occurrence distribution. Detailed experimental results will be presented in the next section.

\section{Experimental Evaluation}
\label{eval}
In this section, we present the practical performance of DaLI through four proposed inference attacks. Specifically, we conduct extensive experiments to address the following key inquiries (KeyIQs):
\begin{itemize}
    \item \textit{KeyIQ1:} Is dataset pruning truly as effective as previously anticipated, ensuring a balance between utility and efficiency while also safeguarding privacy?
    \item \textit{KeyIQ2:} How does the risk of privacy inference change across different pruning methods?
    \item \textit{KeyIQ3:} Can we develop an effective single metric to measure privacy inference risk across different methods and settings, thereby facilitating privacy-level evaluations across different pruning methods?
    \item \textit{KeyIQ4:} How do the attacks fare when the adversary possesses varying degrees of knowledge?
    \item \textit{KeyIQ5:} Can we explore dedicated defense strategies to mitigate the inference effectiveness?
\end{itemize}

\noindent\textbf{Overview. }Specifically, we first present the statistical rationale behind the four proposed threshold attacks and validate the performance of DaLI when the adversary possesses the most knowledge of the pruning process. Following this, we propose a unified evaluation metric, the Brimming score, which effectively measures the privacy risk of different methods under various settings. In addition, we introduce four more challenging scenarios, each based on different levels of adversary knowledge (see Section \ref{partial_know})
We will show that DaLI remains effective even in the most challenging scenario. Finally,
we make a preliminary attempt on the defense side.

\subsection{Experimental Setup}
\noindent {\bf Pruning Methods.}
We use \textbf{twelve} popular pruning methods: DeepFool\cite{ducoffe2018adversarial}, 
Contextual Diversity (Co.Div.)\cite{agarwal2020contextual},
Cal\cite{margatina2021active}, Glister (Glist.)~\cite{killamsetty2021glister}, Forgetting (Forgt.)\cite{toneva2018empirical}, GraNd\cite{paul2021deep}, 
Uncertainty (Unc.)\cite{coleman2020selection}, Craig\cite{mirzasoleiman2020coresets}, GradMatch (G.M.)\cite{killamsetty2021grad}, Submodular (SubM.)\cite{iyer2013submodular}, Herding (Herd.)\cite{welling2009herding} and kCenterGreedy (kCent)\cite{farahani2009facility}. They can be classified into seven types. Detailed taxonomy is given in Appendix \ref{taxonomy}.

\noindent {\bf Datasets.}
In our experiments, we utilize three commonly used datasets: MNIST, CIFAR10, and CIFAR100 for evaluation. Related details can be found in Appendix \ref{dataset}.

\noindent {\bf Hyperparameter Setup. }
We use four pruning fractions with significant spans: 0.2, 0.4, 0.6, and 0.8 and use attack success rate (ASR) for evaluation. Details of other hyperparameters and the evaluation metric are in Appendix \ref{hyper_app}.

\begin{table*}[htbp]
    \centering
 \small 
    \begin{threeparttable}[t]
    \begin{adjustbox}{max width=\textwidth}
    \begin{tabular}{c|c|c|c|c|c|c|c|c|c|c|c|c|c|c}
        \hline
        \hline
        Dataset & Frac. & Thres. & Cal& Co.Div.& Craig& DeepF.& Forgt.& Glist.& G.M.& GraNd& Herd.& kCent.& SubM.& Unc. \\
        \hline
        \multirow{16.5}{*}{MNIST} & \multirow{3}{*}{0.2} 
        & WhoDis & \cellcolor{red!55}53.39 & 52.54 & 51.95 & \cellcolor{blue!45}50.00 & 52.56 & 51.03 & 50.62 & \cellcolor{red!25}52.77 & 51.91 & 51.33 & \cellcolor{blue!15}50.46 & 52.26\\
        \cline{3-15}
        & & CumDis & \cellcolor{red!25}70.51 & 62.80 & 62.00 & 58.62 & \cellcolor{red!55}95.59 & 58.21 & \cellcolor{blue!15}55.16 & 68.98 & \cellcolor{blue!45}52.97 & 56.61 & 55.61 & 62.88 \\
        \cline{3-15}
        & & SpiDis & \cellcolor{red!25}73.73 & \cellcolor{blue!45}54.64 & 65.84 & 63.66 & \cellcolor{red!55}98.65 & 68.55 & 72.67 & 72.39 & \cellcolor{blue!15}57.58 & 63.51& 70.61 & 66.22 \\
        \cline{3-15}
        & & ArraDis & 62.00 & 62.41 & 60.74 & 58.00 & \cellcolor{red!55}79.35 & 60.09 & \cellcolor{blue!45}51.83 & 63.21 & 59.56 & \cellcolor{blue!15}58.43 & \cellcolor{red!25}65.60 & 59.58 \\
        \cline{2-15}
        & \multirow{3}{*}{0.4} 
        & WhoDis & \cellcolor{red!25}55.43 & 54.36 & 53.44 & 53.52 & \cellcolor{blue!45}50.22 & 51.30 & 51.42 & \cellcolor{red!55}56.97 & 54.57 & 53.24 & \cellcolor{blue!45}50.00 & 54.66 \\
        \cline{3-15}
        & & CumDis & 66.78 & 61.69 & 60.20 & 55.83 & \cellcolor{red!55}85.01 & \cellcolor{blue!15}54.49 & \cellcolor{blue!45}52.76 & \cellcolor{red!25}75.00 & 68.36 & 58.04 & 63.76 & 63.18 \\
        \cline{3-15}
        & & SpiDis & 81.63 & 82.29 & 73.71 & 71.26 & \cellcolor{red!25}96.24 & \cellcolor{blue!15}62.30 & \cellcolor{blue!45}56.29 & 83.45 & \cellcolor{red!55}100.00 & 64.47 & 70.76 & 80.00 \\
        \cline{3-15}
        & & ArraDis & 73.23 & 74.11 & 69.07 & 68.64 & \cellcolor{red!25}86.18 & \cellcolor{blue!15}64.78 & \cellcolor{blue!45}52.14 & 74.66 & \cellcolor{red!55}95.65 & 67.00 & 61.66 & 72.12 \\
        \cline{2-15}
        & \multirow{3}{*}{0.6} 
        & WhoDis & 58.29 & 54.63 & 55.92 & 57.02 & \cellcolor{red!55}59.68 & 54.84 & \cellcolor{blue!45}50.00 & 57.25 & \cellcolor{blue!15}51.26 & 54.41 & 53.89 & \cellcolor{red!25}58.56 \\
        \cline{3-15}
        & & CumDis & 66.05 & 62.15 & 60.42 & 63.66 & \cellcolor{red!55}85.31 & \cellcolor{blue!45}51.00 & 53.21 & 65.11 & \cellcolor{blue!15}52.18 & 56.14 & 61.08 & \cellcolor{red!25}66.48 \\
        \cline{3-15}
        & & SpiDis & 86.17 & 82.01 & 73.81 & 82.81 & 86.03 & 72.22 & \cellcolor{blue!45}54.42 & 80.84 & \cellcolor{red!55}100.00 & \cellcolor{blue!15}66.04 & 86.57 & \cellcolor{red!25}88.08 \\
        \cline{3-15}
        & & ArraDis & 76.51 & \cellcolor{blue!15}71.04 & 71.83 & 80.87 & \cellcolor{red!25}82.33 & 76.68 & \cellcolor{blue!45}51.27 & 76.33 & \cellcolor{red!55}100.00 & 63.71 & 74.22 & 78.12 \\
        \cline{2-15}
        & \multirow{3}{*}{0.8} 
        & WhoDis & 56.52 & \cellcolor{blue!45}51.47 & \cellcolor{red!25}61.30 & 54.13 & 55.99 & 57.07 & \cellcolor{blue!15}52.98 & \cellcolor{red!55}63.37 & 53.28 & 56.18 & 53.23 & 60.35 \\
        \cline{3-15}
        & & CumDis & 57.13 & 60.79 & 56.24 & \cellcolor{red!25}60.83 & \cellcolor{blue!45}50.54 & 55.99 & 54.02 & \cellcolor{red!55}66.78 & 57.92 & 56.15 & \cellcolor{blue!15}52.68 & 61.22 \\
        \cline{3-15}
        & & SpiDis & 81.32 & 76.16 & 67.65 & 62.14 & \cellcolor{red!55}100.00 & \cellcolor{blue!15}58.63 & \cellcolor{blue!45}51.94 & \cellcolor{red!25}87.46 & \cellcolor{red!55}100.00 & 61.25 & 64.29 & 80.06 \\
        \cline{3-15}
        & & ArraDis & 65.84 & 67.71 & 64.63 & 62.24 & \cellcolor{red!55}100.00 & 61.80 & \cellcolor{blue!45}50.68 & \cellcolor{red!25}72.90 & 62.21 & 61.37 & \cellcolor{blue!15}58.96 & 68.83 \\
        \hline
        \hline
        \multirow{16.5}{*}{CIFAR10} 
        & \multirow{3}{*}{0.2} 
        & WhoDis & \cellcolor{red!25}52.05 & \cellcolor{blue!45}50.06 & 50.84 & 50.48 & \cellcolor{blue!15}50.15 & 50.51 & 50.24 & \cellcolor{red!55}53.11 & 51.86 & 50.75 & 51.26 & 51.07\\
        \cline{3-15}
        & & CumDis & 58.84 & 52.75 & \cellcolor{blue!15}51.78 & \cellcolor{blue!45}51.53 & 57.51 & 53.01 & 52.14 & \cellcolor{red!55}67.71 & \cellcolor{red!25}60.34 & 53.10 & 56.69 & 57.31 \\
        \cline{3-15}
        & & SpiDis & \cellcolor{red!55}67.04 & \cellcolor{blue!15}57.07 & 61.60 & 58.00 & \cellcolor{blue!45}50.13 & 60.48 & \cellcolor{red!25}65.52 & 69.81 & 57.31 & 59.32 & 60.77 & 63.72 \\
        \cline{3-15}
        & & ArraDis & \cellcolor{red!25}61.43 & 57.62 & 55.75 & 57.11 & \cellcolor{blue!15}52.29 & 58.31 & \cellcolor{blue!45}50.54 & \cellcolor{red!55}61.71 & 57.71 & 55.60 & 57.73 & 59.22 \\
        \cline{2-15}
        & \multirow{3}{*}{0.4} 
        & WhoDis & 54.38 & 52.23 & 53.39 & 52.51 & \cellcolor{red!25}54.43 & 51.82 & \cellcolor{blue!45}51.03 & \cellcolor{red!55}57.95 & 52.89 & 51.32 & \cellcolor{blue!15}51.25 & 54.33 \\
        \cline{3-15}
        & & CumDis & 61.86 & 55.25 & 59.79 & 53.87 & 56.97 & \cellcolor{blue!45}50.97 & 56.70 & \cellcolor{red!55}77.63 & 60.56 & 52.12 & \cellcolor{blue!15}51.14 & \cellcolor{red!25}62.90 \\
        \cline{3-15}
        & & SpiDis & 83.69 & 73.17 & 78.91 & 69.52 & \cellcolor{red!55}86.94 & 73.53 & 82.79 & \cellcolor{red!25}86.30 & 75.90 & \cellcolor{blue!45}59.46 & \cellcolor{blue!15}60.53 & 81.48 \\
        \cline{3-15}
        & & ArraDis & 74.25 & 64.76 & 68.03 & 64.91 & \cellcolor{red!55}81.10 & 71.31 & \cellcolor{blue!45}51.18 & \cellcolor{red!25}76.31 & 66.64 & 59.85 & \cellcolor{blue!15}57.35 & 72.05 \\
        \cline{2-15}
        & \multirow{3}{*}{0.6} 
        & WhoDis & 57.03 & 54.13 & 58.63 & 57.38 & 51.81 & 56.08 & \cellcolor{blue!15}50.18 & \cellcolor{red!55}63.01 & 51.98 & 52.21 & \cellcolor{blue!45}51.23 & \cellcolor{red!25}59.21 \\
        \cline{3-15}
        & & CumDis & 62.23 & 60.96 & 67.25 & 63.64 & \cellcolor{blue!15}52.41 & 55.46 & 52.62 & \cellcolor{red!55}78.55 &56.20 & 53.45 & \cellcolor{blue!45}50.20 & \cellcolor{red!25}68.57 \\
        \cline{3-15}
        & & SpiDis & 88.55 & 73.24 & 86.78 & \cellcolor{red!25}90.45 & 65.38 & 86.05 & \cellcolor{blue!45}53.57 & 89.30 & \cellcolor{red!55}100.00 & 67.74 & \cellcolor{blue!15}57.14 & 87.80 \\
        \cline{3-15}
        & & ArraDis & 79.67 & 68.82 & 79.75 & 78.55 & 67.98 & \cellcolor{red!55}84.56 & \cellcolor{blue!45}51.49 & \cellcolor{red!25}81.23 & 81.82 & \cellcolor{blue!15}57.46 & 61.48 & 78.67 \\
        \cline{2-15}
        & \multirow{3}{*}{0.8} 
        & WhoDis & 61.24 & 56.74 & 65.23 & 63.58 & 55.76 & 54.82 & \cellcolor{blue!45}50.29 & \cellcolor{red!55}69.81 & \cellcolor{red!25}66.79 & 52.54 & \cellcolor{blue!15}52.48 & 63.09 \\
        \cline{3-15}
        & & CumDis & 62.99 & 57.05 & 65.92 & 64.97 & 59.91 & 57.89 & 54.35 & \cellcolor{red!55}75.25 & \cellcolor{red!25}66.67 & \cellcolor{blue!15}52.30 & \cellcolor{blue!45}52.18 & 63.39 \\
        \cline{3-15}
        & & SpiDis & \cellcolor{red!25}82.65 & 64.60 & \cellcolor{red!55}90.81 & 81.38 & 81.89 & \cellcolor{blue!15}61.81 & \cellcolor{blue!45}50.49 & \cellcolor{red!55}90.81 & 70.76 & 73.13 & 68.00 & 80.41 \\
        \cline{3-15}
        & & ArraDis & 71.46 & 64.88 & \cellcolor{red!25}74.62 & 72.71 & 70.90 & \cellcolor{blue!15}57.43 & \cellcolor{blue!45}51.82 & \cellcolor{red!55}75.05 & 69.92 & 59.21 & 60.73 & 73.68 \\
        \hline
        \hline
        \multirow{16.5}{*}{CIFAR100} & \multirow{3}{*}{0.2} 
        & WhoDis & 51.01 & 51.51 & \cellcolor{blue!15}50.24 & \cellcolor{blue!45}50.10 & 50.64 & \cellcolor{red!55}55.69 & 51.18 & \cellcolor{red!25}53.24 & 51.26 & 51.23 & 50.77 & 50.97 \\
        \cline{3-15}
        & & CumDis & 55.19 & \cellcolor{red!25}59.14 & \cellcolor{blue!15}54.30 & \cellcolor{blue!45}54.00 & 55.39 & \cellcolor{red!55}62.98 & 58.26 & 58.15 & 57.41 & 56.62 & 55.17 & 56.07 \\
        \cline{3-15}
        & & SpiDis & 61.86 & 63.93 & 61.16 & \cellcolor{blue!15}59.39 & 59.72 & \cellcolor{red!55}93.37 & 67.42 & \cellcolor{red!25}69.59 & 68.59 & \cellcolor{blue!45}59.36 & 67.14 & 62.89 \\
        \cline{3-15}
        & & ArraDis & 60.25 & 59.86 & \cellcolor{blue!15}55.17 & \cellcolor{blue!45}54.81 & 58.33 & 84.81 & 53.10 & \cellcolor{red!55}63.76 & \cellcolor{red!25}61.02 & 59.64 & 59.47 & 58.10 \\
        \cline{2-15}
        & \multirow{3}{*}{0.4} 
        & WhoDis & \cellcolor{red!55}52.24 & 50.91 & 50.61 & 50.95 & 50.43 & \cellcolor{red!25}51.94 & \cellcolor{blue!45}50.21 & 51.97 & 50.93 & 51.84 & \cellcolor{blue!15}50.24 & 51.49 \\
        \cline{3-15}
        & & CumDis & 52.28 & \cellcolor{blue!45}50.46 & 51.73 & 54.92 & \cellcolor{red!25}55.93 & 51.52 & \cellcolor{blue!15}50.72 & \cellcolor{red!55}58.53 & 53.74 & 50.96 & 52.34 & 54.78 \\
        \cline{3-15}
        & & SpiDis & 78.34 & 65.12 & 70.00 & 70.00 & 61.20 & \cellcolor{red!55}85.07 & \cellcolor{blue!45}50.33 & 79.31 & \cellcolor{red!25}82.05 & 68.97 & \cellcolor{blue!15}60.32 & 66.90 \\
        \cline{3-15}
        & & ArraDis & 68.12 & \cellcolor{blue!15}56.16 & 59.63 & 61.96 & 56.42 & 69.73 & \cellcolor{blue!45}52.85 & \cellcolor{red!55}74.22 & \cellcolor{red!25}73.36 & 64.85 & 61.44 & 63.29 \\
        \cline{2-15}
        & \multirow{3}{*}{0.6} 
        & WhoDis & 51.71 & \cellcolor{blue!15}50.48 & 52.06 & 53.30 & \cellcolor{red!25}54.26 & 52.53 & \cellcolor{blue!45}50.29 & \cellcolor{red!55}59.53 & 51.45 & 51.30 & 50.23 & 53.29 \\
        \cline{3-15}
        & & CumDis & 51.50 & \cellcolor{blue!45}50.15 & 53.20 & 55.58 & \cellcolor{red!25}56.94 & 52.74 & \cellcolor{blue!15}50.46 & \cellcolor{red!55}60.34 & 52.00 & 51.25 & 51.13 & 56.14 \\
        \cline{3-15}
        & & SpiDis & 68.57 & 60.61 & 71.79 & 75.93 & \cellcolor{red!25}83.77 & 68.75 & \cellcolor{blue!45}58.16 & \cellcolor{red!55}88.11 & 75.00 & 66.67 & \cellcolor{blue!15}58.33 & 78.57 \\
        \cline{3-15}
        & & ArraDis & 66.85 & 61.33 & 63.95 & 70.88 & \cellcolor{red!25}74.42 & 68.62 & \cellcolor{blue!45}55.56 & \cellcolor{red!55}78.14 & 56.83 & 62.20 & \cellcolor{blue!15}55.96 & 66.32 \\
        \cline{2-15}
        & \multirow{3}{*}{0.8} 
        & WhoDis & 56.89 & \cellcolor{blue!45}50.01 & 57.03 & \cellcolor{red!25}57.55 & 50.32 & 52.68 & 50.44 & \cellcolor{red!55}68.78 & 50.33 & 53.91 & \cellcolor{blue!15}50.29 & 55.65 \\
        \cline{3-15}
        & & CumDis & 56.90 & 50.95 & \cellcolor{red!25}56.45 & 57.16 & \cellcolor{blue!15}50.26 & 53.52 & 50.45 & \cellcolor{red!55}73.04 & 50.57 & 53.28 & \cellcolor{blue!45}50.23 & 55.74 \\
        \cline{3-15}
        & & SpiDis & \cellcolor{red!25}79.10 & 65.63 & 71.43 & 70.33 & 61.90 & 62.83 & \cellcolor{blue!45}55.06 & \cellcolor{red!55}91.14 & \cellcolor{blue!15}56.00 & 56.96 & 63.89 & 75.86\\
        \cline{3-15}
        & & ArraDis & \cellcolor{red!25}67.31 & 54.02 & 62.27 & 65.25 & \cellcolor{blue!45}50.15 & 64.67 & \cellcolor{blue!15}51.63 & \cellcolor{red!55}75.96 & 52.48 & 60.10 & 52.62 & 64.69 \\
        \hline
        \hline
    \end{tabular}
    \end{adjustbox}
    \caption{Main attack results under different setups. In each row, the method suffering the most/second most severe privacy risk is marked in dark red/light red. Conversely, the safest/second safest method is marked in dark blue/light blue.}
    \vspace{-3mm}
    \label{tab:main1}
\end{threeparttable}
\end{table*}

\subsection{Evidence for Conducting Attack}
\label{evidence}
To facilitate the subsequent discussion on attacks in scenarios of partial knowledge and our defense attempts, we first present the evidence and intuition behind the two proposed attacks (WhoDis and ArraDis). The evidence for the remaining two attacks will be provided in Appendix \ref{appendix_evidence}.

\begin{figure}[h]
    \centering
    \begin{minipage}[b]{.49\linewidth}
        \centering
        \includegraphics[width=\linewidth]{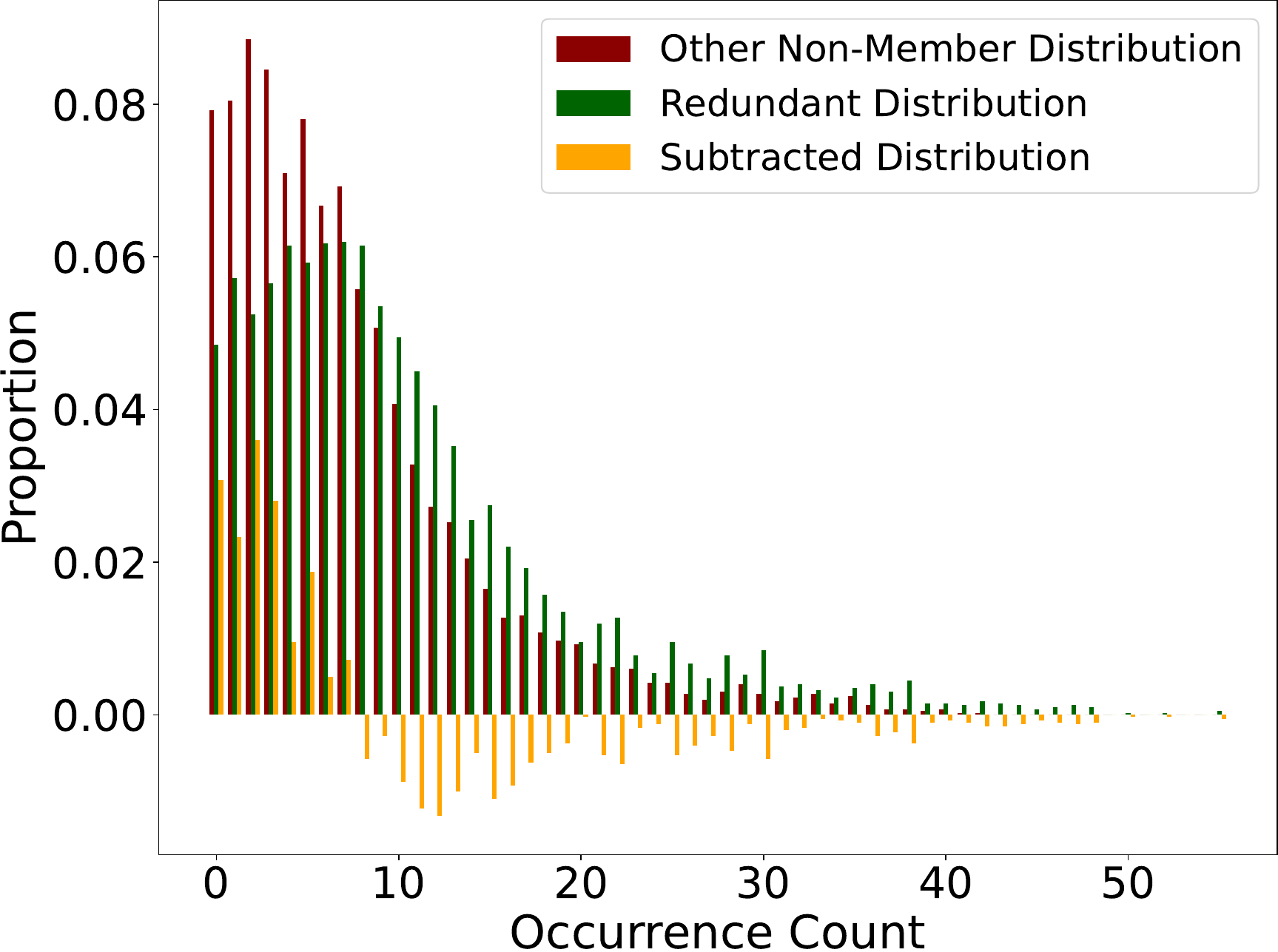}
        \subcaption{WhoDis-Craig}
    \end{minipage}%
    \hfill 
    \begin{minipage}[b]{.49\linewidth}
        \centering
        \includegraphics[width=\linewidth]{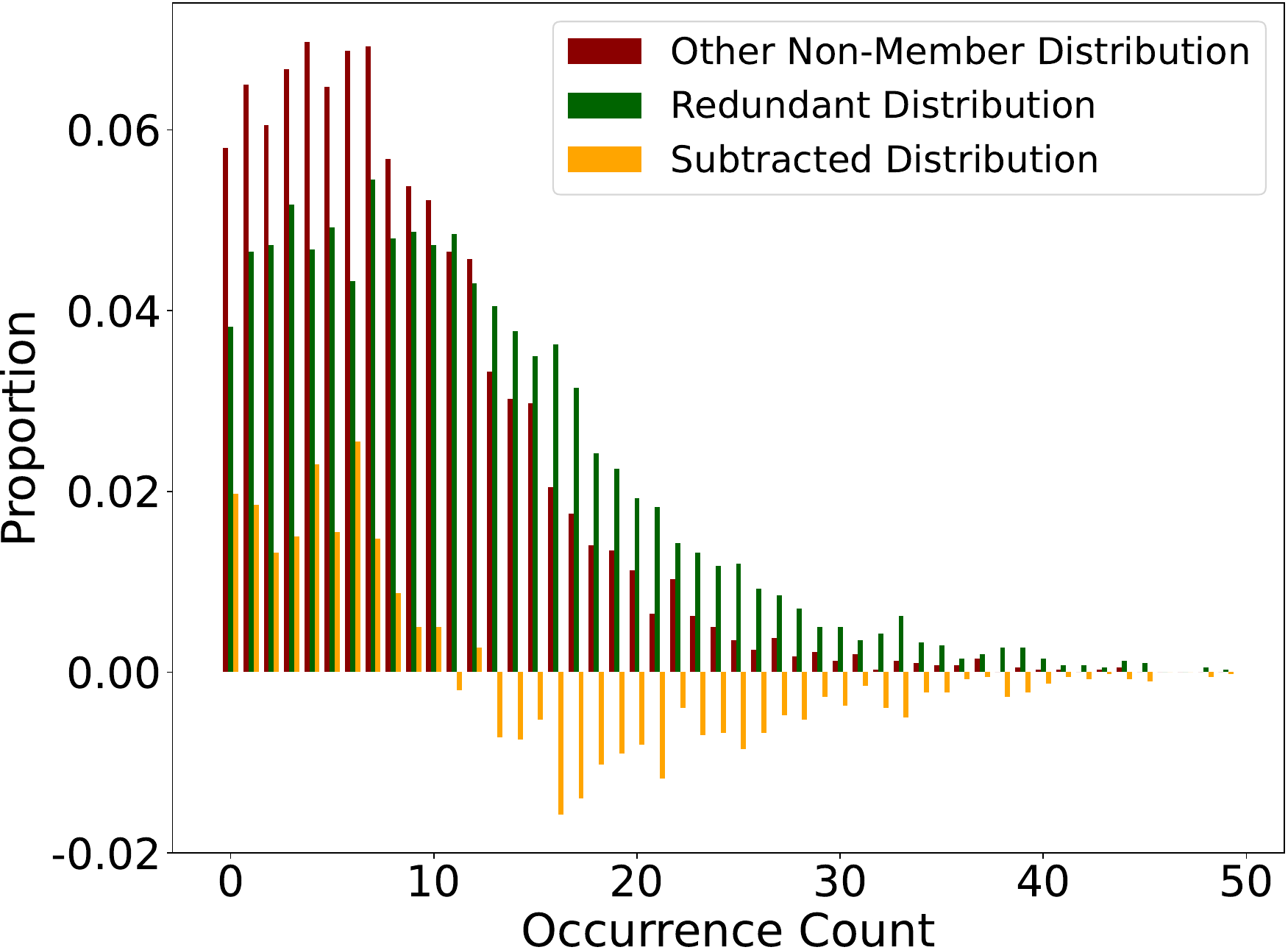}
        \subcaption{WhoDis-DeepF.}
    \end{minipage}%
    \hfill 
    \begin{minipage}[b]{.49\linewidth}
        \centering
        \includegraphics[width=\linewidth]{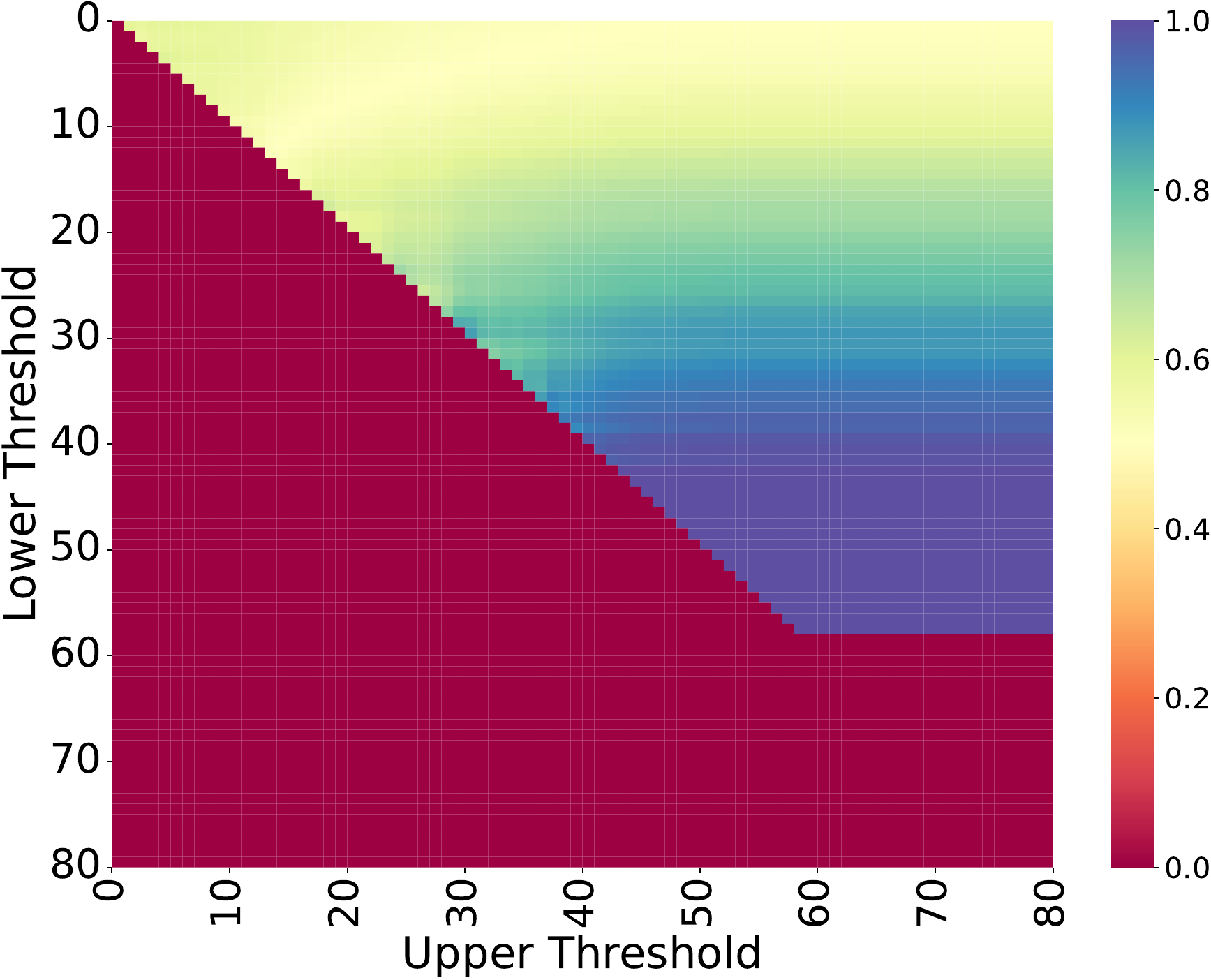}
        \subcaption{ArraDis-Craig}
    \end{minipage}
    \hfill 
    \begin{minipage}[b]{.49\linewidth}
        \centering
        \includegraphics[width=\linewidth]{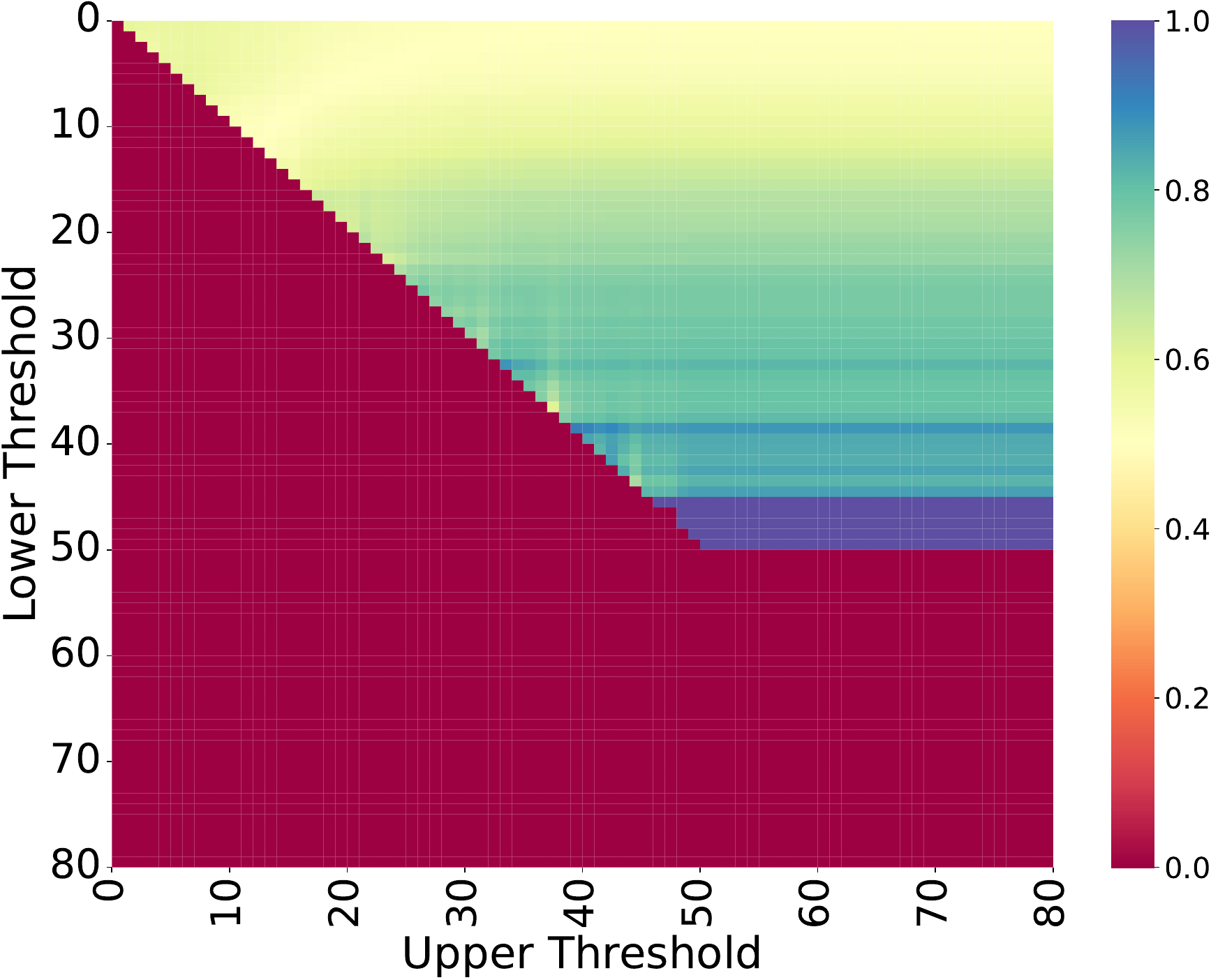}
        \subcaption{ArraDis-DeepF.}
        
    \end{minipage}%
    \caption{Data Lineage Inference vulnerability of different pruning methods under WhoDis and ArraDis. The dataset used here is CIFAR10, the pruning fraction is set to be 0.8, the shadow batch size is 80 and the victim batch size is 100.}
    \label{fig:evidence}
\end{figure}

\noindent {\bf WhoDis.}
As shown in Figures \ref{fig:evidence}(a) and \ref{fig:evidence}(b) (results for other pruning methods can be found in Appendix \ref{remaining_results}), they exhibit a general pattern: the occurrence distribution of the redundant set is more heavy-tailed, while that of other non-members is more concentrated, with the majority of samples having smaller occurrence counts. 
We plotted the difference between the two distributions at different occurrence counts in these figures, as shown by the orange color. It can be observed that there is a specific occurrence count where the difference is predominantly positive on its left side and predominantly negative on its right side. Therefore, if this point or a point near it can be accurately identified as the threshold, the attack can be successfully executed.

\noindent {\bf ArraDis.}
For ArraDis, we seek more fine-grained information from the occurrence distribution. Specifically, we examine the attack success rate when inferring that all data within any given interval of the two distributions belong to one type of data (redundant or other). Results are shown in Figures \ref{fig:evidence}(c) and \ref{fig:evidence}(d).
The deeper the color (blue), the higher the attack success rate. The attack success rate is zero in the lower right corner of the matrix because the upper threshold can not be smaller than the lower one.

It can be observed that for different pruning methods, a threshold range exists where the attack success rate is exceptionally high when the thresholds are set within this region. The corresponding regions differ for each method, providing evidence that ArraDis can conduct an attack successfully and can also be used to design an evaluation metric among different methods.
We will introduce our proposed Brimming score based on this observation later on.

\subsection{Attack Performance}
\subsubsection{Attack Under Full Knowledge}
We first attempt to use DaLI
under the scenario where the adversary has complete knowledge (i.e., knowing both the pruning method and fraction). In this scenario, we primarily aim to provide a negative answer to \textit{KeyIQ1} and a detailed response to \textit{KeyIQ2}. Results are shown in Table \ref{tab:main1}. 
From the results, we can make the following general observations:

\begin{figure}[h]
    \centering
        \includegraphics[width=\linewidth]{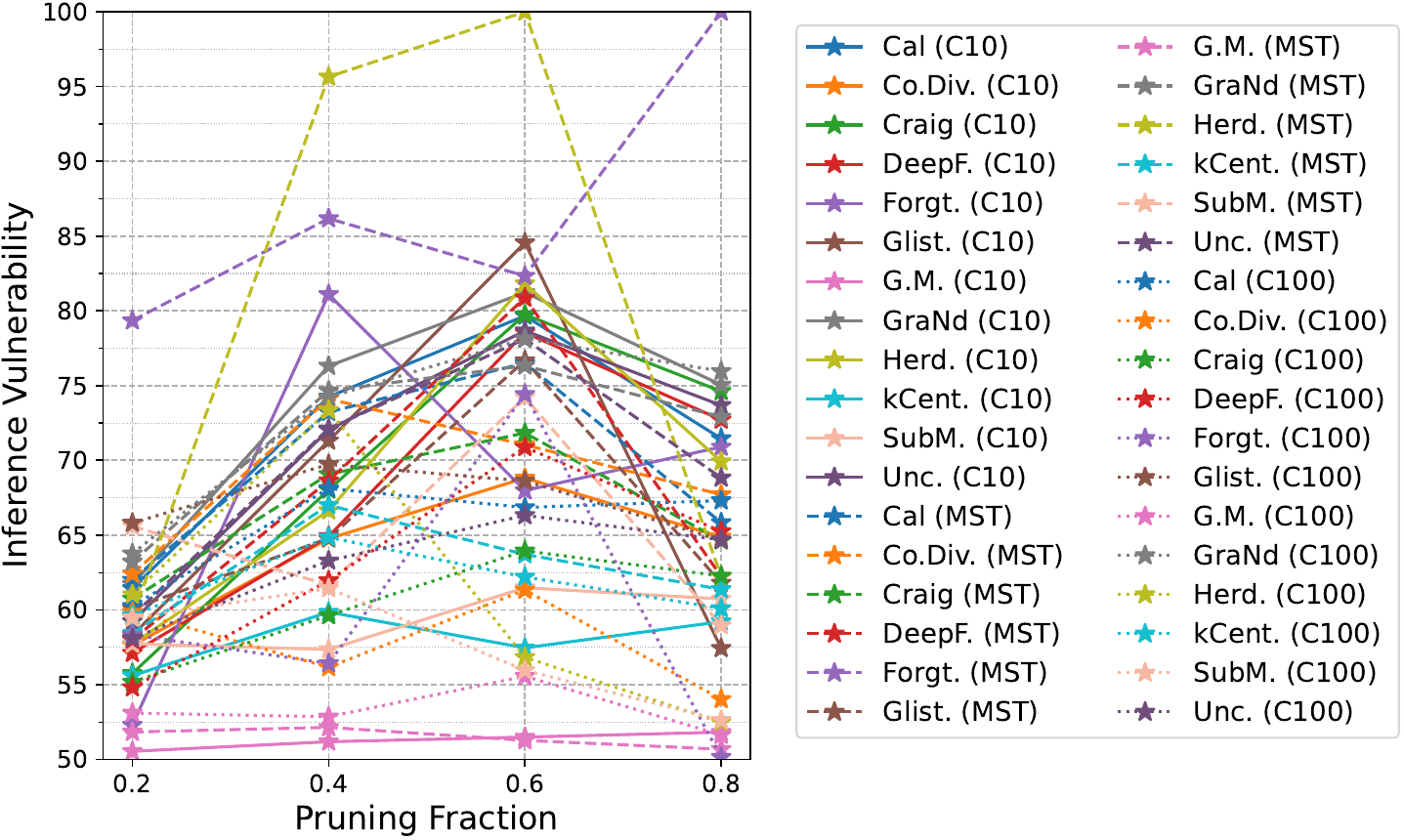}
    \caption{The influence of pruning fraction of a specific pruning method's data lineage inference vulnerability. ArraDis is used here for comparison.}
    \vspace{-3mm}
    \label{fig:fraction}
\end{figure}

\begin{figure*}[h]
    \captionsetup{list=false}
    \centering
    \begin{minipage}[b]{.33\linewidth}
        \centering
        \includegraphics[width=\linewidth]{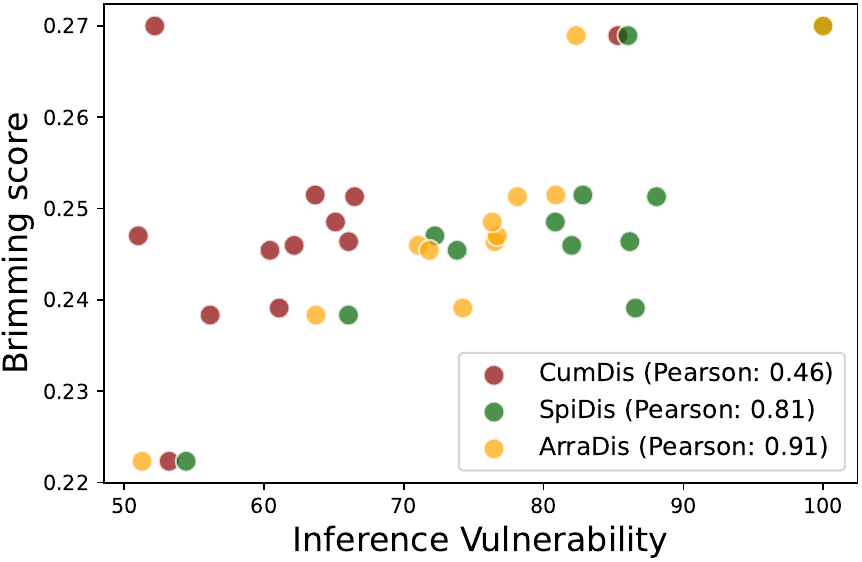}
        \subcaption{MNIST}
    \end{minipage}%
    \hfill 
    \begin{minipage}[b]{.33\linewidth}
        \centering
        \includegraphics[width=\linewidth]{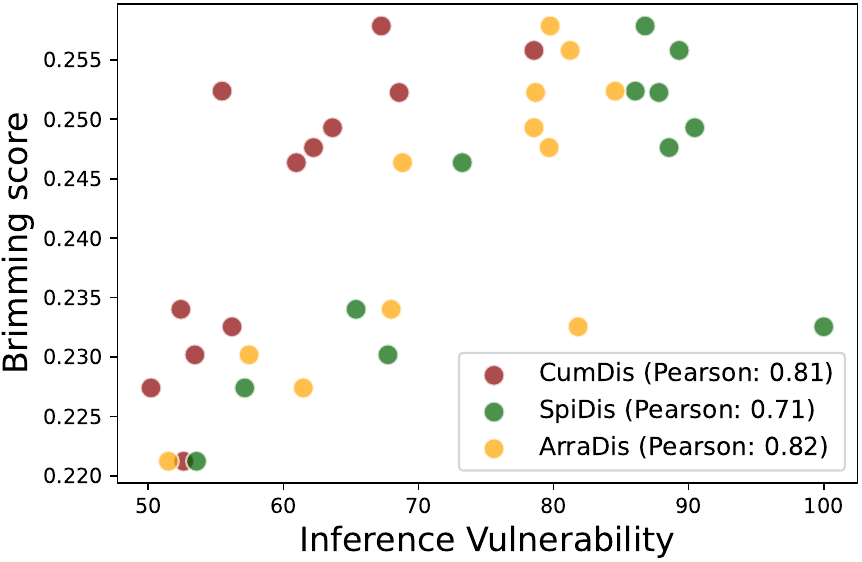}
        \subcaption{CIFAR10}
        
    \end{minipage}%
    \hfill 
    \begin{minipage}[b]{.33\linewidth}
        \centering
        \includegraphics[width=\linewidth]{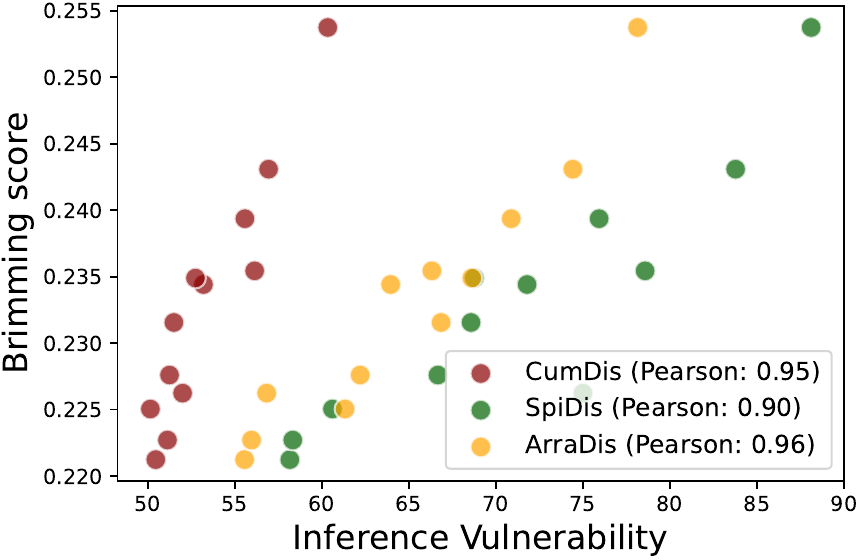}
        \subcaption{CIFAR100}
        
    \end{minipage}
    \vspace{-0.1mm}
    \begin{minipage}[b]{\linewidth}
    \centering
     \small 
    \begin{threeparttable}[t]
        \begin{tabular}{c|c|c|c|c|c|c|c|c|c|c|c|c}
            \hline
            \hline
            Dataset & Cal & Co.Div. & Craig & DeepF. & Forgt. & Glist. & G.M. & GraNd & Herd. & kCent. & SubM. & Unc. \\
            \hline
            MNIST & 0.2463 & 0.2460 & 0.2454 & 0.2515 & \cellcolor{red!25}0.2689 & 0.2470 & \cellcolor{blue!45}0.2223 & 0.2485 & \cellcolor{red!55}0.2700 & \cellcolor{blue!15}0.2383 & 0.2391 & 0.2513 \\
            CIFAR10 & 0.2476 & 0.2464 & \cellcolor{red!25}0.2558 & 0.2493 & 0.2340 & 0.2524 & \cellcolor{blue!45}0.2212 & \cellcolor{red!55}0.2578 & 0.2325 & 0.2302 & \cellcolor{blue!15}0.2274 & 0.2523 \\
            CIFAR100 & 0.2316 & 0.2250 & 0.2344 & 0.2394 & \cellcolor{red!25}0.2431 & 0.2349 & \cellcolor{blue!45}0.2212 & \cellcolor{red!55}0.2537 & 0.2262 & 0.2276 & \cellcolor{blue!15}0.2227 & 0.2354 \\
            \hline
            \hline
        \end{tabular}
        \end{threeparttable}
    \subcaption*{(d) The Brimming score of 12 different pruning methods under the pruning fraction of 0.6.}
    \label{table:metrics}
\end{minipage}
    \caption{The Pearson correlation coefficient between inference vulnerability (ASR) 
    and the Brimming score. Pruning fraction is set to 0.6. Three attacks (CumDis, SpiDis, and ArraDis) are used. A strong positive correlation can be observed.} 
    \vspace{-3mm}
    \label{fig:brimming}
\end{figure*}
\noindent {\bf Within the same dataset, the comparison of privacy risks between different pruning methods remains relatively stable across various pruning fractions.}  For instance, with the MNIST dataset, Forgt. consistently exhibits the highest privacy risk across different fractions. While for the CIFAR10 and CIFAR100 datasets, GraNd is generally the method with the highest privacy risk under different fractions. For all three datasets, G.M. is the safest method. This suggests that \textit{the privacy risk of dataset pruning is largely due to the interaction between the intrinsic characteristics of the dataset and the inherent properties of the pruning method}. Since this paper primarily focuses on the privacy risks introduced by dataset pruning, we consider exploring this intrinsic relationship as future work.

\noindent {\bf Within the same method, the degree of privacy risks varies under different pruning fractions. } To make it more clear, we present the results across different datasets and pruning fractions in Figure \ref{fig:fraction}. We observe that most methods experience the highest privacy risk at a pruning fraction of 0.6. The risks at fractions of 0.4 and 0.8 are similar and both lower than at 0.6, while the privacy risk is minimal at a fraction of 0.2. From a practical standpoint, to better balance efficiency and utility, an appropriate pruning fraction (e.g., around 0.6) is often chosen for the service since a higher fraction may not significantly improve efficiency, while a lower fraction may drastically reduce utility. This further underscores the threat that dataset pruning poses to data privacy. Therefore, finding a pruning method that can effectively protect data privacy at commonly used pruning fractions is an urgent and important issue. 

\vspace{-3mm}
\subsubsection{Brimming Score}
Given the trends and phenomena observed regarding different pruning methods, we can clearly pose the following question: Can we propose a unified single metric to measure the privacy risk of different methods? 
To address this issue, which is consistent with \textit{KeyIQ3}, we propose a metric named the Brimming score. The design and inspiration for this metric come from the results shown in Figure \ref{fig:evidence}(c) and Figure \ref{fig:evidence}(d). As previously analyzed (see Section \ref{evidence}), the deeper the blue section rises from the bottom, the more severe the privacy risk, visually analogous to the level of water brimming in a glass. The Brimming score can be seen as the average ASR across all intervals. It offers a fine-grained assessment of the overall susceptibility of a pruning method to being compromised. The formal definition of the Brimming score is as follows: 
\begin{equation}
\mathcal{B}_{\textrm{s}} = \frac{1}{\eta}
\sum_{p=0}^{\upsilon-1} \sum_{q=p+1}^{\upsilon}
R_{\textrm{cdr}}(p,q),
\end{equation}
where \(\upsilon=\frac{|Q_{\textrm{v}}|}{\zeta_{\textrm{v}}}\), \(R_{\textrm{cdr}}(\cdot,\cdot) \) is the bivariate CDR of the victim datapool, \(\eta = \frac{1}{2} \upsilon  \left( \upsilon + 1 \right)\) is used to average the weight of each interval (derivation details are in Appendix \ref{brimming}).

\begin{figure}[h]
    \centering
    \begin{minipage}[b]{.49\linewidth}
        \centering
        \includegraphics[width=\linewidth]{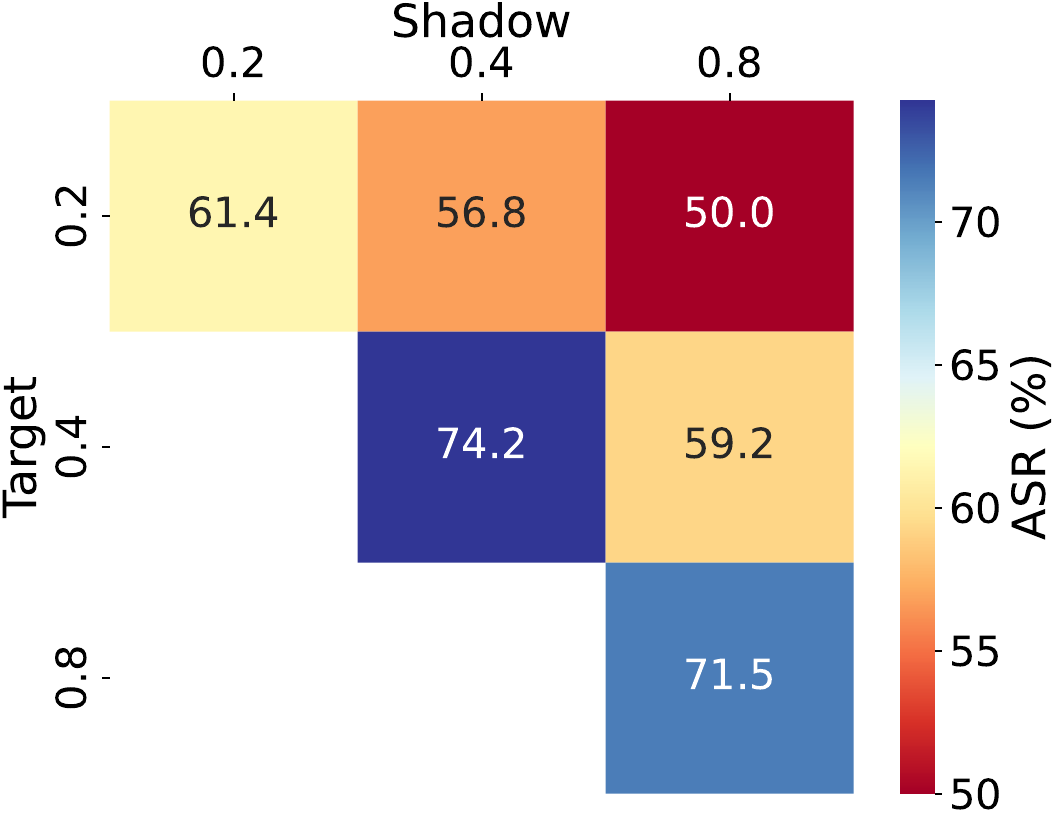}
        \subcaption{CIFAR10-Cal}
        
    \end{minipage}%
    \hfill
    \begin{minipage}[b]{.49\linewidth}
        \centering
        \includegraphics[width=\linewidth]{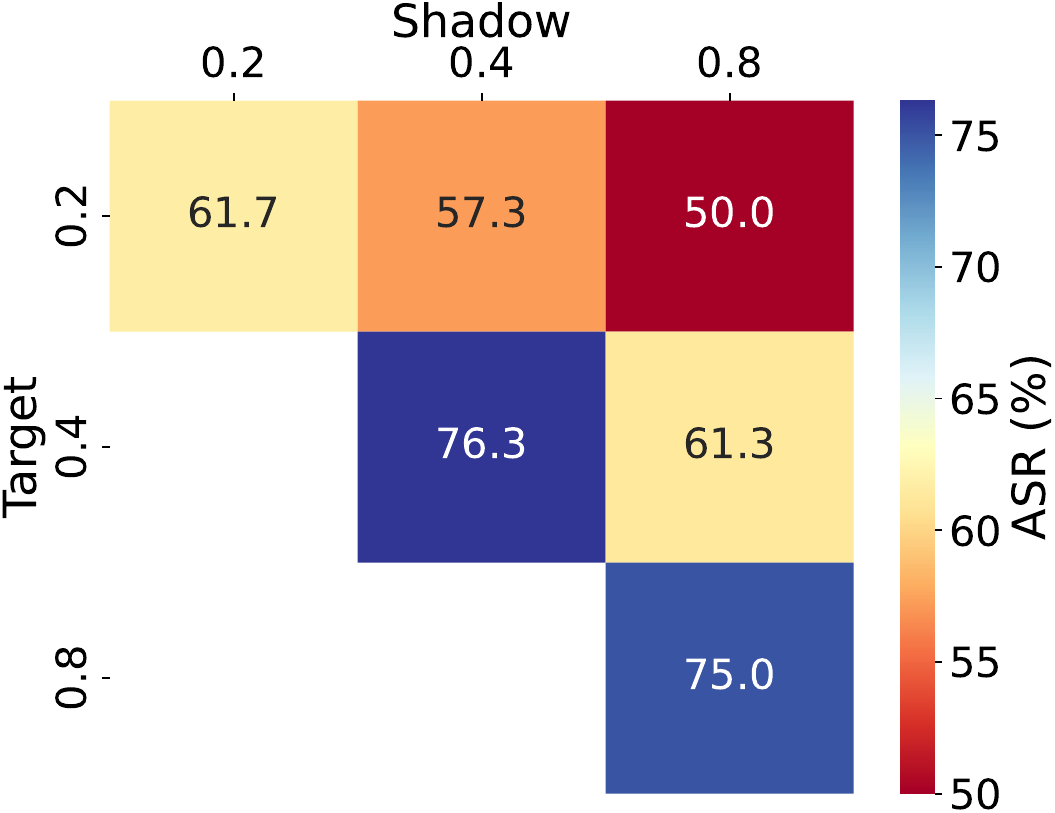}
        \subcaption{CIFAR10-GraNd}
        
    \end{minipage}
    \caption{ASR without knowing the pruning fraction (x-axis: shadow fraction, y-axis: ground truth fraction).}
    \vspace{-3mm}
    \label{fig:attack2}
\end{figure}

Next, experiments are designed to verify the effectiveness of the Brimming score, specifically whether it shows a strong positive correlation with the outcomes of different attacks. The results are shown in Figure \ref{fig:brimming}. We calculate the score for different datasets and pruning methods at a pruning fraction of 0.6 and compute the Pearson correlation coefficient between the score and the ASR of the three different attacks. As seen in Figure \ref{fig:brimming}, the Brimming score exhibits a strong positive correlation with almost all results, effectively representing the privacy risk of different pruning methods. Some specific numerical results are shown in Figure \ref{fig:brimming}(d). Similar to Table \ref{tab:main1}, we marked the methods with the most, second most severe privacy risk and the safest, second safest methods using different colors. It can be observed that the comparative relationships between different methods are well-preserved and consistent with the attack success rates. In Appendix \ref{usecase}, we also provide a use case of the Brimming score to show its effectiveness in the evaluation of different pruning methods across various setups.

\subsection{Attack under Partial Knowledge}
\label{partial_know}
In this section, we consider four other realistic scenarios where the adversary's knowledge of the pruning process is somewhat incomplete, consistent with \textit{KeyIQ4}.

\subsubsection{Unknown Pruning Fraction}
\label{un_frac}
In this scenario, we first conduct attacks in a simple manner: the adversary simply assumes a fraction from a set of potential values. Later, we will show that the adversary can estimate a fraction nearly identical to the ground-truth fraction using the mark-recapture method~\cite{southwood2009ecological,chao2001applications}, indicating that knowledge of the fraction is not necessary for conducting inference.

\noindent\textbf{Simply Guessing.}
In this experiment, we considered three different fraction settings: 0.2, 0.4, and 0.8. The adversary may assume any of these fractions for their pruning process (both on the shadow and victim datapool). We used three pruning methods (Cal, Glister, and GraNd) on CIFAR10. Results are shown in Figure \ref{fig:attack2} (results for other datasets are in Appendix \ref{remaining_results}). From the results, the effectiveness of DaLI sees different degrees of decline under various settings. Some potential patterns are evident: when the difference between the ground truth fraction and the fraction assumed by the adversary is small (e.g., the ground truth fraction is 0.2, and the adversary assumes a fraction of 0.4), considerable attack performance is still observed. However, when the difference is large, the inference becomes less effective. This indicates that protecting the information about the pruning fraction can somewhat safeguard data privacy. However, this protection is still unreliable. 
For instance, the adversary first views one of its shadow datapool as a victim datapool and evaluates the performance using other shadow datapools across various fraction combinations. The adversary can find that assuming a relatively smaller pruning fraction often yields better results. Leveraging this insight as prior knowledge, the adversary can enhance the effectiveness of the actual attack. 

\begin{figure}[h]
    \centering
        \includegraphics[width=0.8\linewidth]{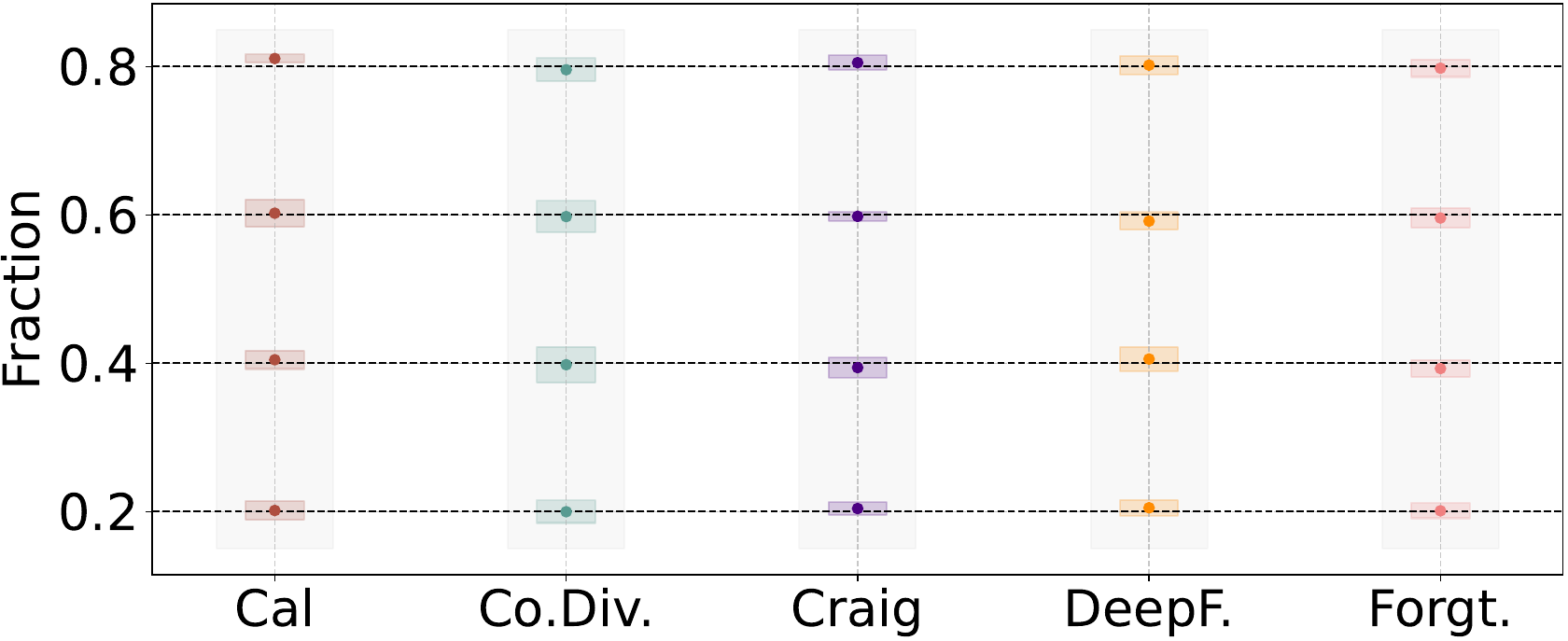}
    \caption{The estimated fraction via mark-recapture.}
    \vspace{-3mm}
    \label{fig:recap}
\end{figure}

\noindent\textbf{Mark-Recapture. } The mark-recapture method, or capture-release method, is a statistical approach widely used in demography and biology to estimate population sizes~\cite{krebs1989ecological}. In our scenario, the adversary can manually mark a certain amount of data before data collection. After the selected set is released, the adversary counts the number of marked data points within it and calculates their proportion relative to the total marked data. This serves as the estimated pruning fraction. Figure \ref{fig:recap} presents the estimated results. We observe that the estimated fraction is nearly identical to the ground-truth value, with generally mild fluctuations. This demonstrates the effectiveness of mark-recapture in estimating the fraction.

\begin{table}[h!]
\centering
 \small 
    \begin{threeparttable}[t]
\begin{adjustbox}{max width=\columnwidth}
\begin{tabular}{cccccc}
\toprule
\toprule
\textbf{Target} & \textbf{Shadow} & \textbf{ASR Drop} & \textbf{Target} & \textbf{Shadow} & \textbf{ASR Drop} \\
\midrule
 Cal-0.2 & GraNd-0.4 & 7.65\% & Cal-0.4 & GraNd-0.8 & 17.02\% \\
\cmidrule(r){1-3} \cmidrule(l){4-6}
Craig-0.2 & Forgt.-0.4 & 5.14\% & Craig-0.4 & Glist.-0.8 & 19.27\% \\
\cmidrule(r){1-3} \cmidrule(l){4-6}
Glist.-0.2 & Cal-0.4 & 6.36\% & Glist.-0.4 & Forgt.-0.8 & 13.25\% \\
\cmidrule(r){1-3} \cmidrule(l){4-6}
GraNd-0.2 & Craig-0.4 & 7.26\% & GraNd-0.4 & Craig-0.8 & 18.41\% \\
\bottomrule
\bottomrule
\end{tabular}
\end{adjustbox}
\caption{ASR Drop for adversaries without knowing the pruning method and the pruning fraction.}
\vspace{-4mm}
\label{fig:asr}
\end{threeparttable}
\end{table}

\subsubsection{Unknown Pruning Method} 
In this scenario, the adversary can only choose one pruning method from some potential methods to carry out the inference. We use Cal and Glister as the ground truth pruning methods (used by the service provider) and use the other three pruning methods (Co.Div., GraNd, kCent.) as the shadow pruning methods employed by the adversary. Experiments were conducted on three datasets with four different pruning fractions (Frac. in short). Results under CIFAR10 are shown in Figure \ref{fig:unknown_method} (results for other datasets are given in Appendix \ref{remaining_results}). The results indicate that the impact of the pruning method is seemingly less significant than that of the pruning fraction. In most cases, the adversary can still exhibit extraordinary attack capabilities (with a success rate exceeding 90\%). Additionally, overall, the privacy risk is most severe when the pruning fraction is 0.6, which is consistent with the previous conclusion. Also, the results vary with different combinations of ground truth and shadow pruning methods. This further demonstrates the interactions between different pruning methods, suggesting some underlying intrinsic connections among various pruning methods.

\begin{figure}[ht]
    \centering
    \begin{minipage}[b]{.49\linewidth}
        \centering
        \includegraphics[width=\linewidth]{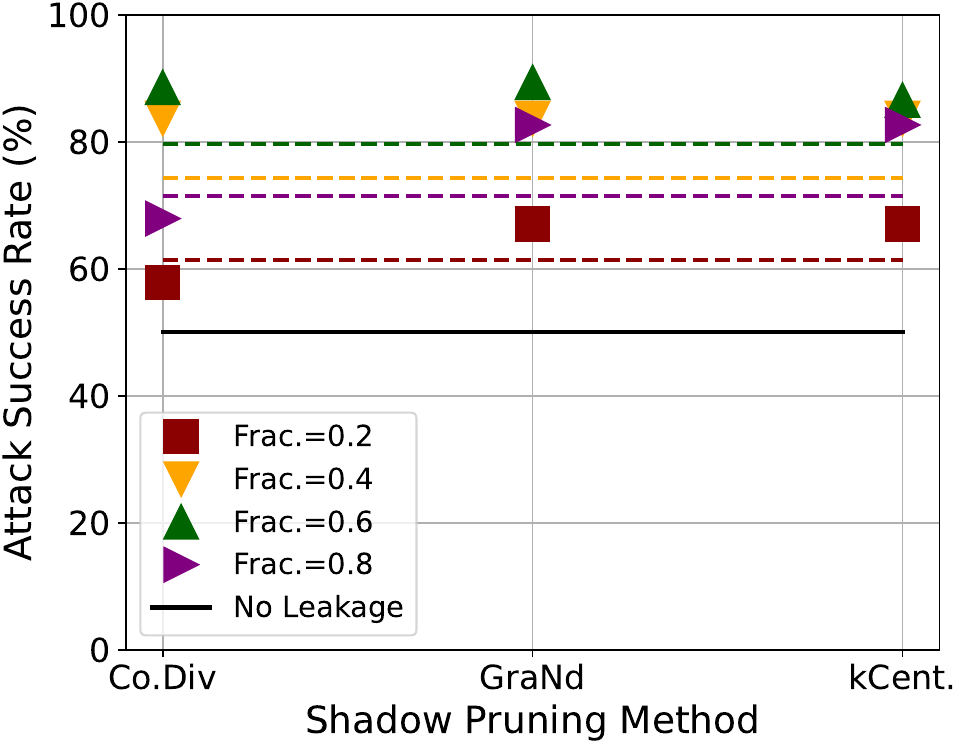}
        \subcaption{CIFAR10-Cal}
        
    \end{minipage}%
    \hfill
    \begin{minipage}[b]{.49\linewidth}
        \centering
        \includegraphics[width=\linewidth]{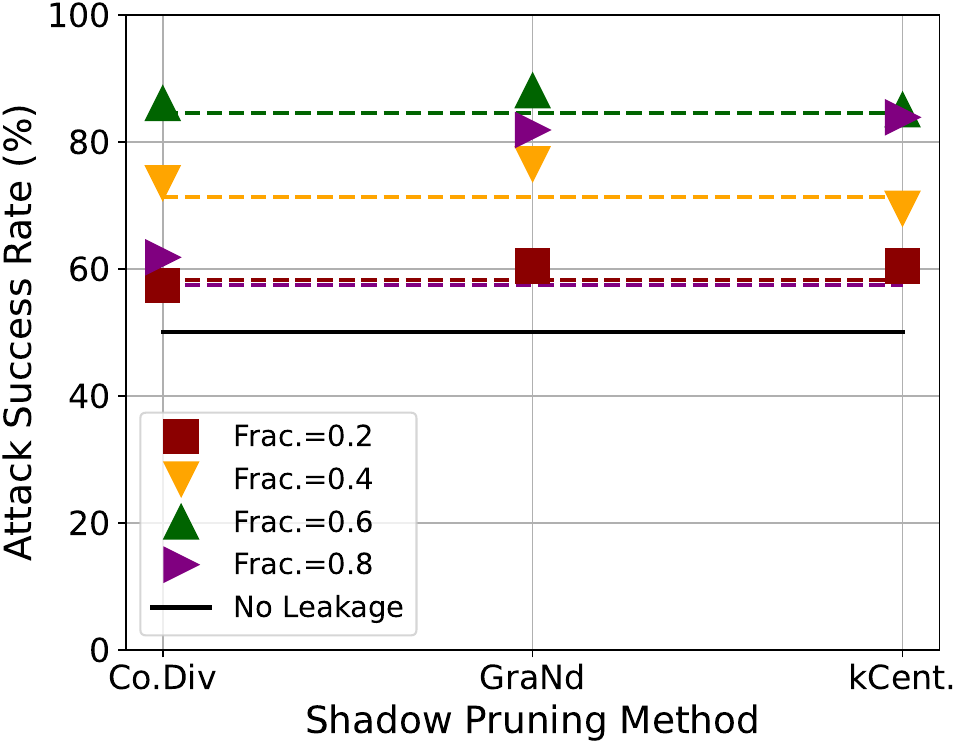}
        \subcaption{CIFAR10-Glist.}
        
    \end{minipage}%
    \caption{Attack success rate (ArraDis) when the pruning method is unknown to the adversary. The colored dashed lines are the results when the pruning method is known.}
    \vspace{-5mm}
    \label{fig:unknown_method}
\end{figure}

\begin{table*}[t!]
    \centering
    \small 
    \begin{threeparttable}[t]
        \begin{tabular}{c|c|c|c|c|c|c|c|c|c|c|c|c}
            \hline
            \hline
            Pruning Method & Cal & Co.Div. & Craig & DeepF. & Forgt. & Glist. & G.M. & GraNd & Herd. & kCent. & SubM. & Unc. \\
            \hline
            w/o. ReDoMi & 59.89 & 50.88 & 54.50 & 58.61 & 56.70 & 56.46 & 50.92 & 71.23 & 65.54 & 53.34 & 51.05 & 59.85 \\
            w. ReDoMi & 50.98 & 50.15 & 51.76 & 50.22 & 55.48 & 52.75 & 51.08 & 56.64 & 50.88 & 50.16 & 50.45 & 50.15 \\
            \hline
            \hline
        \end{tabular}
    \end{threeparttable}
    \caption{Attack success rate (WhoDis) of 12 different pruning methods with (w.) and without (w/o.) using ReDoMi.}
    \vspace{-4mm}
    \label{tab:defense}
\end{table*}

\subsubsection{Unknown both Pruning Method and Pruning Fraction}
As discussed in Section \ref{un_frac}, the pruning fraction can be easily estimated via mark-recapture. Thus, the scenario where both the pruning method and fraction are unknown reduces to a scenario where only the pruning method is unknown, allowing the adversary to still conduct effective inference. However, to provide further insights and more comprehensively evaluate the effectiveness of DaLI, we consider the unrealistic scenario: the adversary simply assumes a method and fraction for inference. In this experiment, without loss of generality, we consider four methods for the victim datapool (Cal, Craig, Glist., GraNd). Two ground truth fractions (0.2, 0.4) and adversary assumed fractions (0.4, 0.8) are considered here. In this experiment, the combinations of methods and fractions are fixed for fair comparison. In Table \ref{fig:asr}, we represent the combination of method and fraction as method-fraction (e.g., Cal-0.2).
We calculated the attack success rate drop (ASR Drop) across all experimental setups, which is defined as
\begin{equation}
\text{ASR Drop} = \frac{\text{ASR}_{\text{known}} - \text{ASR}_{\text{unknown}}}{\text{ASR}_{\text{known}}},
\end{equation}
where \(\text{ASR}_{\text{known}}\) is the ASR when both the pruning method and fraction are known, and \(\text{ASR}_{\text{unknown}}\) is the ASR when the adversary does not know the pruning method and fraction.

Table \ref{fig:asr} shows the results under CIFAR10 (results for other datasets are given in Appendix \ref{remaining_results}). We can observe that regardless of the target pruning method and shadow pruning method when the pruning fraction combination is fixed, the overall change in ASR Drop is small. Specifically, when the (target fraction, shadow fraction) is set to (0.2, 0.4), the overall ASR Drop is around 6\%, whereas when set to (0.4, 0.8), the overall ASR Drop is around 17\%. There are occasional instances where the drop is less noticeable or more pronounced. 
These observations suggest that even when both the pruning method and pruning fraction are unknown, an adversary can still successfully execute an inference. Although the inference performance will be weakened, as discussed in Section \ref{un_frac}, to facilitate a more effective attack, the adversary can initially assume a smaller pruning fraction, which often yields better results.

\subsubsection{Inference can still be successful without any shadow datapool}
\label{no_aux}
We further consider a scenario where the adversary does not rely on auxiliary data sharing the same underlying distribution as the victim data to build shadow datapools. This means they cannot use thresholds derived from shadow distributions to infer membership and have access only to the victim occurrence distribution. Recall insights provided in Figures \ref{fig:evidence}(a) and \ref{fig:evidence}(b) and the conclusion in Section \ref{evidence}. Since the pattern is consistent across different datasets and methods, the adversary can easily develop an intuitive method for inference. Specifically, the adversary can simply assume samples above the middle position in the occurrence distribution belong to the redundant set, while samples below the middle position belong to other non-members.
Results using WhoDis on CIFAR10 are shown in Table \ref{fig:no_shadow} (results for other datasets are in Appendix \ref{remaining_results}).

\begin{table}[h]
\centering
 \small 
    \begin{threeparttable}[t]
\begin{adjustbox}{max width=\columnwidth}
\begin{tabular}{ccccc}
\toprule
\toprule
\textbf{Fraction} & \textbf{DeepF.} & \textbf{Glist.} & \textbf{GraNd} & \textbf{Unc.} \\
\midrule
0.6 &  56.24(-1.14)  & 55.35(-0.73) & 60.24(-2.77)  &  57.89(-1.32)\\
\cmidrule(r){1-5}
0.8 & 57.11(-6.47) & 53.59(-1.23) & 62.32(-7.49) & 56.93(-6.16) \\
\bottomrule
\bottomrule
\end{tabular}
\end{adjustbox}
\end{threeparttable}
\caption{Attack success rate without shadow datapools.}
\vspace{-3mm}
\label{fig:no_shadow}
\end{table}

It can be observed that even without the auxiliary data, the adversary can still design simple schemes for privacy inference based on the pattern found before. This further highlights the privacy risks inherent in the carefully designed process of dataset pruning.

\section{Discussion of Potential Defense Strategies}
Due to the differences between DCMI and traditional model-centric membership inference, defending against DCMI is not straightforward. For DCMI defense, we do not want the participation of any downstream model. Consequently, almost all previously proposed defense techniques that have theoretical guarantees, such as Differentially Private SGD~\cite{abadi2016deep}, are unsuitable for our setup since they are proposed to get downstream models. 
Also, we do not want to add new samples or remove samples from the selected set, as it has been determined through a carefully designed pruning process based on service requests.
Moreover, after implementing a defense strategy, the performance of the downstream model trained with the selected set should not degrade significantly. Therefore, it is preferable not to modify the selected set.

Based on these, we make an initial attempt to defend against DCMI, which is consistent with \textit{KeyIQ5}. We propose a defense strategy named \underline{Red}undant-\underline{O}ther Non-Member \underline{Mi}xing (ReDoMi).
Our inspiration comes from a technique called mixup\cite{zhang2017mixup}, originally used to alleviate model memorization and sensitivity to adversarial examples. Mixup has already been used as a defense strategy in previous works~\cite{chen2021enhanced,li2021membership}. We made appropriate modifications to apply it to the DCMI scenario. Specifically, after the pruning process is completed, the service provider merges the samples from the redundant set with those from other non-members in the pixel space while retaining their original labels. We denote samples from the redundant set as \((x^{\textrm{victim}}_{\textrm{red}},y^{\textrm{victim}}_{\textrm{red}}) \in D^{\textrm{victim}}_{\textrm{red}}\), and those from other non-members as \((x^{\textrm{victim}}_{\textrm{non}},y^{\textrm{victim}}_{\textrm{non}}) \in D^{\textrm{victim}}_{\textrm{non}}\). A formal definition of ReDoMi is as follows:
\begin{equation}
\tilde{x}^{\textrm{victim}}_{\textrm{red}} = \lambda x^{\textrm{victim}}_{\textrm{red}} + (1 - \lambda)x^{\textrm{victim}}_{\textrm{non}},
\end{equation}
\begin{equation}
\tilde{y}^{\textrm{victim}}_{\textrm{red}} = y^{\textrm{victim}}_{\textrm{red}},
\end{equation}
where \(\lambda \in [0,1]\) follows a Beta distribution with both parameters equal to \(\gamma \in 
 (0,\infty)\)~\cite{zhang2017mixup}.

Furthermore, we need to determine the correspondence between each redundant data and other non-members. Specifically, consider each data in the above two sets has its own identifier \( \{ a^{\text{red}}_\textrm{i} \}_{\textrm{i}=1}^\textrm{n} \) and \( \{ a^{\text{non}}_\textrm{j} \}_{\textrm{j}=1}^\textrm{m} \), respectively, where \(\textrm{n}\) is the scale of the redundant set
and \(\textrm{m}\) is the scale of other non-members. We define a pairing function \( f_{\textrm{p}}: \{1, \ldots, \max(\textrm{n}, \textrm{m})\} \to \{ (a^{\text{red}}_\textrm{i}, a^{\text{non}}_\textrm{j}) \} \). For \( \textrm{k} \in \{1, \ldots, \min(\textrm{n}, \textrm{m})\} \), we have:
\begin{equation}
f_{\textrm{p}}(\textrm{k}) = (a^{\text{red}}_\textrm{k}, a^{\text{non}}_\textrm{k}),
\end{equation}
while for \( \textrm{k} \in \{\min(\textrm{n}, \textrm{m}) + 1, \ldots, \max(\textrm{n}, \textrm{m})\} \), we have:
\begin{equation}
f_{\textrm{p}}(\textrm{k}) = \begin{cases} 
(a^{\text{red}}_\textrm{k}, a^{\text{non}}_{((\textrm{k}-1)\bmod\textrm{m})+1}) & \text{if } \textrm{n} > \textrm{m}, \\
(a^{\text{red}}_{((\textrm{k}-1)\bmod\textrm{n})+1}, a^{\text{non}}_\textrm{k}) & \text{if } \textrm{n} < \textrm{m}. \\
\end{cases} 
\end{equation}
This definition ensures that when \( \textrm{n} \neq \textrm{m} \), the larger set's remaining elements will be paired with the starting elements of the smaller set, repeating as necessary. ReDoMi enhances the redundant distribution by performing linear interpolations with other non-members, thereby reducing the difference between the two distributions. Furthermore, since the deterministic correspondence between the redundant set and other non-members is only visible to the service provider, it can easily restore the redundant data, ensuring the continued usability of the mixed data. As shown in Table \ref{tab:defense} (\(\gamma=0.05\), visualization of occurrence distributions under defense can be found in Appendix \ref{remaining_results}), it is evident that ReDoMi successfully brings adversaries obstacles when conducting inference, effectively mitigating the privacy risk. However, it is notable that this defense method is heuristic and lacks theoretical guarantees. Thus, more rigorous approaches are still needed, which is left for future work.

\section{Conclusion}
In this paper, we introduce a novel attack perspective in ML systems called Data-Centric Membership Inference (DCMI) and propose the first attack framework DaLI for this task, which includes four threshold-based attack methods based on the carefully designed occurrence distributions. Extensive experiments demonstrate DaLI's effectiveness under different real world conditions. 
Additionally, we introduce a metric called the Brimming score, which effectively measures the privacy risk of different pruning methods. Finally, we make an initial attempt to defend against DCMI by proposing a defense method named ReDoMi. 

% % use section* for acknowledgment
% \ifCLASSOPTIONcompsoc
%   % The Computer Society usually uses the plural form
%   \section*{Acknowledgments}
% \else
%   % regular IEEE prefers the singular form
% \fi




%\bibliographystyle{IEEEtran}


% \end{thebibliography}

% \bibliographystyle{plain}
\bibliographystyle{IEEEtran}
\bibliography{reference}

\newpage

\appendices
% \begin{appendices}
\section{Pruning Method Taxonomy}
\label{taxonomy}
Without loss of generality, we choose 12 popular dataset pruning methods in our experiments. These methods can be classified into 7 types\cite{guo2022deepcore}:
\begin{itemize}
    \item \textit{Decision boundary based methods.} Methods of this type assume that data near the decision boundary are harder to separate and, thus, can be chosen for the selected set for model training. DeepFool (DeepF.)~\cite{ducoffe2018adversarial} and Contrastive Active Learning (Cal)~\cite{margatina2021active} are used in our experiments.
    \item \textit{Bi-level optimization based methods.} Methods of this type consider the selection of \(D_{\textrm{sel}}\) as the outer objective and the optimization of model parameters on \(D_{\textrm{sel}}\) as the inner objective, among which Glister (Glist.)~\cite{killamsetty2021glister} is used in our paper.
    \item \textit{Error based methods.} Methods of this type assume that samples with more contributions on loss or gradient during model training are of more importance. Forgetting (Forgt.)~\cite{toneva2018empirical} and GraNd~\cite{paul2021deep} are used in our work.
    \item \textit{Uncertainty based method.} This type of method assumes that samples with lower confidence are more important in model training. We choose Least Confidence (LC)~\cite{coleman2020selection} to represent this type and use the method type (Unc.) to refer to it.
    \item \textit{Gradient matching based methods.}  This type of method expect the gradients produced by \(D_{\textrm{tra}}\) can be replaced by those produced by \(D_{\textrm{sel}}\) with minimal difference. We use two methods of this type, named Craig~\cite{mirzasoleiman2020coresets} and GradMatch (G.M.)~\cite{killamsetty2021grad}.
    \item \textit{Submodularity based methods.} A submodular function\cite{iyer2013submodular} is a set function whose value has the property of diminishing returns, meaning the incremental gain from adding an element to a set decreases as the set becomes larger, thus can be naturally used in dataset pruning. We use Facility Location (FL)~\cite{iyer2013submodular} of this type and use the method type (SubM.) to refer to it.
    \item \textit{Geometry based methods.} This type of method assumes that samples with lower confidence are more important in model training. We choose Contextual Diversity (Co.Div.)~\cite{agarwal2020contextual}, Herding (Herd.)~\cite{welling2009herding} and kCenterGreedy (kCent.)~\cite{farahani2009facility} to represent this type.
\end{itemize}

\section{Details of Datasets}
\label{dataset}
\noindent\textbf{MNIST. }This dataset comprises a set of 70,000 grayscale images of handwritten digits, each sized 28\(\times\)28 pixels. These images are categorized into 10 classes, representing digits from 0 to 9, with each class having 7,000 images. The dataset is divided into 60,000 training images and 10,000 test images.

\noindent\textbf{CIFAR10. }This dataset contains a diverse set of 60,000 small, 32\(\times\)32 pixel color images categorized into 10 classes, each represented by 6,000 images. It is organized into 50,000 training images
and 10,000 test images.

\noindent\textbf{CIFAR100. }Similar to CIFAR10, this dataset contains a diverse set of 60,000 small, 32\(\times\)32 pixel color images, while images in this dataset are categorized into 100 classes. It is organized into 50,000 training images and 10,000 test images.

Without loss of generality, we randomly select 20,000 samples from each original dataset to serve as the adversary's auxiliary data, while the remaining samples are designated as the service provider's data. Both of them are then randomly split into two equal parts: one half is used as a candidate for dataset pruning, and the other half is used as other non-members. Each shadow pruning process is conducted on 8,000 randomly sampled data from the candidate samples. 

\section{Hyperparameter Setup}
\label{hyper_app}
Unless otherwise specified, 32 shadow datapools are used, the victim batch size $\zeta_{\textrm{v}}$ is set to 2,000 for CIFAR10 and CIFAR100, and to 2,500 for MNIST. The shadow victim batch size $\zeta_{\textrm{s}}$ is set to 800. Relative ablation studies on the batch size scale can be found in Appendix \ref{batch_size}. We evaluate the inference efficacy using the attack success rate (ASR), which measures the proportion of data in a target dataset for which we successfully inferred membership. Unless otherwise stated, all experiments were repeated three times. 

\begin{figure}
    \centering
        \includegraphics[width=\linewidth]{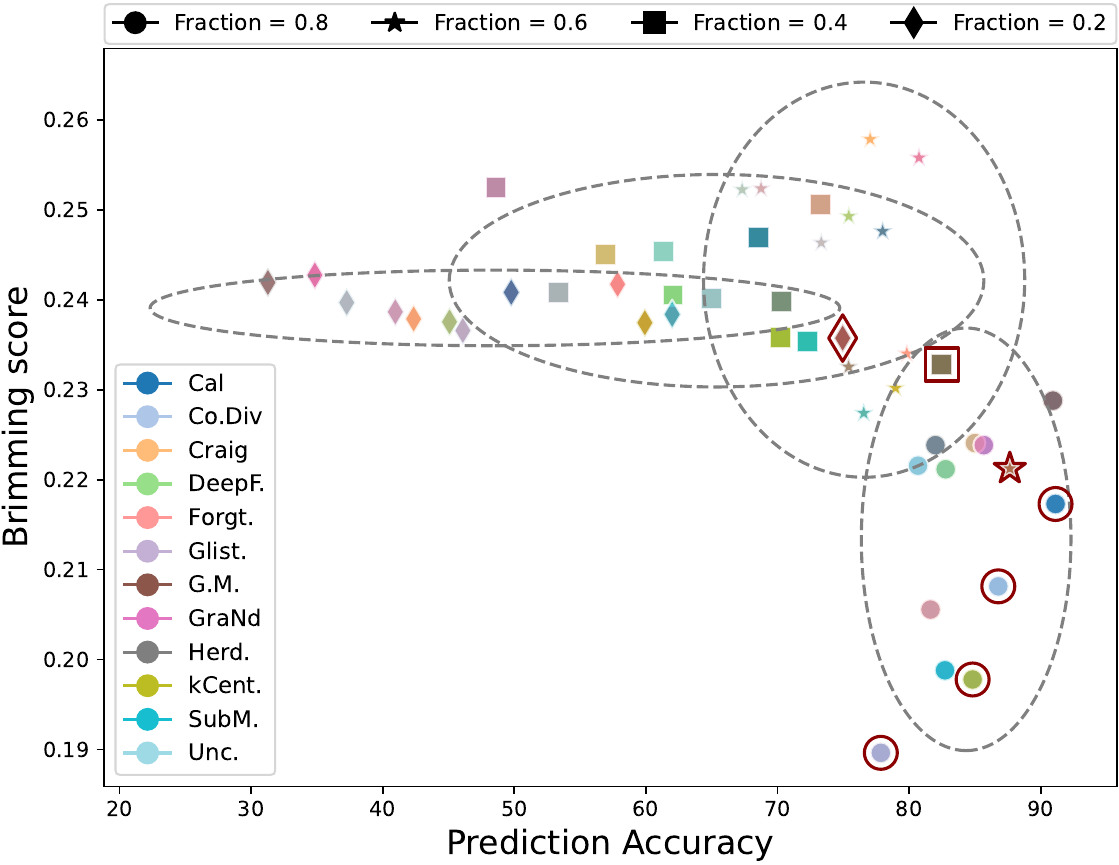}
    \vspace{-4mm}
    \caption{The efficacy that model gains by training on the selected dataset produced by different pruning method v.s. the data lineage inference vulnerability of different pruning method. The Perato Front under different fractions are marked with the corresponding hollow pattern.}
    \vspace{-5mm}
    \label{fig:guide}
\end{figure}

\section{Definition of CDF and CCDF}
\label{CDF_def}
\begin{definition}[CDF and CCDF of Shadow Datapool] 
Given a shadow datapool and the occurrence distribution of its redundant set \(\mathbb{S}_{\textrm{red}}(\textrm{i})(\cdot)\) and other non-members \(\mathbb{S}_{\textrm{non}}(\textrm{i})(\cdot)\), the cumulative distribution of them are functions \(F^{\textrm{cdf}}_{\textrm{red}}(\textrm{i})(\cdot)\) and \(F^{\textrm{cdf}}_{\textrm{non}}(\textrm{i})(\cdot)\), respectively: 
\begin{equation}
\begin{aligned}
&F^{\textrm{cdf}}_{\textrm{red}}(\textrm{i})(t) = \sum_{t' \leq t} \mathbb{S}^{\textrm{shadow}}_{\textrm{red}}(\textrm{i})(t') \\
&= \sum_{t' \leq t} \sum_{s=1}^{|Q_{\textrm{s}}|} \mathbb{I}(x_s \in D^{\textrm{shadow}}_{\textrm{red}}(\textrm{i}))\mathbb{I}(t_s = t'), 
\end{aligned}
\label{cdf_red}
\end{equation}
\begin{equation}
\begin{aligned}
&F^{\textrm{cdf}}_{\textrm{non}}(\textrm{i})(t) = \sum_{t' \leq t} \mathbb{S}^{\textrm{shadow}}_{\textrm{non}}(\textrm{i})(t') \\
&= \sum_{t' \leq t} \sum_{s=1}^{|Q_{\textrm{s}}|} \mathbb{I}(x_s \in D^{\textrm{shadow}}_{\textrm{non}}(\textrm{i}))\mathbb{I}(t_s = t'). 
\label{cdf_non}
\end{aligned}
\end{equation}
\begin{figure*}[ht]
    \centering
    \begin{minipage}[b]{.32\linewidth}
        \centering
\includegraphics[width=\linewidth]{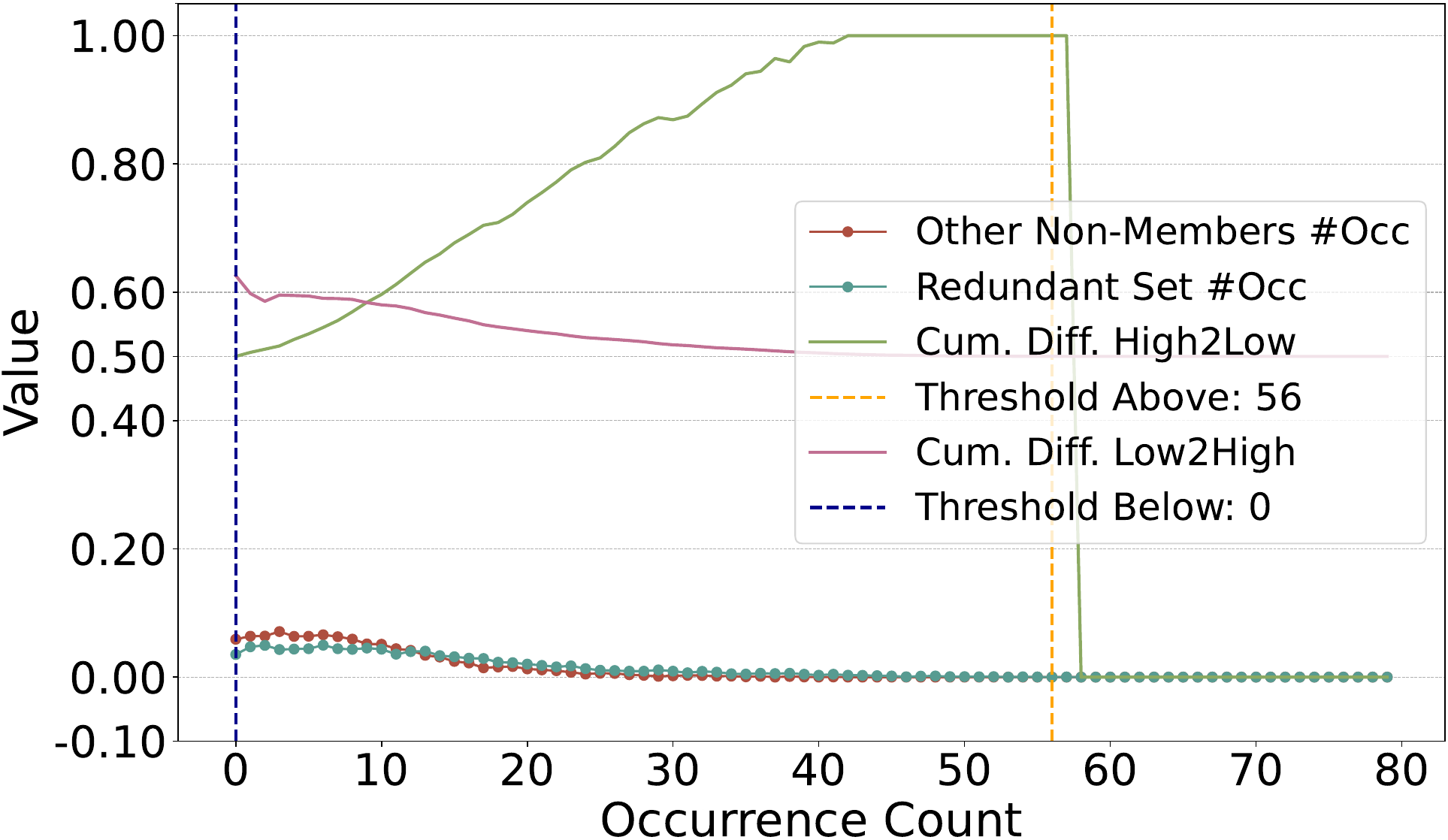}
        \subcaption{CumDis-Cal}
    \end{minipage}%
    \hfill 
    \begin{minipage}[b]{.32\linewidth}
        \centering
        \includegraphics[width=\linewidth]{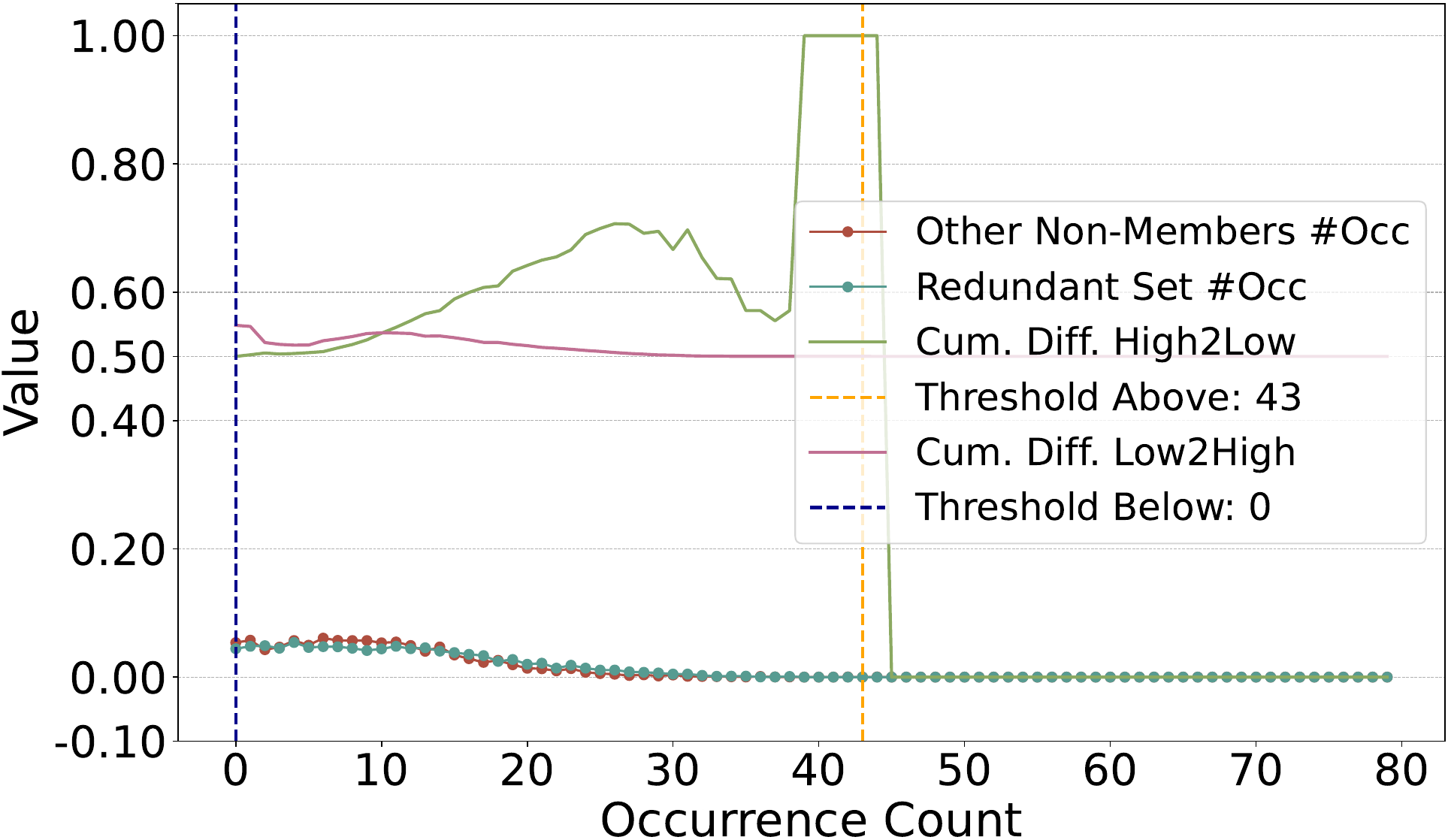}
        \subcaption{CumDis-Co.Div.}
    \end{minipage}%
    \hfill 
    \begin{minipage}[b]{.32\linewidth}
        \centering
        \includegraphics[width=\linewidth]{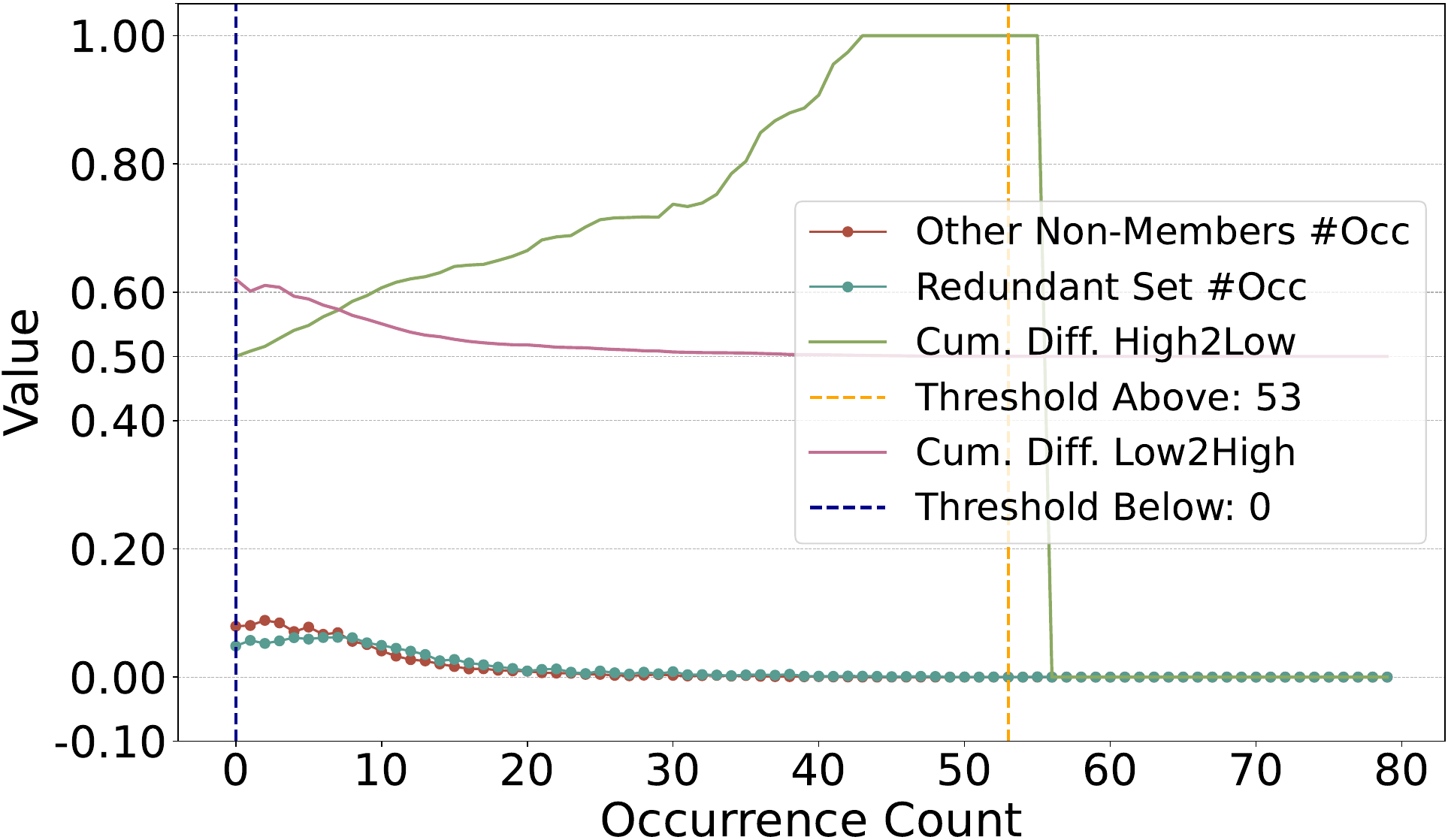}
        \subcaption{CumDis-Craig}
        
    \end{minipage}
    \hfill
    \begin{minipage}[b]{.32\linewidth}
        \centering
        \includegraphics[width=\linewidth]{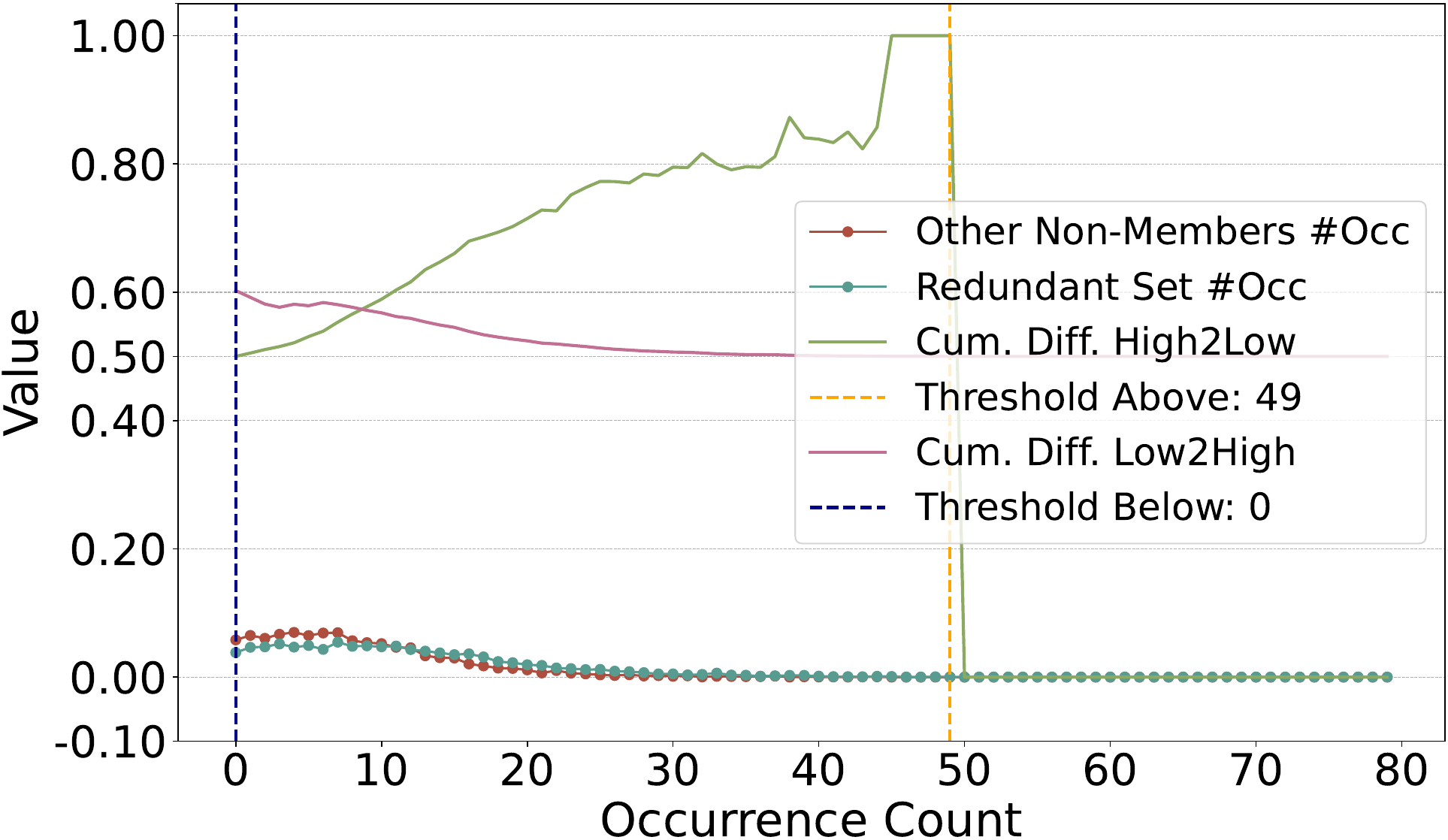}
        \subcaption{CumDis-DeepF.}
    \end{minipage}
    \hfill
    \begin{minipage}[b]{.32\linewidth}
        \centering
        \includegraphics[width=\linewidth]{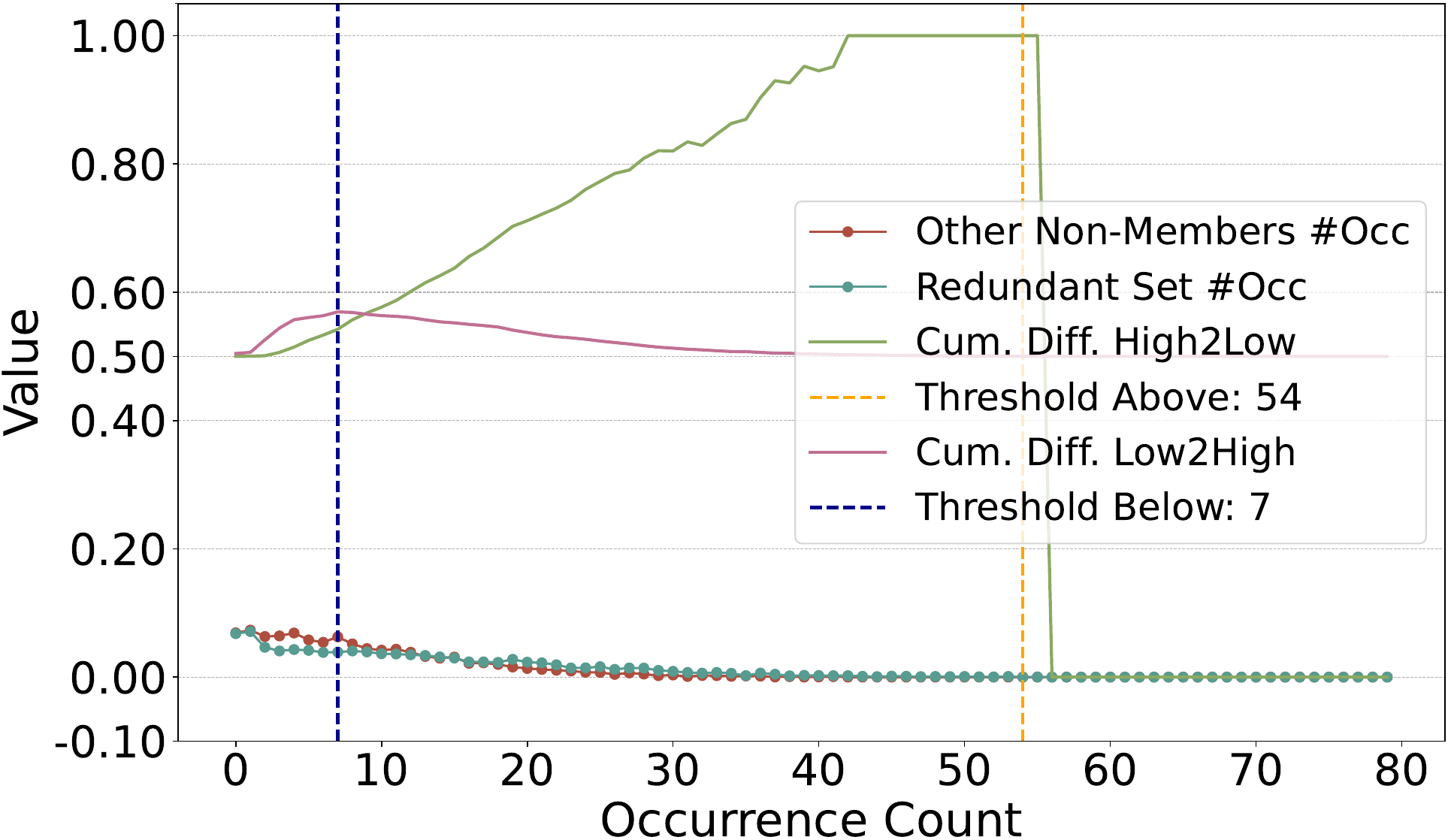}
        \subcaption{CumDis-Forgt.}
        
    \end{minipage}
    \hfill
    \begin{minipage}[b]{.32\linewidth}
        \centering
        \includegraphics[width=\linewidth]{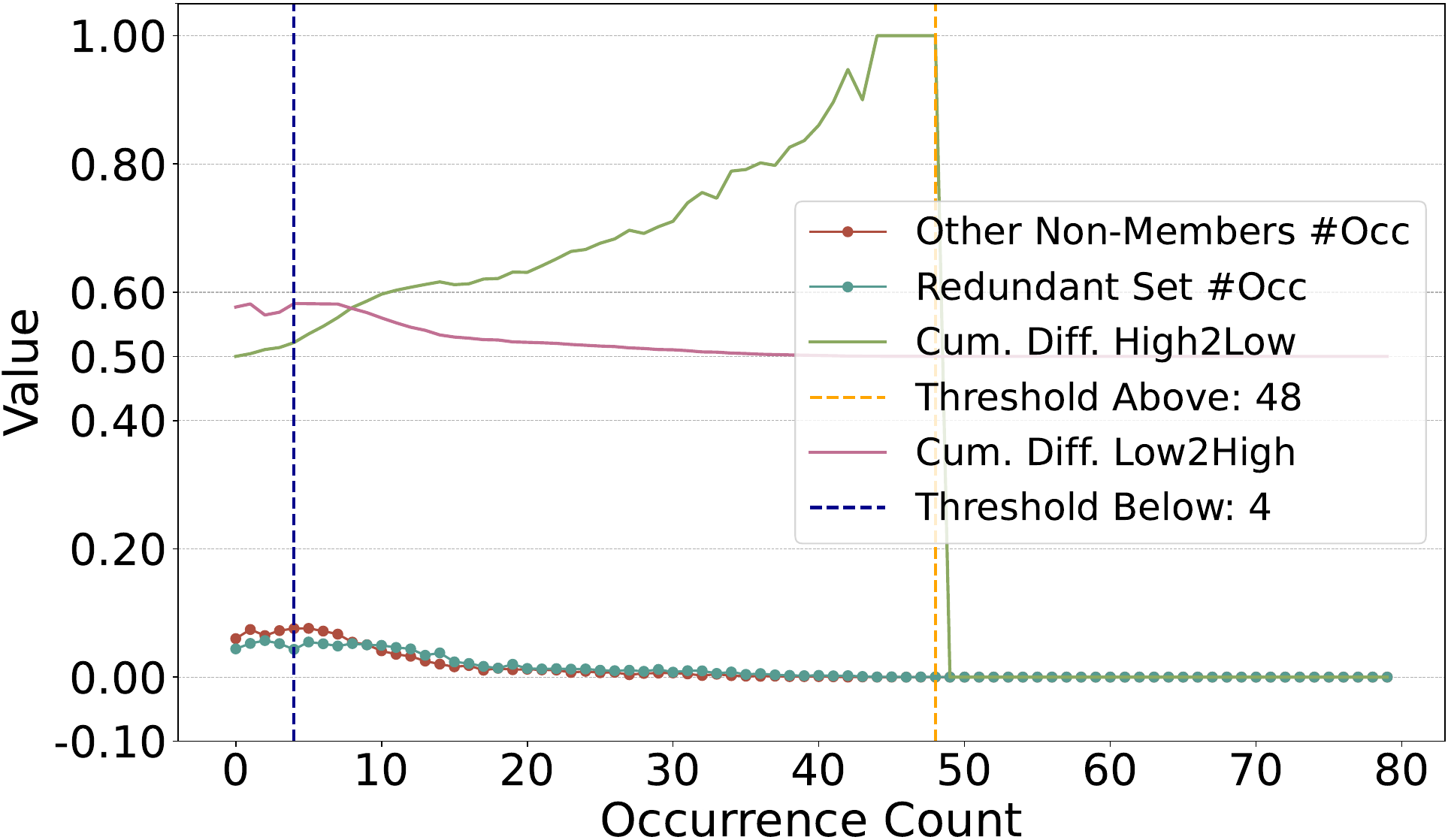}
        \subcaption{CumDis-Glist.}
        
    \end{minipage}
    \hfill
    \begin{minipage}[b]{.32\linewidth}
        \centering
        \includegraphics[width=\linewidth]{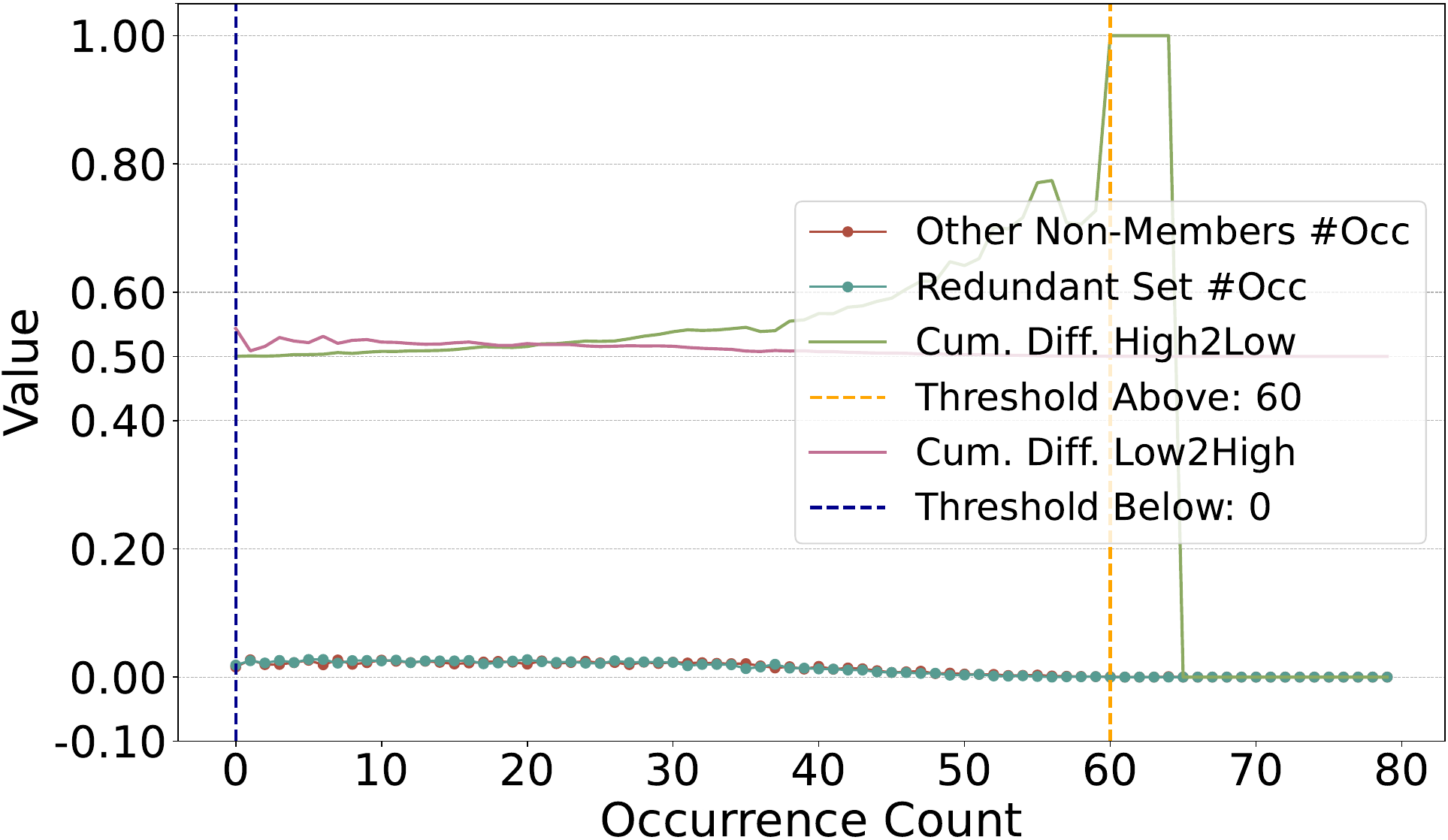}
        \subcaption{CumDis-G.M.}
        
    \end{minipage}
    \hfill
    \begin{minipage}[b]{.32\linewidth}
        \centering
        \includegraphics[width=\linewidth]{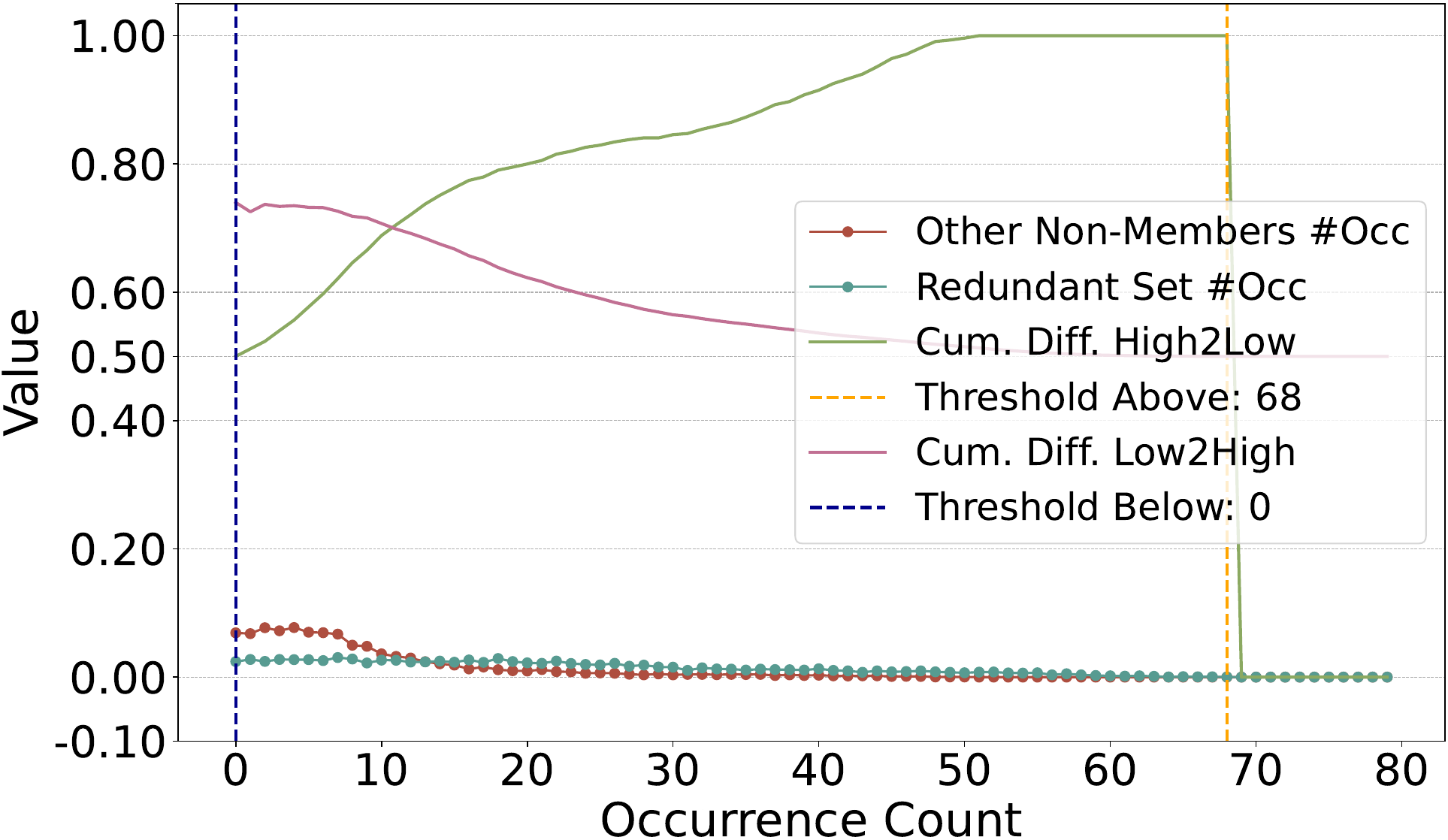}
        \subcaption{CumDis-GraNd}
    \end{minipage}
    \hfill
    \begin{minipage}[b]{.32\linewidth}
        \centering
        \includegraphics[width=\linewidth]{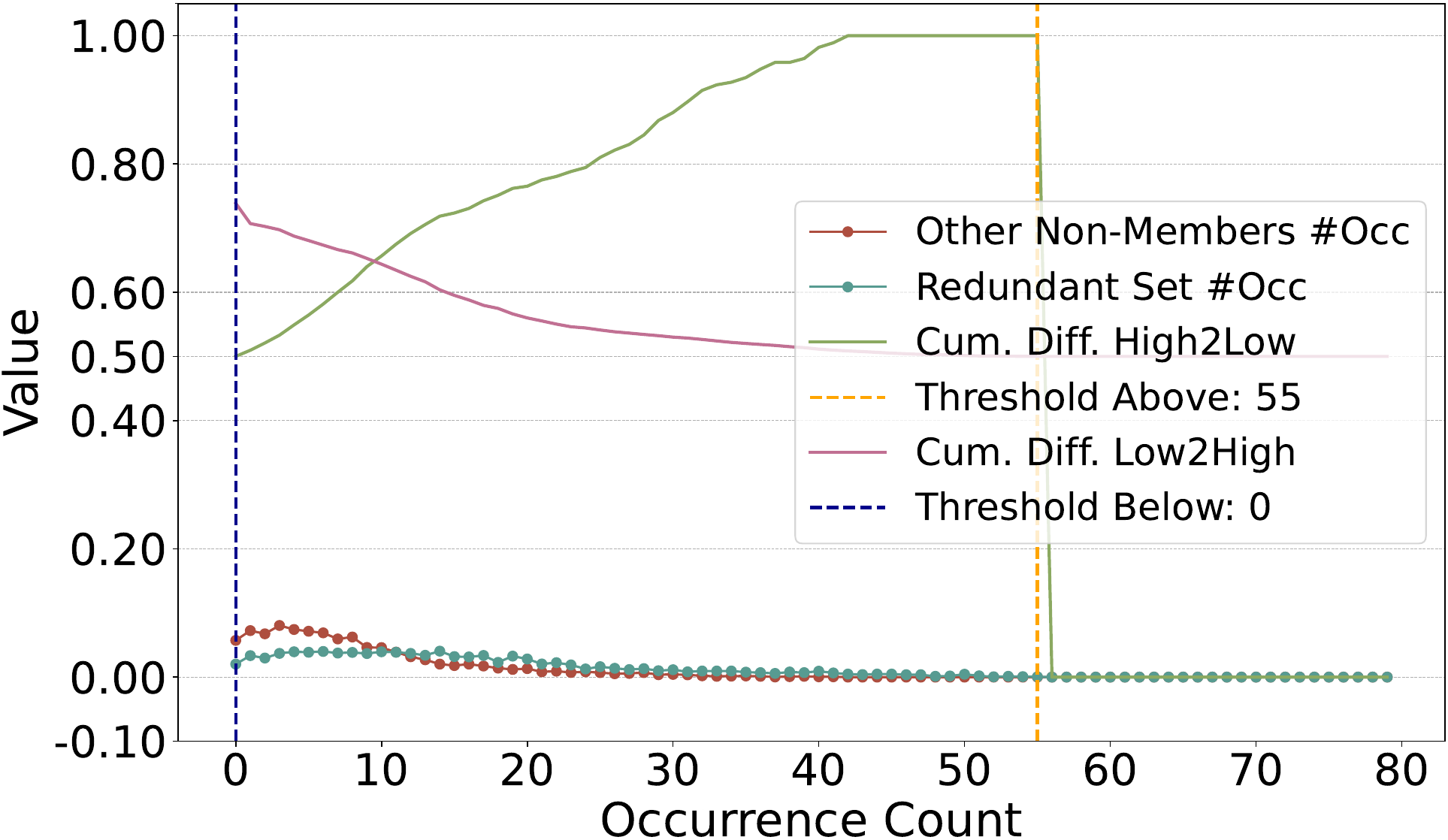}
        \subcaption{CumDis-Herd.}
        
    \end{minipage}
    \hfill
    \begin{minipage}[b]{.32\linewidth}
        \centering
        \includegraphics[width=\linewidth]{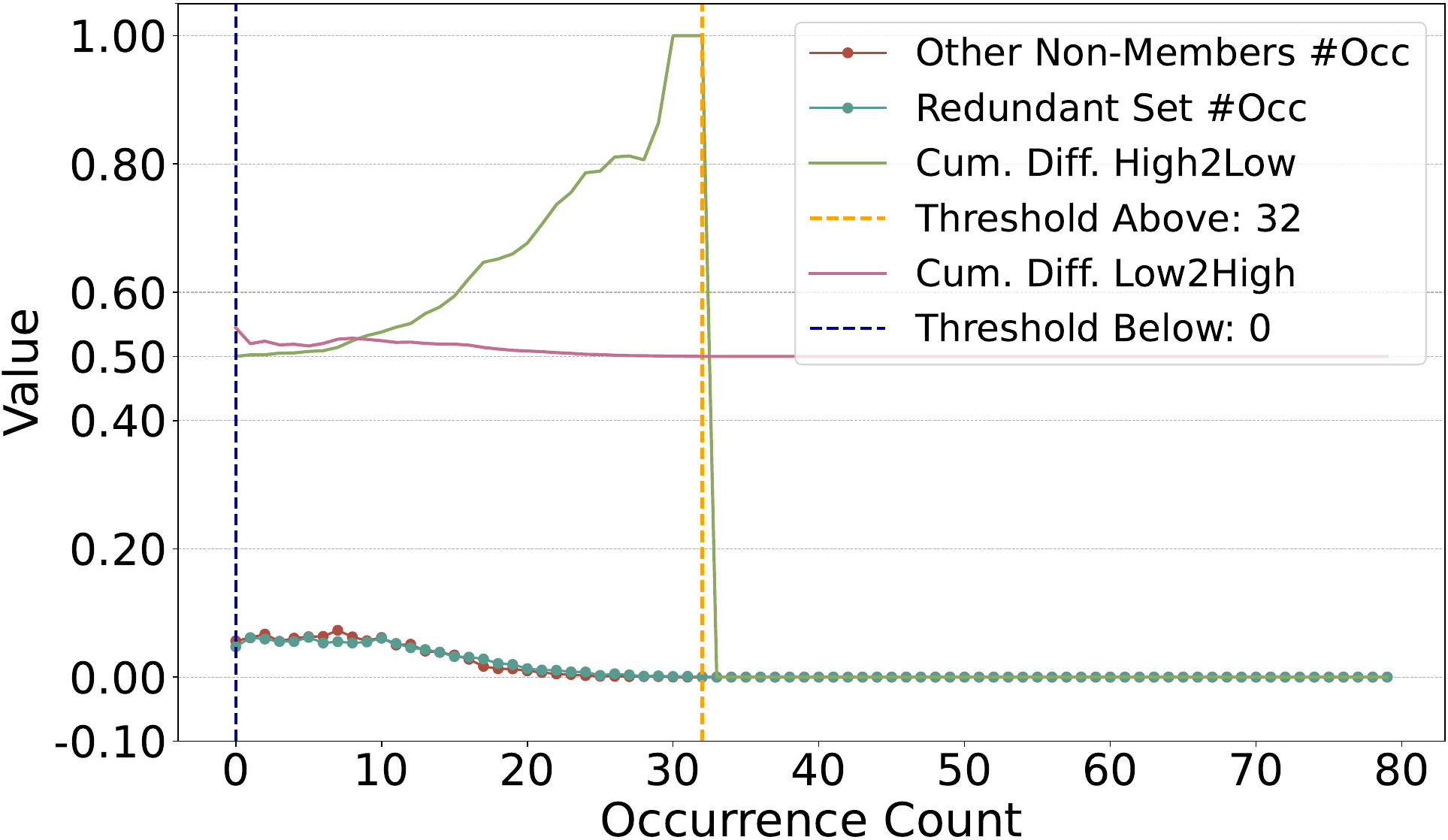}
        \subcaption{CumDis-kCent.}
    \end{minipage}
    \hfill
    \begin{minipage}[b]{.32\linewidth}
        \centering
        \includegraphics[width=\linewidth]{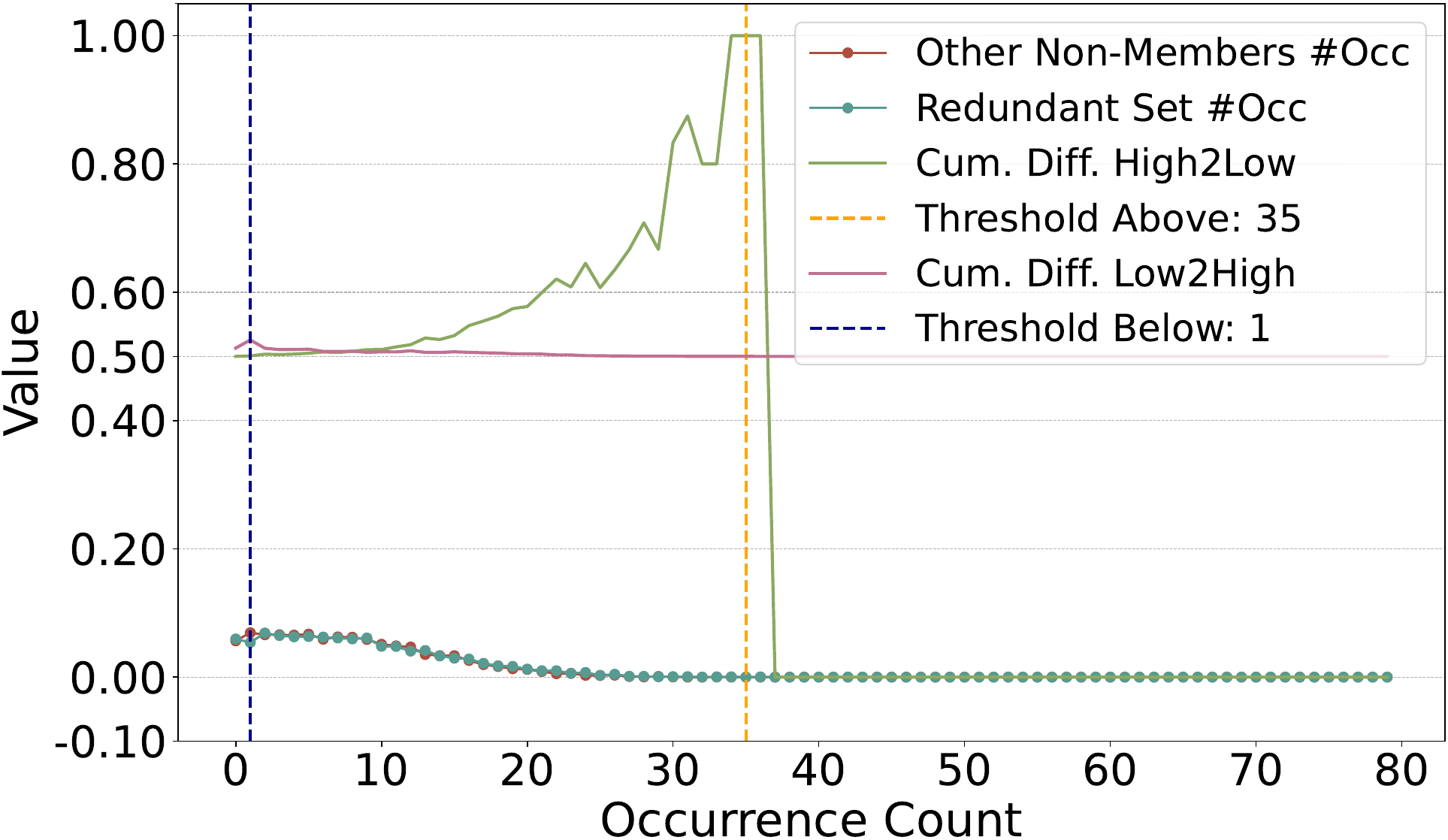}
        \subcaption{CumDis-SubM.}
        
    \end{minipage}
    \hfill
    \begin{minipage}[b]{.32\linewidth}
        \centering
        \includegraphics[width=\linewidth]{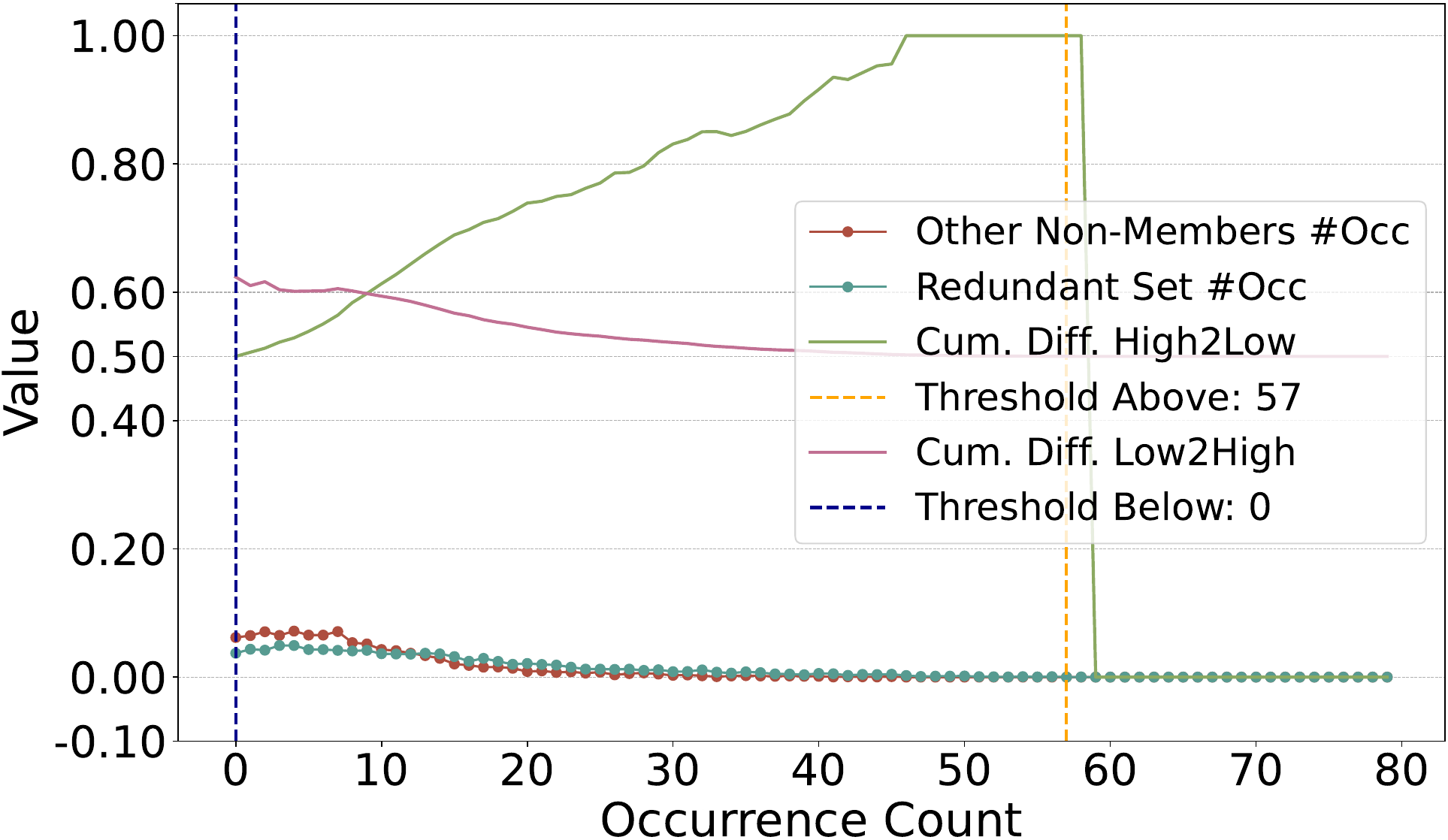}
        \subcaption{CumDis-Unc.}
        
    \end{minipage}
    \caption{Data Lineage Inference vulnerability of different pruning methods under CumDis. The dataset used here is CIFAR10, the pruning fraction is set to be 0.8, the shadow batch size is 80 and the victim batch size is 100.}
    \vspace{-3mm}
    \label{fig:evidence_app}
\end{figure*}
The complementary cumulative distribution function of the shadow redundant set and other non-members are functions \(F^{\textrm{ccdf}}_{\textrm{red}}(\textrm{i})(\cdot)\) and \(F^{\textrm{ccdf}}_{\textrm{non}}(\textrm{i})(\cdot)\), respectively: 
\begin{equation}
\begin{aligned}
&F^{\textrm{ccdf}}_{\textrm{red}}(\textrm{i})(t) = \sum_{t' > t} \mathbb{S}^{\textrm{shadow}}_{\textrm{red}}(\textrm{i})(t') \\
&= \sum_{t' > t} \sum_{s=1}^{|Q_{\textrm{s}}|} \mathbb{I}(x_s \in D^{\textrm{shadow}}_{\textrm{red}}(\textrm{i}))\mathbb{I}(t_s = t'),
\label{ccdf_red}
\end{aligned}
\end{equation}
\begin{equation}
\begin{aligned}
&F^{\textrm{ccdf}}_{\textrm{non}}(\textrm{i})(t) = \sum_{t' > t} \mathbb{S}^{\textrm{shadow}}_{\textrm{non}}(\textrm{i})(t') \\
&= \sum_{t' > t} \sum_{s=1}^{|Q_{\textrm{s}}|} \mathbb{I}(x_s \in D^{\textrm{shadow}}_{\textrm{non}}(\textrm{i}))\mathbb{I}(t_s = t').
\label{ccdf_non}
\end{aligned}
\end{equation} 
\end{definition}

\section{Derivation Detail of Brimming Score}
\label{brimming}
We start with the double summation:
   \[
   N = \sum_{p=0}^{\frac{|Q_{\textrm{v}}|}{\zeta_{\textrm{v}}}-1} \sum_{q=p+1}^{\frac{|Q_{\textrm{v}}|}{\zeta_{\textrm{v}}}} 1.
   \]
This sum counts the number of pairs \((p, q)\) where \(0 \leq p < q \leq \frac{|Q_{\textrm{v}}|}{\zeta_{\textrm{v}}}\).

\begin{figure*}[ht]
    \centering
    \begin{minipage}[b]{.32\linewidth}
        \centering
        \includegraphics[width=\linewidth]{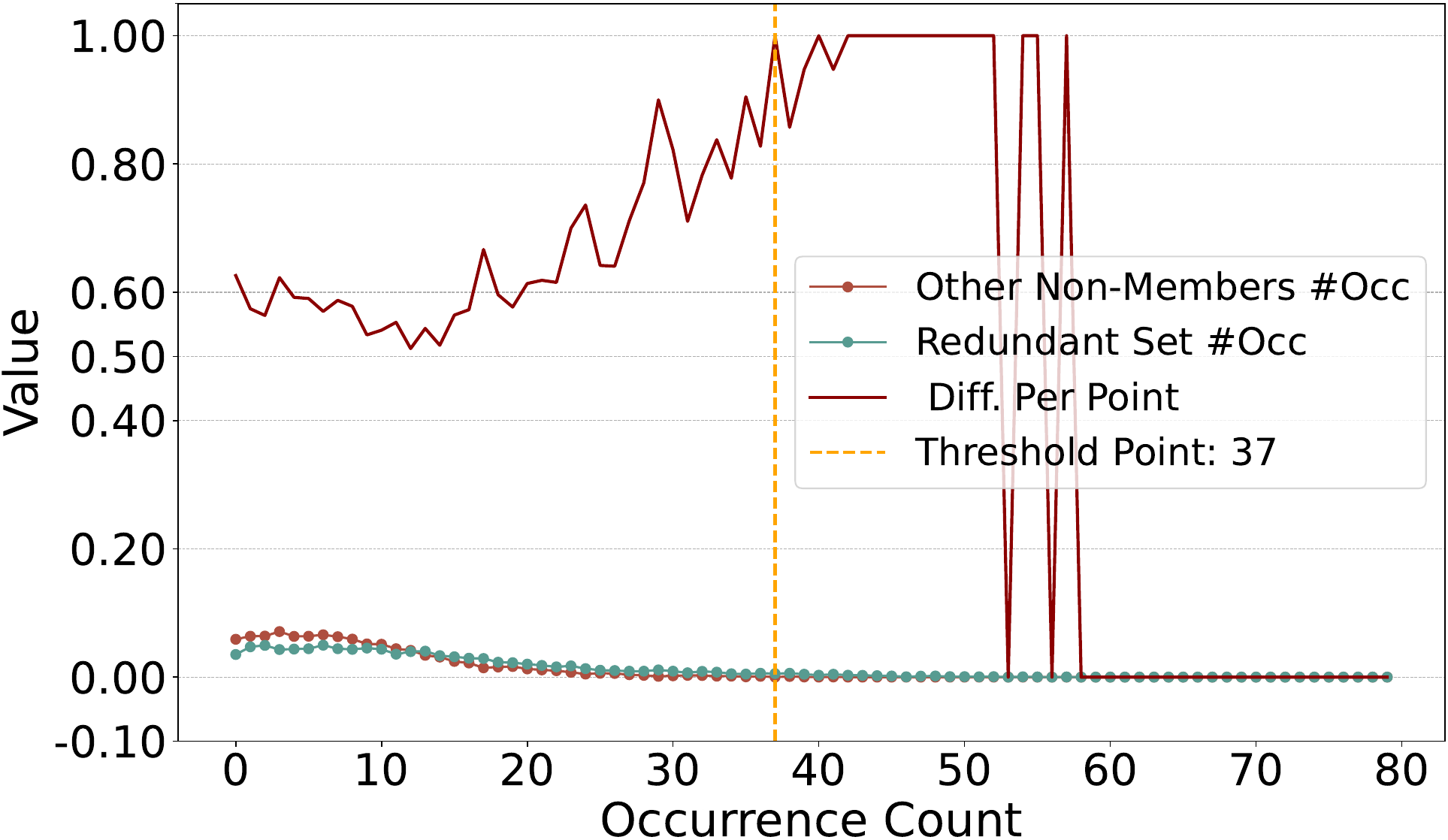}
        \subcaption{SpiDis-Cal}
        
    \end{minipage}%
    \hfill
    \begin{minipage}[b]{.32\linewidth}
        \centering
        \includegraphics[width=\linewidth]{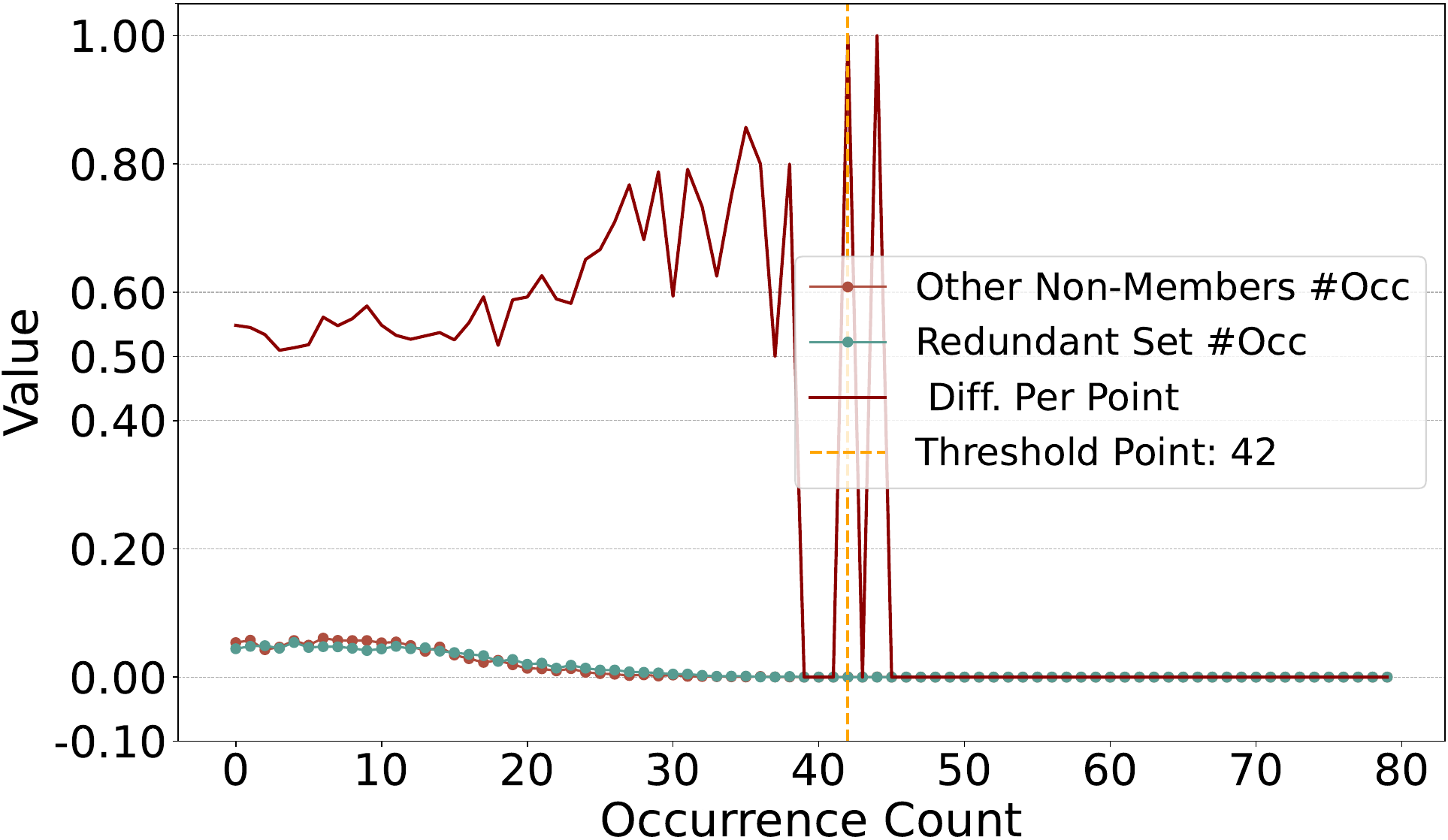}
        \subcaption{SpiDis-Co.Div.}
        
    \end{minipage}%
    \hfill
    \begin{minipage}[b]{.32\linewidth}
        \centering
        \includegraphics[width=\linewidth]{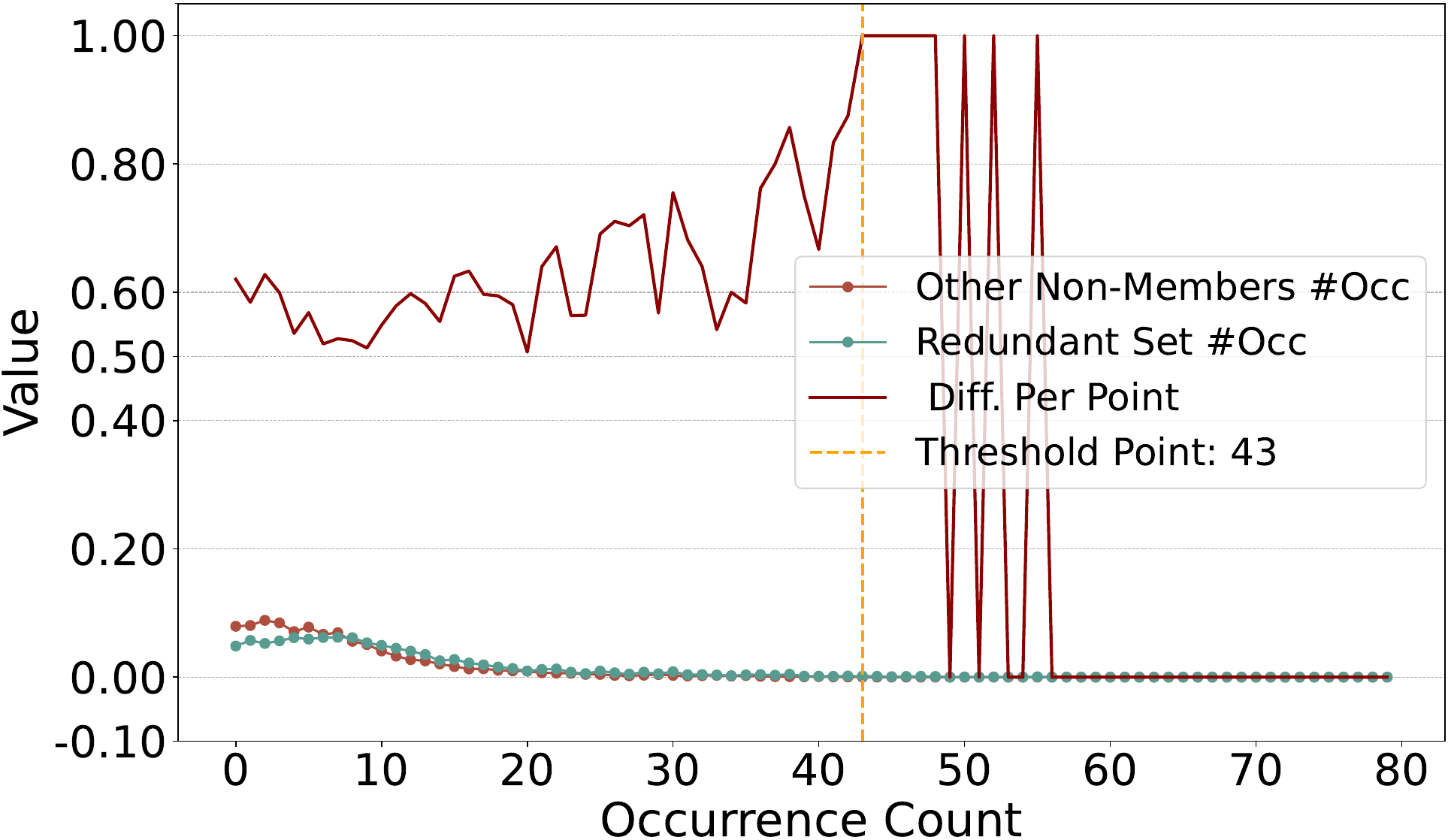}
        \subcaption{SpiDis-Craig}
        
    \end{minipage}%
    \hfill
    \begin{minipage}[b]{.32\linewidth}
        \centering
        \includegraphics[width=\linewidth]{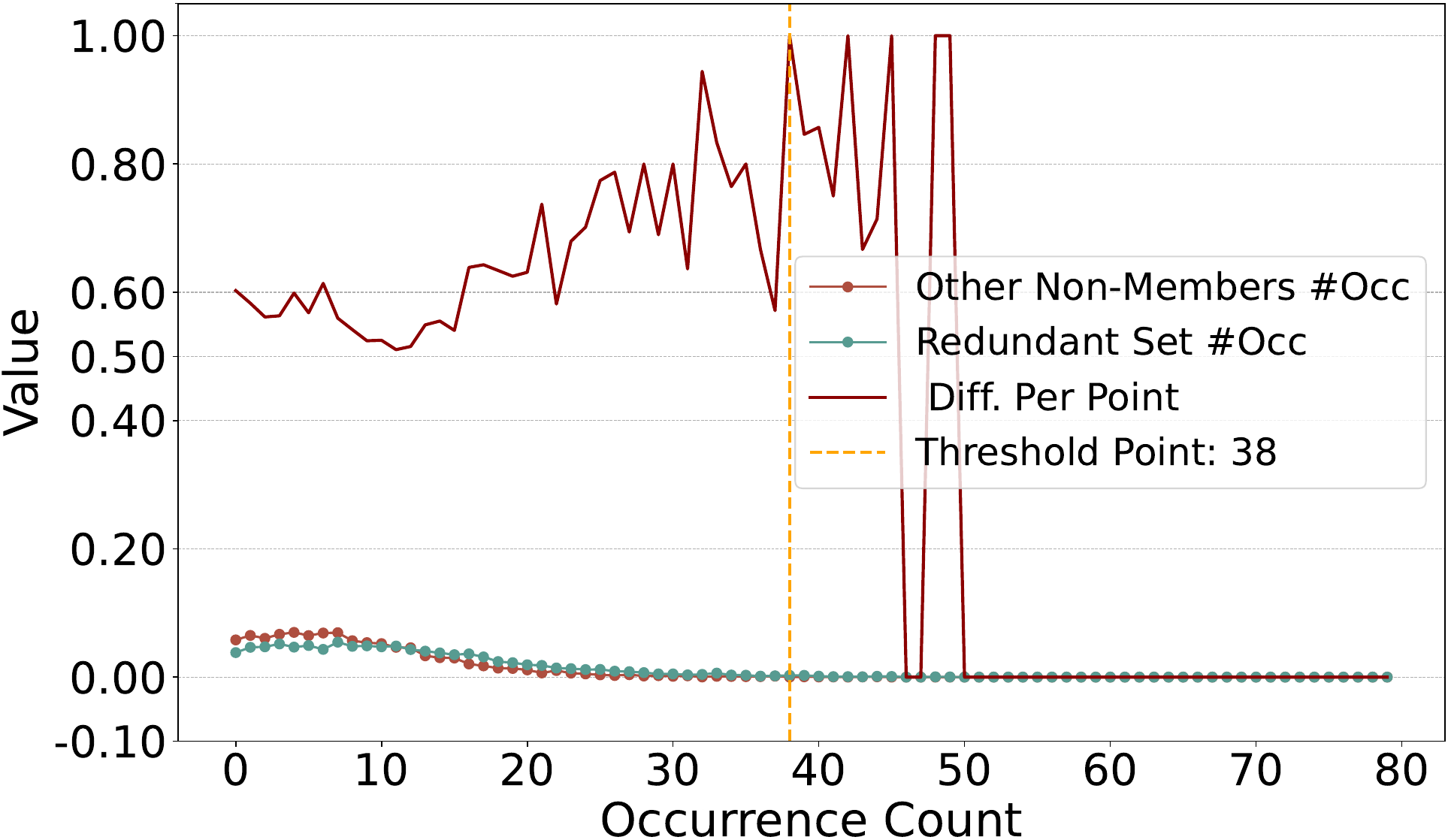}
        \subcaption{SpiDis-DeepF.}
        
    \end{minipage}%
    \hfill
    \begin{minipage}[b]{.32\linewidth}
        \centering
        \includegraphics[width=\linewidth]{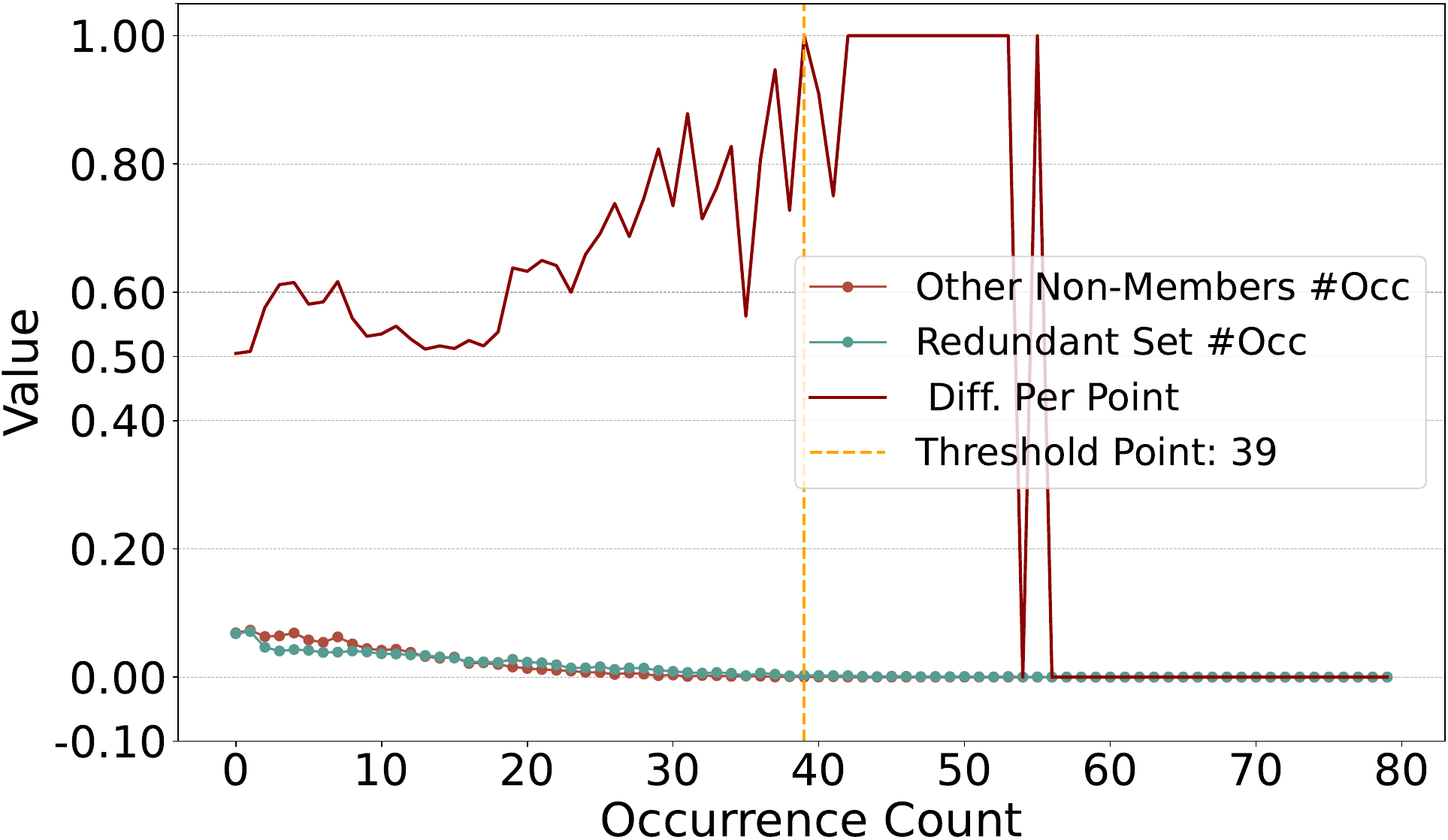}
        \subcaption{SpiDis-Forgt.}
        
    \end{minipage}%
    \hfill
    \begin{minipage}[b]{.32\linewidth}
        \centering
        \includegraphics[width=\linewidth]{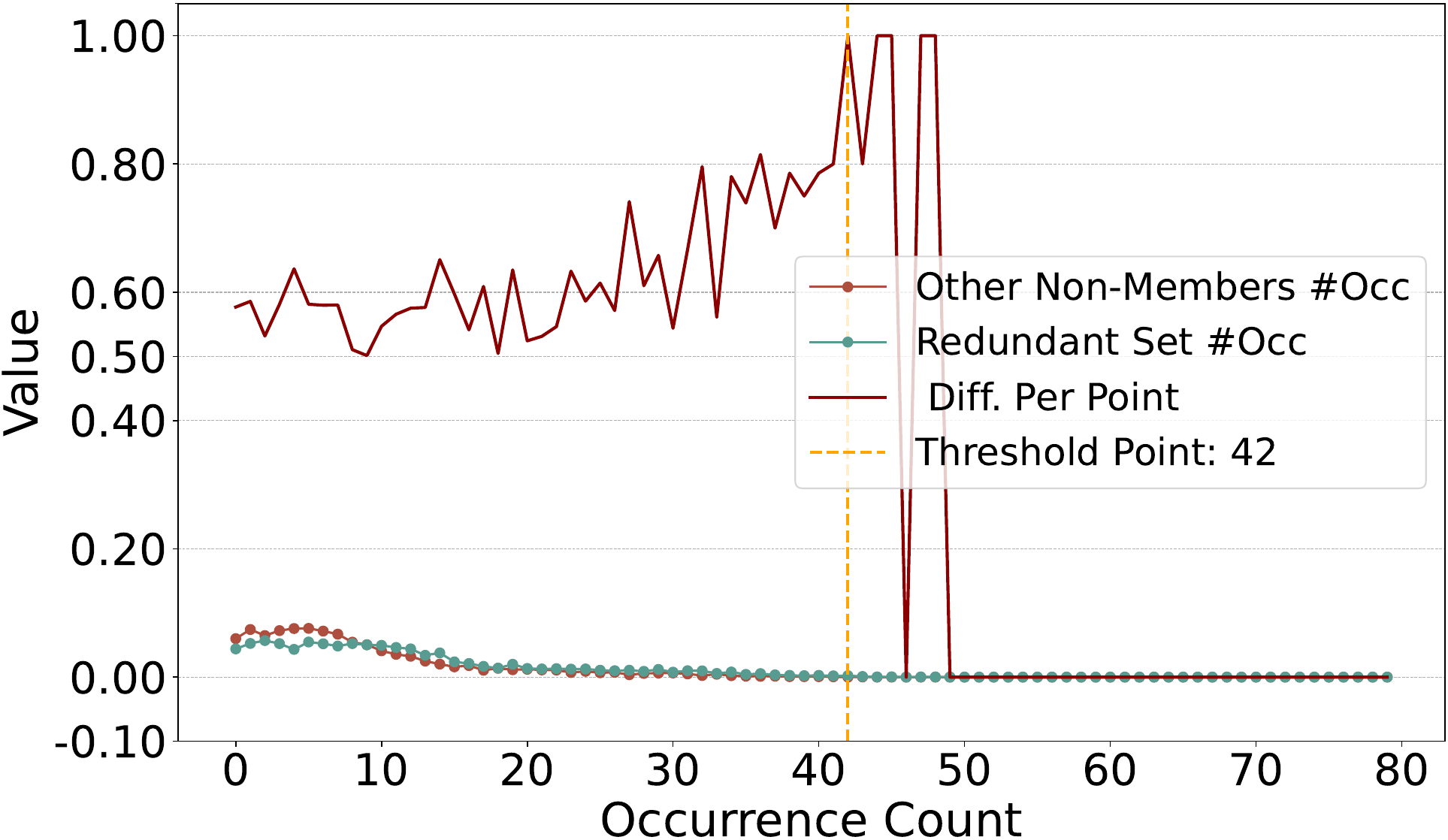}
        \subcaption{SpiDis-Glist.}
        
    \end{minipage}%
    \hfill
    \begin{minipage}[b]{.32\linewidth}
        \centering
        \includegraphics[width=\linewidth]{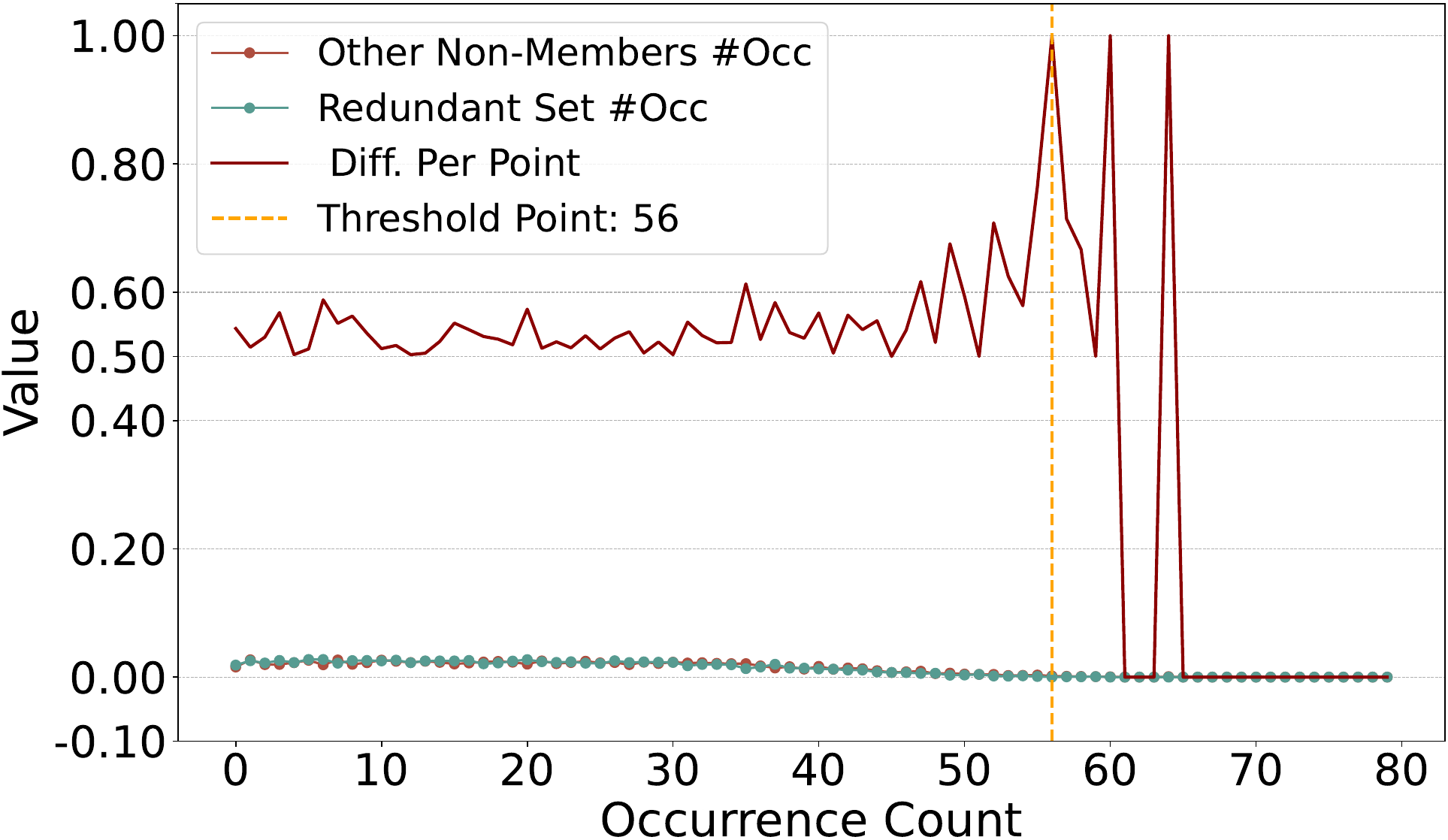}
        \subcaption{SpiDis-G.M.}
        
    \end{minipage}%
    \hfill
    \begin{minipage}[b]{.32\linewidth}
        \centering
        \includegraphics[width=\linewidth]{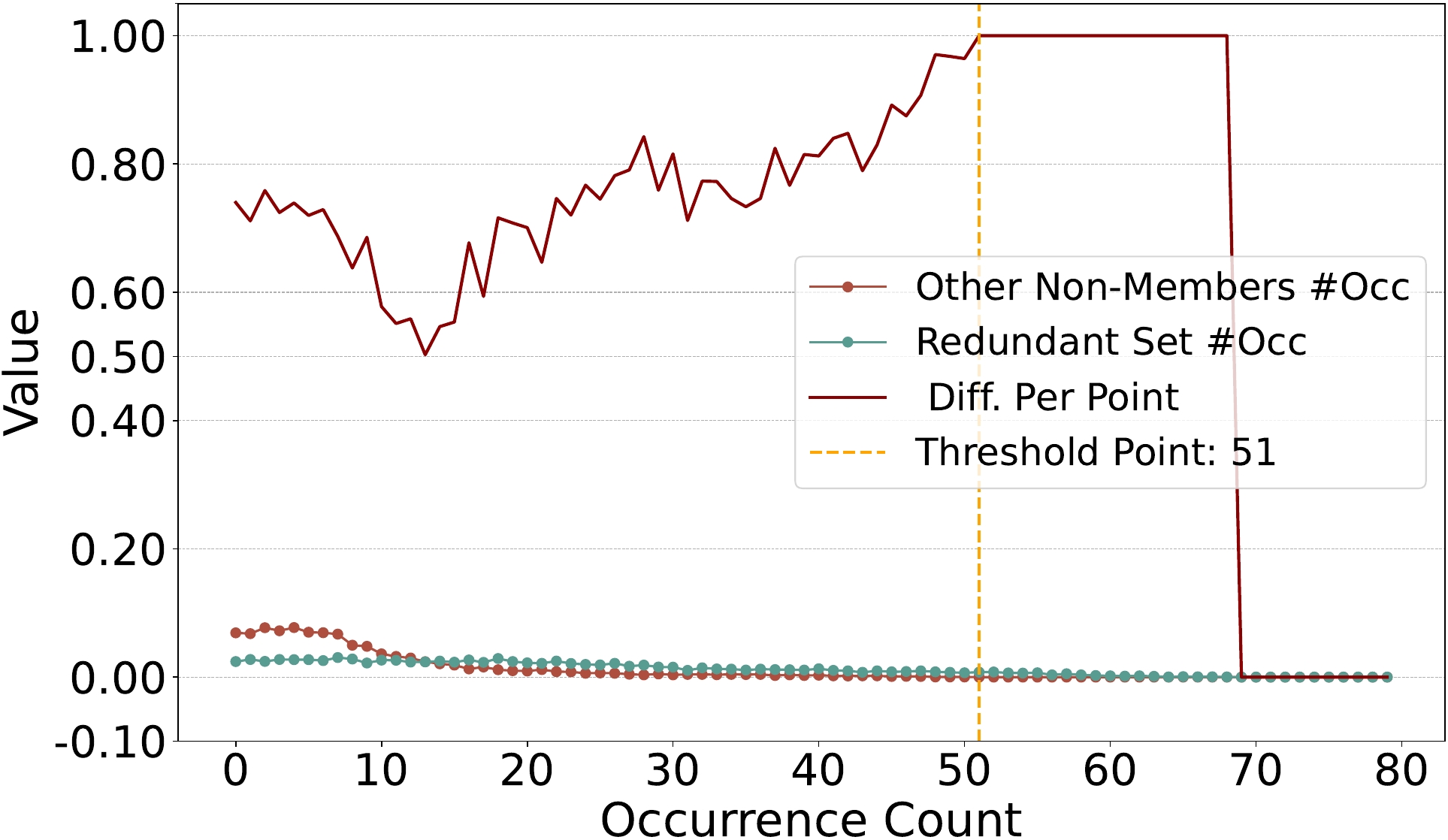}
        \subcaption{SpiDis-GraNd}
        
    \end{minipage}%
    \hfill
    \begin{minipage}[b]{.32\linewidth}
        \centering
        \includegraphics[width=\linewidth]{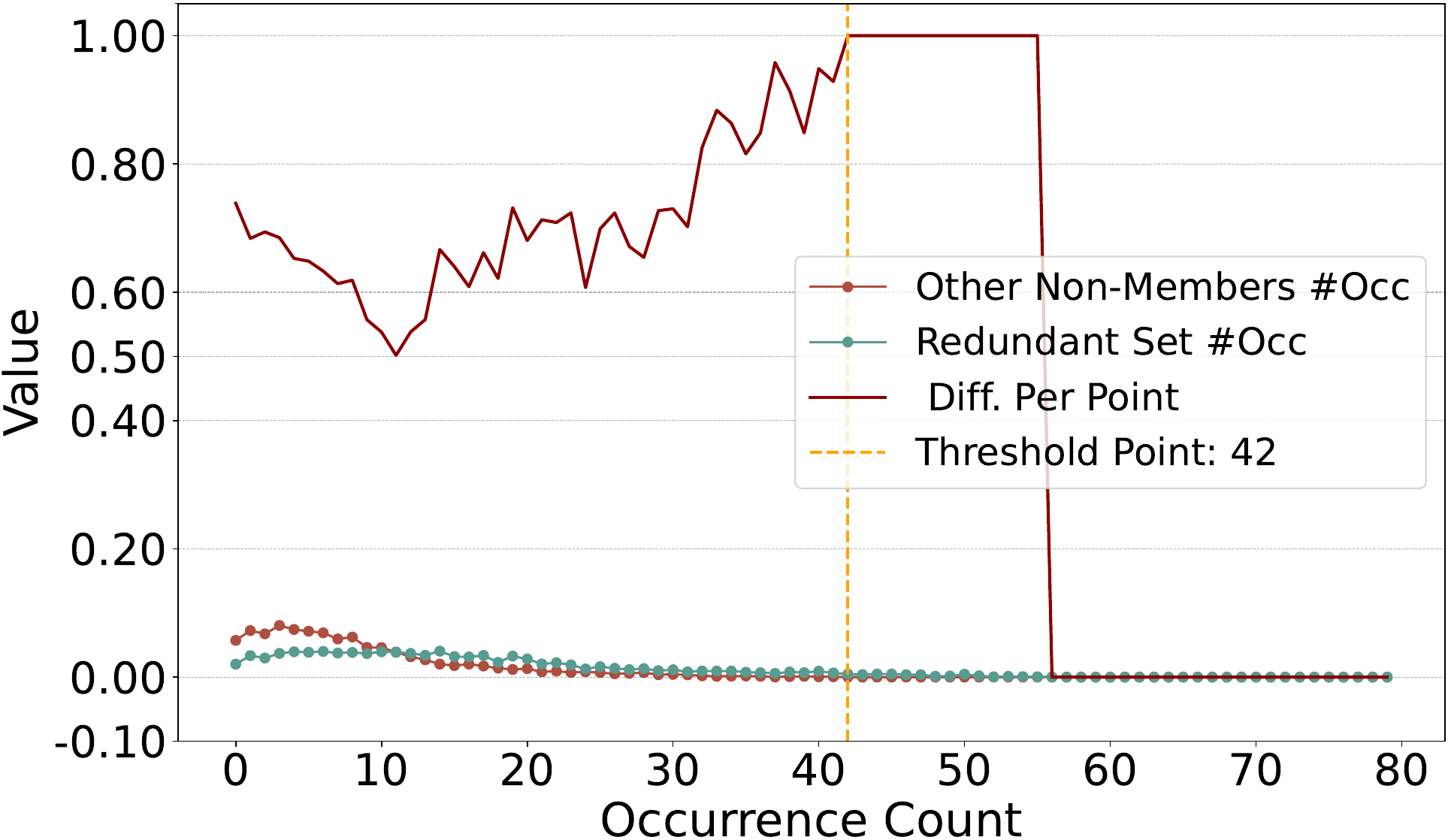}
        \subcaption{SpiDis-Herd.}
        
    \end{minipage}%
    \hfill
    \begin{minipage}[b]{.32\linewidth}
        \centering
        \includegraphics[width=\linewidth]{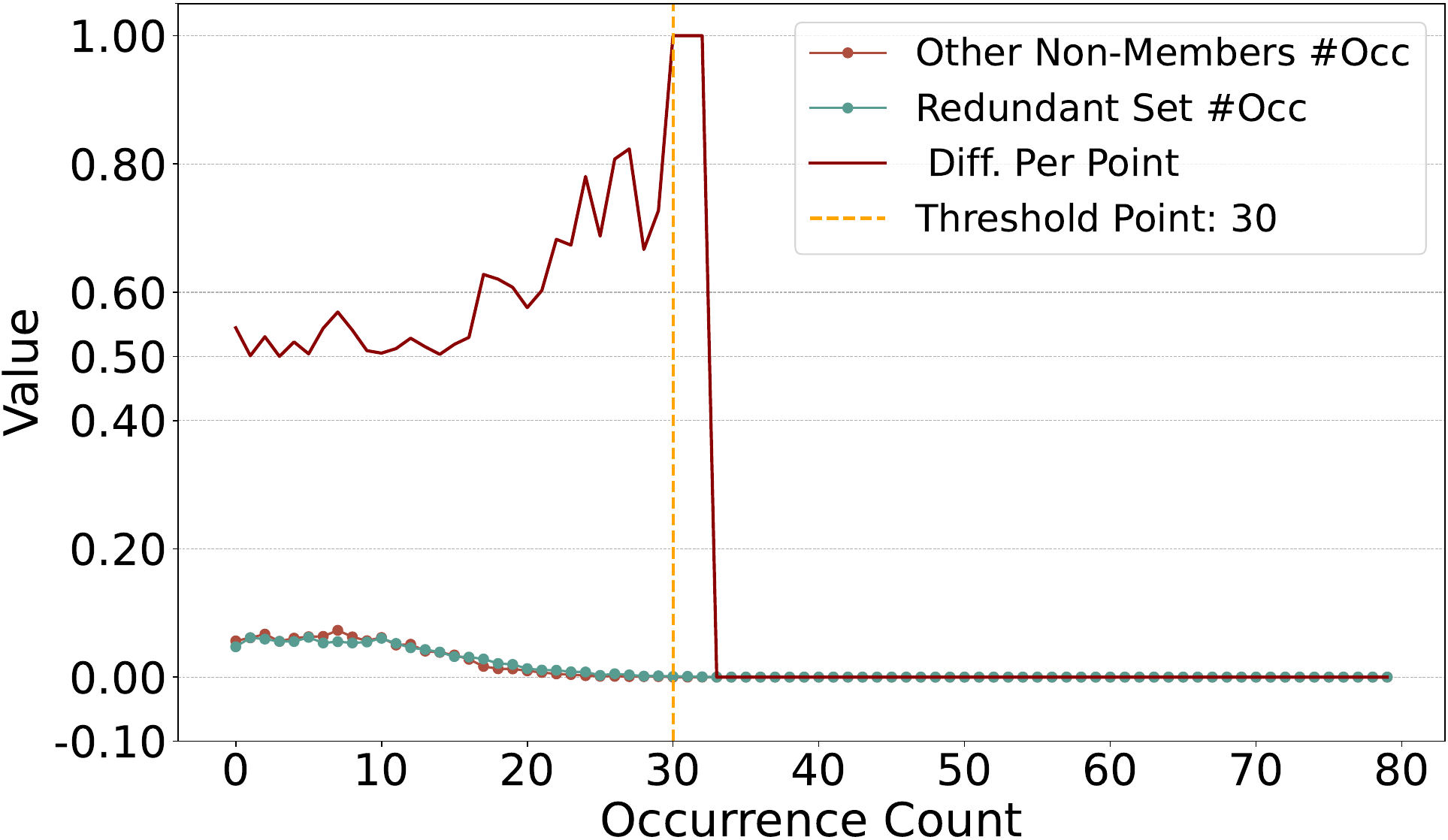}
        \subcaption{SpiDis-kCent.}
        
    \end{minipage}%
    \hfill
    \begin{minipage}[b]{.32\linewidth}
        \centering
        \includegraphics[width=\linewidth]{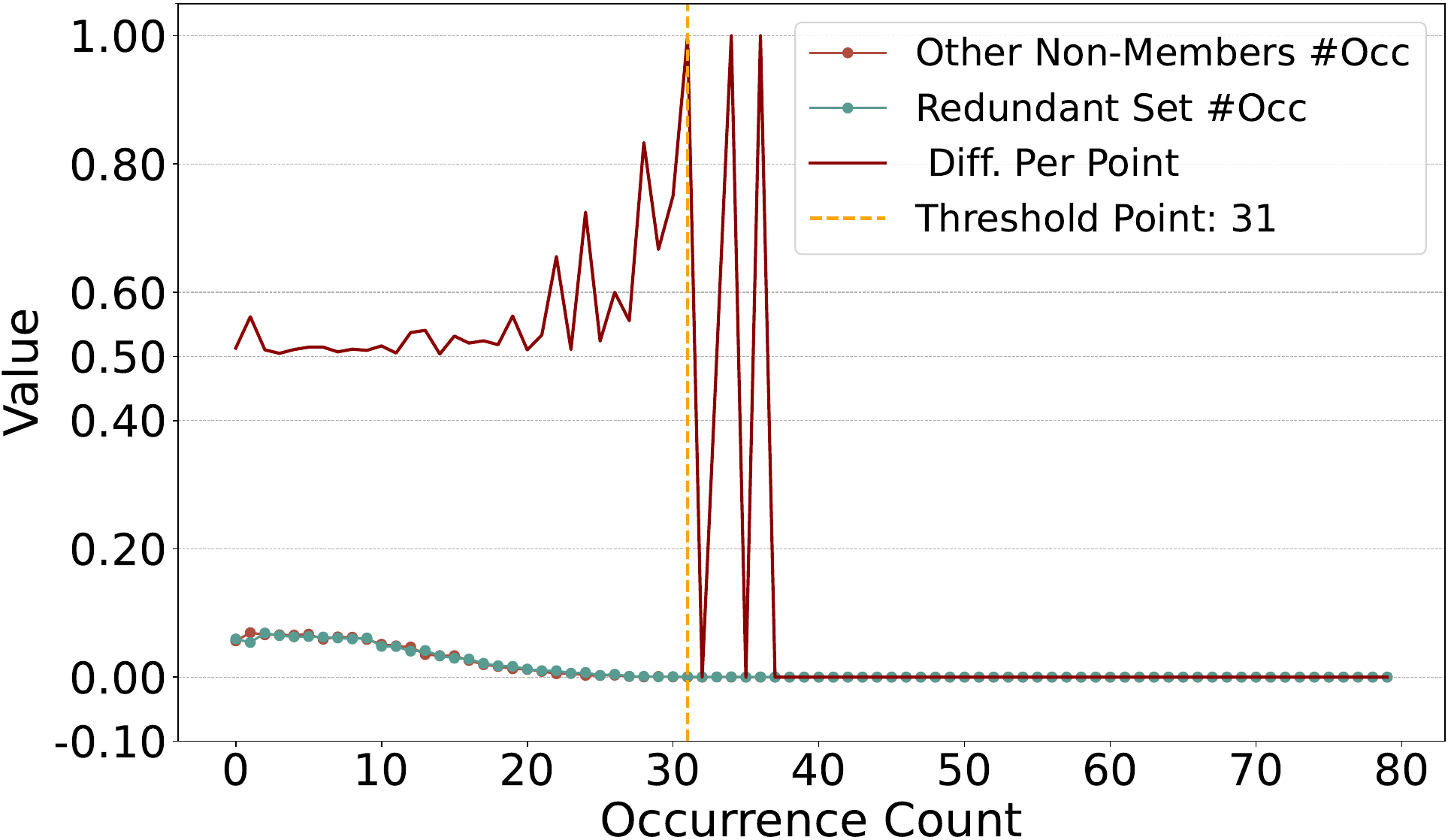}
        \subcaption{SpiDis-SubM.}
        
    \end{minipage}%
    \hfill
    \begin{minipage}[b]{.32\linewidth}
        \centering
        \includegraphics[width=\linewidth]{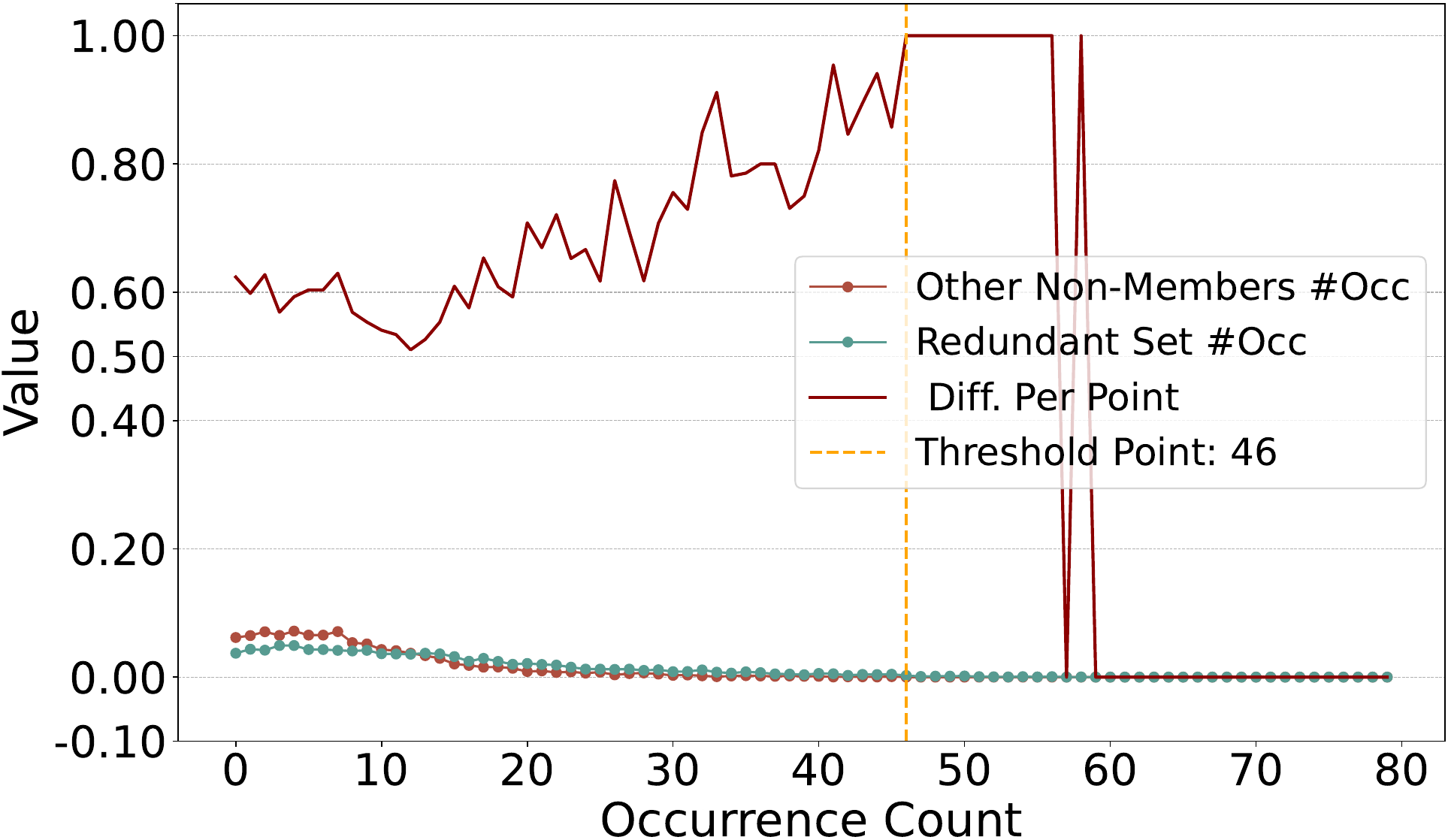}
        \subcaption{SpiDis-Unc.}
        
    \end{minipage}%
    \caption{Data Lineage Inference vulnerability of different pruning methods under SpiDis. The dataset used here is CIFAR10, the pruning fraction is set to be 0.8, the shadow batch size is 80 and the victim batch size is 100.}
    \label{fig:evidence_app}
\end{figure*}
For a fixed \(p\), the inner sum:
   \[
   \sum_{q=p+1}^{\frac{|Q_{\textrm{v}}|}{\zeta_{\textrm{v}}}} 1
   \]
counts the number of terms from \(q = p+1\) to \(q = \frac{|Q_{\textrm{v}}|}{\zeta_{\textrm{v}}}\). The number of such terms is:
   \[
   \frac{|Q_{\textrm{v}}|}{\zeta_{\textrm{v}}} - (p + 1) + 1 = \frac{|Q_{\textrm{v}}|}{\zeta_{\textrm{v}}} - p.
   \]
Substituting this back into the outer sum:
   \[
   N = \sum_{p=0}^{\frac{|Q_{\textrm{v}}|}{\zeta_{\textrm{v}}}-1} \left(\frac{|Q_{\textrm{v}}|}{\zeta_{\textrm{v}}} - p \right).
   \]
   This sum can be split into two separate sums:
   \[
   N = \sum_{p=0}^{\frac{|Q_{\textrm{v}}|}{\zeta_{\textrm{v}}}-1} \left(\frac{|Q_{\textrm{v}}|}{\zeta_{\textrm{v}}}\right) - \sum_{p=0}^{\frac{|Q_{\textrm{v}}|}{\zeta_{\textrm{v}}}-1} p.
   \]
\begin{figure*}[ht!]
    \centering
    \begin{minipage}[b]{.32\linewidth}
        \centering
        \includegraphics[width=\linewidth]{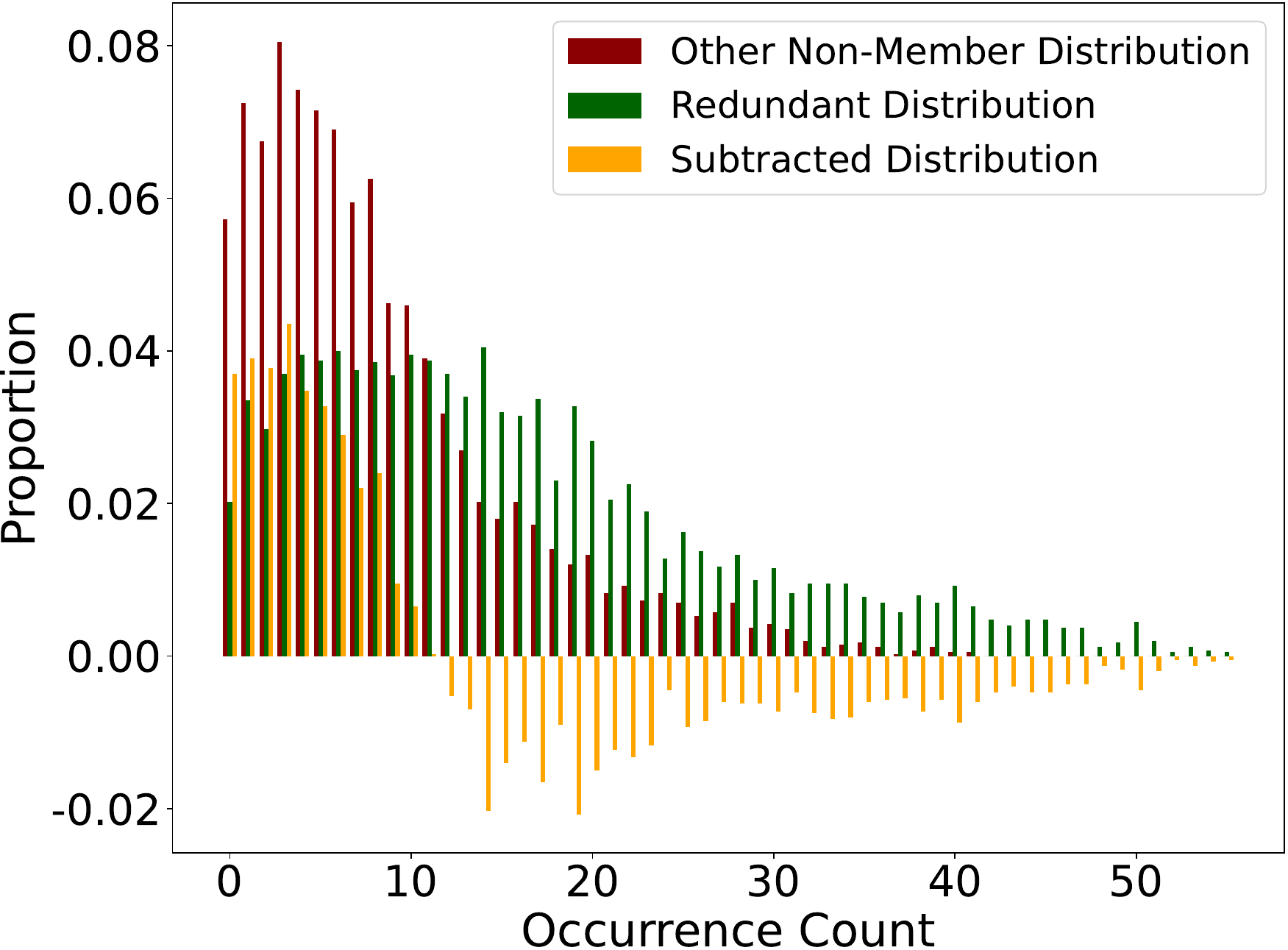}
        \subcaption{\(\zeta_{\textrm{v}}\)=100, \(\zeta_{\textrm{s}}\)=80}
    \end{minipage}
    \hfill 
    \begin{minipage}[b]{.32\linewidth}
        \centering
        \includegraphics[width=\linewidth]{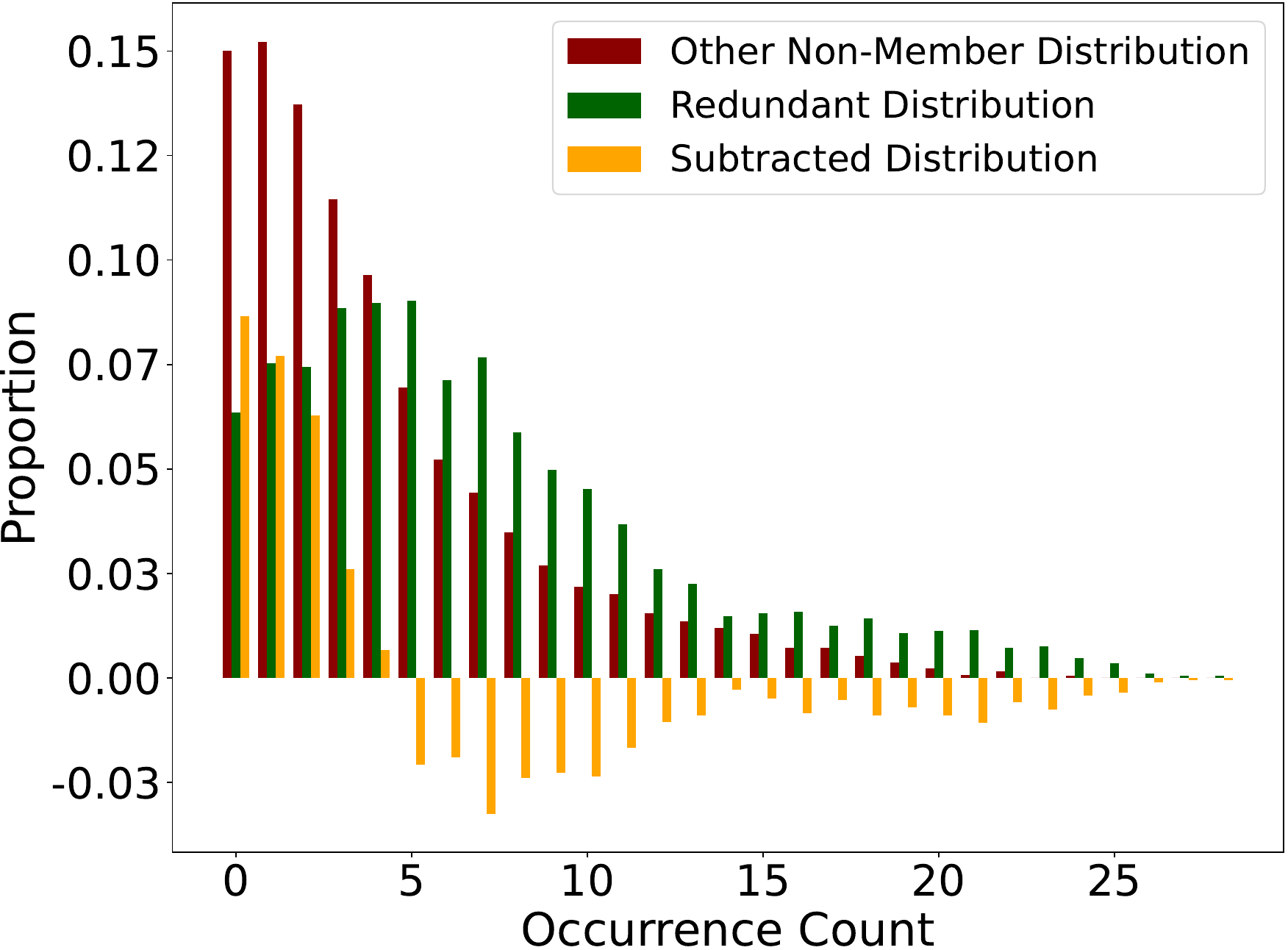}
        \subcaption{\(\zeta_{\textrm{v}}\)=200, \(\zeta_{\textrm{s}}\)=160}
    \end{minipage}
    \hfill 
    \begin{minipage}[b]{.32\linewidth}
        \centering
        \includegraphics[width=\linewidth]{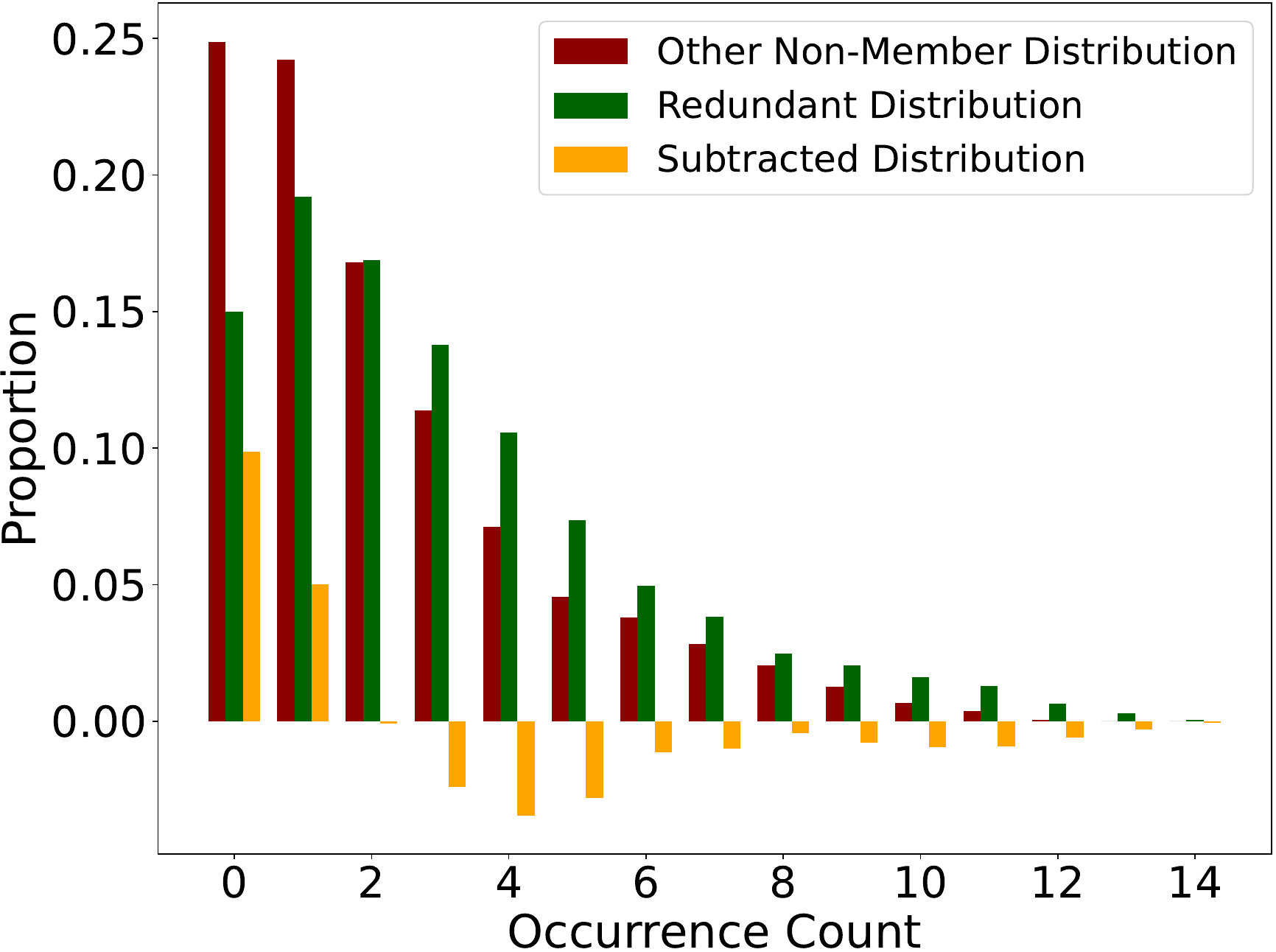}
        \subcaption{\(\zeta_{\textrm{v}}\)=400, \(\zeta_{\textrm{s}}\)=200}
    \end{minipage}
    \hfill 
    \begin{minipage}[b]{.32\linewidth}
        \centering
        \includegraphics[width=\linewidth]{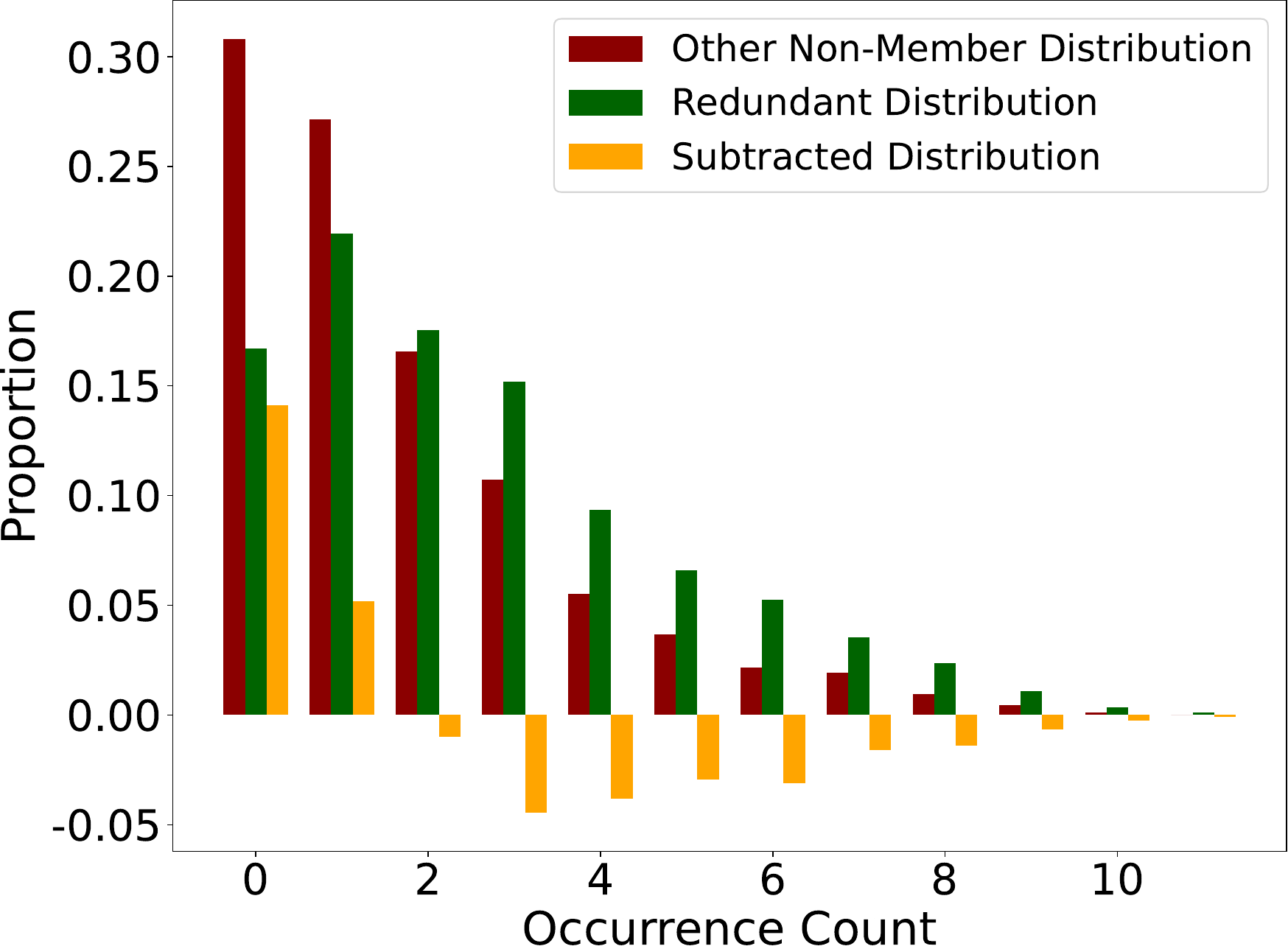}
        \subcaption{\(\zeta_{\textrm{v}}\)=500, \(\zeta_{\textrm{s}}\)=400}
    \end{minipage}
    \hfill 
    \begin{minipage}[b]{.32\linewidth}
        \centering
        \includegraphics[width=\linewidth]{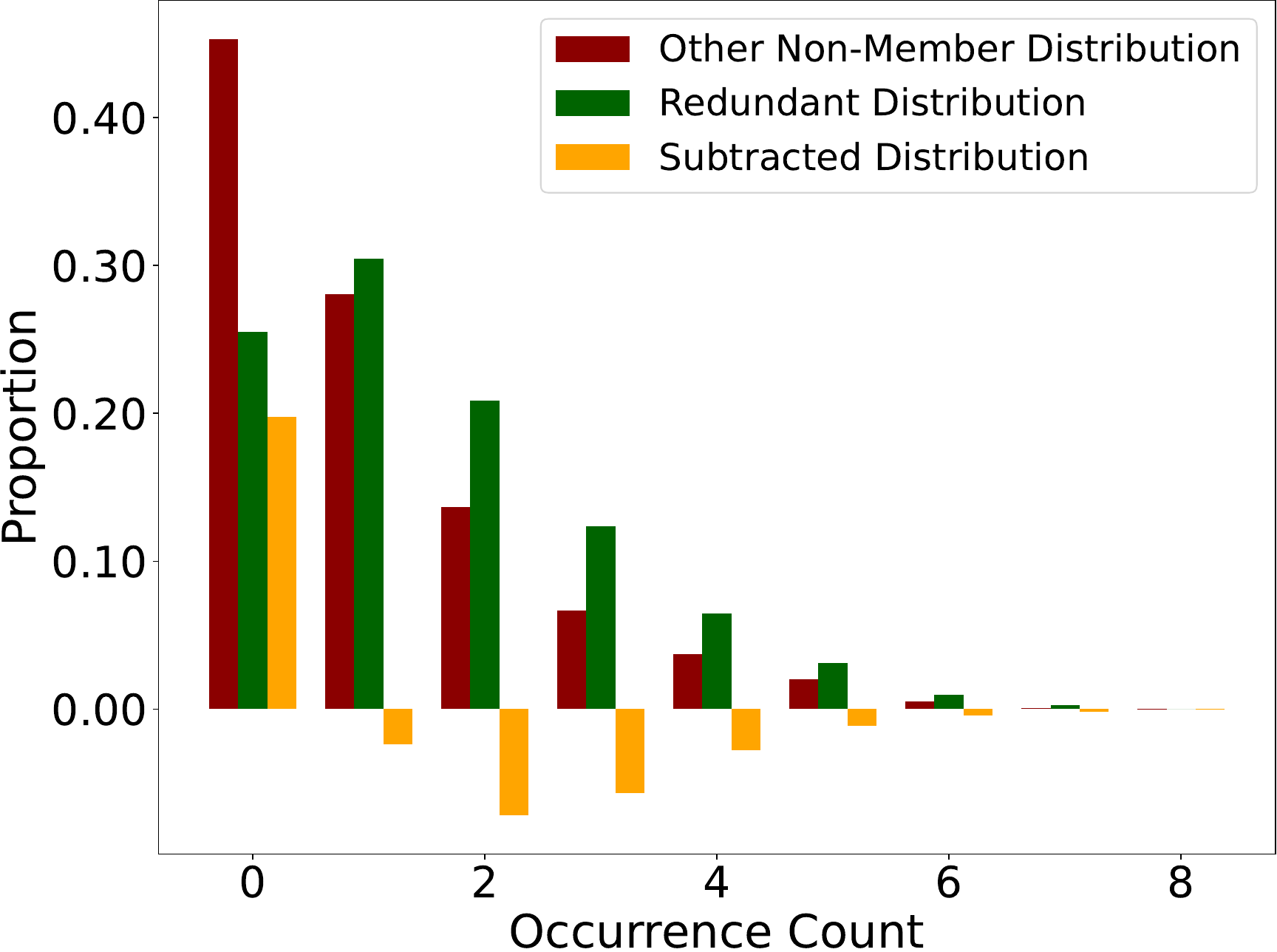}
        \subcaption{\(\zeta_{\textrm{v}}\)=800, \(\zeta_{\textrm{s}}\)=400}
    \end{minipage}
    \hfill 
    \begin{minipage}[b]{.32\linewidth}
        \centering
        \includegraphics[width=\linewidth]{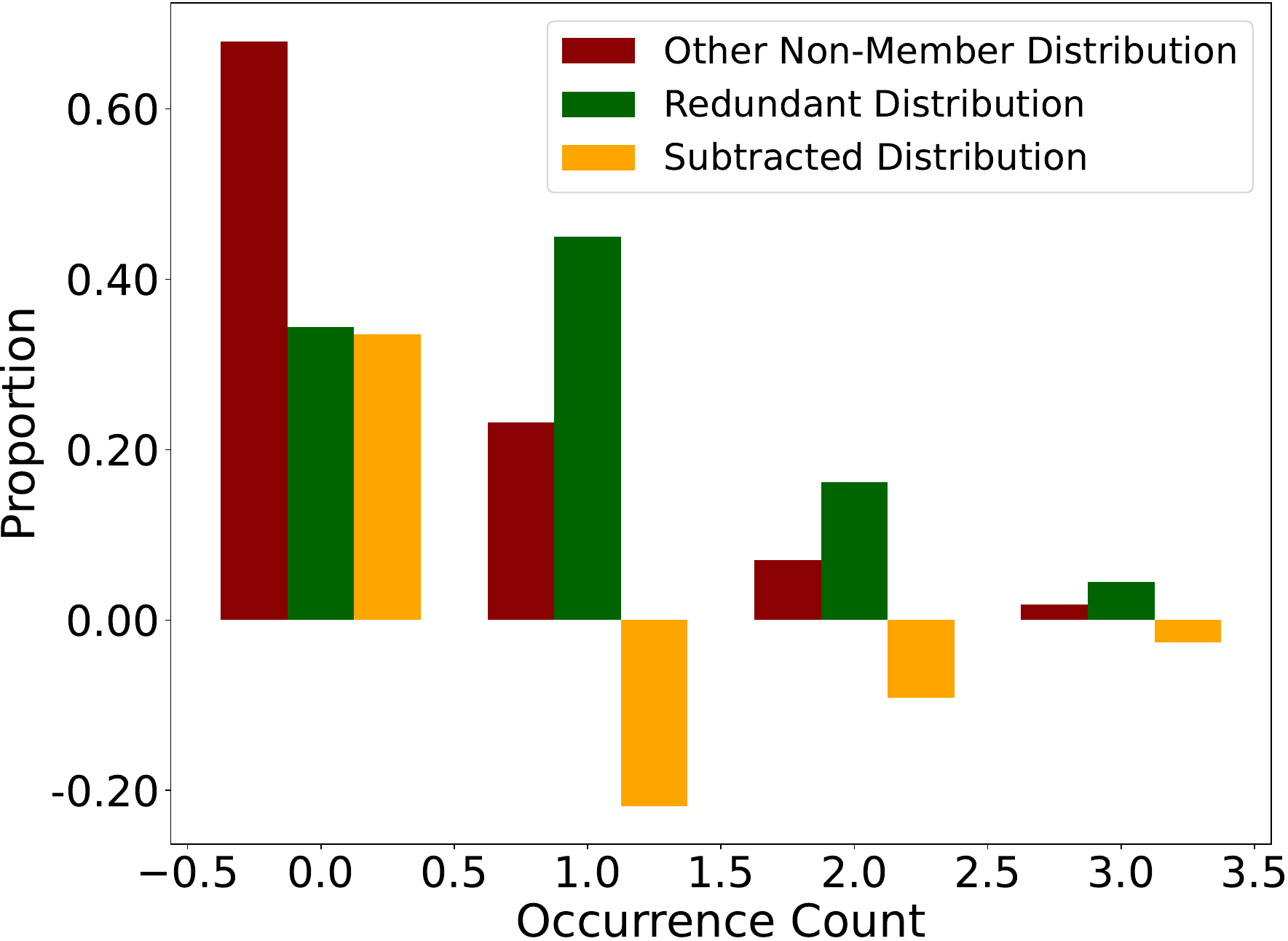}
        \subcaption{\(\zeta_{\textrm{v}}\)=2000, \(\zeta_{\textrm{s}}\)=800}
    \end{minipage}
    \vspace{0.3cm}
    \begin{minipage}[b]{\linewidth}
    \centering
     \small 
    \renewcommand{\arraystretch}{0.5}
    \renewcommand\tabcolsep{5.5pt}
    \begin{threeparttable}[t]
        \begin{tabular}{c|c|c|c|c|c|c}
            \hline
            \hline
            Attack & (a) & (b) & (c) & (d) & (e) & (f) \\
            \hline
            WhoDis & 65.54 & 58.14 & 54.49 & 57.05 & 59.89 & 66.79  \\
            CumDis & 83.96 & 72.77 & 63.23 & 64.84 & 64.00 & 66.67 \\
            ArraDis & 95.35 & 90.70 & 95.24 & 73.81 & 73.33 & 70.76  \\
            SpiDis & 100.00 & 82.86 & 77.61 & 70.97 & 78.57 & 69.92  \\
            \hline
            \hline
        \end{tabular}
    \end{threeparttable}
    \subcaption*{(g) Attack success rate under different batch size.}
    \label{table:metrics}
\end{minipage}
    \caption{Occurrence distributions and their difference when using different batch size.}
    \label{fig:batch_size}
\end{figure*}
\begin{figure*}[h!]
    \centering
    \begin{minipage}[b]{.16\linewidth}
        \centering
        \includegraphics[width=\linewidth]{figures/sp_final_Craig_occ_dis_dif_0.8_groupSize_100.pdf}
        \subcaption{Craig (w/o.)}
    \end{minipage}
    \hfill
    \begin{minipage}[b]{.16\linewidth}
        \centering
        \includegraphics[width=\linewidth]{figures/sp_final_DeepFool_occ_dis_dif_0.8_groupSize_100.pdf}
        \subcaption{DeepF. (w/o.)}
        
    \end{minipage}
    \hfill
    \begin{minipage}[b]{.16\linewidth}
        \centering
        \includegraphics[width=\linewidth]{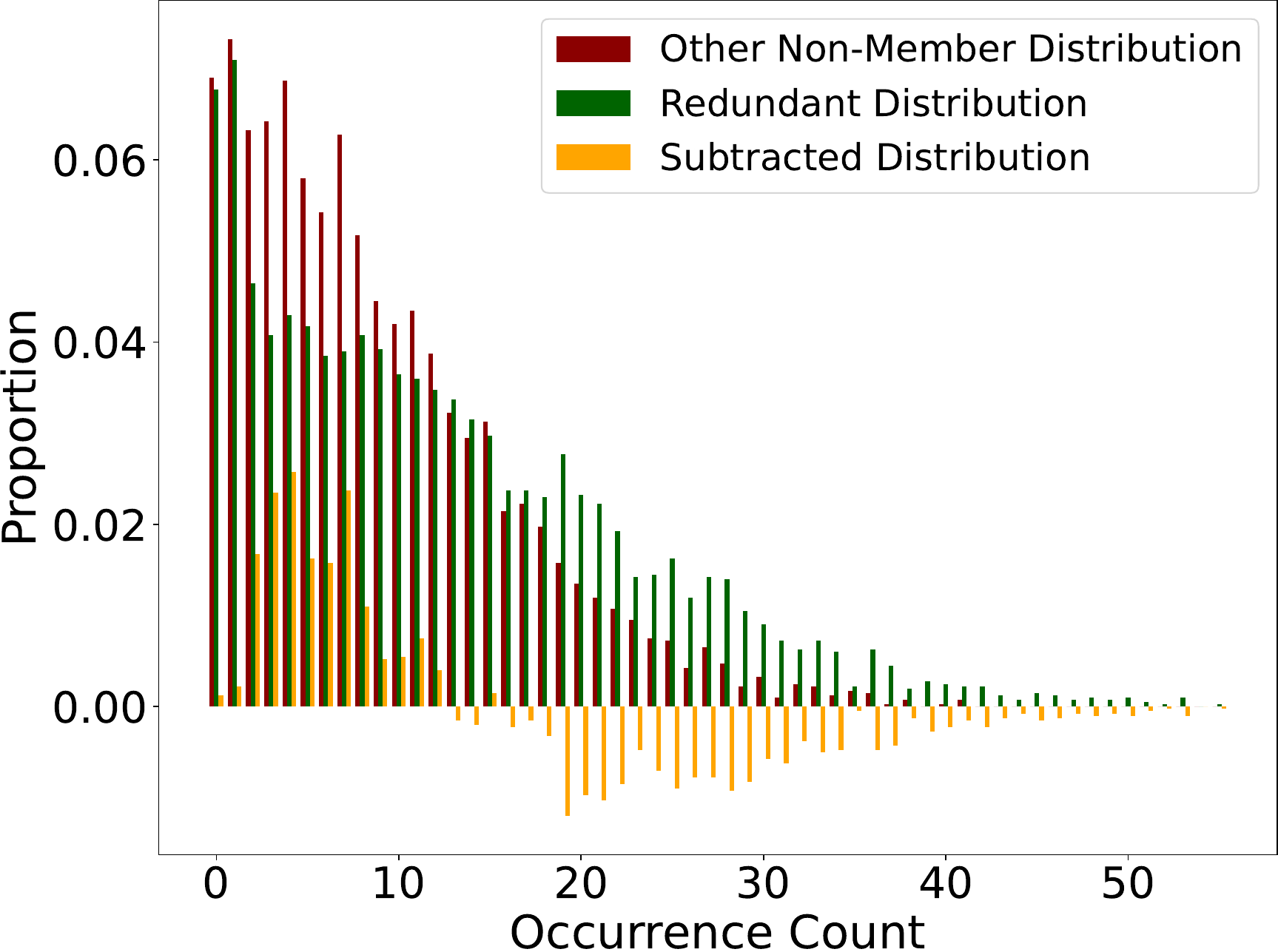}
        \subcaption{Forgt. (w/o.)}
        
    \end{minipage}%
    \hfill
    \begin{minipage}[b]{.16\linewidth}
        \centering
        \includegraphics[width=\linewidth]{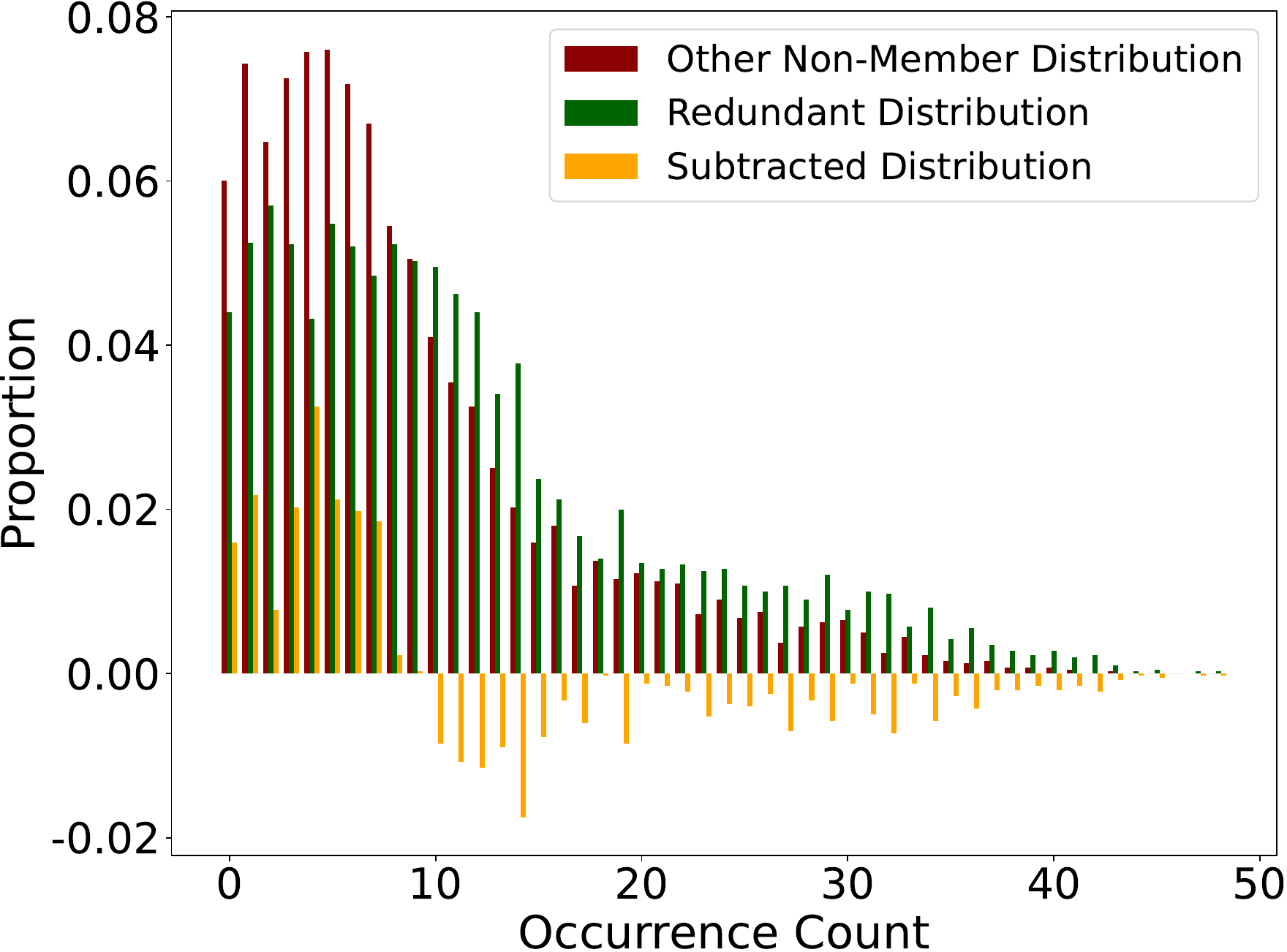}
        \subcaption{Glist. (w/o.)}
        
    \end{minipage}
    \hfill
    \begin{minipage}[b]{.16\linewidth}
        \centering
        \includegraphics[width=\linewidth]{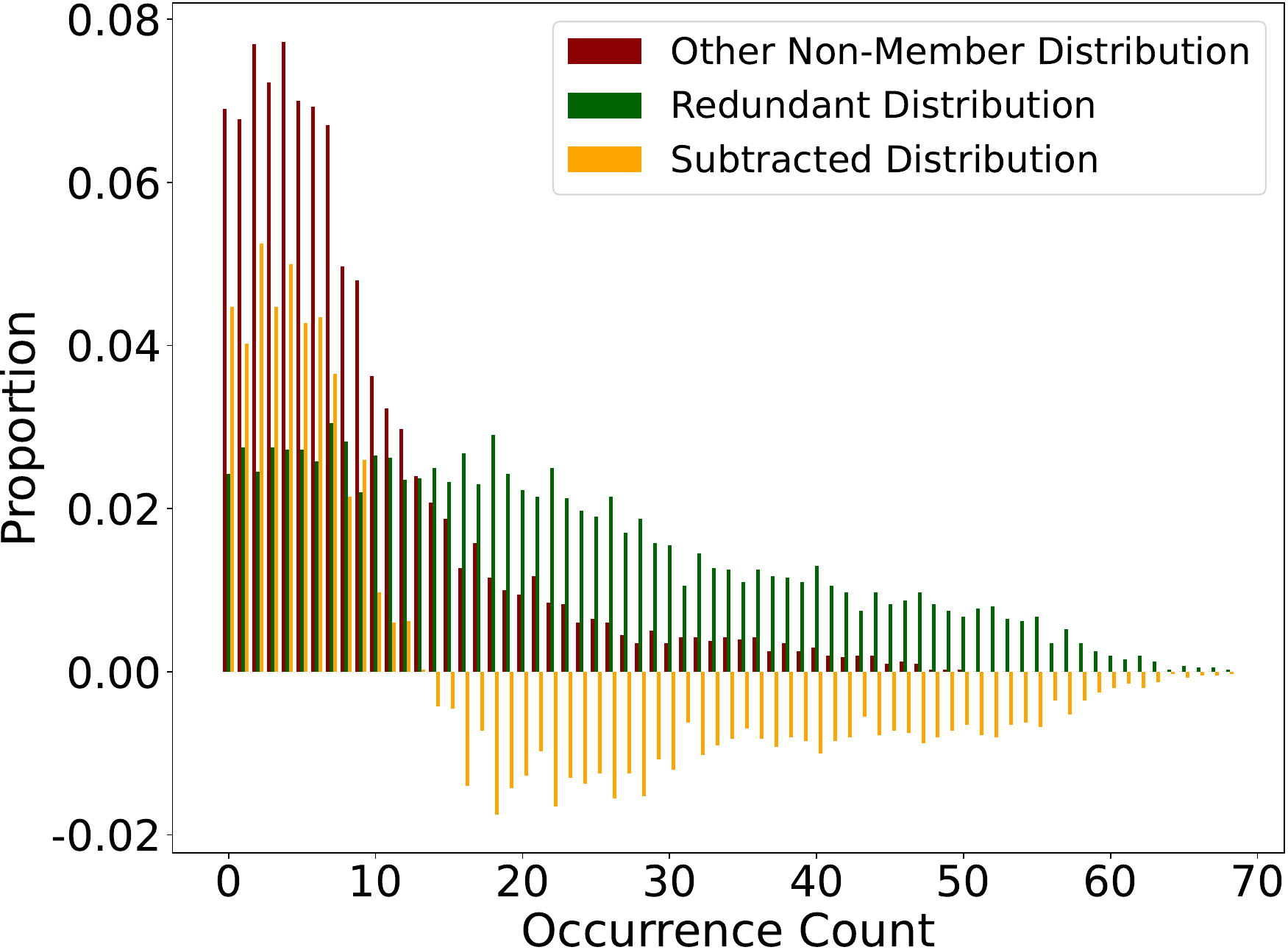}
        \subcaption{GraNd (w/o.)}
        
    \end{minipage}
    \hfill
    \begin{minipage}[b]{.16\linewidth}
        \centering
        \includegraphics[width=\linewidth]{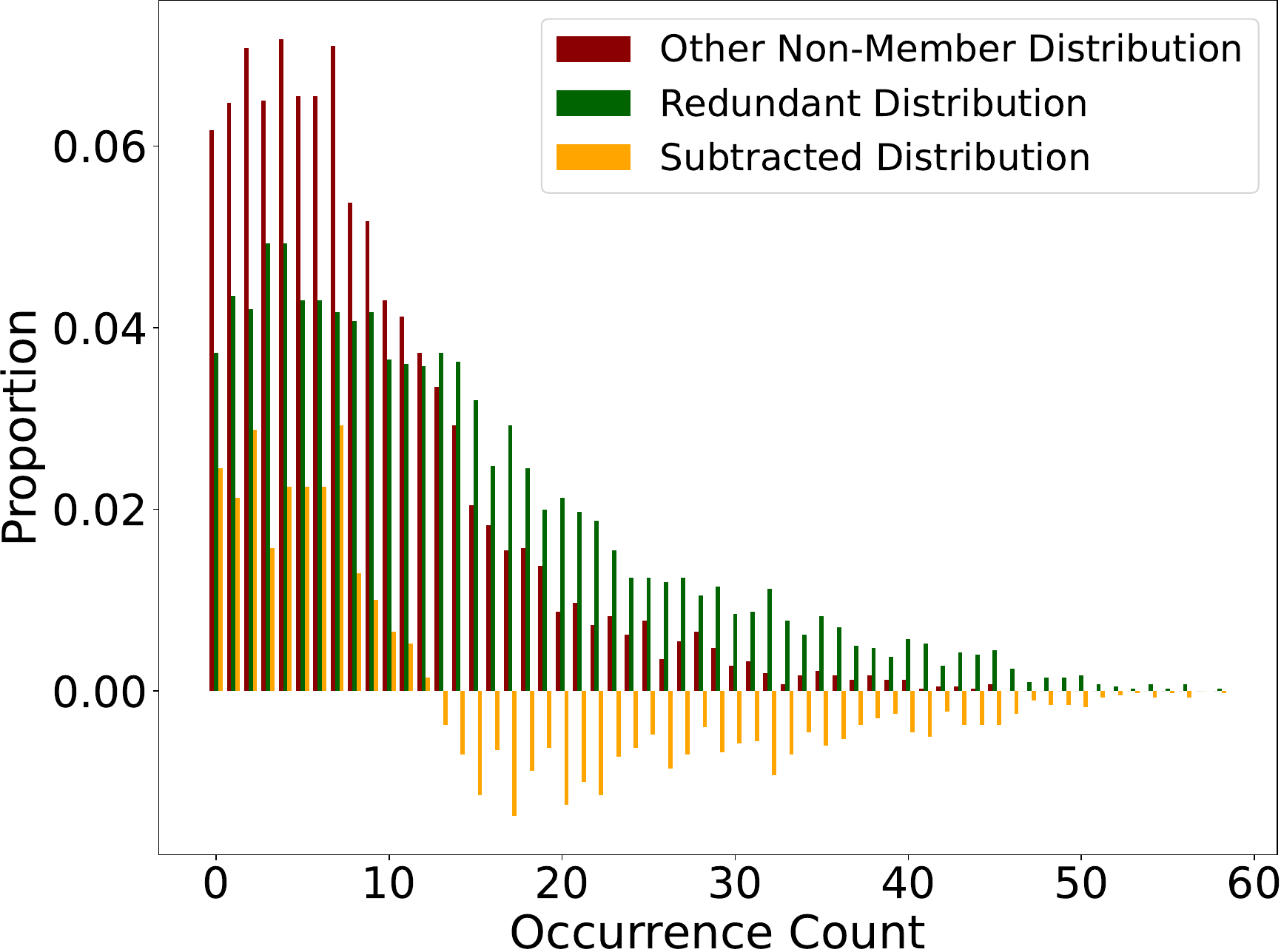}
        \subcaption{Unc. (w/o.)}
        
    \end{minipage}
    \hfill
        \begin{minipage}[b]{.16\linewidth}
        \centering
        \includegraphics[width=\linewidth]{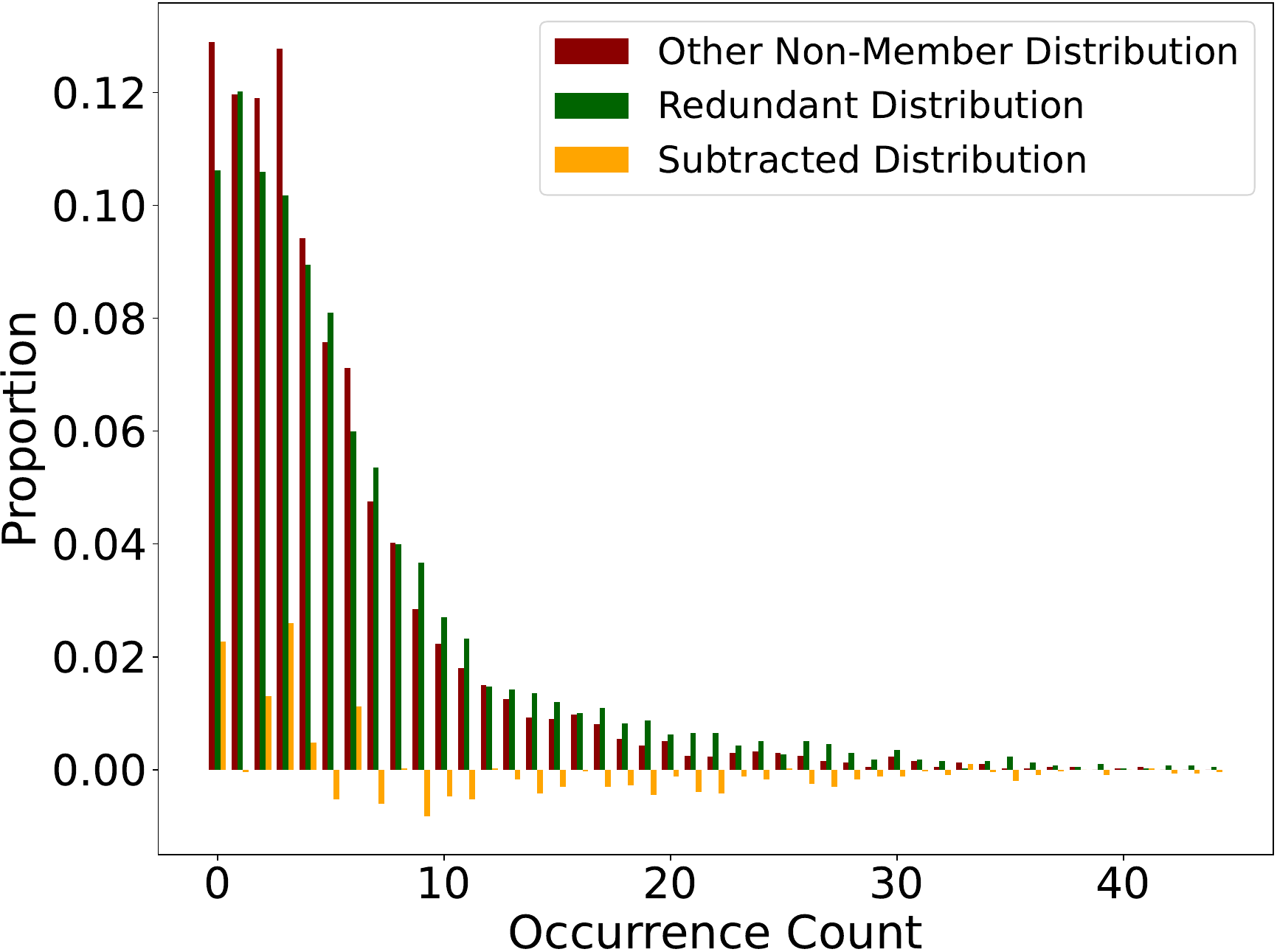}
        \subcaption{Craig (w.)}
    \end{minipage}
    \hfill
    \begin{minipage}[b]{.16\linewidth}
        \centering
        \includegraphics[width=\linewidth]{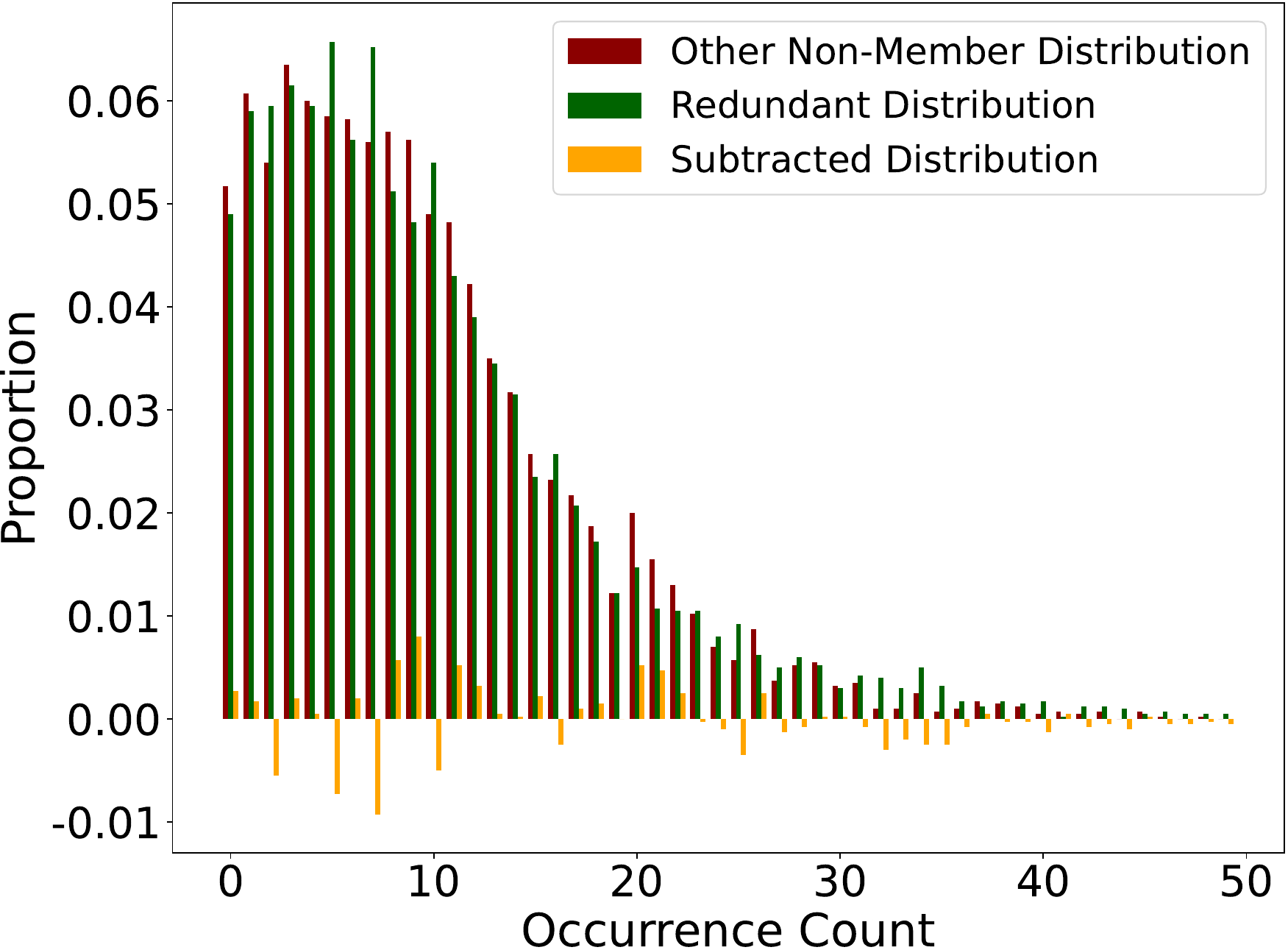}
        \subcaption{DeepF. (w.)}
        
    \end{minipage}
    \hfill
    \begin{minipage}[b]{.16\linewidth}
        \centering
        \includegraphics[width=\linewidth]{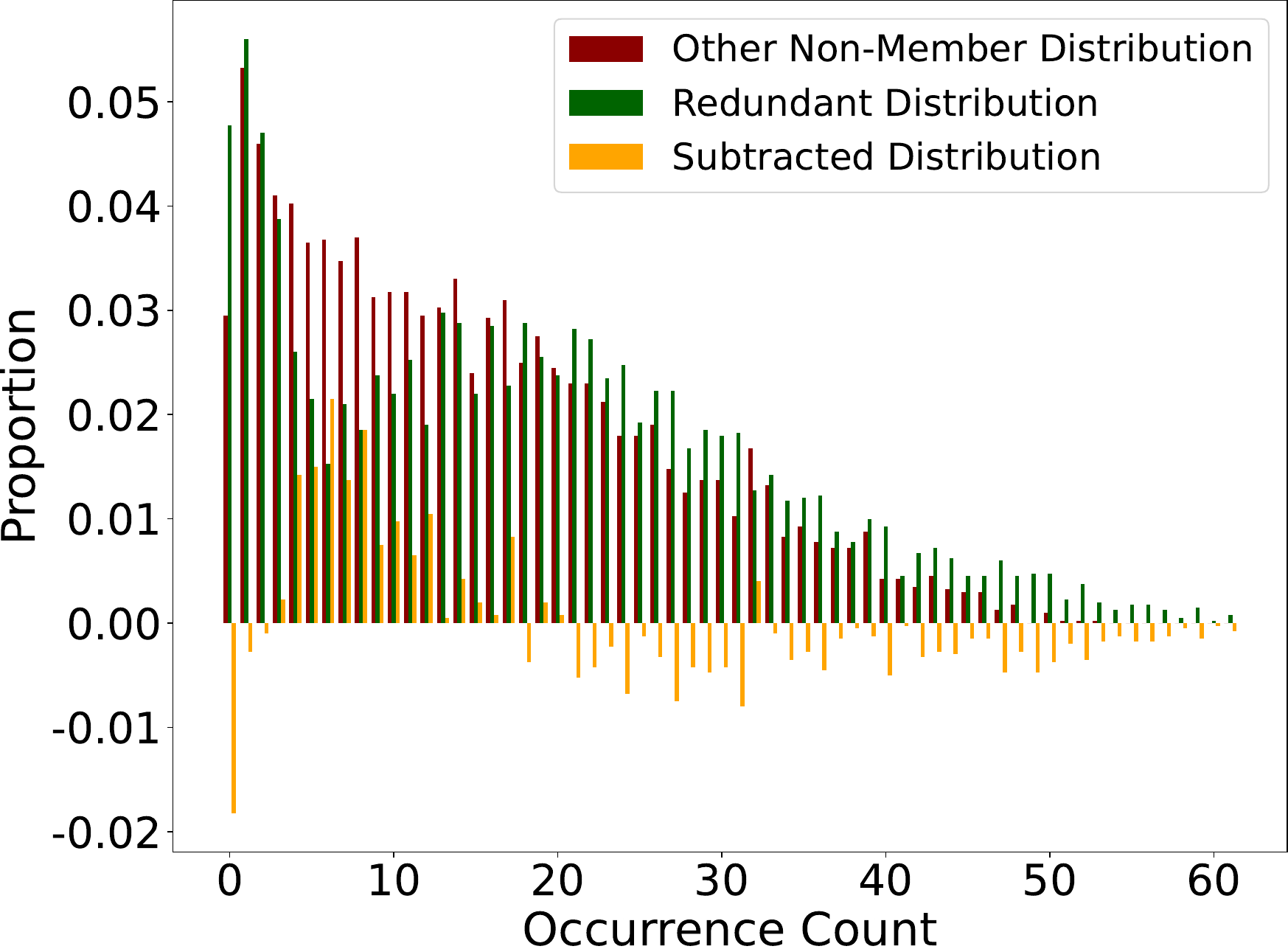}
        \subcaption{Forgt. (w.)}
        
    \end{minipage}%
    \hfill
    \begin{minipage}[b]{.16\linewidth}
        \centering
        \includegraphics[width=\linewidth]{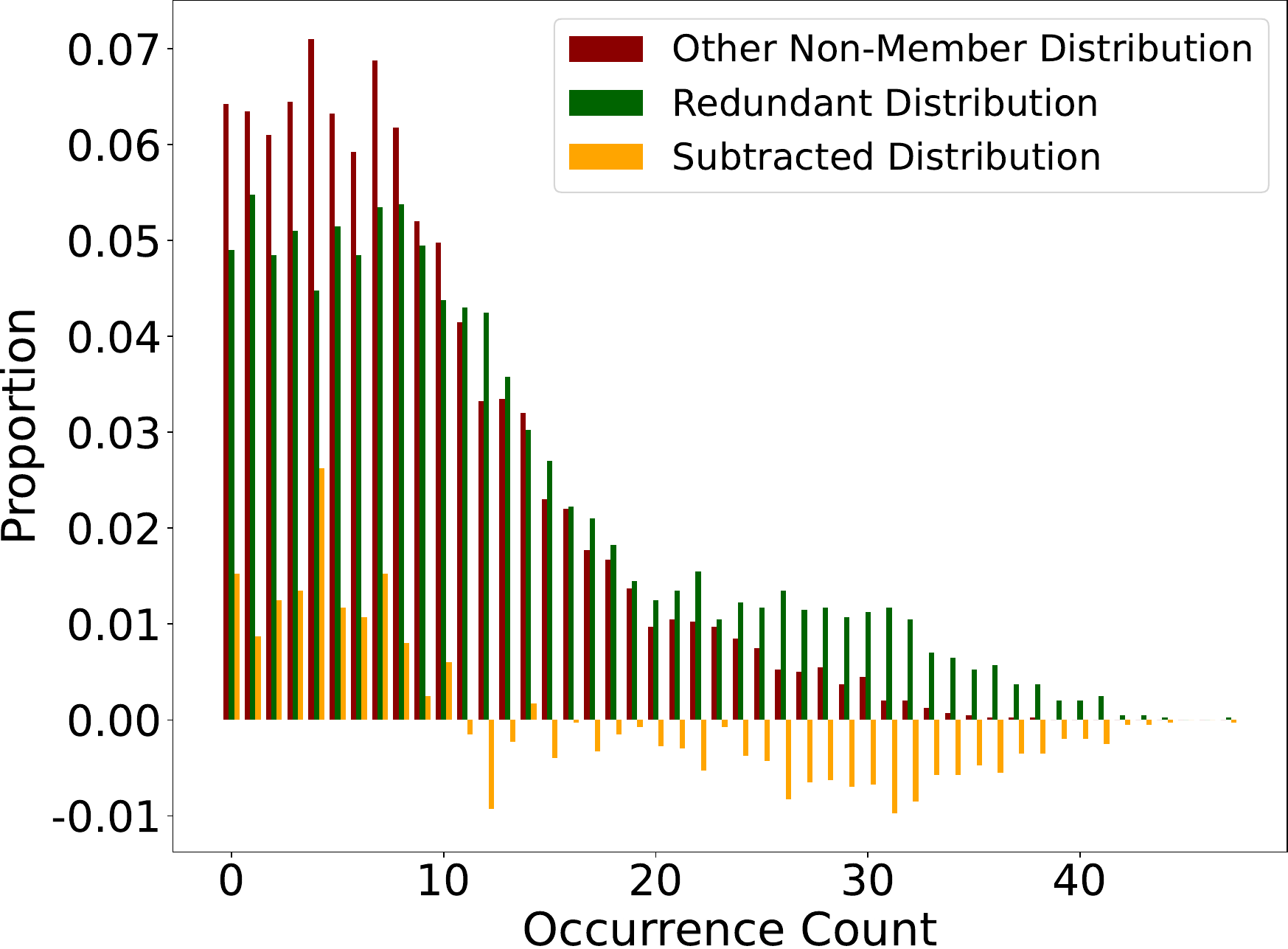}
        \subcaption{Glist. (w.)}
        
    \end{minipage}
    \hfill
    \begin{minipage}[b]{.16\linewidth}
        \centering
        \includegraphics[width=\linewidth]{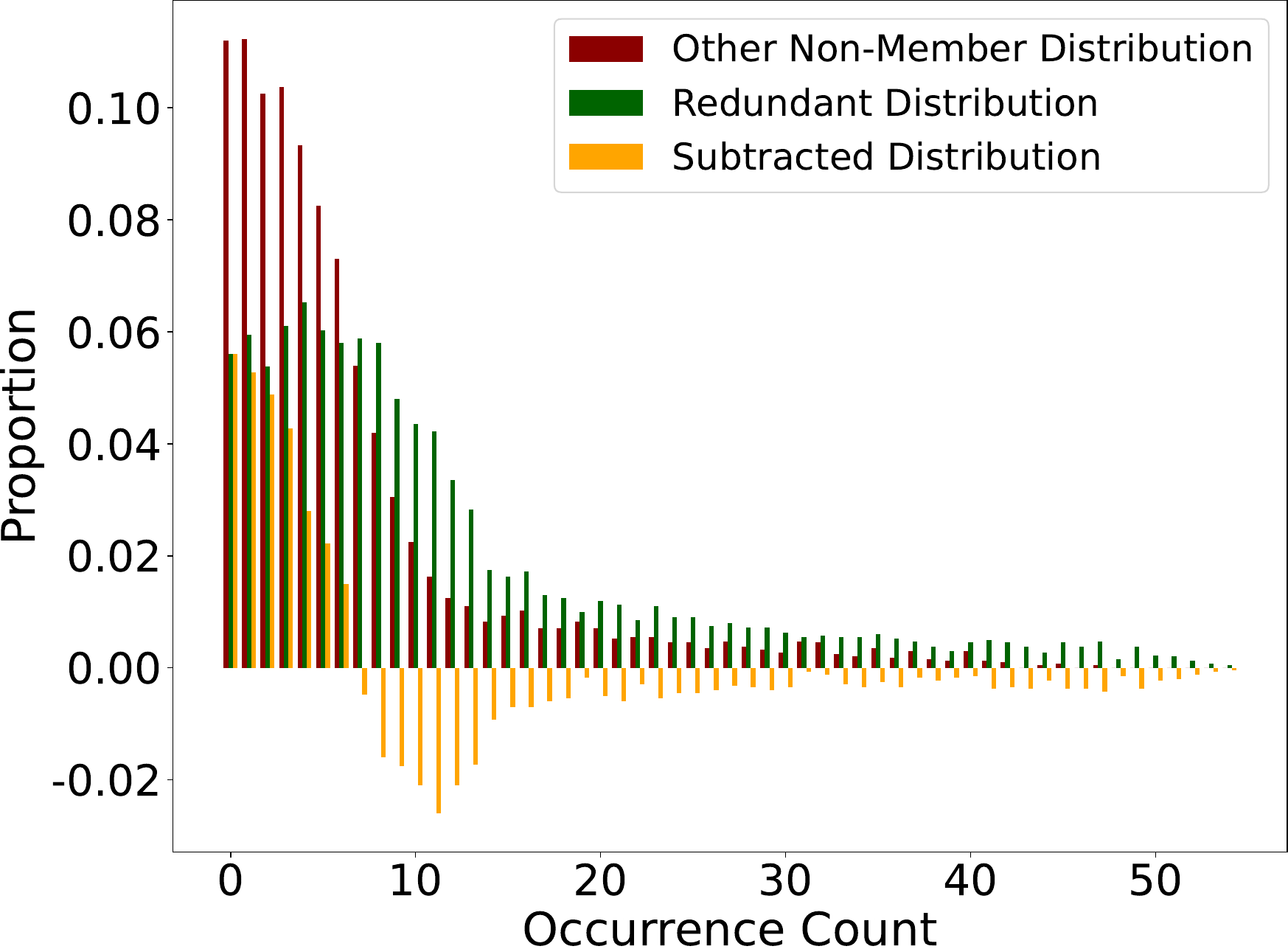}
        \subcaption{GraNd (w.)}
        
    \end{minipage}
    \hfill
    \begin{minipage}[b]{.16\linewidth}
        \centering
        \includegraphics[width=\linewidth]{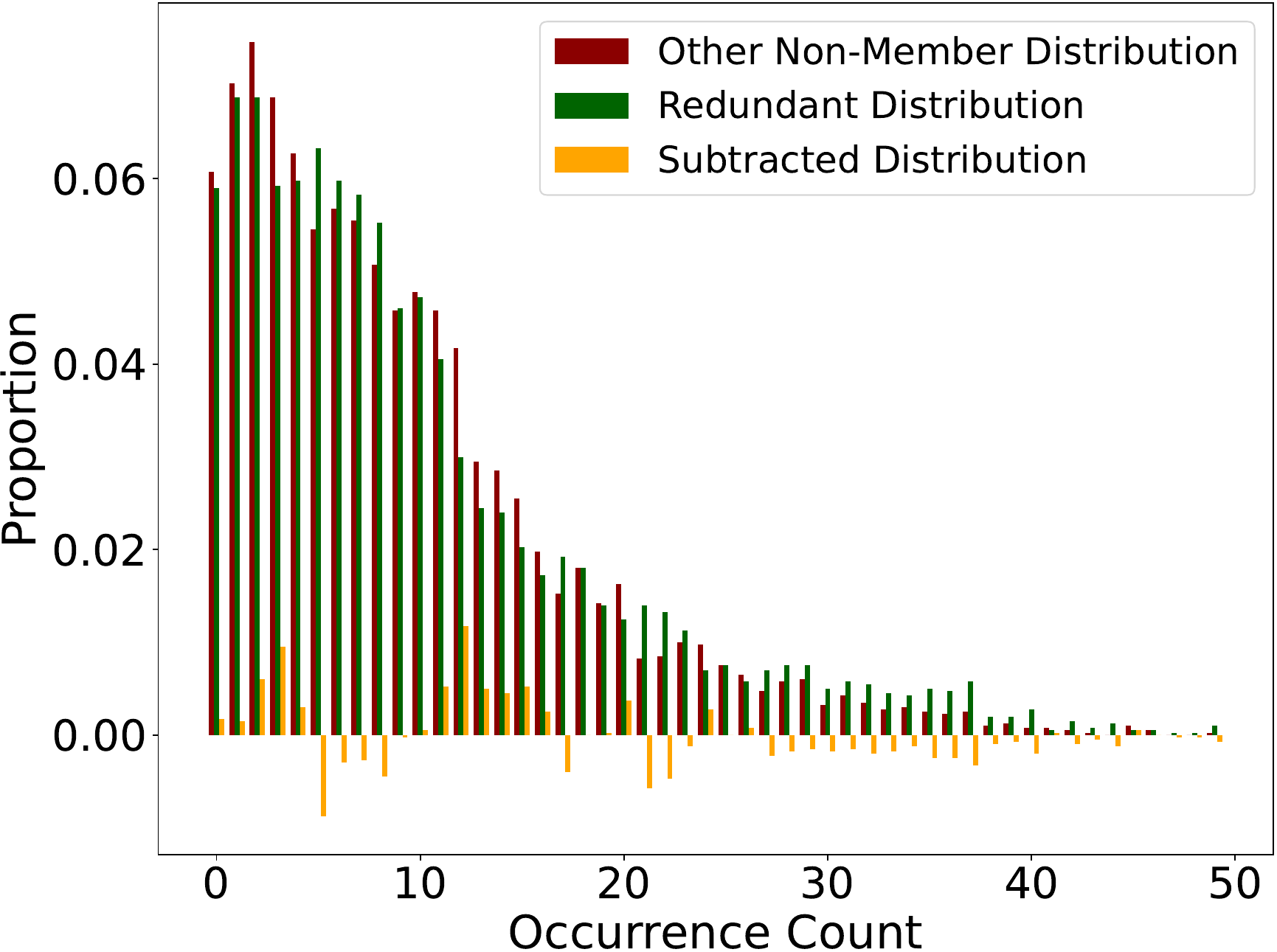}
        \subcaption{Unc. (w.)}
        
    \end{minipage}
    \hfill
        \begin{minipage}[b]{.16\linewidth}
        \centering
        \includegraphics[width=\linewidth]{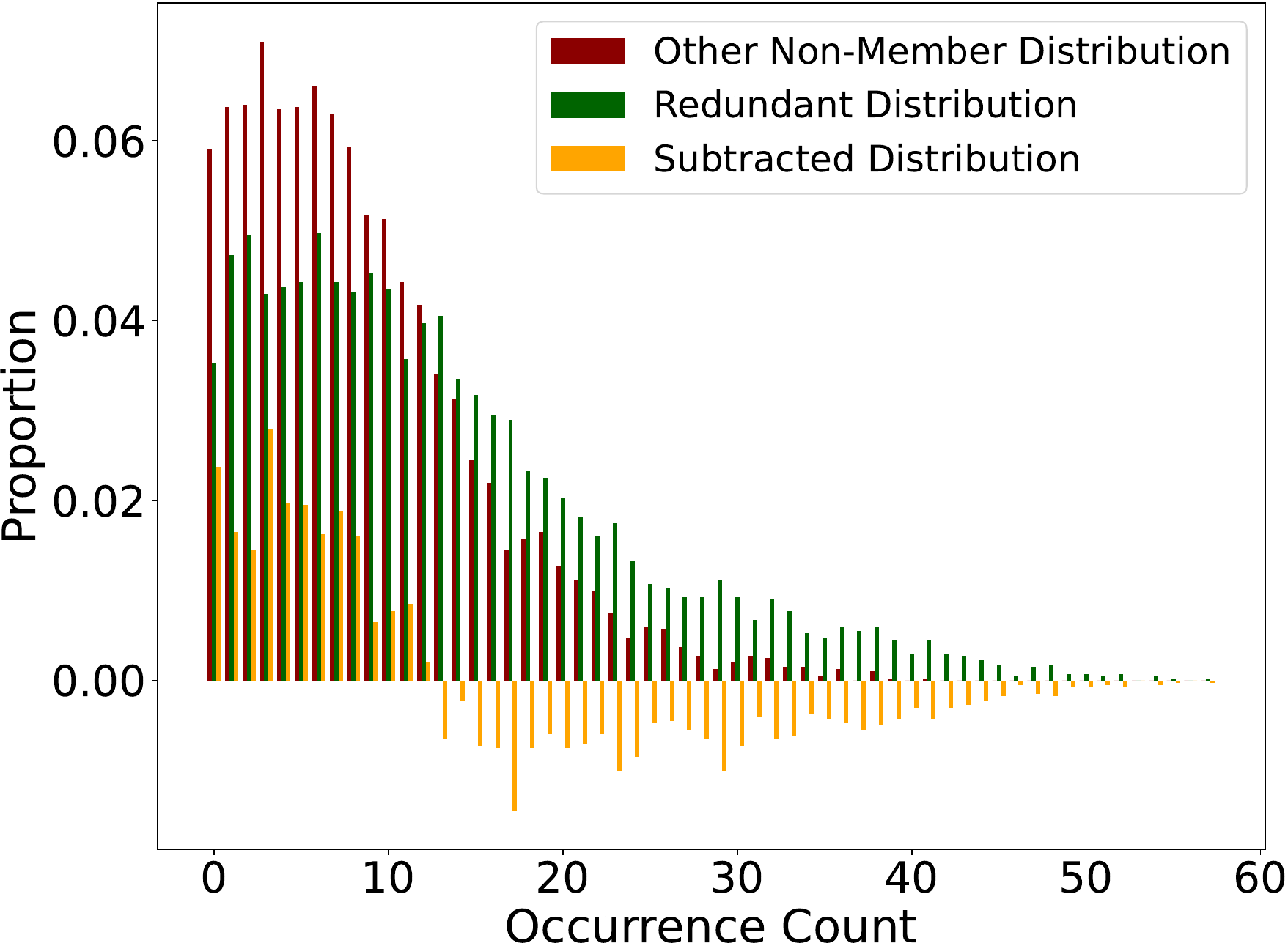}
        \subcaption{Cal (w/o.)}
    \end{minipage}
    \hfill
    \begin{minipage}[b]{.16\linewidth}
        \centering
        \includegraphics[width=\linewidth]{figures/sp_final_Herding_occ_dis_dif_0.8_groupSize_100.pdf}
        \subcaption{Herd. (w/o.)}
        
    \end{minipage}
    \hfill
    \begin{minipage}[b]{.16\linewidth}
        \centering
        \includegraphics[width=\linewidth]{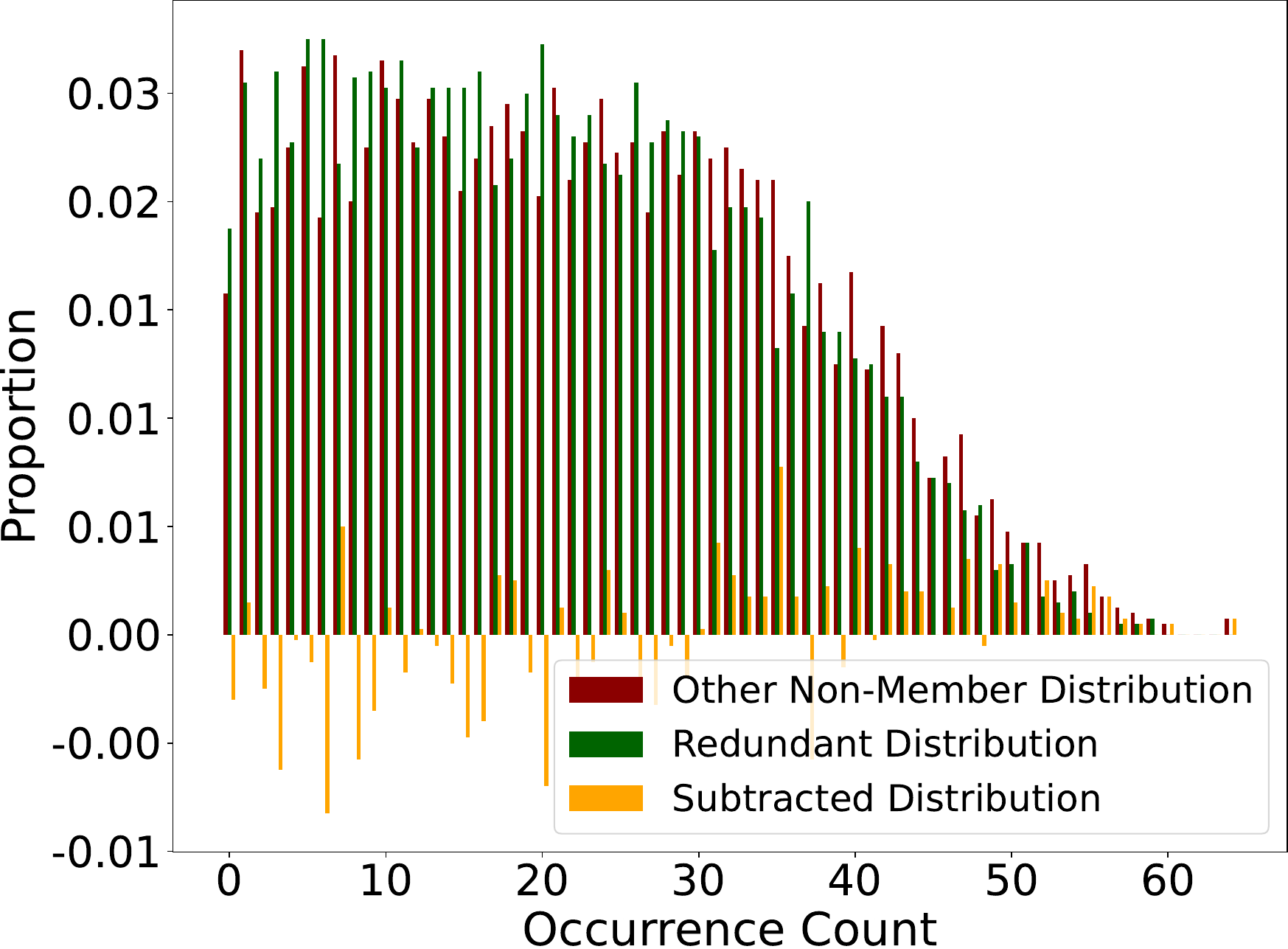}
        \subcaption{G.M. (w/o.)}
        
    \end{minipage}
    \hfill
    \begin{minipage}[b]{.16\linewidth}
        \centering
        \includegraphics[width=\linewidth]{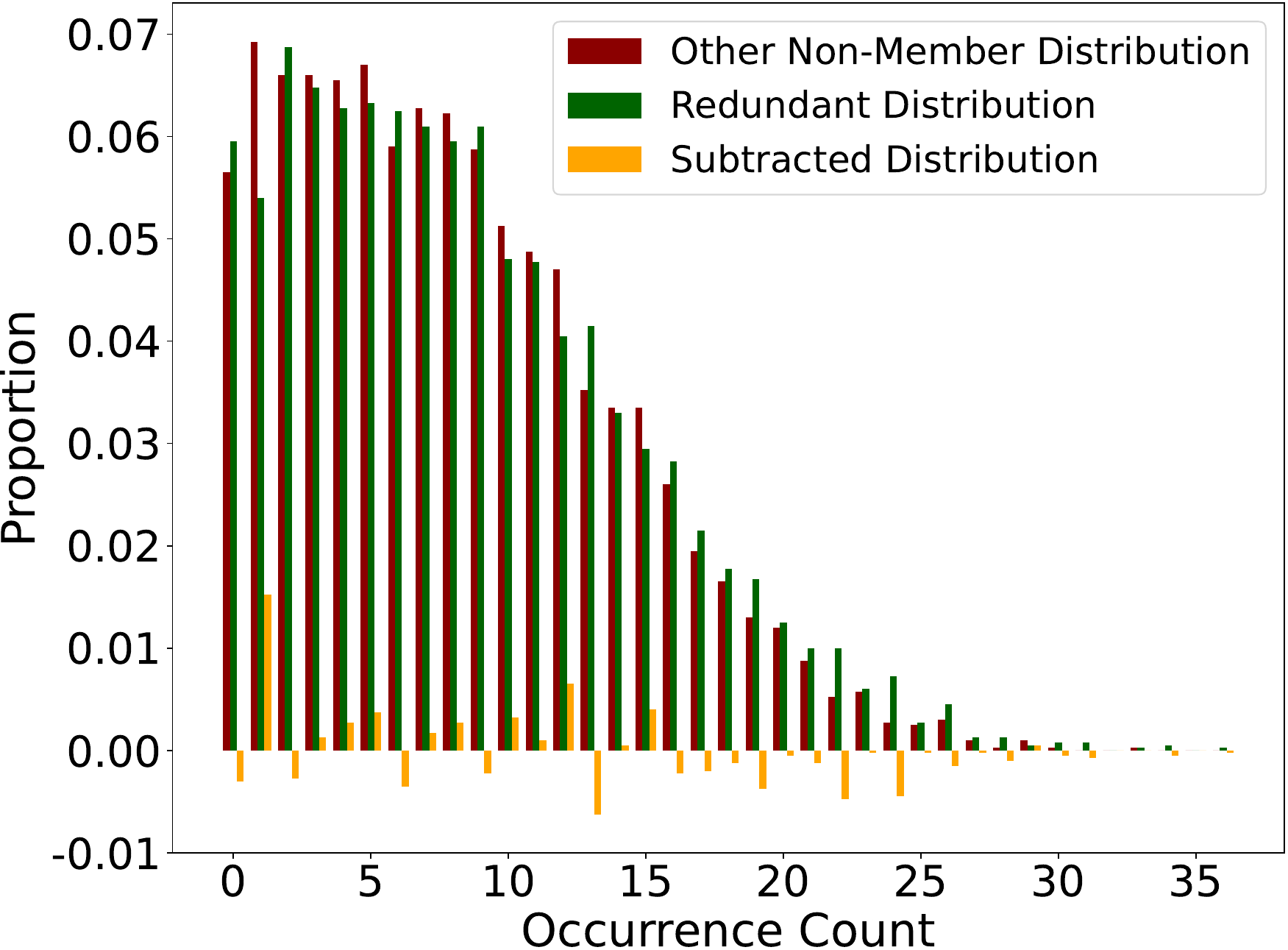}
        \subcaption{SubM. (w/o.)}
        
    \end{minipage}
    \hfill
    \begin{minipage}[b]{.16\linewidth}
        \centering
        \includegraphics[width=\linewidth]{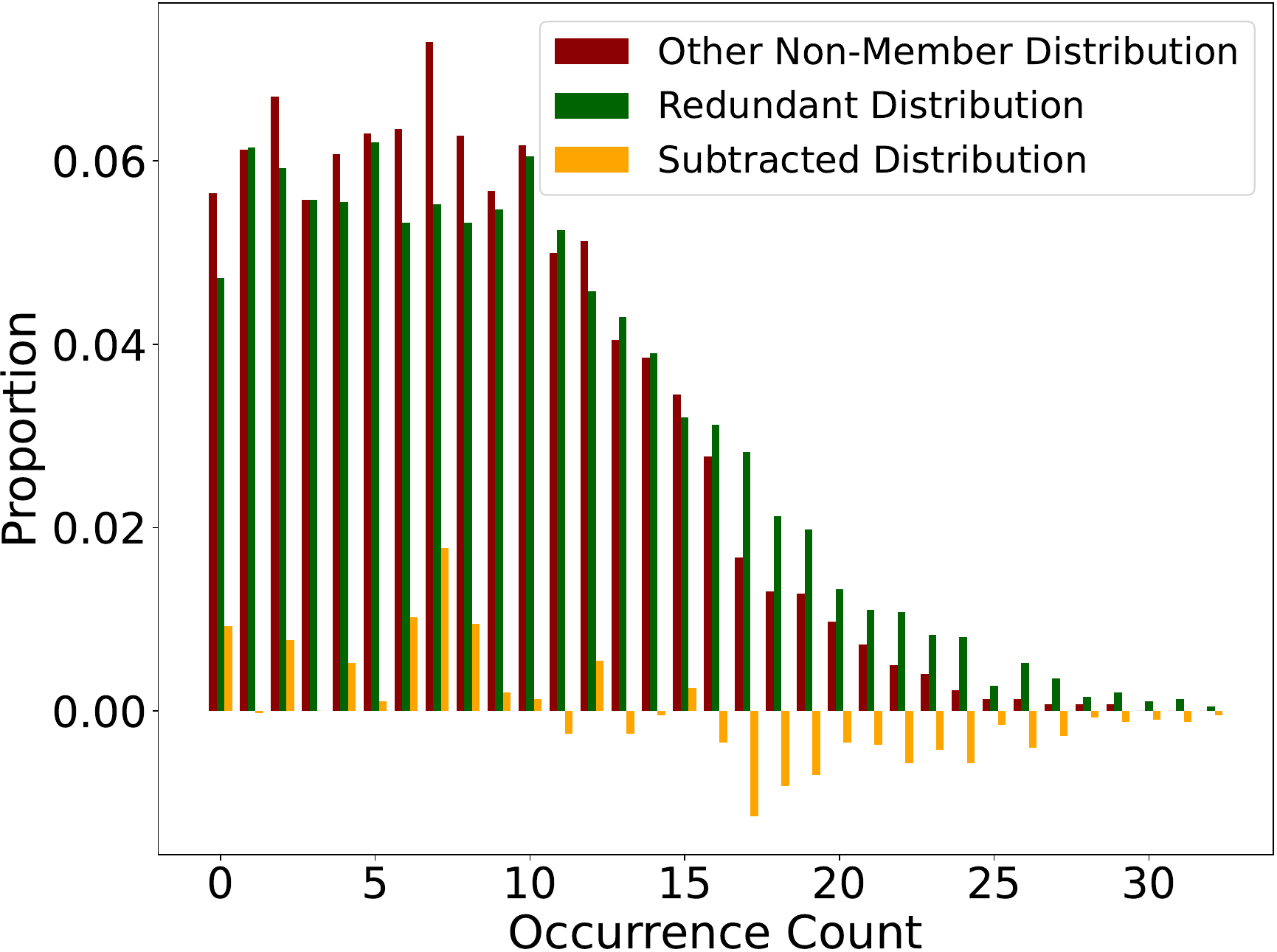}
        \subcaption{kCent. (w/o.)}
        
    \end{minipage}
    \hfill
    \begin{minipage}[b]{.16\linewidth}
        \centering
        \includegraphics[width=\linewidth]{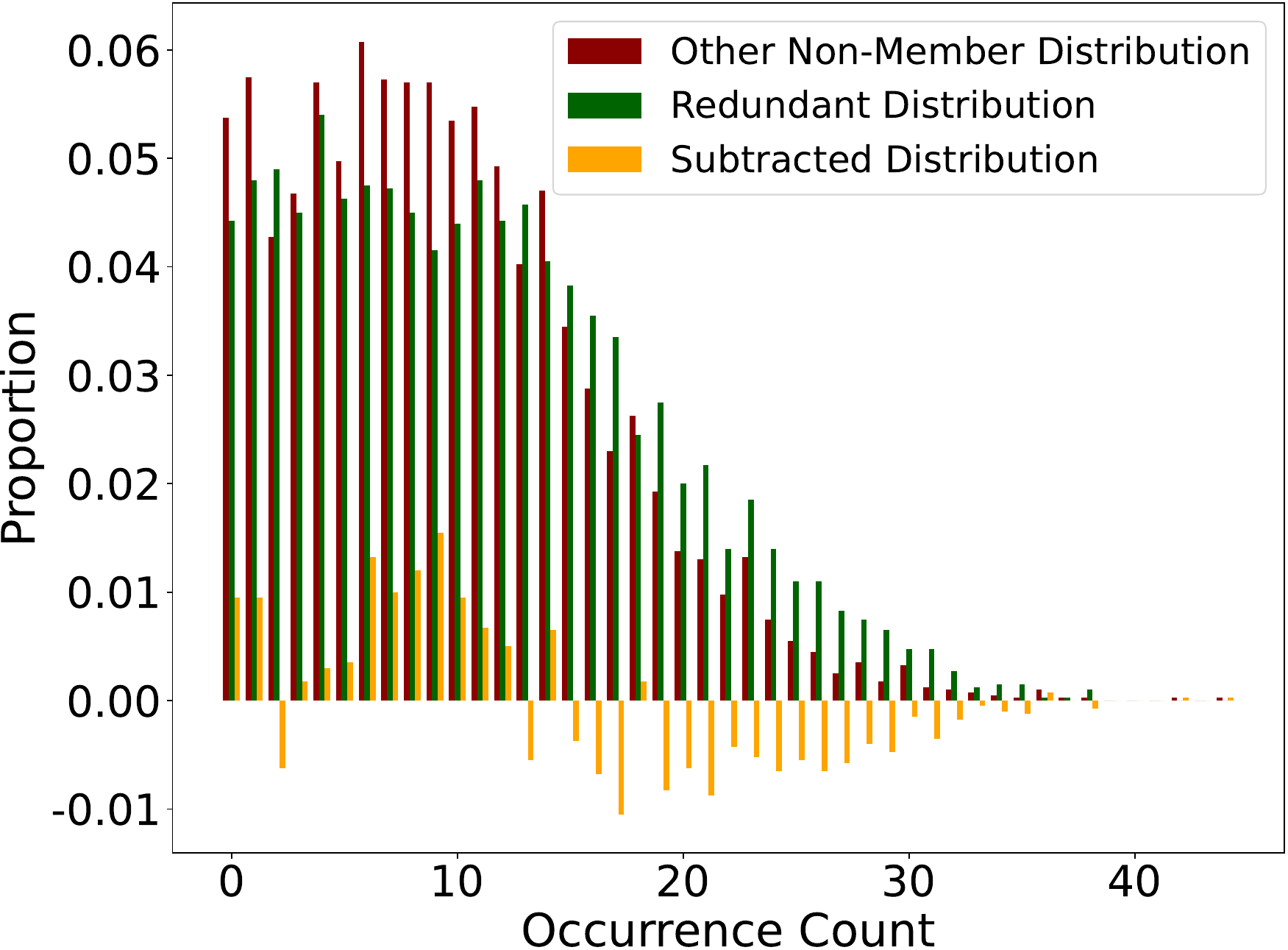}
        \subcaption{Co.Div. (w/o.)}
        
    \end{minipage}
    \hfill
        \begin{minipage}[b]{.16\linewidth}
        \centering
        \includegraphics[width=\linewidth]{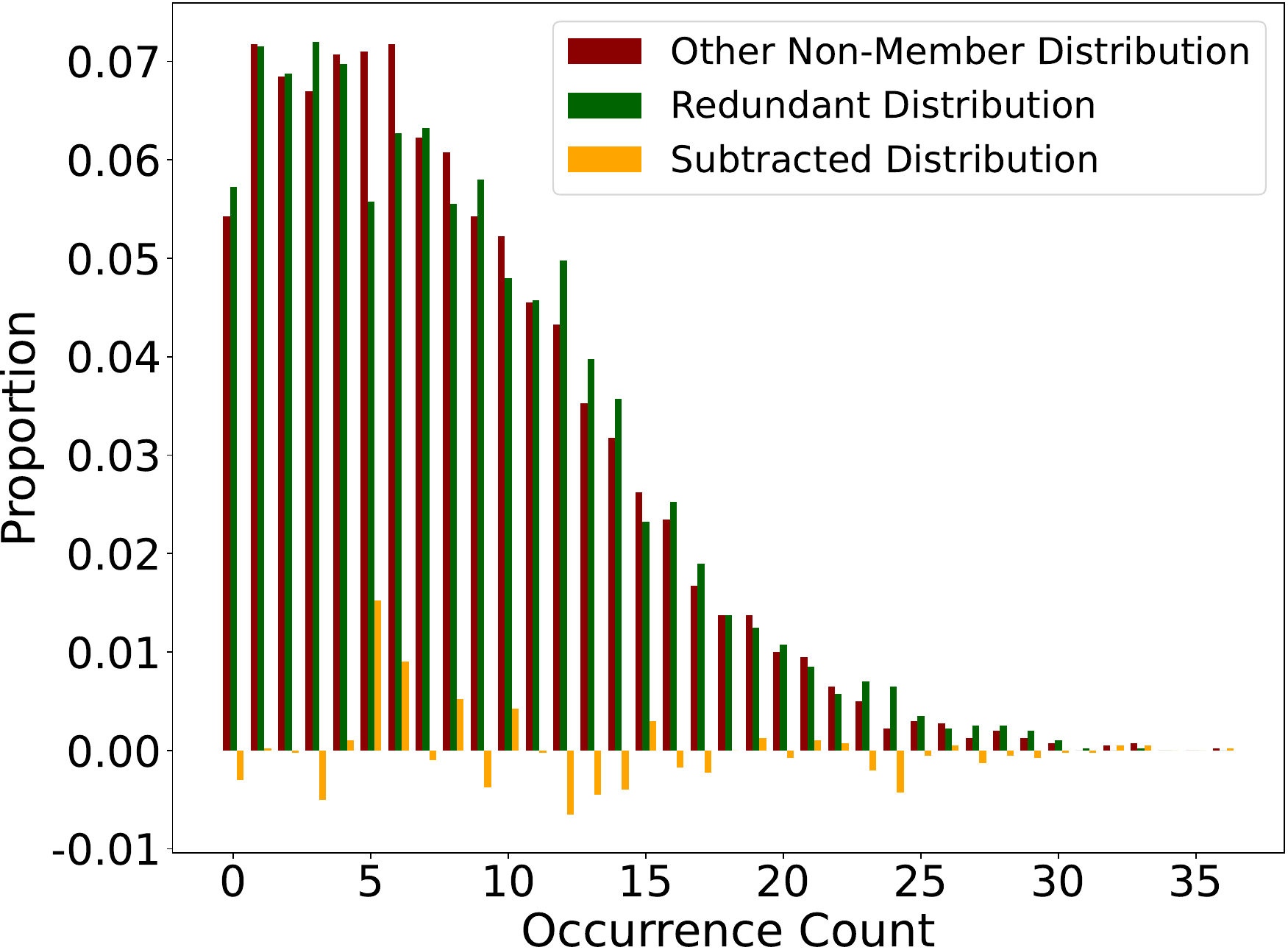}
        \subcaption{Cal (w.)}
    \end{minipage}
    \hfill
    \begin{minipage}[b]{.16\linewidth}
        \centering
        \includegraphics[width=\linewidth]{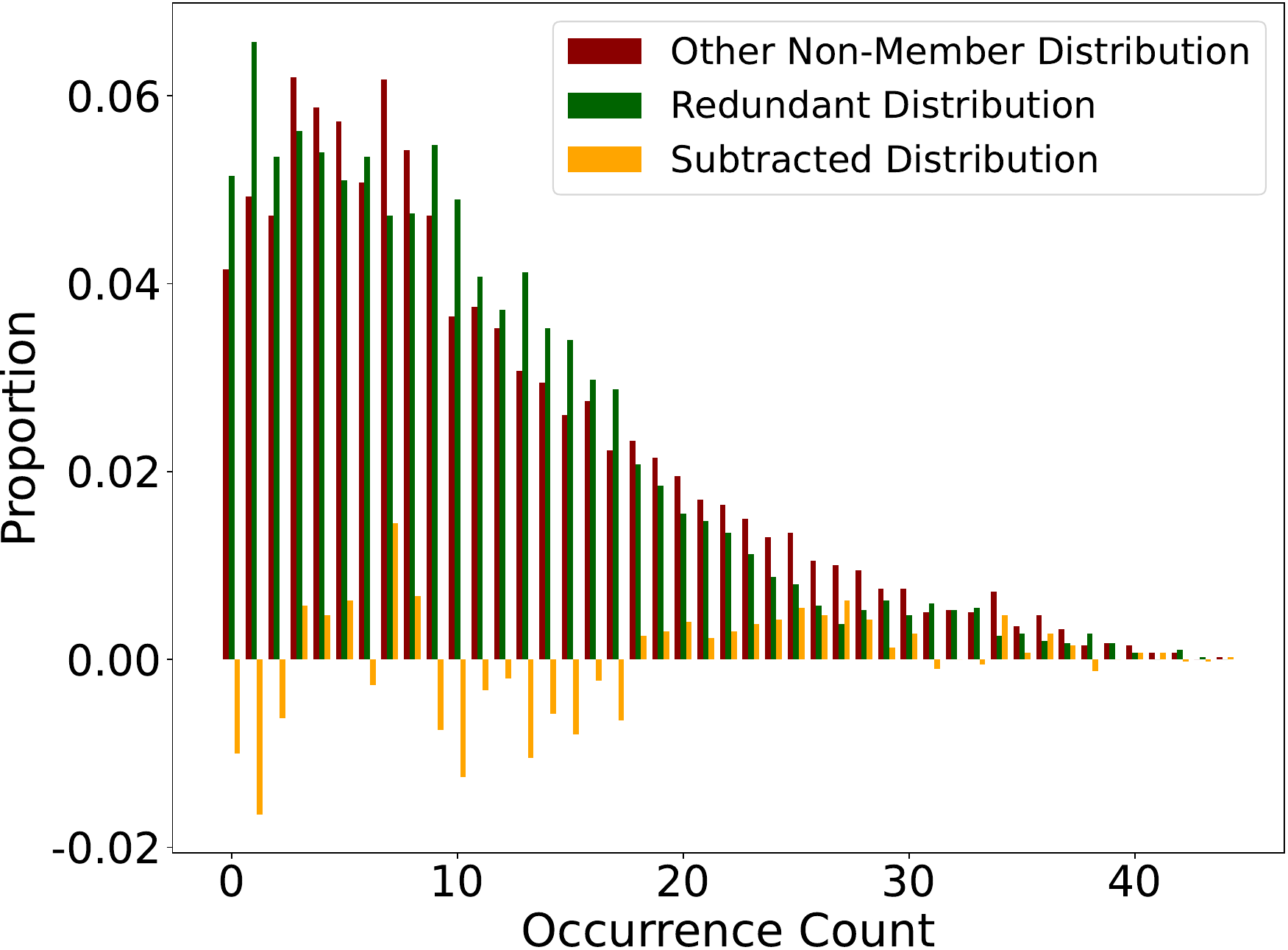}
        \subcaption{Herd. (w.)}
        
    \end{minipage}
    \hfill
    \begin{minipage}[b]{.16\linewidth}
        \centering
        \includegraphics[width=\linewidth]{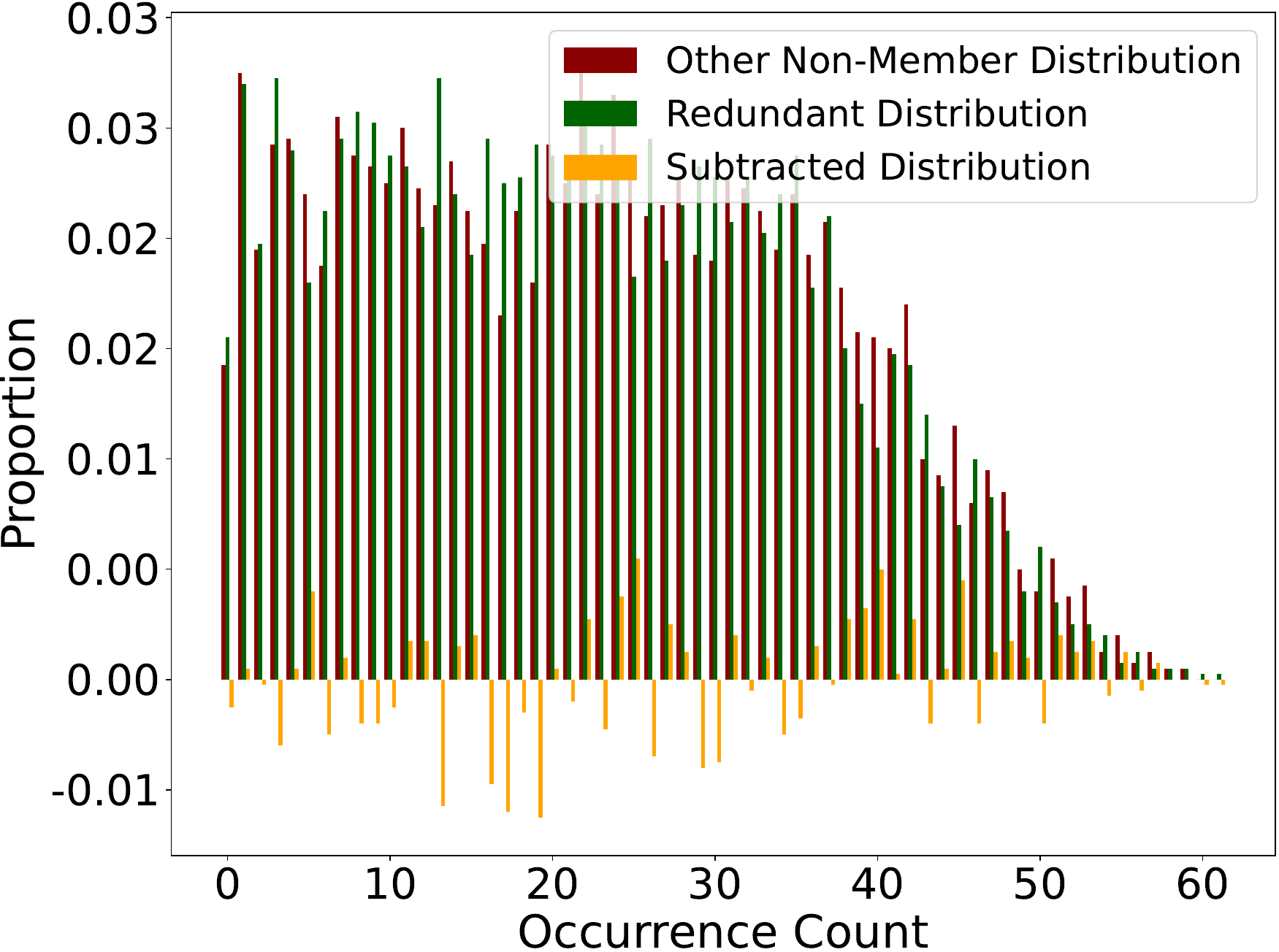}
        \subcaption{G.M. (w.)}
        
    \end{minipage}
    \hfill
    \begin{minipage}[b]{.16\linewidth}
        \centering
        \includegraphics[width=\linewidth]{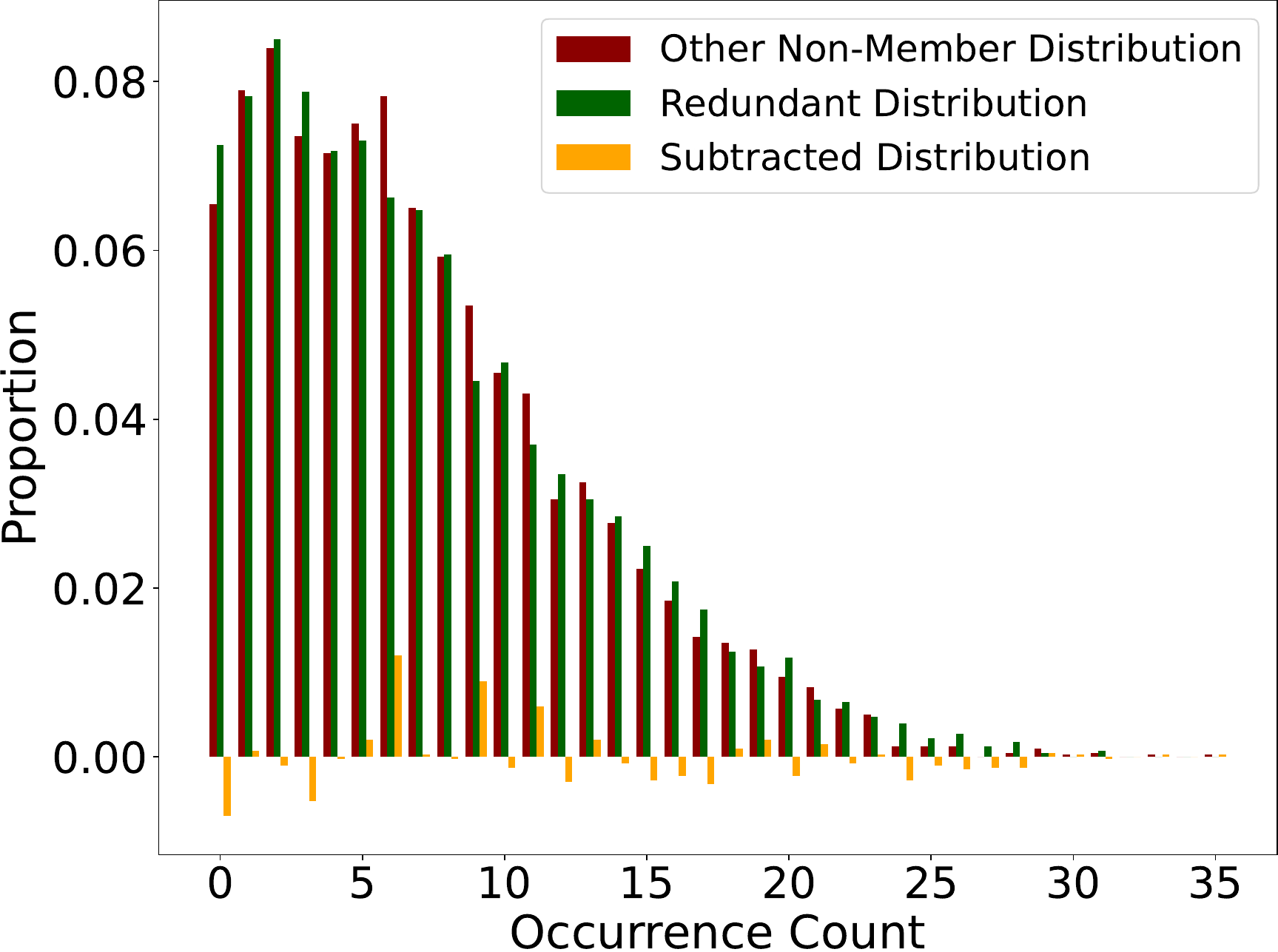}
        \subcaption{SubM (w.)}
        
    \end{minipage}
    \hfill
    \begin{minipage}[b]{.16\linewidth}
        \centering
        \includegraphics[width=\linewidth]{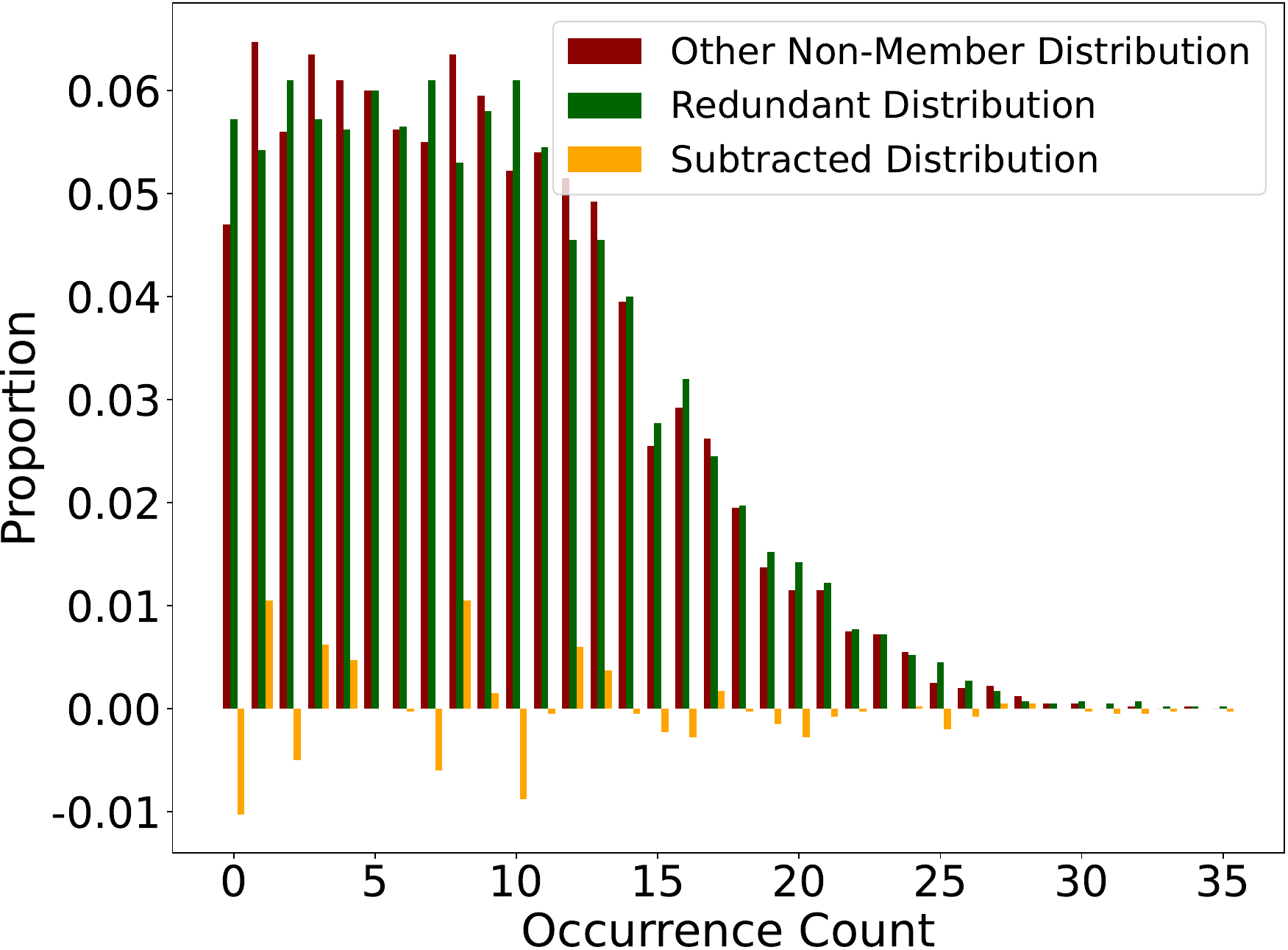}
        \subcaption{kCent. (w.)}
        
    \end{minipage}%
    \hfill
    \begin{minipage}[b]{.16\linewidth}
        \centering
        \includegraphics[width=\linewidth]{figures/sp_final_ContextualDiversity_occ_dis_dif_0.8_groupSize_100_defense.pdf}
        \subcaption{Co.Div. (w.)}
        
    \end{minipage}
    \caption{The occurrence distribution and their difference of all the twelve pruning methods with (w.) and without (w/o.) the proposed ReDoMi defense technique. The pruning fraction is set to 0.8 and dataset is CIFAR10. \(\zeta_{\textrm{v}}\) and \(\zeta_{\textrm{s}}\) are set to 100 and 80, respectively.}
    \vspace{-3mm}
    \label{fig:defense_remain}
\end{figure*}
The first sum is a constant term \(\frac{|Q_{\textrm{v}}|}{\zeta_{\textrm{v}}}\) added \(\frac{|Q_{\textrm{v}}|}{\zeta_{\textrm{v}}}\) times:
   \[
   \sum_{p=0}^{\frac{|Q_{\textrm{v}}|}{\zeta_{\textrm{v}}}-1} \left(\frac{|Q_{\textrm{v}}|}{\zeta_{\textrm{v}}}\right) = \left(\frac{|Q_{\textrm{v}}|}{\zeta_{\textrm{v}}}\right)^{2}.
   \]
   The second sum is the sum of the first \(\frac{|Q_{\textrm{v}}|}{\zeta_{\textrm{v}}}\) integers:
   \[
   \sum_{p=0}^{\frac{|Q_{\textrm{v}}|}{\zeta_{\textrm{v}}}-1} p = \frac{\left(\frac{|Q_{\textrm{v}}|}{\zeta_{\textrm{v}}} - 1\right) \left(\frac{|Q_{\textrm{v}}|}{\zeta_{\textrm{v}}}\right)}{2}.
   \]
   Now combine the two parts:
   \[
   \eta = \left(\frac{|Q_{\textrm{v}}|}{\zeta_{\textrm{v}}}\right)^{2} - \frac{1}{2}\left(\frac{|Q_{\textrm{v}}|}{\zeta_{\textrm{v}}}\right) \left(\frac{|Q_{\textrm{v}}|}{\zeta_{\textrm{v}}} - 1\right).
   \]
Therefore, the final simplified expression for \( \eta \) is:
\[
\eta = \frac{1}{2} \left( \frac{|Q_{\textrm{v}}|}{\zeta_{\textrm{v}}} \right) \left( \frac{|Q_{\textrm{v}}|}{\zeta_{\textrm{v}}} + 1 \right).
\]

\begin{figure*}[h!]
    \centering
    \begin{minipage}[b]{.19\linewidth}
        \centering
        \includegraphics[width=\linewidth]{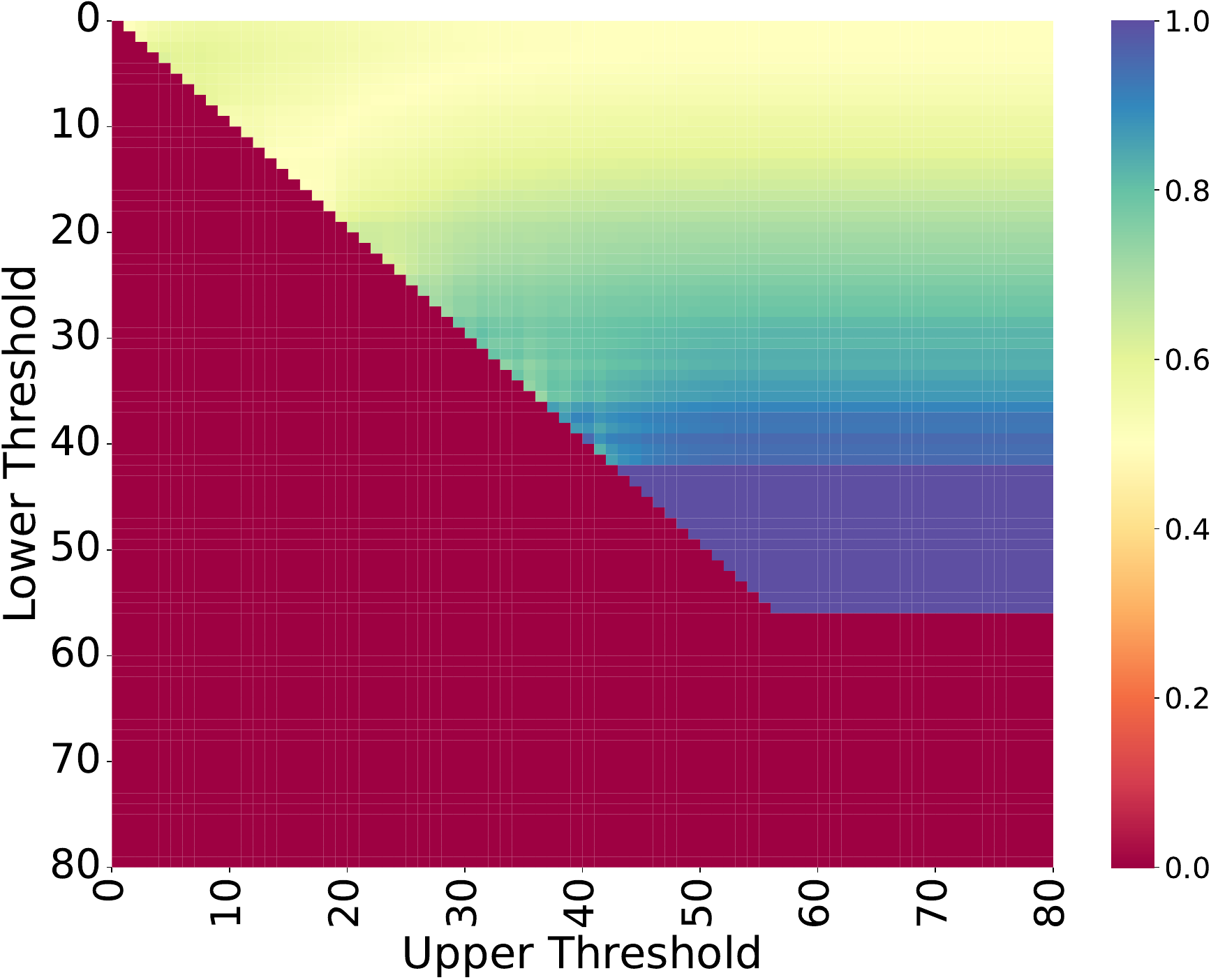}
        \subcaption{ArraDis-Forgt.}
        %\label{fig:mutag-big_stu}
    \end{minipage}%
    \hfill
    \begin{minipage}[b]{.19\linewidth}
        \centering
        \includegraphics[width=\linewidth]{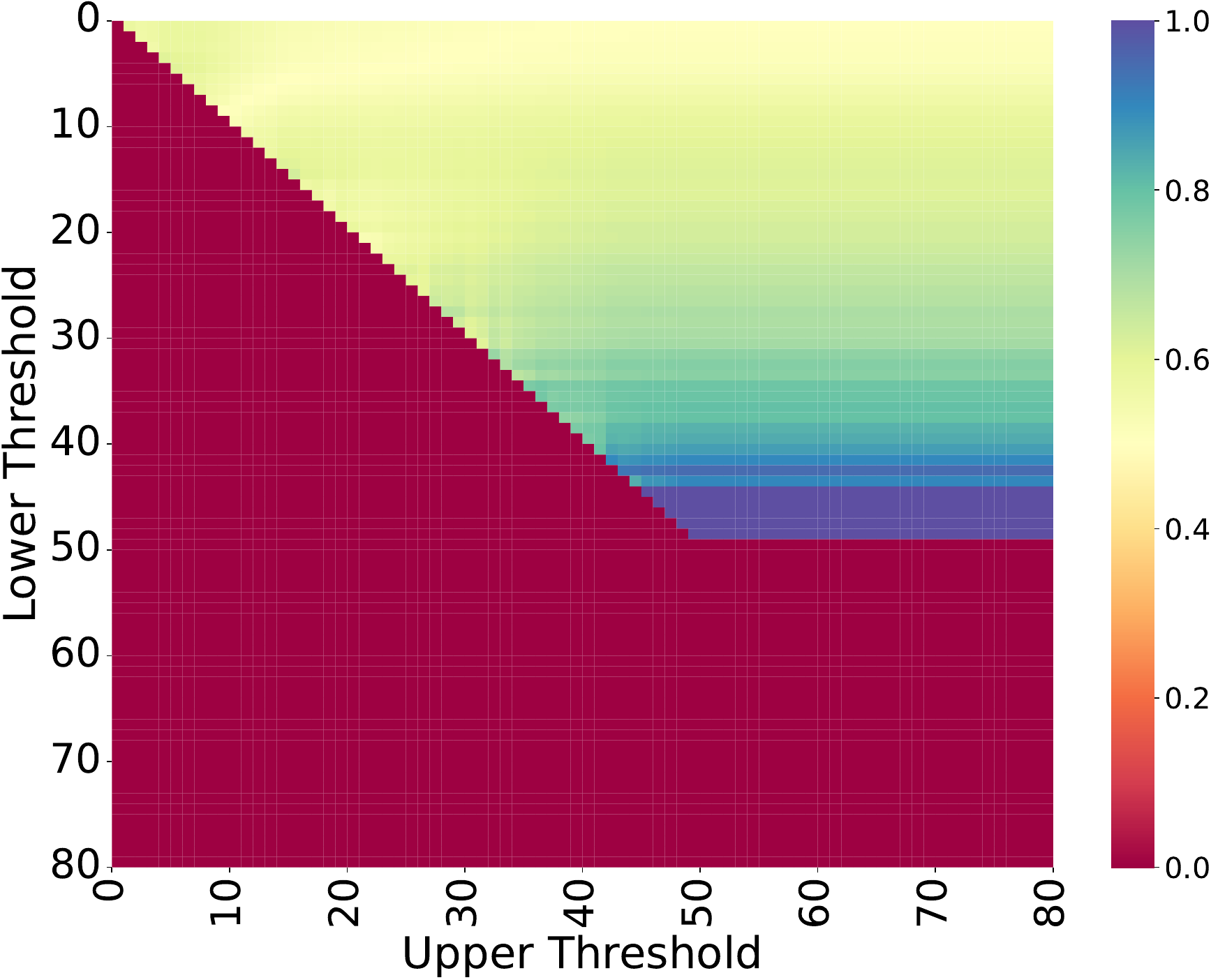}
        \subcaption{ArraDis-Glister}
        
    \end{minipage}%
    \hfill
    \begin{minipage}[b]{.19\linewidth}
        \centering
        \includegraphics[width=\linewidth]{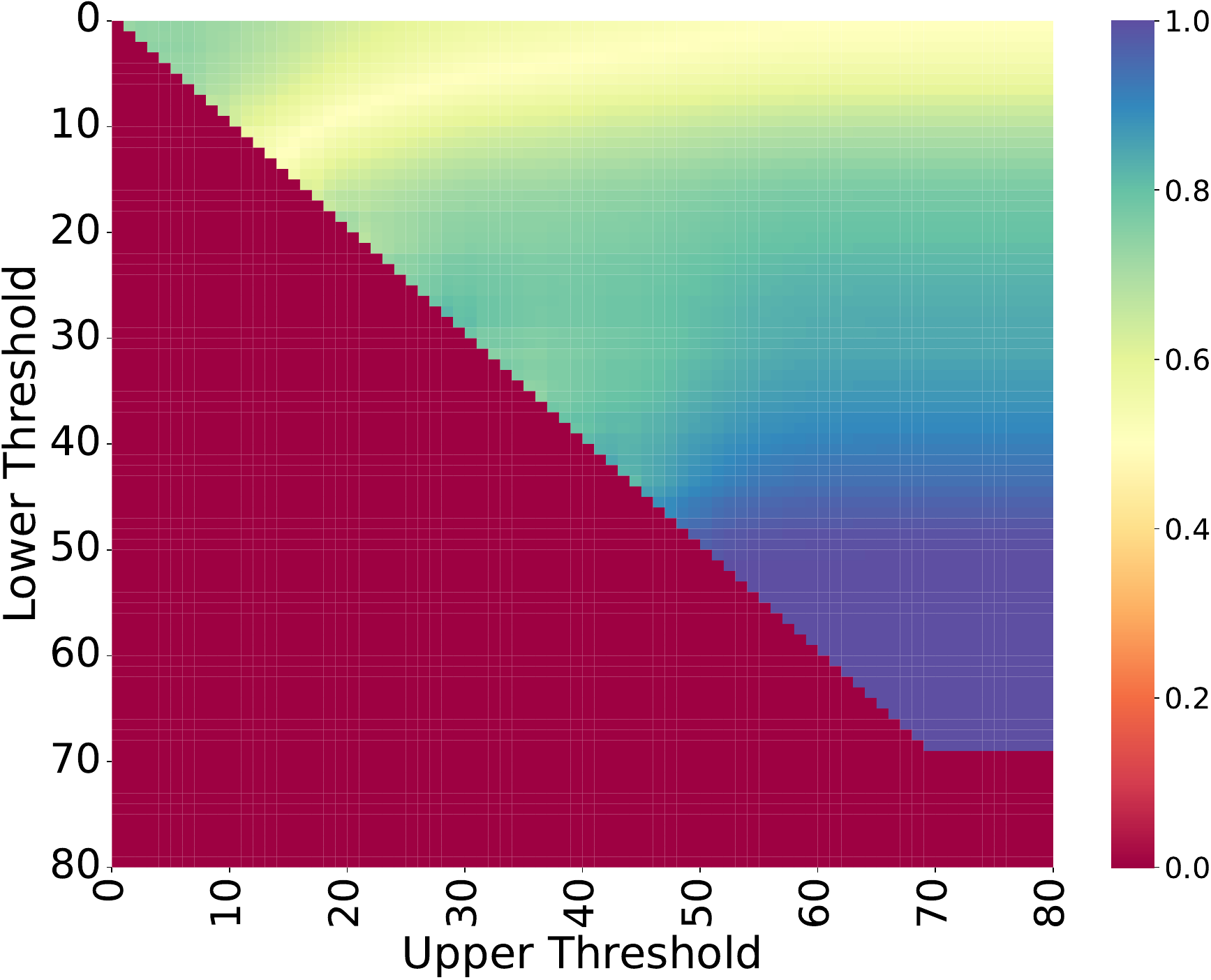}
        \subcaption{ArraDis-GraNd}
        
    \end{minipage}%
    \hfill
    \begin{minipage}[b]{.19\linewidth}
        \centering
        \includegraphics[width=\linewidth]{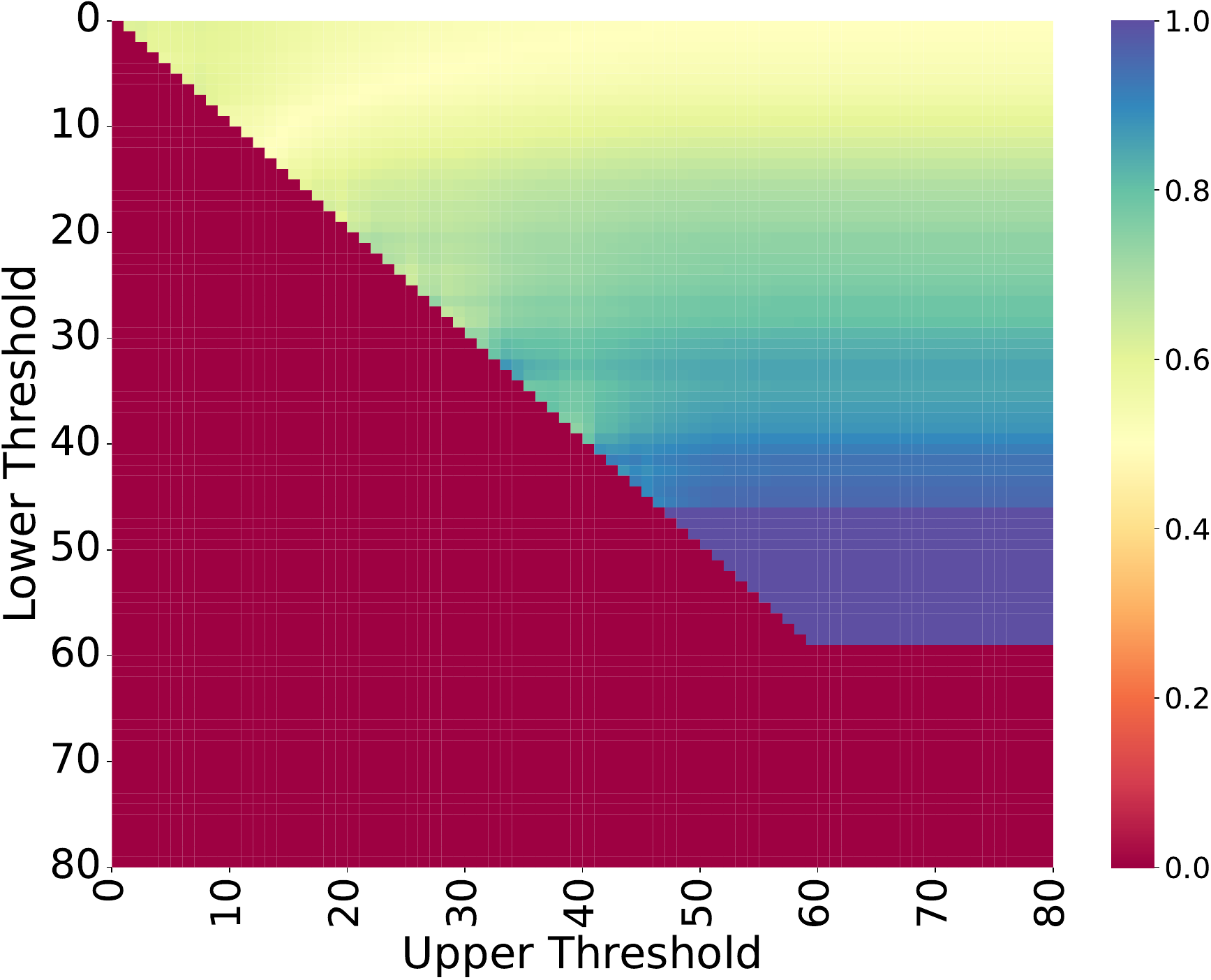}
        \subcaption{ArraDis-Unc.}
        
    \end{minipage}%
    \hfill
    \begin{minipage}[b]{0.19\linewidth}
        \centering
        \includegraphics[width=\linewidth]{figures/sp_final_Cal_differences_interval_0.8.pdf}
        \subcaption{ArraDis-Cal}
    \end{minipage}%
    \hfill
    \begin{minipage}[b]{.19\linewidth}
        \centering
        \includegraphics[width=\linewidth]{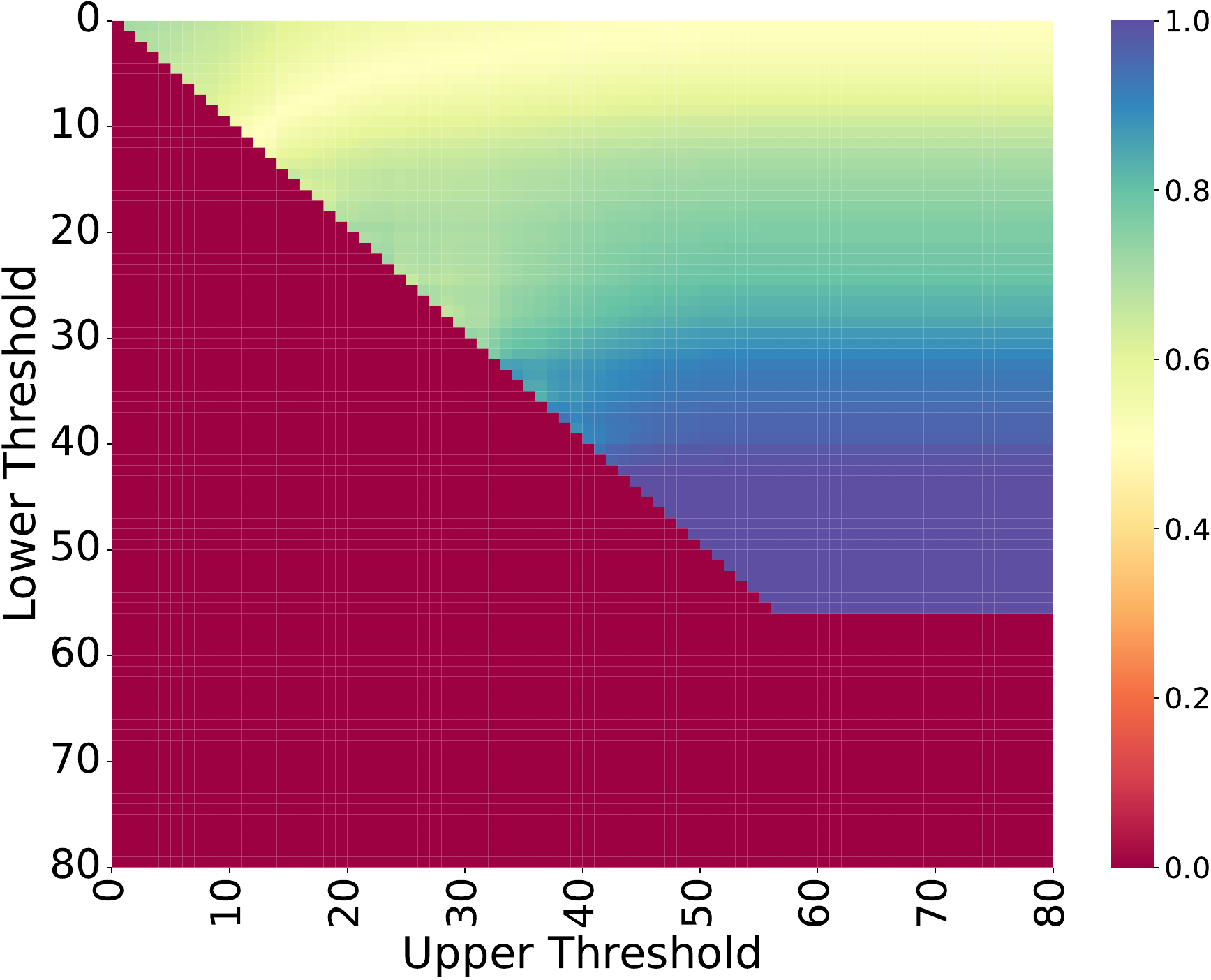}
        \subcaption{ArraDis-Herd.}
        
    \end{minipage}%
    \hfill
    \begin{minipage}[b]{.19\linewidth}
        \centering
        \includegraphics[width=\linewidth]{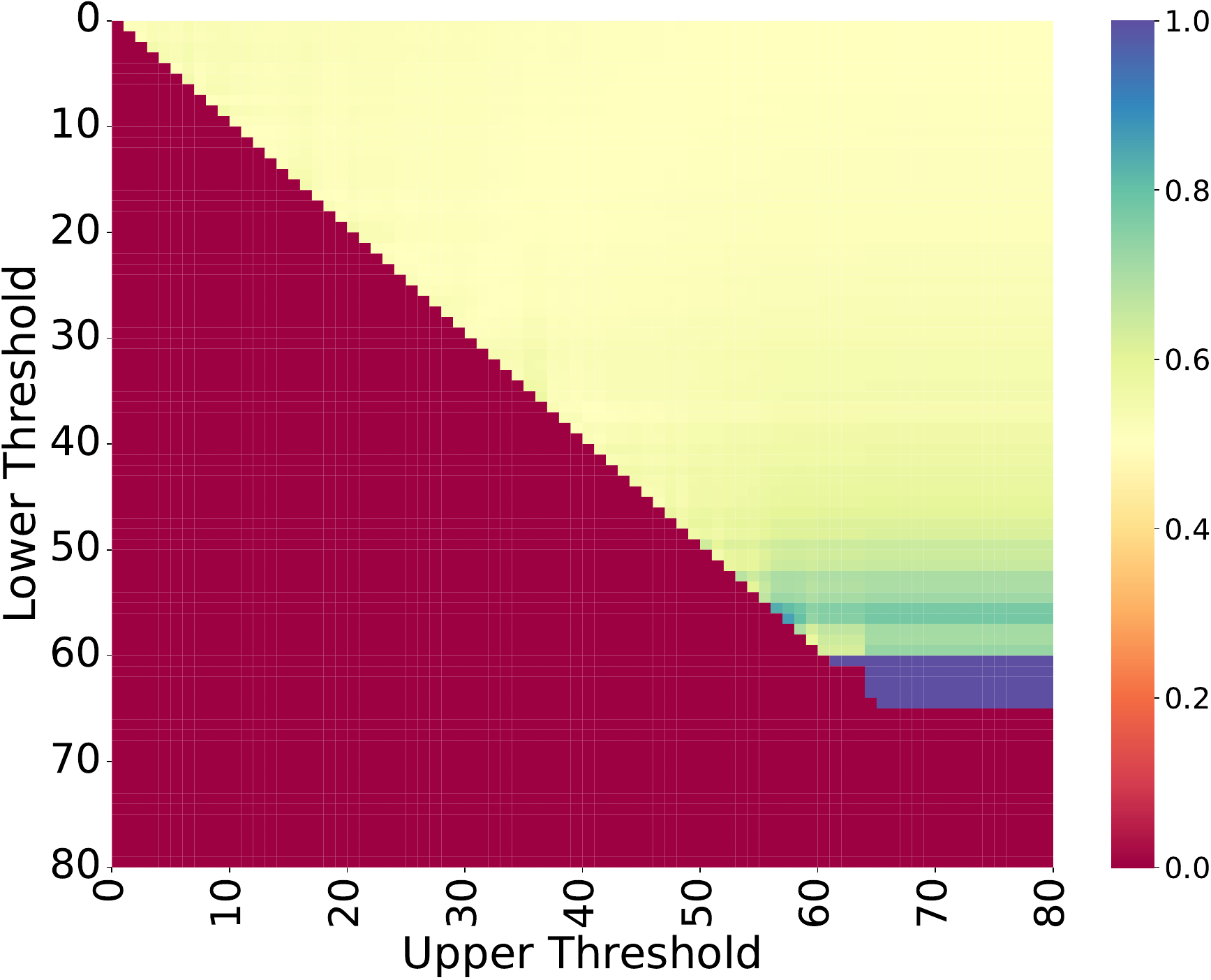}
        \subcaption{ArraDis-G.M.}
        
    \end{minipage}%
    \hfill
    \begin{minipage}[b]{.19\linewidth}
        \centering
        \includegraphics[width=\linewidth]{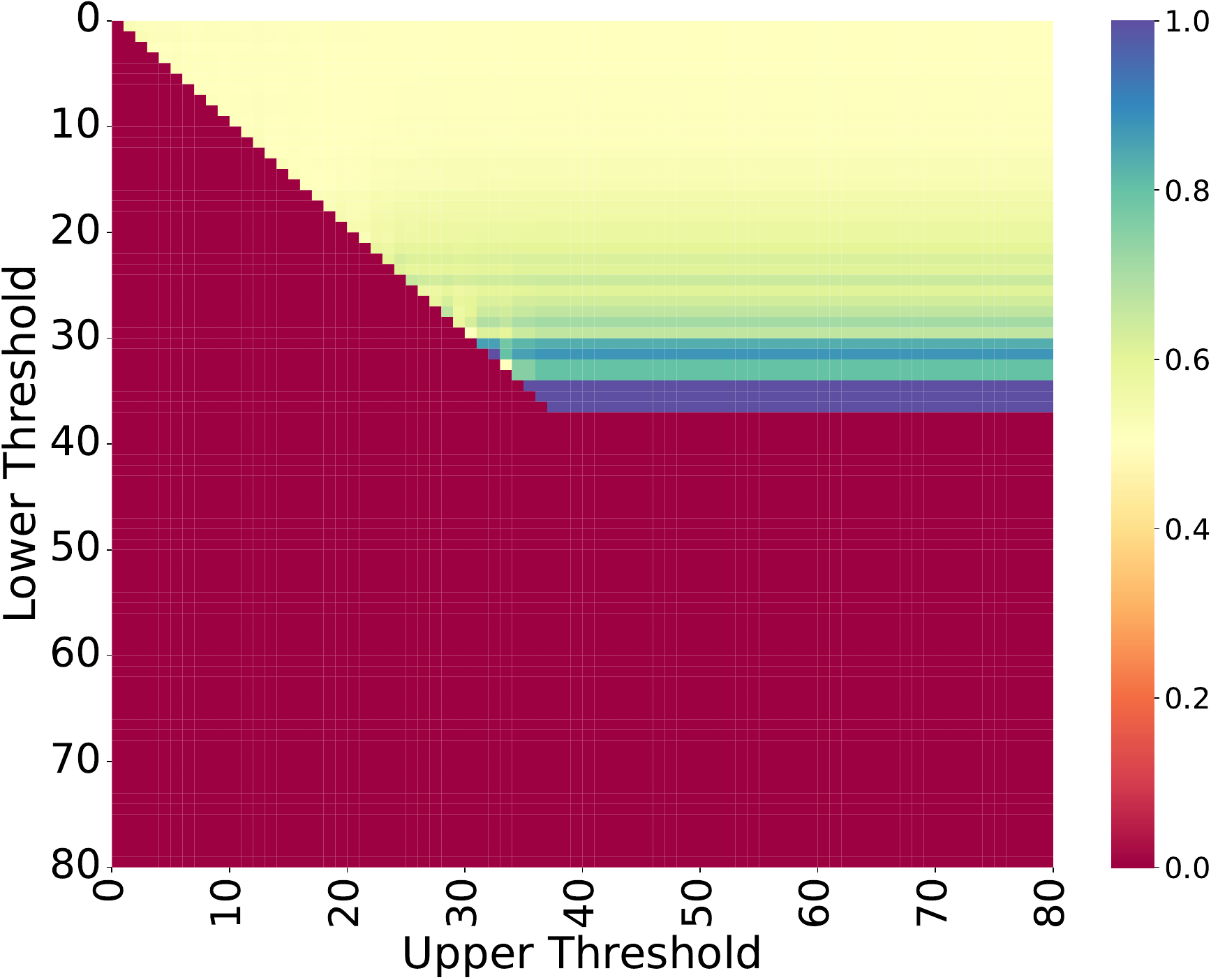}
        \subcaption{ArraDis-SubM.}
        
    \end{minipage}%
    \hfill
    \begin{minipage}[b]{.19\linewidth}
        \centering
        \includegraphics[width=\linewidth]{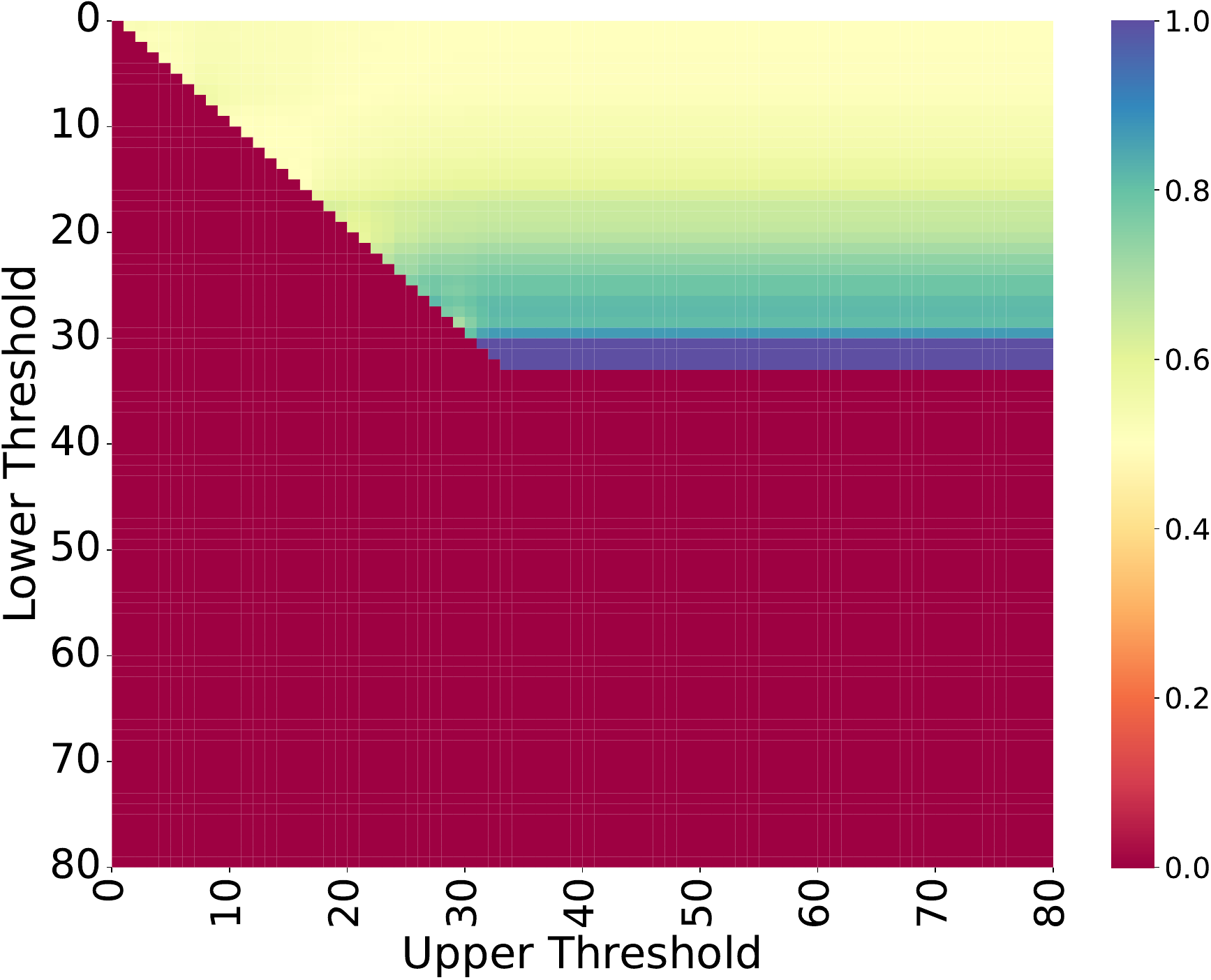}
        \subcaption{ArraDis-kCent.}
        
    \end{minipage}%
    \hfill
    \begin{minipage}[b]{.19\linewidth}
        \centering
        \includegraphics[width=\linewidth]{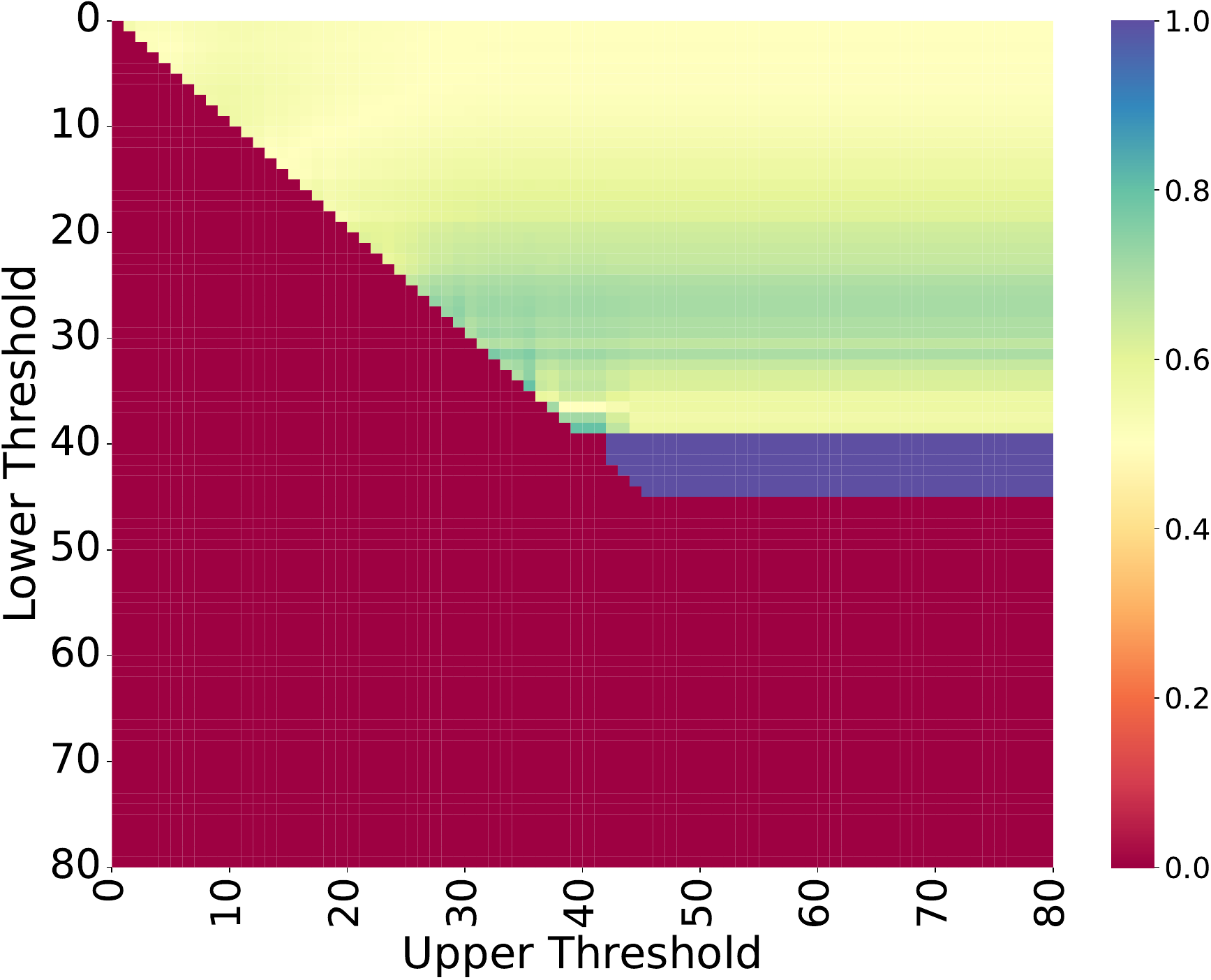}
        \subcaption{ArraDis-Co.Div.}
        
    \end{minipage}%
    \caption{The ArraDis results of the remaining six pruning methods. The pruning fraction is set to 0.8 and dataset is CIFAR10. \(\zeta_{\textrm{v}}\) and \(\zeta_{\textrm{s}}\) are set to 100 and 80, respectively.}
    \label{fig:brim_remain}
\end{figure*}
\begin{figure*}[h]
    \centering
    \begin{minipage}[b]{0.24\linewidth}
        \centering
        \includegraphics[width=\linewidth]{figures/cal_cifar10_att2_line.pdf}
        \subcaption{MNIST-Cal}
        %\label{fig:mutag-big_stu}
    \end{minipage}%
    \hfill
    \begin{minipage}[b]{.24\linewidth}
        \centering
        \includegraphics[width=\linewidth]{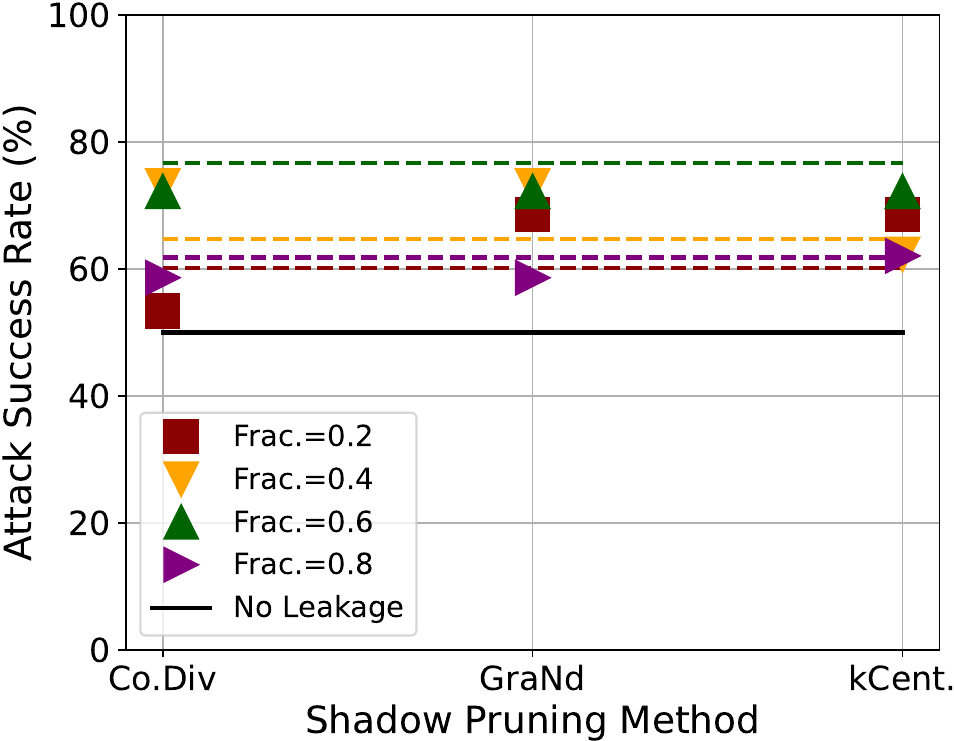}
        \subcaption{MNIST-Glist.}
        
    \end{minipage}%
    \hfill
    \begin{minipage}[b]{.24\linewidth}
        \centering
        \includegraphics[width=\linewidth]{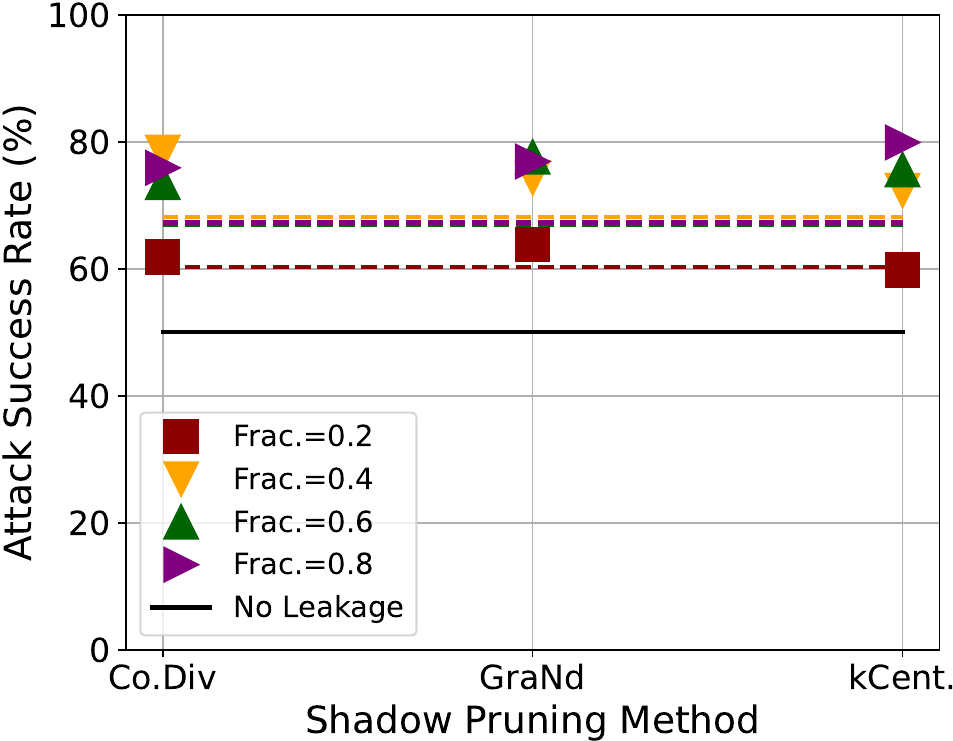}
        \subcaption{CIFAR100-Cal}
        
    \end{minipage}%
    \hfill
    \begin{minipage}[b]{.24\linewidth}
        \centering
        \includegraphics[width=\linewidth]{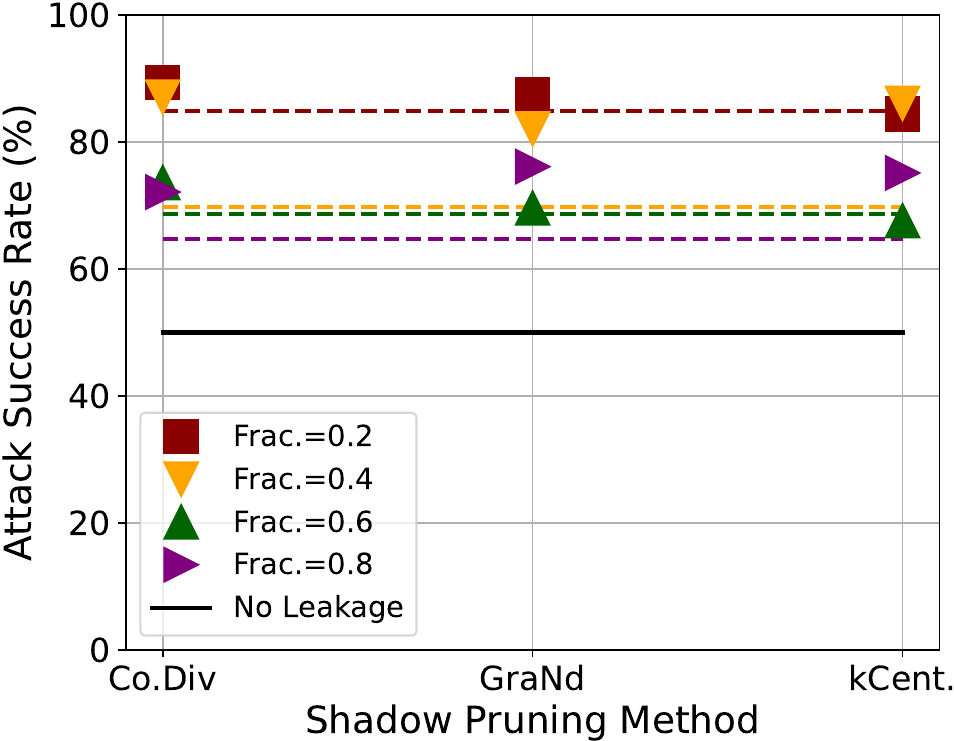}
        \subcaption{CIFAR100-Glist.}
        
    \end{minipage}%
    \caption{Attack success rate (ArraDis) when the adversary has no knowledge of pruning method.}
    \label{fig:unknown_method_appendix}
\end{figure*}
\begin{figure*}[ht]
    \centering
    \begin{minipage}[b]{.33\linewidth}
        \centering
        \includegraphics[width=\linewidth]{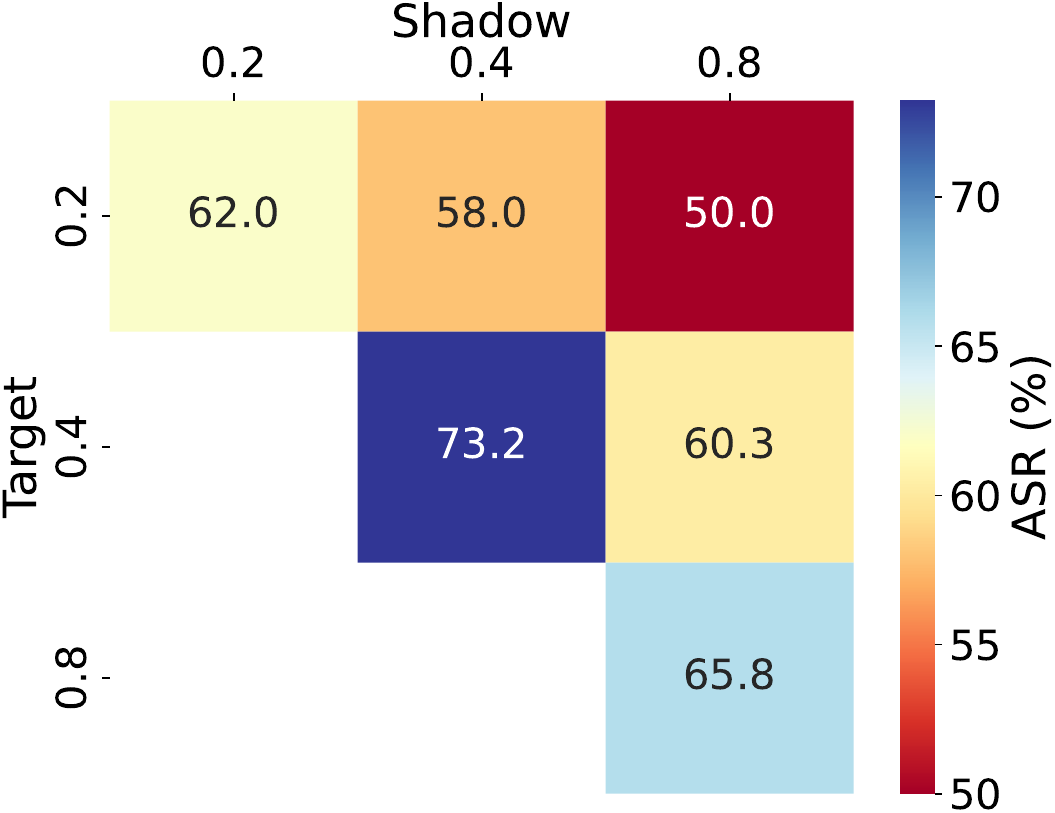}
        \subcaption{MNIST-Cal}
    \end{minipage}%
    \hfill
    \begin{minipage}[b]{.33\linewidth}
        \centering
        \includegraphics[width=\linewidth]{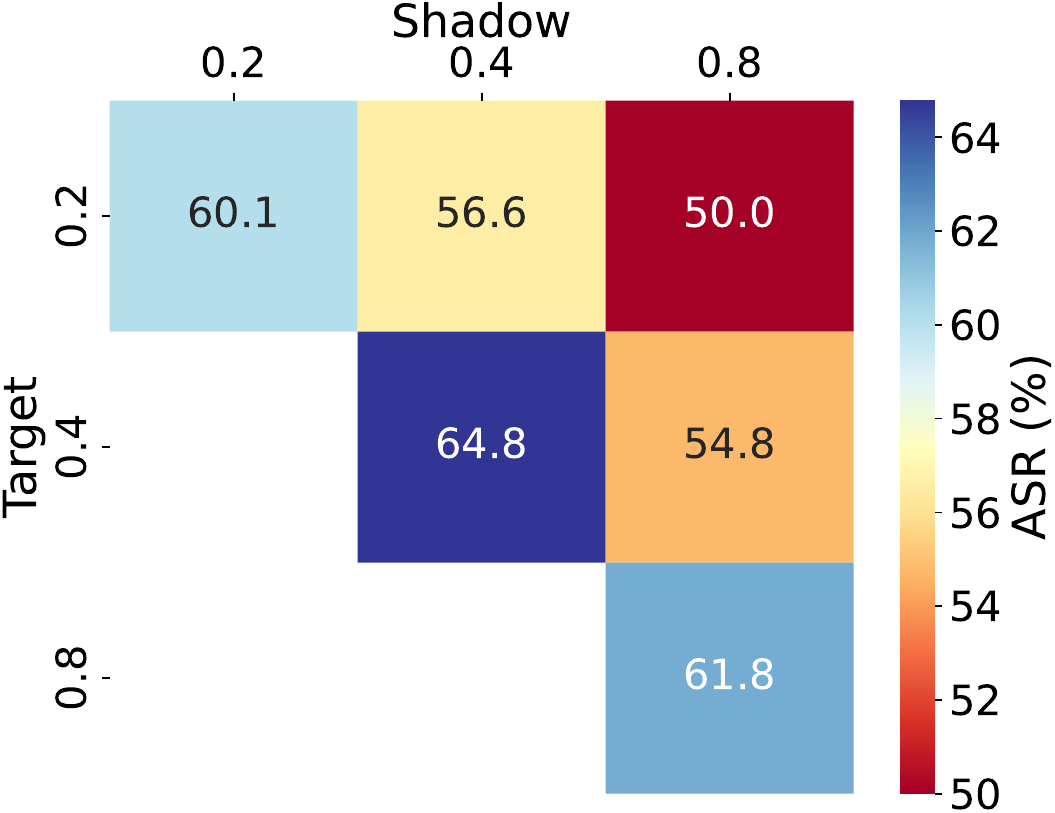}
        \subcaption{MNIST-Glist.}
        
    \end{minipage}%
    \hfill
    \begin{minipage}[b]{.33\linewidth}
        \centering
        \includegraphics[width=\linewidth]{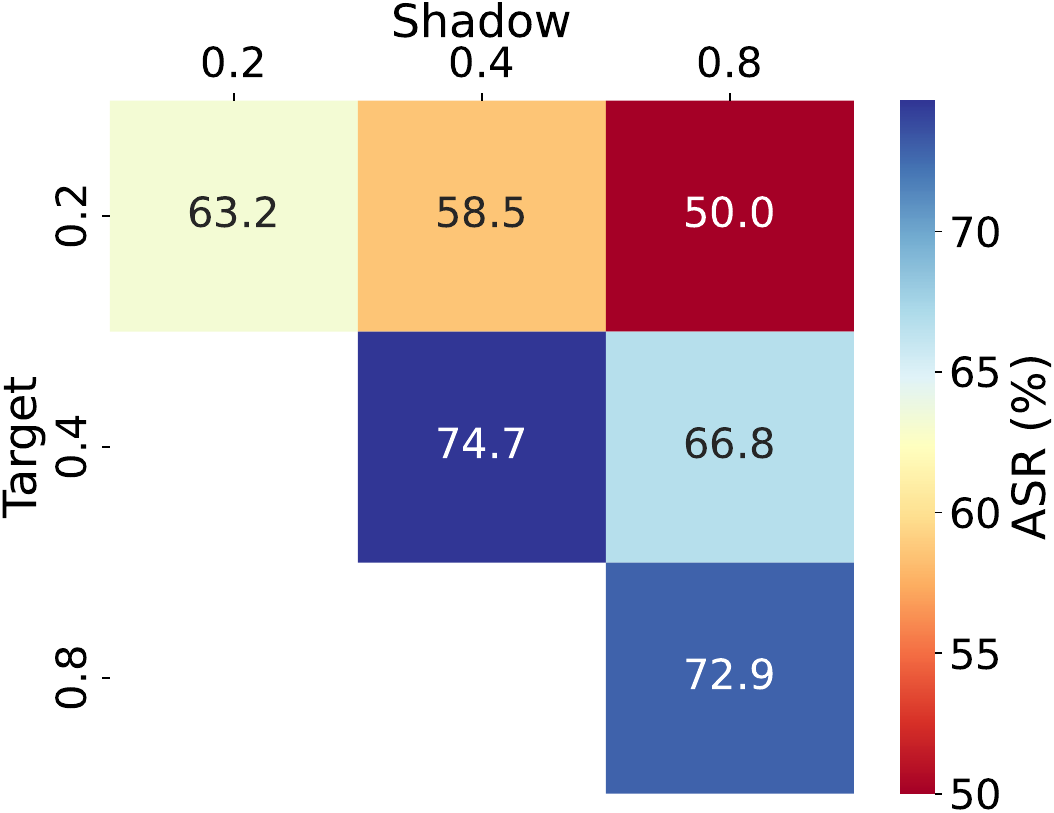}
        \subcaption{MNIST-GraNd}
        
    \end{minipage}%
    \hfill
    \begin{minipage}[b]{.33\linewidth}
        \centering
        \includegraphics[width=\linewidth]{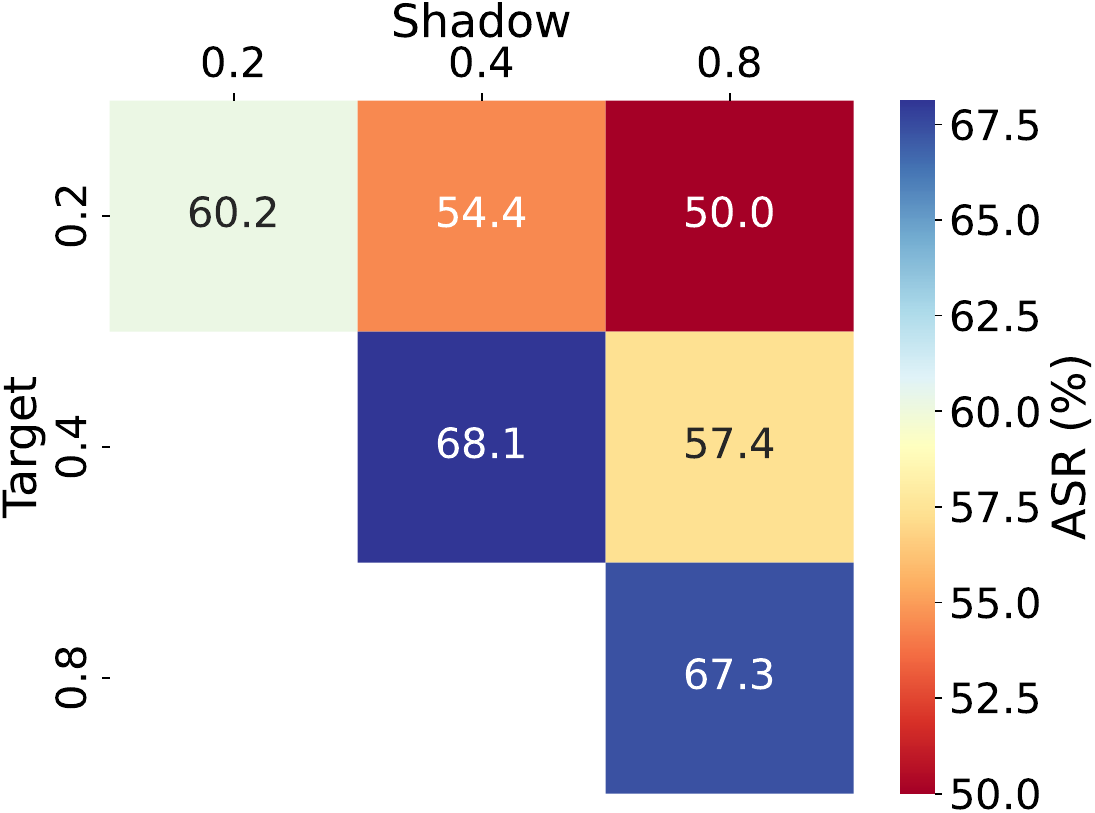}
        \subcaption{CIFAR100-Cal}
        
    \end{minipage}%
    \hfill
    \begin{minipage}[b]{.33\linewidth}
        \centering
        \includegraphics[width=\linewidth]{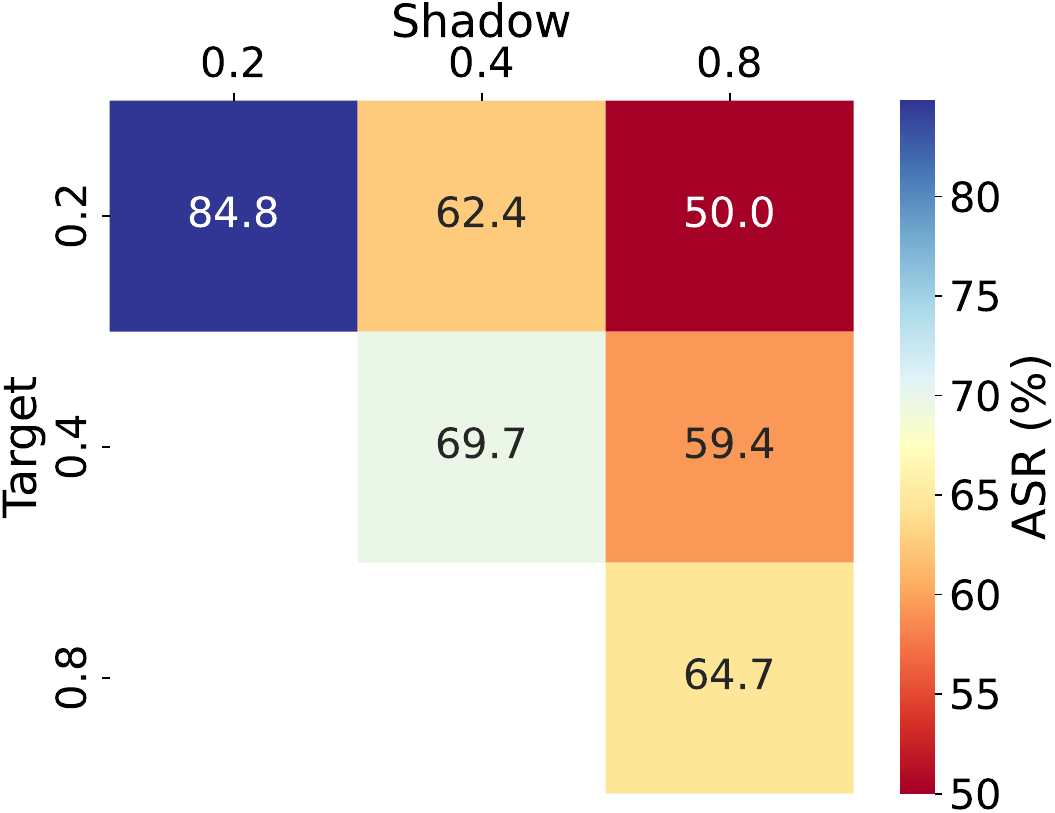}
        \subcaption{CIFAR100-Glist.}
        
    \end{minipage}%
    \hfill
    \begin{minipage}[b]{.33\linewidth}
        \centering
        \includegraphics[width=\linewidth]{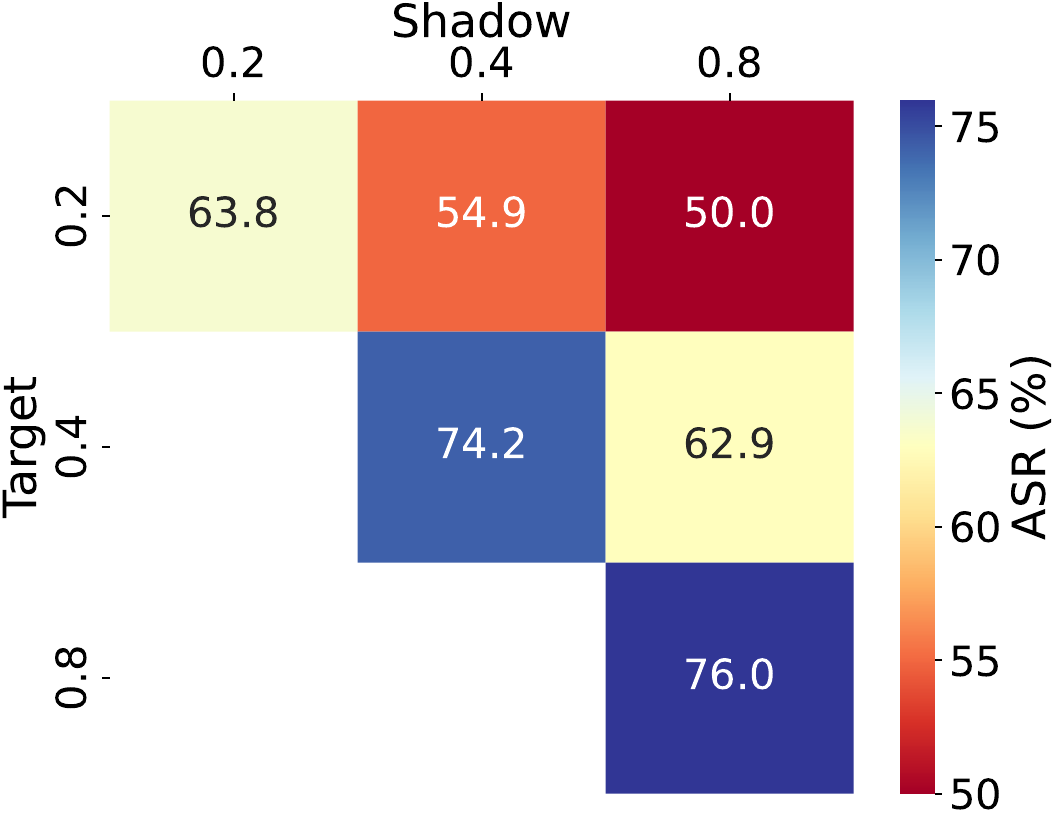}
        \subcaption{CIFAR100-GraNd}
        
    \end{minipage}%
    % \vspace{-3mm}
    \caption{ASR without knowing the pruning fraction (x-axis: shadow fraction, y-axis: ground truth fraction).}
    \vspace{-3mm}
    \label{fig:attack2_appendix}
\end{figure*}
\section{Influence of Different Batch Size}
\label{batch_size}
As shown in Figure \ref{fig:batch_size}, we set \(\zeta_{\textrm{v}}\) to six different values (i.e., 100, 200, 400, 500, 800, and 2,000). The pruning method used here is Herd., and the dataset is CIFAR10. It can be observed that as the batch size increases, the difference between the two occurrence distributions follows the same trend. In other words, when the occurrence count is approximately within the first quarter of its range, other non-members appear more frequently. Beyond this range, data from the redundant set become more prevalent. Regarding the attack success rate, although there is a decline when the batch size is set to a larger value compared to a smaller one, the results remain satisfactory. Considering the trade-off between performance and efficiency, setting batch size to a larger value is a better choice.

\vspace{-2mm}
\section{Evidence for Conducting CumDis and SpiDis}
\label{appendix_evidence}
\noindent {\bf CumDis. }
For CumDis, we leverage the differences in the CDFs and the CCDFs at various occurrence counts between the redundant set and other non-members. As shown in Figures \ref{fig:evidence_app}(a) to Figure \ref{fig:evidence_app}(b), these figures present three key pieces of information:

\noindent\textbf{Info 1}: The occurrence distribution of the redundant set (Redundant Set \#Occ) and the occurrence distribution of  other non-members (Other Non-Members \#Occ) 
(same as in Figures \ref{fig:evidence}(a) and \ref{fig:evidence}(b)).

\noindent\textbf{Info 2}: (1). The attack success rate assuming all samples that have occurrence count \(t\) higher than a certain threshold belongs to a certain data type (Cum. Diff. High2Low). The attack success rate drops to zero when reaching a certain value is because there are no samples present \(t\) higher than that value (see footnote \ref{fn1}). (2). The attack success rate assuming all samples lower than a certain occurrence count \(t\) belong to one data type (Cum. Diff. Low2High). 

\noindent\textbf{Info 3}: When the adversary knows the two cumulative distributions (CDF and CCDF) of the victim datapool, the two ground truth thresholds (Threshold Below and Threshold Above) found by CumDis.

From \textbf{Info 1}, it can be observed that the two occurrence distributions show significant differences under different pruning methods. \textbf{Info 2} provides even more clear information, directly indicating that different pruning methods consistently exhibit strong cumulative distribution differences at different occurrence counts thus can result in high attack success rate. Additionally, from \textbf{Info 3}, we understand that the thresholds derived from CumDis can successfully pinpoint positions with higher attack success rates.

\noindent {\bf SpiDis.}
For SpiDis, we further attempt to identify the points within the entire occurrence distribution where the attack success rate is highest, specifically where the difference between the two distributions (Redundant Set \#Occ and Other Non-Members \#Occ) is the greatest at a particular occurrence count \(t\). As shown in Figures \ref{fig:evidence_app}(a) to \ref{fig:evidence_app}(l), we present the attack success rate based on the differences at various \(t\) values between the two distributions (Diff. Per Point). If the value at a certain position is 0, it indicates that there are no samples present at that position. It can be observed that, except for positions with no samples observed, there are clearly high success rates across the entire distribution, with many positions showing an attack success rate of 100\%. This information demonstrates the feasibility of SpiDis.

\section{Use Case of the Brimming score}
\label{usecase}
After validating the effectiveness of the Brimming score in Figure \ref{fig:brimming}, we now conduct a demonstrative experiment to provide further explanation. In Figure \ref{fig:guide}, we use the CIFAR10 dataset as an example and calculate the generalization performance of models trained using selected sets obtained from all 12 methods under 4 pruning fractions. The generalization performance is plotted on the x-axis, while the corresponding Brimming score is plotted on the y-axis. Different symbols represent different fractions, and different colors represent different dataset pruning methods. A large dashed ellipse encloses the approximate range of all results for the same fraction. We calculate the Pareto front\footnote{The Pareto front is a concept from multi-objective optimization that represents a set of solutions where no solution can be improved on one objective without degrading another objective. It is the boundary separating the feasible region of optimal trade-offs from the rest of the solution space, highlighting the most efficient solutions in terms of multiple criteria.} of all results under the same pruning fraction, represented by hollow symbols of the corresponding shapes.

Given the goal of achieving high generalization performance while ensuring good privacy, the Pareto front here is selected from samples located in the lower right corner for different fractions. From Figure \ref{fig:guide}, we can see that in terms of generalization performance, the differences between methods are smaller when the pruning fraction is relatively high (e.g., 0.6, 0.8). However, from the data privacy perspective, their Brimming score show significant differences. In contrast, when the pruning fraction is low, the resulting model generalization performance varies significantly between different methods, while the differences among the Brimming scores are small compared to those under higher fractions. This indicates that existing dataset pruning methods exhibit poor stability and robustness, failing to ensure utility at smaller fractions. It also further highlights the difficulty of simultaneously achieving efficiency, utility, and privacy protection. In this experiment, the fraction of 0.8 relatively achieves the latter two goals better, but the fraction is too high to guarantee efficiency. When the fraction is 0.6, it better balances utility and efficiency, but the privacy risk increases. As the fraction further decreases, it provides better privacy protection and efficiency, but the model utility significantly drops. We urge the AI security community and the AI efficiency community to focus on this issue and strive to propose dataset pruning methods that can better balance the aforementioned three objectives.

\section{Additional Experimental Results}
\label{remaining_results}
In this section, we present additional experiments results that are not included in the main manuscript.

\noindent\textbf{Defense Occurrence Distribution Outcomes of All Pruning Methods. }As shown in Figure \ref{fig:defense_remain}, we compare the occurrence distribution of the redundant set and other non-members with and without using the proposed defense technique ReDoMi. It is evident that the distributions are more similar when ReDoMi is applied, which hinders the adversary and reduces the effectiveness of the inference.

\noindent\textbf{Outcomes When Unknown Pruning Method Under MNIST and CIFAR100. } As shown in Figure \ref{fig:unknown_method_appendix}, similar to Figure \ref{fig:unknown_method}, we present the outcomes when the adversary does not know the pruning method. The setup is the same as in Figure \ref{fig:unknown_method}.
\begin{table}[h!]
\centering
 \small 
    \begin{threeparttable}[t]
\begin{adjustbox}{max width=\columnwidth}
\begin{tabular}{ccccccc}
\toprule
\toprule
\textbf{Dataset} & \textbf{Target} & \textbf{Shadow} & \textbf{ASR Drop} & \textbf{Target} & \textbf{Shadow} & \textbf{ASR Drop} \\
\midrule
\multirow{5}{*}{MNIST} & Cal-0.2 & GraNd-0.4 & 7.50\% & Cal-0.4 & GraNd-0.8 & 22.67\% \\
\cmidrule(r){2-4} \cmidrule(l){5-7}
& Craig-0.2 & Forgt.-0.4 & 2.62\% & Craig-0.4 & Glister-0.8 & 24.28\% \\
\cmidrule(r){2-4} \cmidrule(l){5-7}
& Glister-0.2 & Cal-0.4 & 16.53\% & Glister-0.4 & Forget.-0.8 & 22.33\% \\
\cmidrule(r){2-4} \cmidrule(l){5-7}
& GraNd-0.2 & Craig-0.4 & 9.57\% & GraNd-0.4 & Craig-0.8 & 23.37\% \\
\midrule
\midrule
\multirow{5}{*}{CIFAR100} & Cal-0.2 & GraNd-0.4 & 14.92\% & Cal-0.4 & GraNd-0.8 & 22.15\% \\
\cmidrule(r){2-4} \cmidrule(l){5-7}
& Craig-0.2 & Forgt.-0.4 & 15.25\% & Craig-0.4 & Glister-0.8 & 19.68\% \\
\cmidrule(r){2-4} \cmidrule(l){5-7}
& Glister-0.2 & Cal-0.4 & 6.36\% & Glister-0.4 & Forget.-0.8 & 13.25\% \\
\cmidrule(r){2-4} \cmidrule(l){5-7}
& GraNd-0.2 & Craig-0.4 & 9.19\% & GraNd-0.4 & Craig-0.8 & 28.40\% \\
\bottomrule
\bottomrule
\end{tabular}
\end{adjustbox}
\caption{ASR Drop for adversaries without knowing the pruning method and the pruning fraction.}
\vspace{-5mm}
\label{fig:asr_app}
\end{threeparttable}
\end{table}

\noindent\textbf{Outcomes When Unknown Pruning Fraction Under MNIST and CIFAR100. }
As shown in Figure \ref{fig:attack2_appendix}, similar to Figure \ref{fig:attack2}, we present the outcomes when the adversary does not know the pruning fraction and instead assumes one. The setup is the same as in Figure \ref{fig:attack2}.

\begin{table}[h]
\centering
 \small 
    \begin{threeparttable}[t]
\begin{adjustbox}{max width=\columnwidth}
\begin{tabular}{cccccccc}
\toprule
\toprule
\textbf{Dataset} & \textbf{Fract.} & \textbf{Craig} & \textbf{DeepF.} & \textbf{Forgt.} & \textbf{Glister} & \textbf{GraNd} & \textbf{Unc.} \\
\midrule
\multirow{3}{*}{MNIST} & 0.6 & 54.67(-1.25) & 56.85(-0.17) & 57.52(-2.16) & 54.02(-0.82) & 56.41(-0.84) & 56.97(-1.59) \\
\cmidrule(r){2-8}
& 0.8 & 53.02(-8.28) & 52.75(-1.38) & 56.00(+0.01) & 51.81(-5.26) & 58.16(-5.21) & 55.64(-4.71) \\
\midrule
\multirow{3}{*}{CIFAR100} & 0.6 & 52.67(+0.61) & 53.08(-0.22) & 53.74(-0.52) & 52.52(-0.01) & 57.30(-2.23) & 52.95(-0.34) \\
\cmidrule(r){2-8}
& 0.8 & 52.46(-4.57) & 53.55(-4.00) & 50.06(-0.26) & 52.51(-0.17) & 60.80(-7.98) & 53.30(-2.35) \\
\bottomrule
\bottomrule
\end{tabular}
\end{adjustbox}
\end{threeparttable}
\caption{Attack success rate without shadow datapools.}
\vspace{-4mm}
\label{fig:no_shadow_app}
\end{table}
\noindent\textbf{Outcomes of ArraDis on Other Pruning Methods. }As shown in Figure \ref{fig:brim_remain}, similar to Figure \ref{fig:evidence}, we present the outcomes of ArraDis on ten other pruning methods using the same experimental setup. It is evident that the last four methods are significantly harder to attack than the other eight methods, as indicated by the much smaller region with high attack success rates (blue region), especially for GradMatch (G.M.). This trend aligns with our outcomes shown in Table \ref{tab:main1} and serves as evidence for the effectiveness of the Brimming score.

\noindent\textbf{Outcomes when both pruning method and fraction are unknown. } Results for MNIST and CIFAR100 are shown in Table~\ref{fig:asr_app}. The setup is the same as in Table \ref{fig:asr}.

\noindent\textbf{Outcomes without shadow datapool.} Results under this scenario for MNIST and CIFAR100 are shown in Table~\ref{fig:no_shadow_app}. The setup is the same as in Table \ref{fig:no_shadow}.

% \end{appendices}
\end{document}